\documentclass[master,       
               oneside,        
               BCOR=10mm,       
               ngerman, english, 
                english,
               ]{GAUBM}

\usepackage{import}
\usepackage{tikz}
\usetikzlibrary{positioning}
\usepackage[onehalfspacing]{setspace}
\usepackage{mathptmx,lmodern,amsmath,amssymb, float, latexsym,exscale}
\usepackage{mathtools}
\usepackage{interval}
\usepackage{amsfonts}
\usepackage[utf8]{inputenc}
\usepackage[T1]{fontenc}
\usepackage[top=2.5cm, bottom=2cm, left=2.5cm, right=2.5cm]{geometry}
\usepackage{romannum}
\usepackage{url}
\usepackage{graphicx}
\usepackage{subcaption}
\usepackage{wrapfig}
\usepackage{textcomp}
\usepackage{pifont}
\usepackage{dsfont}
\usepackage [labelfont={bf,bf},font={small},
  labelsep=colon]{caption}
\newtagform{fn}{(}{)\footnotemark}

\usepackage{tikz} 
\usepackage[nointegrals]{wasysym}
\usepackage{eurosym}
\usepackage[toc,page]{appendix}
\usepackage{color}
\usepackage{xcolor}
\usepackage{enumitem}
\usepackage{wrapfig}
\usepackage{sidecap}
\usepackage{times}
\usepackage{multirow}
\usepackage{svg}
\usepackage{float}
\usepackage{booktabs}
\usepackage{verbatim}
\usepackage{sudoku}
\usepackage[version=4,arrows=pgf]{mhchem}
\usepackage{textcomp}
\usepackage[column=Q]{cellspace}
\newcommand{\uproman}[1]{\uppercase\expandafter{\romannumeral#1}}

\usepackage{caption}
\usepackage{wasysym}%
\usepackage{cprotect}
\usepackage{physics}
\usepackage{amsthm}
\floatstyle{plaintop}
\restylefloat{table}
\usepackage[backend=biber,sorting=none,sortlocale=auto,bibencoding=auto, giveninits, url=false]{biblatex}
\bibliography{sources}
\AtEveryBibitem{\clearfield{pages}}
\AtEveryBibitem{\clearfield{month}}
\AtEveryBibitem{\clearfield{day}}
\AtEveryBibitem{\clearfield{issue}}

\usepackage{algorithm}
\usepackage[noend]{algpseudocode}
\makeatletter
\def\BState{\State\hskip-\ALG@thistlm}
\makeatother

\usepackage{setspace}  
\usepackage[main=english, ngerman]{babel}     
\usepackage{calc}      
\usepackage[T1]{fontenc}       
\usepackage[autostyle=true,german=quotes]{csquotes}

\usepackage{lmodern} 
\usepackage{booktabs}                      
\usepackage{longtable}                     
\usepackage{array}                         

\usepackage{textcomp,gensymb}

\usepackage[pdfstartview=FitH,      
            breaklinks=true,        
            bookmarksopen=true,     
            bookmarksnumbered=true  
            ]{hyperref}
\usepackage[capitalise]{cleveref}

\numberwithin{equation}{section}
\newtheorem{prop}{Proposition}[section]

\newtheorem{theorem}{Theorem}[section]
\newtheorem{corollary}{Corollary}[section]
\newtheorem{lemma}[theorem]{Lemma}

\theoremstyle{definition}
\newtheorem{definition}{Definition}[section]
\newtheorem*{remark}{Remark}
\newtheorem*{example}{Example}

\newcommand{\Eqref}[1]{Eq.~\eqref{#1}}

\DeclareTOCStyleEntry[numwidth=3em]{tocline}{figure}

\begin{document}

\ThesisAuthor{Lars Torbj\o rn}{Stutzer}
\PlaceOfBirth{-}
\ThesisTitle{Stochastic Calculus Approach to Thermodynamics of Jump Processes}{}
\FirstReferee[Dr. Alja\v{z} Godec\\{\small \textit{Max-Planck-Institut f\"ur Multidisziplin\"are Naturwissenschaften, G\"ottingen}}]{Dr. Alja\v{z} Godec}
\Institute{Max-Planck-Institut f\"ur Multidisziplin\"are Naturwissenschaften}
\SecondReferee{Prof. Dr. Peter Sollich\\{\small \textit{Institut f\"ur Theoretische Physik, G\"ottingen}}}
\ThesisBegin{11}{11}{2024}
\ThesisEnd{20}{02}{2025}
\frontmatter
\maketitle
\cleardoublepage

\begin{abstract}
    Stochastic thermodynamics is the field of study relating fluctuations in stochastic systems to thermodynamic quantities. The total entropy production (EP), is central to the thermodynamic classification of systems. Non-equilibrium systems manifestly all have non-zero EP and therefore impose an "arrow of time". 
    Thermodynamic inequalities are lower bounds on the total EP and are especially useful when only parts of systems are operationally accessible. We use a stochastic calculus approach to directly derive and generalise three classes of inequalities for Markov jump processes using correlations of path observables, e.g., currents and densities. Our theoretical predictions are compared with simulations, where a good agreement is observed. The thermodynamic bounds we investigate include the thermodynamic uncertainty relation (TUR), thermodynamic transport bound (TB), and thermodynamic correlation bound (CB). We provide insight into the saturation conditions for these bounds and to what degree saturation can be achieved. Additionally, for the TUR and TB, we show how the bounds are related, which includes identifying a diffusion coefficient for jump dynamics. 
    Comparisons are drawn between the TUR and TB for relaxation and stationary processes in biologically relevant settings. Specifically, calmodulin folding dynamics and secondary active transport, where differences in long-time relaxation and convergence are observed. For a systematic way to construct models, we formulate two methods to drive systems out of equilibrium without changing the stationary probability distribution. 
\end{abstract}

\cleardoublepage
\onehalfspacing
\tableofcontents

\mainmatter

\newpage

\chapter{Introduction}

Non-equilibrium systems are present in various aspects of everyday life, especially when it comes to biological systems \cite{NonEqLivingSystGnesotto, NonEqBiologyXiaona, NonEqBiologyColombo, SotoSepulveda, ChemSystOutOfEqEsch, ProtometabolismNader, LandscapeFluxTheoryWang, NonEqEnzymaticNijemeisland, IntermolecularForcesLeckband}. These are generally driven by some form of energy consumption, e.g., by ATP hydrolysis \cite{NonEqLivingSystGnesotto, NonEqBiologyXiaona, NonEqEnzymaticNijemeisland}. Unsurprisingly, this has led to a broad interest in the understanding of systems far from equilibrium. \textcolor{black}{Due to the large versatility of {non-equilibrium} systems, ranging from the cosmic scale \cite{Pietroni2009} to the realm of quantum physics \cite{Allahverdyan2014}, a unifying method to describe non-equilibrium systems is yet to be developed. Instead, various fields of non-equilibrium physics have been developed to describe specific systems \cite{Pietroni2009, Allahverdyan2014, Seifert_2012}. In the last decade, the field of stochastic thermodynamics has gained significant traction \cite{Seifert_2012, dieball2024thermodynamic, DirectTUR, TURArbitraryInit, TURBiomolecular, TURTimeDependentDriving, SeifertStochasticTD} as a field of non-equilibrium physics to relate observable fluctuations to thermodynamical quantities in noisy systems coupled to one or more heat baths \cite{Seifert_2012, SeifertStochasticTD}.} 

Determining the thermodynamic cost of a non-equilibrium process is of great interest \cite{dieball2024perspectivetimeirreversibilitysystems, Seifert_2012}, e.g., in a molecular motor where ATP hydrolysis turns chemical energy into mechanical work \cite{Boksenbojm_2009}. A central quantity in determining the thermodynamic cost is the total entropy production \cite{Seifert_2012}. From classical thermodynamics, it is known that the stationary entropy production is proportional to the heat flux from the system to the bath \cite{Seifert_2012, dieball2024thermodynamic}. 

In most cases, however, quantifying the exact amount of produced entropy is a task that proves extremely difficult, if not impossible. It requires the knowledge of  \emph{all} dissipative degrees of freedom, i.e., \textit{all} degrees of freedom which are not in equilibrium \cite{DirectTUR, dieball2024perspectivetimeirreversibilitysystems}. For instance, experimental setups generally cannot capture all information due to finite resolution \cite{SeifertPartiallyAccessibleNetworks}. To mitigate this, lower bounds on entropy production have been  studied intensively in stochastic thermodynamics in the last decade \cite{TURArbitraryInit, TURBiomolecular, TURTimeDependentDriving, DirectTUR, KineticTUR, VuUnderdampedTUR, KwonUnderdampedTUR, LeeUnderdanmpedTUR, EspositoUnderdampedTUR, CrutchfieldUnderdampedTUR, dieball2024thermodynamic, BoundsCorrelationTimes}. These are referred to as thermodynamic bounds and generally consist of trade-off relations between entropy production and, e.g., correlations between observables \cite{BoundsCorrelationTimes}. Additionally, they can be seen as extensions of the second law of thermodynamics, stating that the entropy production is non-negative and therefore indicates an "arrow of time" \cite{Seifert_2012, SeifertEntropyProd, DirectTUR, dieball2024thermodynamic, ThreeFacesI, SEIFERT2018176}. One of the most notable classes of bounds are thermodynamic uncertainty relations (TUR), where the precision of currents bounds the total entropy production \cite{DirectTUR, KineticTUR, TURArbitraryInit}.

Recent work has shown the strength of proving thermodynamic bounds using a stochastic calculus approach for continuous space dynamics \cite{dieball2024thermodynamic, DirectTUR, CrutchfieldUnderdampedTUR}. Contrary to other proof methods, the stochastic calculus approach is not limited to specific bounds. Additionally, it requires no information beyond the dynamics of the system. Other approaches to prove thermodynamic bounds found in the literature use quantities and methods commonly encountered in, e.g., information theory \cite{Shiraishi2021} and large deviation theory \cite{TURTimeDependentDriving, Gingrich2016, Proesmans_2017}. We refer to these latter proof methods as "indirect" methods, while the stochastic calculus approach yields a "direct" method, as explained below.

A key ingredient often missing in the "indirect" methods is the insight into the saturation of the thermodynamic bounds \cite{DirectTUR}. To deduce the thermodynamic cost from the bounds, we want the latter to be as sharp as possible. Hence, there is valuable knowledge to be gained from the information on how the thermodynamic bounds saturate, e.g., the dissipative degrees of freedom. Directness in the form of using equations of motion together with basic inequalities - in this case, the Cauchy-Schwarz inequality - has the advantage of being straightforward, while also yielding valuable key insight into the saturation \cite{DirectTUR, dieball2024thermodynamic}.

Discrete dynamics in the form of Markov jump processes are a special case of continuous space dynamics in an energy landscape with sharp barriers \cite{Moro1995}. If the continuous space dynamics have a time-scale separation and the time-scale at which the system is observed is sufficiently large, then only the slow dynamics (corresponding to the jumps between long-lived metastable states) are observed \cite{Moro1995}. With an appropriate spatial coarse-graining, the continuous space dynamics can be approximated by \emph{discrete} states and jumps between those. It is expected that a stochastic calculus approach for discrete systems should be applicable as in the continuous case, since the former is the appropriate limit of the latter \cite{DirectTUR}.

This leads us to the main research question to be answered in this thesis: Is there a "direct" method based on a stochastic calculus approach for Markov jump processes that, similar to the continuous dynamics, allows for direct proofs of thermodynamic bounds? Moreover, we aim to explore the versatility of the "direct" method as well as the saturation of the thermodynamic bounds. Included in this exploration are the formulations and proofs of thermodynamic bounds that only are proven for continuous space dynamics \cite{dieball2024thermodynamic, BoundsCorrelationTimes}.  We also strive to use the same ansatz as in Ref.~\cite{Kaiser2017} to develop a method of driving specific parts of a system to create consistent comparisons. This method allows for systematic investigations of how the thermodynamic bounds perform at various distances from equilibrium.

The structure of the thesis is as follows. In Ch.~\ref{ch:methods},  fundamental mathematical and physical concepts are introduced. This is followed by Ch.~\ref{ch:advmeth}, where more advanced topics are presented, including a short overview of thermodynamic bounds and various methods found in the literature to prove them. Additionally, the stochastic calculus approach is introduced, and it is shown how it may be used to evaluate correlations. This part is complemented by a systematic approach to driving systems out of equilibrium while keeping the underlying stationary probability density invariant. Subsequently, in Ch.~\ref{ch:directProofs}, the direct proof of the thermodynamic bounds is shown. A comparison of the bounds follows in Ch.~\ref{ch:ComparisonBounds}. We conclude in Ch.~\ref{ch:conclusion} with a discussion of the presented work and an outlook on possible extensions to be made in the future.



\chapter{Methods \label{ch:methods}}

In this chapter, we  provide the mathematical and physical background relevant for the thesis. Since we will discuss the use  of stochastic calculus of MJP for thermodynamic inference, it is in some sense the complement to the work on continuous space systems in Refs. \cite{DirectTUR, DieballCoarseGraining, DieballCurrentVariance, dieball2024thermodynamic}. To properly compare the two settings, we will introduce various ways of describing them. In the discrete case, this will be with the master equation in Sec.~\ref{sec:ME} and the path measure in Sec.~\ref{sec:Path}. Both descriptions will be relevant as, e.g., motivation for building the stochastic calculus description for MJP in Ch. \ref{ch:advmeth}. To properly compare the results from this thesis with the continuous space description, we also briefly touch on the stochastic calculus description of continuous systems in the form of the Langevin equation in Sec.~\ref{sec:continuous}. We conclude the chapter by giving a brief overview of the example systems considered throughout the thesis, including some remarks on their biological relevance.

\section{Discrete State Space and Local Detailed Balance}

Before going into the details on probabilities, stochastic processes, and entropy production, we need to specify the state spaces considered. In this thesis, we are interested in discrete state spaces. These are relevant for describing a wide range of systems, such as conformations of biological molecules, e.g., calmodulin \cite{CalmodulinStigler, FirstPassageRick, TURTimeDependentDriving} and transport of molecules through cell membranes \cite{Berlaga2021, Berlaga2022, Berlaga2022_2}. In general, such discrete systems are coarse-grained representations of high-dimensional continuous state spaces \cite{EspositoCoarseGraining, LocalDetailedBalanceAcross, dieball2024perspectivetimeirreversibilitysystems}. While many interesting results have been found for continuous systems, e.g., thermodynamic bounds \cite{DirectTUR, CrutchfieldUnderdampedTUR, dieball2024thermodynamic, BoundsCorrelationTimes} and (co)variances of observables \cite{DieballCurrentVariance}, a generalisation to, or adaptation for, MJP is sometimes missing. This work fills this gap and provides a comparison with the continuous case where relevant. 

It proves useful to describe discrete systems using directed graphs\footnote{In the literature, this is sometimes also referred to as a network.} $\mathcal{G}=(\Omega, \mathcal{V})$, where each node $i\in \Omega\subset\mathds{N}$ represents a state, and an edge $i\to j\in \mathcal{V}$ between two states $i$ and $j$ exists if the system allows a direct transition from $i$ to $j$. Directed graphs are in general not symmetric, meaning that the existence of an edge $i\to j$ does not imply the existence of the reversed edge $j \to i$. These  asymmetric systems give rise to multiple physical and mathematical problems, such as steady states not being unique \cite{Schnakenberg},  infinite entropy production, and the absence of a thermodynamically consistent description \cite{ThreeFacesI}. Hence, this thesis will only deal with symmetric directed graphs.


Lastly, we will only tackle finite graphs, i.e., graphs with a finite number of states $|\Omega|<\infty$. To be more precise, when defining probabilities and operators on graphs, we can express them as vectors and matrices. Furthermore, when defining observables in Ch. \ref{ch:directProofs}, there is no constraint on the functions entering beyond requiring them to be bounded. This changes if we consider graphs of infinite countable states, where these functions have to be, e.g., $L^2(\mathcal{G}, \mathds{P})$ square-summable functions if variances are to be calculated for some probability measure $\mathds{P}$ \cite{Dunford1971}. As most biological systems which we ultimately aim to describe are finite, we choose to describe the dynamics of finite systems. However, in Sec.~\ref{sec:comparisonNESS} we will extend the model introduced in Sec.~\ref{sec:SecondaryTransport} to a countable infinite-sized system to be able to make appropriate comparisons. 

\textcolor{black}{Fundamental to stochastic thermodynamics is the assumption that the thermal baths (often referred to as environment or medium \cite{ThreeFacesI}), as well as degrees of freedom that are not observed completely, are in thermal equilibrium \cite{LocalDetailedBalanceAcross, LocalDetailedBalance, EspositoCoarseGraining, BauerLocalDetailedBalance, StochasticNonLinearKorbel}. 
This leads to the \textit{local detailed balance} (LDB) condition, which states that the entropy flow causing a transition between two states is given by the log ratio of the transition rates between these states \cite{LocalDetailedBalanceAcross}. This is, e.g., necessary for a meaningful first and second law of thermodynamics \cite{ThreeFacesI}. For MJP, local detailed balance states that the internal dynamics of a state are in equilibrium, while the transitions between states, which are the observed part, may be out-of-equilibrium. Hence, the LDB condition is assumed implicitly when using, e.g., the master equation \cite{LocalDetailedBalanceAcross, ThreeFacesI}.}

\section{Master Equation and Probability Fluxes\label{sec:ME}}

Now that the fundamental description of the underlying state space is in place, we can turn to the dynamics. Throughout the thesis, we will only consider time-homogeneous processes. With $p_i(\tau)$ we denote the \textit{probability} of being in a state $i\in\Omega$ at time $\tau$, and $P(j, \tau'| i, \tau)$ is the \textit{conditional transition probability} of being in state $j$ at time $\tau'\geq \tau$ given that the system is in state $i$ at time $\tau$. We will refer to this transition probability as the \textit{propagator}, as it propagates the probability from time $\tau$ to some later time $\tau'$ through \cite{MarkovJumpProcessesBook}
\begin{equation}
    \begin{aligned}[b]
        p_j(\tau') = \sum_{i\in\Omega} P(j, \tau'| i, \tau) p_i(\tau)\,.
    \end{aligned}
    \label{eq:forwardKolmogorov}
\end{equation}
For $\tau' = \tau+\mathrm{d}\tau$, $\mathrm{d}\tau\to0$, and $i\neq j$, the transition probabilities $P(j, \tau+\mathrm{d}\tau| i, \tau)\to r_{ij}(\tau)\mathrm{d}\tau$ are proportional to $\mathrm{d}\tau$ \cite{VanKampen1976}. The proportionality $r_{ij}(\tau) = r_{ij}$ is the \textit{transition rate}\footnote{If $i\to j\notin\mathcal{V}$, then $r_{ij}=0$.} of going from $i$ to $j$, i.e., $r_{ij}\mathrm{d}\tau$ in the limit $\mathrm{d}\tau\to 0$ is the probability of transitioning from $i$ to $j$ in the time-interval $[\tau, \tau+\mathrm{d}\tau]$. Note that these are time-independent due to our assumption of time-homogeneous dynamics. For $i=j$, the conservation of probability enforces 
$P(i, \tau+\mathrm{d}\tau| i, \tau) = 1 - \sum_{\substack{j\in\Omega\\ j\neq i}}P(j, \tau+\mathrm{d}\tau| i, \tau)\to 1 - \mathrm{d}\tau\sum_{\substack{j\in\Omega\\ j\neq i}}r_{ij}$, where the sum goes over all $j\neq i$ with fixed $i$. Defining the exit rate from $i$ as
\begin{equation}
    \begin{aligned}
        r_{ii} = - \sum_{\substack{j\in\Omega\\ j\neq i}} r_{ij}\,,
    \end{aligned}
    \label{eq:exit_rate}
\end{equation}
we can rewrite \Eqref{eq:forwardKolmogorov} and take the limit $\mathrm{d}\tau\to0$ to get the master equation \cite{Schnakenberg, SEIFERT2018176}
\begin{equation}
    \begin{aligned}
        \partial_\tau p_i(\tau) = \sum_{j\in\Omega} r_{ji}p_j(\tau) = \sum_{\substack{j\in\Omega\\ j\neq i}} \left(r_{ji}p_j(\tau) - r_{ij}p_i(\tau)\right)\,.
    \end{aligned}
    \label{eq:MasterEquation}
\end{equation}
This equation is quite intuitive: the change of the probability $\partial_\tau p_i(\tau)$ comes from a probability flux from each state $j\neq i$ into the state $i$ \cite{StochasticProcessesRoss},
\begin{equation}
    \begin{aligned}
        J_{ji}(\tau) = r_{ji}p_j(\tau)\,,
    \end{aligned}
    \label{eq:flux}
\end{equation}
and a probability flux $J_{ij}(\tau)$ leaving state $i$ to $j$. Summing over all $j$, one gets the total probability of entering and leaving the state and hence the change in probability. This naturally leads to the definition of the \textit{net probability flux} between two states $i$ and $j$
\begin{align}
    \mathcal{J}_{ij}(\tau) = r_{ij}p_i(\tau) - r_{ji}p_j(\tau) = - \mathcal{J}_{ji}(\tau)\,,
    \label{eq:NetFlux}
\end{align}
so that \Eqref{eq:MasterEquation} can be written as $\partial_\tau p_i(\tau)= -\sum_{\substack{j\in\Omega\\ j\neq i}}\mathcal{J}_{ij}(\tau)$. It should also be noted that \Eqref{eq:MasterEquation} can be written in a matrix-vector form
\begin{equation}
    \partial_\tau \mathbf{p}(\tau) = \mathbf{L}\mathbf{p}(\tau)\,,
    \label{eq:ME_vector}
\end{equation}
where $\mathbf{p}(\tau)=(p_1(\tau), p_2(\tau), \dots)^T$ is the probability density vector and $\mathbf{L}$ is the master operator, or generator, with entries $(\mathbf{L})_{ji} = r_{ij}$. 
With the assumptions made so far, i.e., a finite irreducible state space and time-homogeneous dynamics, the generator $\mathbf{L}$ has a single unique zero eigenvalue \cite{Schnakenberg, Horn_Johnson_2012}. The remaining eigenvalues have negative real parts that can be ordered such that $0 = \lambda_0 > \mathrm{Re}(\lambda_1)\geq \mathrm{Re}(\lambda_2) \geq \dots$ \cite{Schnakenberg}. Solving \Eqref{eq:ME_vector}, we recover the propagator of the system \cite{StochasticProcessesRoss}
\begin{equation}
    \begin{aligned}
        P(j, \tau'-\tau|i)=P(j, \tau'|i, \tau) = \left(\mathrm{e}^{\mathbf{L}(\tau'-\tau)}\right)_{ji}\,.
    \end{aligned}
    \label{eq:Propagator}
\end{equation}
Notably, the propagator only depends on the time difference \cite{MarkovJumpProcessesBook}. Additionally, we emphasise that these Markovian propagators form a semi-group even when the propagator explicitly depends on time \cite[Sec. 2.3]{StochasticProcessesApplicationsPavliotis}. 

\section{Classification of Systems\label{sec:SystemClassification}}
We formulated the dynamics of the state probability on graphs above. Now we will classify various types of systems and introduce concepts and quantities that will help in this classification. This distinctions are useful, as different systems have fundamentally different attributes. For instance, not all systems are microscopically reversible \cite{Seifert_2012} and systems may explicitly depend on time \cite{TURTimeDependentDriving}. Systems can be divided into three classes: (i) \textit{equilibrium} steady state, (ii) \textit{non-equilibrium} steady state (NESS), and (iii) \textit{transient}. We denote the equilibrium  and NESS probability density vectors with $\mathbf{p}^\mathrm{eq}$ and $\mathbf{p}^\mathrm{s}$, respectively. Unlike the transient $\mathbf{p}(\tau)$, these do not depend on time $\tau$. In other words, both $\mathbf{p}^\mathrm{eq}$ and $\mathbf{p}^\mathrm{s}$ are the eigenvectors to the zero eigenvalue of the respective generators, i.e., $\partial_\tau \mathbf{p}^\mathrm{eq} = \mathbf{0} = \partial_\tau \mathbf{p}^\mathrm{s}$. For a stationary system \Eqref{eq:MasterEquation} implies 
\begin{align}
    \sum_{\substack{j\in\Omega\\ j\neq i}} p_i^\mathrm{s} r_{ij} = \sum_{\substack{j\in\Omega\\ j\neq i}} p_j^\mathrm{s} r_{ji}
    \label{eq:StationaryCondition}
\end{align}
Equilibrium systems satisfy the stronger (global) \textit{detailed balance} (DB) condition \cite{PrincipleDBKlein, NonEqLivingSystGnesotto}
\begin{align}
    p_i^\mathrm{eq}r_{ij} = p_j^\mathrm{eq}r_{ji}\quad \forall i,j\in\Omega\,.
    \label{eq:DetailedBalance}
\end{align}
Equation \eqref{eq:DetailedBalance} is a special case of \Eqref{eq:StationaryCondition}, where the net flux \Eqref{eq:NetFlux} vanishes on each edge.

By transient, we mean all systems where the probability density $\mathbf{p}(\tau)$ depends on time $\tau$. On the one hand, this means that $\partial_\tau \mathbf{p}(\tau) \neq 0$. We will see, e.g., in Sec.~\ref{sec:Correlationproof}, that this generally leads to more complicated formulas and descriptions compared to steady-state systems. On the other hand, transient systems in finite symmetric directed graphs with constant rates will for times $\tau\to \infty$ relax to a steady state
, be it an equilibrium or NESS system, as a consequence of the eigenvalues of $\mathbf{L}$ discussed in the previous section. 


Cycles prove to be a useful tool to classify the stationary properties of a system \cite{SeifertPartiallyAccessibleNetworks}. A cycle is a closed, non-repeating path on the underlying graph with a specific direction. In other words, it is the set of edges $\mathcal{C} = \{x_1\to x_2, x_2\to x_3, \dots, x_{k-1}\to x_k, x_k\to x_1\}$ where $x_j\neq x_l\in\Omega\,\,\forall j\neq l$. 

Edge affinities are the DB-breaking contributions on an edge $i\to j\in \mathcal{V}$, as one defines these as \cite{PietzonkaCurrentFluctuations}
\begin{equation}
    \begin{aligned}
        \mathcal{A}_{ij} = \log(J_{ij} / J_{ji})\,,
    \end{aligned}
    \label{eq:affinities}
\end{equation}
or equivalently $\mathrm{e}^{\mathcal{A}_{ij}}=\frac{r_{ij}p_i}{r_{ji}p_j}$ using \Eqref{eq:flux}. In equilibrium, $\mathcal{A}_{ij} = 0$ for all $i,j$, while $\mathcal{A}_{ij}\neq 0$ for at least one $i,j$ in general for a NESS. Two more observations can be made: (i) the edge affinities are anti-symmetric $\mathcal{A}_{ij} = -\mathcal{A}_{ji}$, and (ii) one needs to know both transition rates as well as the probability to calculate it. Knowing the probability density would make calculating the affinities to determine the nature of the system superfluous, as one can use \Eqref{eq:DetailedBalance}. This is where the cycle affinities $\mathcal{A}_\mathcal{C}$ come in. Instead of comparing the fluxes of an edge with its reverse, this compares a cycle $\mathcal{C}$ with its reversed $\Tilde{\mathcal{C}} = \{x_1 \to x_k, x_{k}\to x_{k-1}, \dots, x_2\to x_1\}$ \cite{Ohga2023, Liang2023}
\begin{equation}
    \begin{aligned}
        \mathcal{A}_\mathcal{C} = \log\left(\frac{\prod_{i\to j\in\mathcal{C}}r_{ij}}{\prod_{i\to j\in\mathcal{C}}r_{ji}}\right) = \sum_{i\to j\in\mathcal{C}}\mathcal{A}_{ij}\,.
    \end{aligned}
    \label{eq:CycleAffinity}
\end{equation}
Since a cycle is a closed loop, the probabilities in \Eqref{eq:affinities} all cancel. Hence, the cycle affinities only depend on the transition rates such that one can deduce violations of DB without inferring $\mathbf{p}^\mathrm{s}$. However, this is a tool we can only use to describe the (eventual) steady state of the system. As shown in App.~\ref{sec:AlternativeDriving}, controlling the edge affinities can be used to systematically drive systems.

\section{Paths, Path Measures, and Ergodicity\label{sec:Path}}

\begin{figure}
    \centering
    \includegraphics[width=0.45\textwidth]{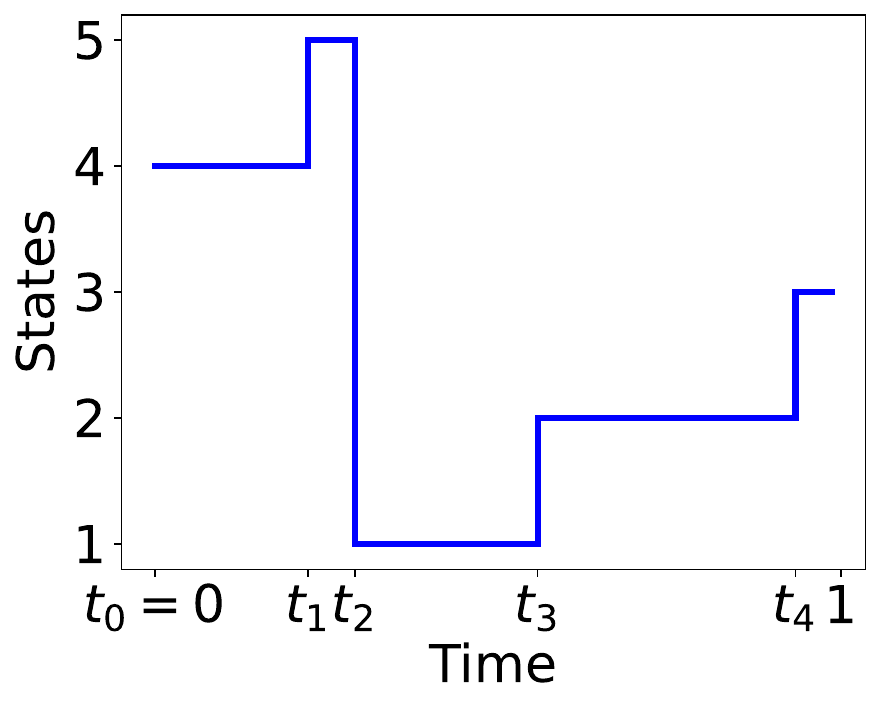}
    \caption[Example of a Path]{Single trajectory in the toy model. This path is $(x_\tau)_{0\leq\tau\leq1}=(4, 0; 5, t_1; 1, t_2; 2, t_3; 3, t_4)$ and has length $t=1$.}
    \label{fig:TrajectoryExample}
\end{figure}

If we prepare a system so it initially is in state $x_0$ at time $t_0=0$, it will over time transition to other states. We refer to such a sequence of states and times of transition between the states we refer to as a \emph{path}. The states in the path are denoted by $x_i\in\Omega$ and the times at which the transition from $x_{i-1}$ to $x_{i}$ take place are denoted by $t_{i}$, $i\geq 1$. With $t_0$ we denote the initial time when the trajectory starts and we set $t_0=0$ without loss of generality. Hence, the total path of length $t$ can be written as $(x_\tau)_{0\leq \tau \leq t} = (x_0, t_0=0; x_1, t_1; x_2, t_2; \dots)$. An example trajectory is shown in Fig.~\ref{fig:TrajectoryExample}. Note that length refers to the total time of the path, not the number of states visited along a path. 

The probability of a path $P_{\mathbf{p}_0}\left[(x_\tau)_{0\leq \tau \leq t}\right] \equiv P_{x_0}\left[(x_\tau)_{0\leq \tau \leq t}\right]p_{x_0}(t_0=0)$ of a given length $t$ evolving from a distribution $\mathbf{p}_0=\mathbf{p}(t_0=0)$ can be written using the transition and exit rates in \Eqref{eq:exit_rate}~\cite{hartichGodecPathProb}
\begin{equation}
    \begin{aligned}
        P_{\mathbf{p}_0}\left[(x_\tau)_{0\leq \tau \leq t}\right] = p_{x_0}(t_0=0)\mathrm{e}^{t_1 r_{x_0x_0}}r_{x_0x_1}\mathrm{e}^{(t_2-t_1)r_{x_1x_1}}r_{x_1x_2}\dots\,,
    \end{aligned}
\end{equation}
where the exponentials account for the time spent in each state before transitioning to the next state. As a state can be visited multiple times along the same path, the path probability can be written in a more compact form \cite{hartichGodecPathProb}
\begin{equation}
    \begin{aligned}
        P_{\mathbf{p}_0}\left[(x_\tau)_{0\leq \tau \leq t}\right] = p_{x_0}(t_0=0)\mathrm{e}^{\sum_{x\in\Omega} \tau_x r_{xx}}\prod_{i,j\in\Omega}\left(r_{ij}\right)^{n_{ij}}\,.
    \end{aligned}
    \label{eq:PathMeasure}
\end{equation}
Here, the integer $n_{ij}$ is the number of times the transition $i\to j$ occurs and $\tau_i$ is the total time spent in state $i$ along the path. Hence, the second part in \Eqref{eq:PathMeasure} is the weight of all transitions while the third part is the weight from waiting in the states. The $n_{ij}$ and $\tau_i$ are extensively studied in Ch.~\ref{ch:advmeth}, where a framework of stochastic calculus for MJP is introduced. 


We may also be interested in an observable $\hat{O}_{x_\tau}$ which is a function of the state $x_\tau$ at time $\tau$. If we have a path $(x_\tau)_{0\leq \tau \leq t}$, we define the time average of the observable as
\begin{align}
    \overline{O}(t) = \frac{1}{t}\int_0^t\mathrm{d}\tau \hat{O}_{x_\tau}\,.
    \label{eq:TimeAverageObservable}
\end{align}
Since the path is a stochastic quantity, this time average will also be a stochastic quantity. In case the system is strongly ergodic we can relate \Eqref{eq:TimeAverageObservable} to the stationary (either equilibrium or NESS) average \cite{Ergodicity1, Kaiser2017, NonEqBiologyColombo, InfiniteErgodicErez, FractionalFeynmanKacCarmi, MarkovJumpProcessesBook, StochasticProcessesApplicationsPavliotis}
\begin{align}
    \langle \hat{O}_n \rangle_\mathrm{s} = \sum_{n\in\Omega} \hat{O}_n p_n^\mathrm{s} \stackrel{\mathrm{a.s.}}{=} \lim_{t\to \infty} \overline{O}(t) = \lim_{t\to \infty} \frac{1}{t}\int_0^t\mathrm{d}\tau \hat{O}_{x_\tau}\,,
    \label{eq:Ergodicity}
\end{align}
where $\stackrel{\mathrm{a.s.}}{=}$ (read as "almost surely equal") is an equality on all non-zero measure sets. Experimentally, this is useful to estimate the steady-state average from single-trajectory measurements as long as these trajectories are sufficiently long. 

\section{Stochastic Thermodynamics and Entropy Production\label{sec:Entropy}}
\textit{Stochastic thermodynamics}, unlike macroscopic thermodynamics, is a framework to describe the energetics of noisy systems coupled to a heat bath arbitrarily far from equilibrium. Stochastic thermodynamics accounts for fluctuations in the description of energetics via, e.g., with fluctuation theorems \cite{SeifertStochasticTD, Jarzynski, Jarzynski2} and thermodynamic bounds \cite{DirectTUR, dieball2024thermodynamic, BoundsCorrelationTimes, ImprovingBoundsCorrelations}. "System" generally refers to the degrees of freedom we are interested in. In this thesis, system refers to the discrete state space considered. On the other hand, in continuous systems it may be the position of a particle or a polymer \cite{dieball2024thermodynamic}. Crucial to stochastic thermodynamics is that the system is coupled to a bath which is in thermal equilibrium and does not get affected by the state of the system \cite{Seifert_2012, SeifertStochasticTD, ThreeFacesI}. Also, the bath is assumed to be independent of any time-dependent forces acting on the system \cite{SeifertStochasticTD}. 

Central to the description of systems out-of-equilibrium is the \textit{total entropy production} $\Delta S_\mathrm{tot}(t)$ in the time-interval $[0, t]$ defined as the average log-ratio \cite{SeifertStochasticTD, Seifert_2012, LocalDetailedBalance, LocalDetailedBalanceAcross, roldan2023martingales},
\begin{equation}
    \begin{aligned}
        \Delta S_\mathrm{tot}(t) = \left\langle\log\left(\frac{P\left[(x_\tau)_{0\leq \tau \leq t}\right]}{\Tilde{P}\left[(\Tilde{x}_\tau)_{0\leq \tau \leq t}\right]}\right) \right\rangle\,,
    \end{aligned}
    \label{eq:total_entropy_production_definition}
\end{equation}
of probabilities of paths $(x_\tau)_{0\leq \tau \leq t}$ and the reversed paths $(\Tilde{x}_\tau)_{0\leq \tau \leq t} = (x_{t-\tau})_{0\leq \tau \leq t}$, with $\Tilde{P}((\Tilde{x}_\tau)_{0\leq \tau \leq t})$
the probability of the reversed path and the average is taken over all paths. This we can understand as a measure of the irreversibility of the system \cite{SeifertStochasticTD, Seifert_2012, LocalDetailedBalance, dieball2024thermodynamic, DirectTUR, ImprovingBoundsCorrelations, SeifertEntropyProd, ThreeFacesI}, since paths and their reversed are equally likely in case the system is \textcolor{black}{microscopically} reversible, i.e., in equilibrium. The log-ratio in \Eqref{eq:total_entropy_production_definition} is referred to as the path-dependent entropy change $\Delta s_\mathrm{tot}\left[(x_\tau)_{0\leq \tau \leq t}\right]$, which is a fluctuating quantity. It obeys the integration fluctuation theorem \cite{ThreeFacesI, SeifertStochasticTD, Seifert_2012},
\begin{align}
    \left\langle e^{\Delta s_\mathrm{tot}\left[(x_\tau)_{0\leq \tau \leq t}\right]}\right\rangle = 1\,,
\end{align}
for all times $t$. Rearranging this, summing (or integrating in case of continuous state spaces) over all paths, and using Jensen's inequality $\langle \mathrm{e}^{-X}\rangle \geq  \mathrm{e}^{\langle- X\rangle}$ we recover the second law of thermodynamics \cite{ThreeFacesI, SeifertStochasticTD}
\begin{equation}
    \begin{aligned}
        \Delta S_\mathrm{tot}(t) = \langle \Delta s_\mathrm{tot}\left[(x_\tau)_{0\leq \tau \leq t}\right]\rangle \geq 0\,.
    \end{aligned}
    \label{eq:secondlaw}
\end{equation}
Hence, the fluctuation theorem can be seen as a generalisation of the second law for noisy systems, since the average becomes deterministic in the limit of infinite system sizes due to the law of large numbers. Furthermore, this allows single trajectories to yield a negative entropy change. To be more precise, in an ensemble there may exist paths with $\Delta s_\mathrm{tot}\left[(x_\tau)_{0\leq \tau \leq t}\right] < 0$.

We may also consider the entropy production $\Dot{S}_\mathrm{tot}(\tau)$ of the system at time $\tau$, which relates to the total entropy production through $\Delta S_\mathrm{tot}(t) = \int_0^t\mathrm{d}\tau \Dot{S}_\mathrm{tot}(\tau)$. The entropy production of a MJP has the form \cite{ThreeFacesI, PietzonkaCurrentFluctuations, SEIFERT2018176, Schnakenberg}
\begin{equation}
    \begin{aligned}
        \Dot{S}_\mathrm{tot}(\tau) = \sum_{\substack{x,y\in\Omega\\x\neq y}}r_{xy}p_x(\tau)\log{\frac{r_{xy}p_x(\tau)}{r_{yx}p_{y}(\tau)}}\,,
    \end{aligned}
\end{equation}
so that the total entropy production can be written as \cite{TURArbitraryInit}
\begin{equation}
    \begin{aligned}[b]
        \Delta S_\mathrm{tot}(t) 
        &= \frac{1}{2}\sum_{\substack{x,y\in\Omega\\x\neq y}}\int_0^t\mathrm{d}\tau(r_{xy}p_x(\tau)-r_{yx}p_y(\tau))\log{\frac{r_{xy}p_x(\tau)}{r_{yx}p_{y}(\tau)}}\,.
    \end{aligned}
    \label{eq:TotalEntropy}
\end{equation}
In NESS Eq.~\eqref{eq:TotalEntropy} will be linear in $t$, while transient systems generally have more complicated dependencies on $t$. While in most cases not useful, it is also possible to write \Eqref{eq:TotalEntropy} in terms of net fluxes \Eqref{eq:NetFlux} and edge affinities \Eqref{eq:affinities} \cite{Schnakenberg}
\begin{equation}
    \begin{aligned}
        \Delta S_\mathrm{tot}(t) &= \frac{1}{2}\sum_{\substack{x,y\in\Omega\\x\neq y}}\int_0^t\mathrm{d}\tau\mathcal{J}_{xy}(\tau)\mathcal{A}_{xy}(\tau)\,.
    \end{aligned}
\end{equation}

\section{Spectral Decomposition\label{sec:Spec}}
In this thesis, we will limit ourselves to systems with diagonalisable generator $\mathbf{L}$ with a bi-orthonormal basis. This assumption greatly restraints systems this applies to, as general driven systems are excluded from it \cite{Hartich_2019}. For systems where no bi-orthonormal basis exists, the calculations become substantially more complicated and would go beyond the scope of this thesis. All examples where we apply spectral decomposition have a bi-orthonormal basis.

Let $\lambda_k$ be the eigenvalues of $\mathbf{L}$ and without loss of generality let $0=\lambda_0\geq \mathrm{Re}(\lambda_1)\geq \mathrm{Re}(\lambda_2)\geq \dots\geq \mathrm{Re}(\lambda_{N-1})$. We use the standard inner product, i.e., for two vectors $\boldsymbol{u},\boldsymbol{v}$ we write $\bra{\boldsymbol{u}}\ket{\boldsymbol{v}} = \overline{\boldsymbol{u}}^T\boldsymbol{v} = \sum_k \overline{u}_kv_k$. The left and right eigenvectors to eigenvalue $\lambda_k$ of $\mathbf{L}$ we denote by $\bra{\boldsymbol{\psi}^L_k}$ and $\ket{\boldsymbol{\psi}^R_k}$. Specifically, this means that $\bra{\boldsymbol{\psi}^L_k}\mathbf{L} = \lambda_k \bra{\boldsymbol{\psi}^L_k}$ and $\mathbf{L}\ket{\boldsymbol{\psi}^R_k} = \lambda_k \ket{\boldsymbol{\psi}^R_k}$. These eigenvectors form the bi-orthonormal basis of the generator. With our assumptions about the system, we can decompose the generator into the bi-orthonormal eigenbasis \cite{Hartich_2019}
\begin{align}
    \mathbf{L} = \sum_{k=0}^{N-1}\lambda_k\ket{\boldsymbol{\psi}^R_k}\bra{\boldsymbol{\psi}^L_k}\,,
\end{align}
where $\bra{\boldsymbol{\psi}^L_k}\ket{\boldsymbol{\psi}^R_l} = \delta_{kl}$. Furthermore, $\bra{i}\ket{\boldsymbol{\psi}^R_0} = p_i^\mathrm{s}$ and $\bra{\boldsymbol{\psi}^L_l}\ket{i} = 1$ for all $i\in\Omega$. In general, we will write the projection of the left (right) eigenvector onto a state $i$ as $\bra{\boldsymbol{\psi}^L_k}\ket{i} = \psi^L_{k,i}$ ($\bra{i}\ket{\boldsymbol{\psi}^R_k} = \psi^R_{k,i}$). The propagator in this decomposition reads
\begin{align}
    P(i,\tau|j) = \bra{i}\mathrm{e}^{\mathbf{L}\tau}\ket{j} = \sum_{k=0}^{N-1}\psi^R_{k,i}\mathrm{e}^{\lambda_k\tau}\psi^L_{k,j}\,.
    \label{eq:SpectralPropagator}
\end{align}

\section{Continuous Description and Langevin Equation\label{sec:continuous}}
So far, we have presented and discussed discrete systems. However, we have to introduce continuous space descriptions to make appropriate comparisons to results in continuous space. Like discrete systems, continuous systems can be described on multiple levels: path-wise description, single-time ensemble description, etc. We limit the discussion to the path-wise description, most commonly known as \textit{Langevin equation}, which is based on stochastic calculus in continuous space.  


The Langevin equation has proven extremely successful in describing various systems and scenarios varying from passive to active \cite{Loewen2020}, single particle to many particle systems \cite{Lizana2010, Quinn2000, zwanzig2001nonequilibrium}, as well as having various applications in finance \cite{Bouchaud1998}. Suppose $\boldsymbol{x}_\tau, \boldsymbol{v}_\tau\in\mathds{R}^d$ are position and velocity in a $d-$dimensional space coupled to a memoryless thermal bath. The underdamped Langevin equation is a \textit{stochastic differential equation}\footnote{While the consensus in the literature is to call this "differential equations", it would be more intuitive to refer to them as "integral equations" \cite{StochasticProcessesApplicationsPavliotis}. However, we will use the established nomenclature.} (SDE) \cite{StochasticProcessesApplicationsPavliotis, gardiner2004handbook},
\begin{equation}
    \begin{aligned}[b]
        \mathrm{d}\boldsymbol{x}_\tau &= \boldsymbol{v}_\tau\mathrm{d}\tau\,,\\
        \boldsymbol{m}\mathrm{d}\boldsymbol{v}_\tau &= \boldsymbol{F}(\boldsymbol{x}_\tau, \tau)\mathrm{d}\tau - \boldsymbol{\gamma}\boldsymbol{v}_\tau\mathrm{d}\tau + \sqrt{2k_\mathrm{B}T\boldsymbol{\gamma}}\mathrm{d}\boldsymbol{W}_\tau\,,
    \end{aligned}
    \label{eq:Langevin}
\end{equation}
describing the evolution of $\boldsymbol{x}_\tau$ and $\boldsymbol{v}_\tau$ using the rules of \textit{stochastic calculus}. Here, $\boldsymbol{F}(\boldsymbol{x}_\tau, \tau)$ is an external force field. The friction $\boldsymbol{\gamma}$ and mass $\boldsymbol{m}$ matrices are $d\cross d$ positive definite and symmetric. Furthermore, $k_\mathrm{B}$ is the Boltzmann constant and $T$ is the temperature of the thermal bath that couples to the system. The stochasticity enters through the $d-$dimensional \textit{Wiener process} $\mathrm{d}\boldsymbol{W}_\tau$ \cite{StochasticProcessesApplicationsPavliotis}. The white noise $\mathrm{d}\boldsymbol{W}_\tau$ is Gaussian distributed with \cite{StochasticProcessesApplicationsPavliotis, gardiner2004handbook}
\begin{equation}
    \begin{aligned}[b]
        \langle \mathrm{d}\boldsymbol{W}_\tau \rangle &= \boldsymbol{0}\,,\\
        \langle (\mathrm{d}\boldsymbol{W}_\tau)_i (\mathrm{d}\boldsymbol{W}_{\tau'})_j\rangle &= \delta(\tau - \tau') \delta_{ij}\mathrm{d}\tau\mathrm{d}\tau'\,,
    \end{aligned}
    \label{eq:WienerStatistics}
\end{equation}
with respective Dirac and Kronecker $\delta$. That is, the noise has zero mean and the variance scales linearly in time $\langle \mathrm{d}\boldsymbol{W}_\tau^2\rangle\sim \mathrm{d}\tau$. Note that we assume the noise to be additive for simplicity here. A commonly used simplification is the so-called overdamped limit of the full dynamics, which includes taking the limit $\boldsymbol{\gamma}^{-1}\boldsymbol{m}\to 0$ of negligible mass compared to friction. The details of this limit, i.e., the exact formulation and conditions, are shown by Pavliotis \cite{StochasticProcessesApplicationsPavliotis} and Wilemski \cite{Wilemski1976}. The resulting \textit{overdamped Langevin equation} reads \cite{StochasticProcessesApplicationsPavliotis, gardiner2004handbook}
\begin{align}
    \mathrm{d}\boldsymbol{x}_\tau &= \boldsymbol{\gamma}^{-1}\boldsymbol{F}(\boldsymbol{x}_\tau, \tau)\mathrm{d}\tau + \sqrt{2\boldsymbol{D}}\mathrm{d}\boldsymbol{W}_\tau\,.
    \label{eq:OverdampedLangevin}
\end{align}
The diffusion matrix $\boldsymbol{D} = k_\mathrm{B}T\boldsymbol{\gamma}^{-1}$ follows the Einstein-Smoluchowski relation \cite{Blickle2007}. As a result of \Eqref{eq:WienerStatistics}, there is no well defined velocity, since $|\mathrm{d}\boldsymbol{W}_\tau/\mathrm{d}\tau| \sim 1 / \sqrt{\mathrm{d}\tau}\xrightarrow{\mathrm{d}\tau\to0}\infty$ and, as a consequence, $|\mathrm{d}\boldsymbol{x}_\tau/\mathrm{d}\tau| \xrightarrow{\mathrm{d}\tau\to0}\infty$ \cite{dieball2024thermodynamic, gardiner2004handbook}.

The difference between stochastic and classical calculus becomes apparent when calculating integrals since the former requires that we decide how to understand the integral \cite{gardiner2004handbook}. For example, given some scalar function $f(\boldsymbol{x}_\tau)$, the product \cite{StochasticProcessesApplicationsPavliotis, CaiThesis}
\begin{equation}
    \begin{aligned}[b]
        f(\boldsymbol{x}_\tau)\mathrm{d}\boldsymbol{x}_\tau = 
        \begin{cases}
            f(\boldsymbol{x}_\tau)(\boldsymbol{x}_{\tau+\mathrm{d}\tau} - \boldsymbol{x}_{\tau}) & \text{(It\^o)}\,,\\
            f\left(\frac{\boldsymbol{x}_{\tau+\mathrm{d}\tau} + \boldsymbol{x}_{\tau}}{2}\right)(\boldsymbol{x}_{\tau+\mathrm{d}\tau} - \boldsymbol{x}_{\tau}) & \text{(Stratonovich)}\,,
        \end{cases}
    \end{aligned}
\end{equation}
can be understood as evaluating the function at the beginning of the interval, in which cases the integral $\int_{\tau=0}^{\tau=t}f(\boldsymbol{x}_\tau)\mathrm{d}\boldsymbol{x}_\tau$ is an It\^o, integral while evaluating it at the mid-point of the considered interval $(\boldsymbol{x}_{\tau+\mathrm{d}\tau} + \boldsymbol{x}_{\tau})/2$ yields the Stratonovich integral \cite{StochasticProcessesApplicationsPavliotis}. We follow the literature by introducing the notation $\int_{\tau=0}^{\tau=t}f(\boldsymbol{x}_\tau)\circ\mathrm{d}\boldsymbol{x}_\tau$ whenever we mean a Stratonovich integral \cite{DirectTUR, dieball2024thermodynamic, LocalDetailedBalance, StochasticProcessesApplicationsPavliotis}.

We make a final remark on the notation used. In physics literature, the overdamped Langevin equation \Eqref{eq:OverdampedLangevin} is often written as
\begin{equation}
    \begin{aligned}
        \Dot{\boldsymbol{x}}_\tau = \boldsymbol{\gamma}^{-1}\boldsymbol{F}(\boldsymbol{x}_\tau, \tau) + \sqrt{2\boldsymbol{D}}\boldsymbol{\xi}_\tau\,,
    \end{aligned}
\end{equation}
where $\boldsymbol{\xi}_\tau$ is a \textit{Gaussian white noise} with covariance $\langle(\boldsymbol{\xi}_\tau)_i(\boldsymbol{\xi}_{\tau'})_j \rangle = \delta_{ij}\delta(\tau - \tau')$ and $\Dot{\boldsymbol{x}}_\tau$ is the temporal derivative of $\boldsymbol{x}_\tau$. As mentioned above, these quantities are, mathematically speaking, only well defined if understood in the sense of \Eqref{eq:OverdampedLangevin}, as, e.g., the Wiener process is nowhere differentiable \cite{Wilemski1976}. 
\section{Overview of (Biological) Example Systems}

In this thesis, we consider three example systems to illustrate the various mathematical and theoretical aspects that are demonstrated. Specifically, these are a toy model with an interesting topology, the calmodulin folding dynamics from Refs.~\cite{FirstPassageRick, CalmodulinStigler}, and the secondary active transport, which is relevant for the trans-membrane movement of molecules. We provide a short introduction into the relevance of the systems considered (where appropriate) in addition to information on how to model them as MJPs.

\subsection{Toy Model\label{sec:toymodel}}
The toy model we study is shown in Fig.~\hyperref[fig:ModelExamples]{\ref*{fig:ModelExamples}a} and consists of multiple cycles. This particular construction gives us freedom on how to drive the system out of equilibrium since we can drive each cycle individually. The way we drive the cycles is presented in Sec.~\ref{sec:DrivingSystems}. Specifically, the model is constructed such that two cycles can have both parallel and anti-parallel overlap. As an example, consider the cycles $\mathcal{C}_1=\{1\to3, 3\to2, 2\to4, 4\to5, 5\to1\}$ and $\mathcal{C}_2=\{1\to2, 2\to4, 4\to3, 3\to1\}$, so that $\mathcal{C}_1\cap\mathcal{C}_2=\{2\to4\}$ and $\mathcal{C}_1\cap\Tilde{\mathcal{C}}_2=\{1\to3\}$. Cases like these are important to consider when systematically breaking detailed balance.

\subsection{Calmodulin Folding Dynamics\label{sec:calmodulin}}
As receptors for $\mathrm{Ca}^{2+}$ signals, the calmodulin proteins play an important role in physiological processes \cite{CalmodulinIntroductionFrits}. This is due to the crucial role $\mathrm{Ca}^{2+}$ plays in, e.g., cellular functions \cite{CalmodulinPrototypicalChin}. Various proteins are receptors for $\mathrm{Ca}^{2+}$ with varying degrees of conformation changes when binding calcium. Calmodulin is a type of protein which has a $\mathrm{Ca}^{2+}$-induced change in conformation \cite{CalmodulinPrototypicalChin}. As the role of $\mathrm{Ca}^{2+}$ receptors in molecules is not understood in detail, calmodulin has been used as a model system to improve that understanding of $\mathrm{Ca}^{2+}$ signaling\cite{CalmodulinIntroductionFrits, CalmodulinPrototypicalChin}. 

To illustrate the importance of calmodulin, we mention some of its biological functions. Firstly, it enters in the regulation of smooth muscle contraction \cite{CalmodulinSmoothMuscleWalsh, CalmodulinSmoothMuscleTansey}. The calmodulin-dependent myosin light-chain kinase is activated by $\mathrm{Ca}^{2+}$, which, through various biological processes leads to the contraction of the muscle \cite{CalmodulinSmoothMuscleWalsh}. Secondly, calmodulin enters the metabolism, e.g., by being involved in the process of splitting off glucose from glycogen \cite{CalmodulinMetabolismNishizawa}. Thirdly, experimental studies have shown evidence that calmodulin plays an important role in synaptic plasticity important for memory \cite{LTPMemoryLynch, CalmodulinMemoryLledo}. This is but a taste of the involvement of calmodulin in biology and physiology, as there are many more applications and functions beyond what is mentioned above \cite{CalmodulinBookEldik, CalmodulinHeartMaier, CalmodulinMyocardial}, hence motivating further investigations.

\begin{figure}
    \centering
    \includegraphics[width=.9\textwidth]{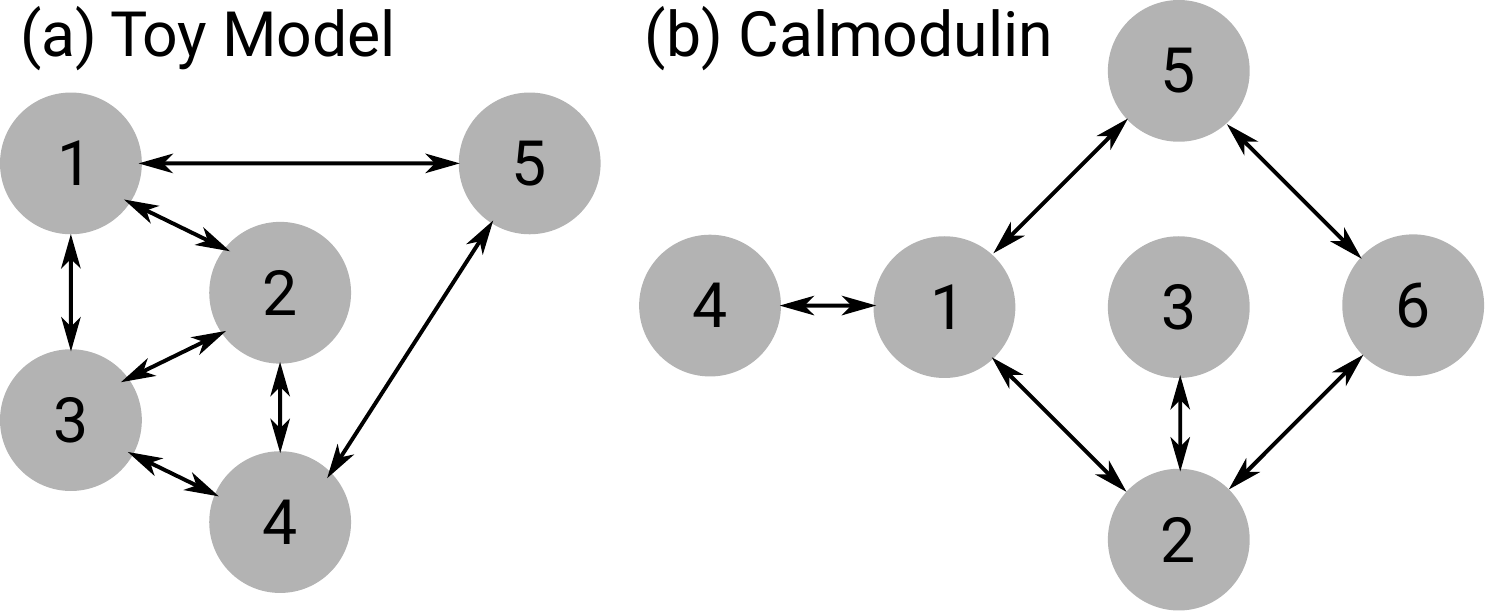}
    \caption[Graph structure of toy and calmodulin model]{Graph structure of the toy model and calmodulin folding graph. The toy model (a) is constructed specifically to allow for interesting features, e.g., cycle overlaps of both orientations. The calmodulin system (b) is a commonly used example which obeys DB \cite{FirstPassageRick, CalmodulinStigler}. The energies and rates are presented in Tab.~\ref{tab:ToyModelEnergies} and \ref{tab:Calmodulin_Rates} for the toy model and calmodulin, respectively.}
    \label{fig:ModelExamples}
\end{figure}

From a physics perspective, the calmodulin folding dynamics is a commonly used example in various fields, such as the study of first passage times \cite{FirstPassageRick} and stochastic thermodynamics \cite{TURTimeDependentDriving}. This is due to the biological relevance mentioned above, but crucially as it can be well approximated by Markov jump dynamics under DB\cite{CalmodulinStigler}. The states represent different conformations of the protein and the transitions are changes of conformation \cite{CalmodulinStigler}. In this work, we use the transition rates from Ref.~\cite{FirstPassageRick}. 
The rates are shown in Tab.~\ref{tab:Calmodulin_Rates} and the topology of the folding network is shown in Fig.~\hyperref[fig:ModelExamples]{\ref*{fig:ModelExamples}b}.

\subsection{Secondary Active Transport\label{sec:SecondaryTransport}}

Biological cells are fundamental to life and cannot function isolated from surrounding influence \cite{Bhowmick2024}. For instance, cells require nutrients to survive \cite{Berlaga2021, Poulsen2015}. To properly perform their functions, various proteins, molecules, and ions need to be able to pass through the cell membrane \cite{Berlaga2021, Berlaga2022, Berlaga2022_2, Poulsen2015, LeVine2016}. Since such molecules in many cases are large, they require an energy input to transport them through the membrane \cite{Berlaga2022}. Additionally, there may be chemical gradients of these going in the opposite direction, hence it may be required to move the molecules against the trans-membrane gradients \cite{Berlaga2022}. This is the reason why we refer to it as "active transport"; an energy input is required for transport. There are, generally speaking, two mechanisms that realise this trans-membrane transport \cite{Berlaga2022_2}: \textit{primary active transport} by, e.g.,  hydrolysis of ATP and \textit{secondary active transport} (SAT) by driving induced by transport of other molecules. In this study, we will use the latter as a motivation for a specific type of graph structure and, hence, we will explain the fundamental principles here. 

\begin{figure}
    \centering
    \includegraphics[width=.9\textwidth]{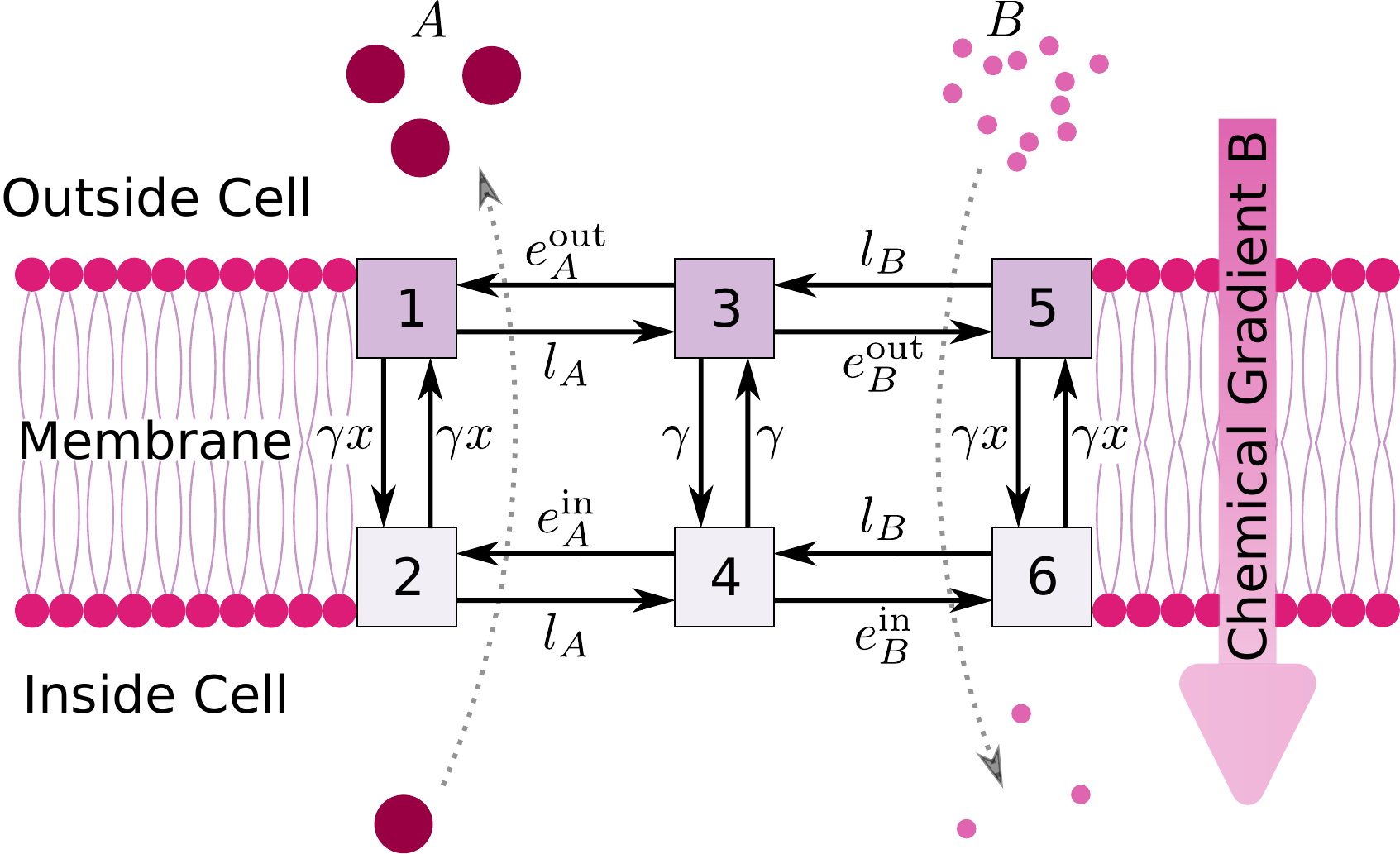}
    \caption[Schematic of secondary active transport]{Schematic representation of an SAT antiporter system reduced to a six-state Markov model. The model corresponds to the one shown in Ref. \cite{Berlaga2022_2}. Species $A$ is the molecule which the cell requires to pass through the membrane and species $B$ represents the driving molecule going along the electrochemical gradient.}
    \label{fig:SecondaryTransportModel}
\end{figure}

The SAT exploits that one type of molecule moving along its electrochemical gradient can influence another type of molecule \cite{Berlaga2021, Berlaga2022, Berlaga2022_2, Poulsen2015, LeVine2016}. This is a well-studied phenomenon, e.g., in the transport of glucose \cite{Poulsen2015}. The SAT can again be divided into two sub-processes, depending on whether the molecule moving along the gradient makes the second type move with or against that direction. These are often referred to as symporters and antiporters, respectively. The protein channels, through which the translocation of the molecules occurs, follow the alternating access model \cite{Weyand2010-em, Berlaga2021}. The general idea is that binding of the molecule on one side of the membrane will lead to the protein changing its conformation to allow the molecule to be released on the other side.

This way of functioning motivates the use of a six-state Markov model shown in Fig.~\ref{fig:SecondaryTransportModel}, which is adapted from Refs. \cite{Berlaga2021, Berlaga2022, Berlaga2022_2}, as an example to simulate a SAT antiporter system. We emphasise that simulations suggest that a Markov model is a good approximation for the dynamics of binding ions \cite{Adelman2016}. The states 1, 3, and 5 correspond to protein channels being open to the outside of the cell with species $A$, no species, or species $B$ present, respectively. Analogously, states 2, 4, and 6 are with the channel being open to the inside of the cell with the same occupation as before. With $e^{\mathrm{out}/\mathrm{in}}_{A/B}$ we denote the rates at which species $A/B$ enter the channel from the outside/inside. These are proportional to the concentration on the outside and inside \cite{Berlaga2021}. The rates of the respective species leaving the channel are denoted by $l_{A/B}$. Following Ref. \cite{Berlaga2022_2}, they are for simplicity assumed to be independent of whether the channel is open to the inside or outside. Lastly, $\gamma$ denotes the conformation rate for opening the tunnel on the opposing side of the membrane. In case there are molecules of either type $A$ or $B$ present, this may or may not inhibit or assist this rate, which is introduced using the dimensionless factor $x$.

The model in Fig.~\ref{fig:SecondaryTransportModel} consist of three cycles $\mathcal{C}_1 = \{1\to2, 2\to4, 4\to3, 3\to1\}$, $\mathcal{C}_2=\{3\to4, 4\to6, 6\to5, 5\to3\}$, and $\mathcal{C}_3=\{1\to2, 2\to4, 4\to6, 6\to5, 5\to3, 3\to1\}$. How strongly this system is driven can be identified in terms of the cycle affinities of the first two cycles
\begin{equation}
    \begin{aligned}[b]
        \mathcal{A}_{\mathcal{C}_1} &= \log\left(\frac{x\gamma^2 l_A e^\mathrm{out}_A}{x\gamma^2 l_A e^\mathrm{in}_A}\right) = \log\left(\frac{e^\mathrm{out}_A}{e^\mathrm{in}_A}\right)\,,\\
        \mathcal{A}_{\mathcal{C}_2} &= \log\left(\frac{e^\mathrm{in}_B}{e^\mathrm{out}_B}\right)\,.
    \end{aligned}
    \label{eq:SecondaryTransportAffinities}
\end{equation}
In this case, the third cycle does not need to be considered, as $\mathcal{A}_{\mathcal{C}_3} = \mathcal{A}_{\mathcal{C}_1} + \mathcal{A}_{\mathcal{C}_2} = \log\left({e^\mathrm{out}_Ae^\mathrm{in}_B}/{e^\mathrm{in}_Ae^\mathrm{out}_B}\right)$. For the system to be in equilibrium, one can read from \Eqref{eq:SecondaryTransportAffinities} that $e^\mathrm{out}_{A/B}=e^\mathrm{in}_{A/B}$. Assuming that the proportionality between rates and concentration outside and inside the cell is the same \cite{Berlaga2021}, this is equivalent to the concentrations of $A$ and $B$ being constant across the cell membrane. 

The reason this simulates antiporters becomes obvious when considering the rates of entering the channel. If $e^\mathrm{out}_{B}/e^\mathrm{in}_{B}\gg e^\mathrm{out}_{A}/e^\mathrm{in}_{A}>1$, then $\mathcal{C}_3$ is sufficiently strongly driven clockwise so that, on average, molecules of type $A$ will pass out of the cell against its chemical gradient. In Sec.~\ref{sec:comparisonNESS}, we show that $\mathcal{A}_{\mathcal{C}_3}<0$ is a necessary but not sufficient condition for the system to behave like an antiporter. While similar constructions can be made for symporters, we choose to focus on this particular construction in this work.

\chapter{Advanced Methods: Stochastic Calculus\label{ch:advmeth}}
Now that the mathematical and physical foundations are in place, we can turn to the more advanced topics relevant to this thesis. To be precise, we will discuss why thermodynamic bounds have become state of the art. Three concrete bounds are introduced and discussed. Similar to Ref. \cite{DirectTUR}, we then emphasise why direct proofs are useful. The chapter is rounded off by introducing a stochastic calculus approach for MJP, as well as a systematic approach to driving systems out of equilibrium without changing the steady state. The latter is the method we use to drive the toy model.

\section{Relevance of Thermodynamic Bounds}


One thing the formulas for the total entropy production in discrete and continuous space  have in common is that they both require the knowledge of the probability density at all times $0\leq\tau\leq t$. Even for NESS, this means that a complete sampling is required, e.g., to evaluate how much energy is needed to keep the process going. In practice, this means that all dissipative degrees of freedom have to be known and, more importantly, accessible \cite{SeifertPartiallyAccessibleNetworks, ThreeFacesI}. However, this is generally not the case. For instance, performing any type of projection will generally lead to a loss of information \cite{SeifertPartiallyAccessibleNetworks}. A type of projection which we will encounter repeatedly in this thesis is observing only one or more states and the transitions between these. If the transitions that are observed are not all driven edges, e.g., when only observing one edge of a driven cycle, then one a priori cannot infer the total dissipation. 

One possible solutions to this challenge are thermodynamic bounds, which are lower bounds on the total entropy production \cite{TURBiomolecular,  dieball2024thermodynamic, DirectTUR}. These can be seen as a generalisation to the second law of thermodynamics, \Eqref{eq:secondlaw}, since they may yield tighter, non-zero lower bounds. In many cases, these consist of trade-off relations between total entropy production and some other quantity, e.g., the precision of a current \cite{DirectTUR}.

The reason why lower bounds are more useful is that they require less knowledge about the systems. They therefore allow to give estimates about the cost of a process from observed subsystems, making them more applicable in experiments probing projections.


\section{Thermodynamic Bounds in Continuous Space\label{sec:thermodynamicbounds}}

In the following, we will present recent results regarding three different thermodynamic bounds and extensions of them: (i) the thermodynamic uncertainty relation, (ii) the transport bound, and (iii) the correlation bound. While (i) has been proven for MJP using methods discussed in Sec.~\ref{sec:AdvantagesDirectProof}, (ii) and (iii) have so far only been established in continuous space. Hence, the bounds will be introduced for continuous dynamics. This will also allow for easier comparison between continuous and discrete dynamics, which we discuss in Ch.~\ref{ch:directProofs}.

\subsection{Thermodynamic Uncertainty Relation\label{sec:TURintro}}

The thermodynamic uncertainty relation (TUR), first introduced by Barrato and Seifert in Ref. \cite{TURBiomolecular}, gives a trade-off between the precision of an observed current and entropy production. Alternatively, one may interpret it as a bound on the cost required to achieve a specific precision of a process. A large variety of TURs are developed to describe various scenarios \cite{TURArbitraryInit, TURTimeDependentDriving, KineticTUR, DirectTUR}. One of the most recognizable versions is the NESS TUR \cite{TURArbitraryInit, TURTimeDependentDriving, KineticTUR, DirectTUR, TighteningUncertaintyPolettini},
\begin{align}
    \frac{\langle {J}_t\rangle ^2}{\mathrm{var}({J}_t)}\leq \frac{\Delta S_\mathrm{tot}(t)}{2}\,,
    \label{eq:continuousTURss}
\end{align}
in terms of the continuous-space current and total entropy production \cite{DirectTUR}
\begin{subequations}
    \begin{alignat}{2}
        J_t &= \int_{\tau=0}^{\tau=t}\boldsymbol{U}(\mathbf{x}_\tau, \tau)\cdot {\circ\mathrm{d}\boldsymbol{x}_\tau}\,,\label{eq:continuousCurrent}\\
        \Delta S_\mathrm{tot}(t) &= t\int \mathrm{d}\boldsymbol{x}\frac{{\boldsymbol{j}(\boldsymbol{x})}^T \boldsymbol{D}^{-1}\boldsymbol{j}(\boldsymbol{x})}{\rho(\boldsymbol{x})}\,. \label{eq:continuousEntropy}  
    \end{alignat}
\end{subequations}
Here, $\boldsymbol{j}(\boldsymbol{x})$ and $\rho(\boldsymbol{x})$ are the probability current and density, respectively. Additionally, $\boldsymbol{U}(\mathbf{x}_\tau, \tau)$ is a vector-valued function that can be freely chosen. For details on these, we refer to Refs. \cite{DirectTUR, Risken1996, ConsistentNumericalSolver}. Note that these quantities are defined for overdamped dynamics, as \Eqref{eq:continuousTURss} does not hold for underdamped dynamics in \Eqref{eq:Langevin} \cite{EspositoUnderdampedTUR}. The first two moments of the current and the current itself Eq.~\eqref{eq:continuousCurrent} are operationally accessible, as the evaluation solely requires knowing the trajectory $(\boldsymbol{x}_\tau)_{0\leq \tau \leq t}$.
Recent work extended this framework to underdamped dynamics \cite{VuUnderdampedTUR, KwonUnderdampedTUR, LeeUnderdanmpedTUR, EspositoUnderdampedTUR, CrutchfieldUnderdampedTUR}. We will not go further into this here and instead refer to the respective references. 

There exist generalisations to transient systems, e.g., where the driving is time-dependent \cite{TURTimeDependentDriving} or when the system initially is not in steady state \cite{DirectTUR, TURArbitraryInit}. The latter will be focused on further requiring the generalisation to transient currents and total entropy production
\begin{subequations}
    \begin{alignat}{2}
        J_t &= \int_{\tau=0}^{\tau=t}\boldsymbol{U}(\mathbf{x}_\tau, \tau)\cdot {\circ\mathrm{d}\boldsymbol{x}_\tau}\,,\label{eq:continuousCurrent2}\\
        \Delta S_\mathrm{tot}(t) &= \int \mathrm{d}\boldsymbol{x}\int_0^t\mathrm{d}\tau\frac{{\boldsymbol{j}(\boldsymbol{x}, \tau)}^T \boldsymbol{D}^{-1}\boldsymbol{j}(\boldsymbol{x}, \tau)}{\rho(\boldsymbol{x}, \tau)}\,,\label{eq:continuousEntropy2}  
    \end{alignat}
\end{subequations}
where $\boldsymbol{j}(\boldsymbol{x}, \tau)$ and $\rho(\boldsymbol{x}, \tau)$ are the time-dependent probability current and density, respectively \cite{DirectTUR}. More precisely, Dieball and Godec recently proved the transient TUR \cite{DirectTUR}
\begin{equation}
    \begin{aligned}
        \frac{\left(t\partial_t\langle {J}_t\rangle - \langle\Tilde{J}_t\rangle \right)^2}{\mathrm{var}({J}_t)}\leq \frac{\Delta S_\mathrm{tot}(t)}{2}\,,
    \label{eq:continuousTURtransient}
    \end{aligned}
\end{equation}
using stochastic calculus (see also Sec.~\ref{sec:continuous}) and the Cauchy-Schwarz inequality for expectation values; motivating the same approach for MJP in this thesis. The modified current,
\begin{equation}
    \begin{aligned}
        \Tilde{J}_t= \int_{\tau=0}^{\tau=t} \tau\partial_\tau\boldsymbol{U}(\boldsymbol{x}_\tau, \tau)\cdot\circ\mathrm{d}\boldsymbol{x}_\tau\,,
    \end{aligned}
    \label{eq:continuousCurrentModified}
\end{equation}
accounts for the change of $\boldsymbol{U}(\boldsymbol{x}_\tau, \tau)$ in time. If $\boldsymbol{U}(\boldsymbol{x}_\tau, \tau) = \boldsymbol{U}(\boldsymbol{x}_\tau)$, Eq.~\eqref{eq:continuousTURtransient} corresponds to a TUR in, e.g., Ref.~\cite{Dechant_2018}. It is also shown in Ref. \cite{DirectTUR} that introducing a density and modified density
\begin{subequations}
    \begin{alignat}{2}
        \rho_t &= \int_{\tau=0}^{\tau=t}V(\boldsymbol{x}_\tau, \tau)\mathrm{d}\tau\,,\\
        \Tilde{\rho}_t &= \int_{\tau=0}^{\tau=t}\tau\partial_\tau V(\boldsymbol{x}_\tau, \tau)\mathrm{d}\tau\,, 
    \end{alignat}
\end{subequations}
where $V(\boldsymbol{x}_\tau, \tau)$ is a scalar function, e.g., a window function \cite{DieballCoarseGraining, DieballCurrentVariance}. Including them in the Cauchy-Schwarz inequality yields the transient correlation TUR \cite{DirectTUR}
\begin{equation}
    \begin{aligned}
        \frac{\left(t\partial_t\langle {J}_t\rangle - \langle\Tilde{J}_t\rangle - a\left((t\partial_t - 1)\langle \rho_t\rangle - \langle \Tilde{\rho}_t\rangle\right) \right)^2}{\mathrm{var}({J}_t)}\leq \frac{\Delta S_\mathrm{tot}(t)}{2}\,.
        \label{eq:continuousCorrelationTURtransient}
    \end{aligned}
\end{equation}
Here, $a\in\mathds{R}$ is a real number. 
The strength of the correlation TUR lies in the particular choice $aV(\boldsymbol{x}_\tau, \tau) = \boldsymbol{U}(\boldsymbol{x}_\tau, \tau)\cdot\boldsymbol{F}(\boldsymbol{x}_\tau) + \nabla\cdot\left(\boldsymbol{D}\boldsymbol{U}(\boldsymbol{x}_\tau, \tau)\right)$, with which \eqref{eq:continuousCorrelationTURtransient} can be saturated \cite{DirectTUR}. Performing this saturation does, however, require knowing $\boldsymbol{F}(\boldsymbol{x}_\tau)$ and $\boldsymbol{D}$, making a lower bound obsolete as Eq.~\eqref{eq:continuousEntropy} can be calculated explicitly. 
This extends the previously known saturation for NESS systems \cite{ImprovingBoundsCorrelations} to possible saturation for transient dynamics arbitrarily far from equilibrium \cite{DirectTUR}. 

For MJP, the state of the art to prove variations of the TUR is discussed in Sec.~\ref{sec:AdvantagesDirectProof}.

\subsection{Thermodynamic Transport Bound}

The transport bound is a recent result applicable to Newtonian, underdamped, and overdamped dynamics \cite{dieball2024thermodynamic}. It is a bound which relates the transport $\langle z_t - z_0\rangle$ of some observable $z_\tau = z(\boldsymbol{x}_\tau, \tau)$ to the total entropy produced. In mathematical terms, it reads
\begin{equation}
    \begin{aligned}
        T\Delta S_\mathrm{tot}(t) \geq \frac{\langle z_t - z_0 - \int_0^t\mathrm{d}\tau\partial_\tau z_\tau\rangle^2}{t\mathcal{D}^z(t)}\,.
    \end{aligned}
    \label{eq:continuousTransport}
\end{equation}
The continuous fluctuation-scale function $\mathcal{D}^z(t)$ describes the short-time fluctuations of $z$ \cite{BoundsCorrelationTimes, dieball2024thermodynamic}. Explicitly, if one defines the displacement $\mathrm{d}z_\tau = z_{\tau + \mathrm{d}\tau} - z_{\tau}$, the fluctuations of $\mathrm{d}z_\tau$ are quantified through $\mathrm{var}(\mathrm{d}z_\tau)\sim\mathrm{d}\tau$, so that $\mathcal{D}^z(t)$ is the time-average of these fluctuations \cite{BoundsCorrelationTimes, dieball2024thermodynamic}
\begin{equation}
    \begin{aligned}
        \mathcal{D}^z(t) = \frac{1}{2t}\int_0^t \mathrm{var}(\mathrm{d}z_\tau) =  \frac{1}{t}\int_0^t\mathrm{d}\tau \langle \left[\nabla_{\boldsymbol{x}}z_\tau\right]^T \frac{\boldsymbol{D}}{k_\mathrm{B}T}\nabla_{\boldsymbol{x}}z_\tau\rangle\,.
    \end{aligned}
    \label{eq:continuousDiffusion}
\end{equation}
Note that this definition is equivalent to the definition in Ref.~\cite{BoundsCorrelationTimes} for stationary systems, i.e., \Eqref{eq:continuousDiffusion} is the extension to transient dynamics \cite{dieball2024thermodynamic}. A more detailed discussion of the fluctuation-scale function $\mathcal{D}^z(t)$ can be found in Ref.~\cite{dieball2024thermodynamic}. 

Comparing the TUR \Eqref{eq:continuousTURtransient} and transport bound \Eqref{eq:continuousTransport} multiple observations can be made. Firstly, while the TUR can be applied in NESS \Eqref{eq:continuousTURss}, the transport bound yields the trivial lower bound $\Delta S_\mathrm{tot}\geq 0$, as there is no transport in a NESS. Secondly, however, \Eqref{eq:continuousTransport} only requires averages, while \Eqref{eq:continuousTURtransient} includes sample-to-sample fluctuation in $\mathrm{var}(J_t)$. For practical purposes, it may therefore often be beneficial to use the transport bound over the TUR, because it requires fewer data to achieve the desired statistical accuracy. Lastly, the naive "translation" of \Eqref{eq:continuousTURtransient} and \Eqref{eq:continuousTransport} from continuous to discrete state spaces proves non-trivial, especially for the latter bound. The TUR depends on time derivatives, which can be identified similarly for discrete states. However, the fluctuation-scale function \Eqref{eq:continuousDiffusion} includes spatial derivatives, which a priori cannot be extended to discrete state spaces. The solution to this is, as may be expected, to identify the diffusion of $z$ on the graph as the difference of the observable in neighbouring sites multiplied by the average activity. The details of this are shown in Sec.~\ref{sec:Transportproof}.

\subsection{Thermodynamic Correlation Bound}
The last bound we present is the correlation bound introduced in Ref.~\cite{BoundsCorrelationTimes}. This bound, in contrast to the TUR and transport bound, is motivated by bounding the "correlation time" of an observable $z$ defined as \cite{BoundsCorrelationTimes}
\begin{equation}
    \begin{aligned}
        \tau^z = \frac{\int_0^\infty\mathrm{d}\tau\mathrm{cov}_\mathrm{s}(z_\tau, z_0)}{\mathrm{var}_\mathrm{s}(z)}\,,
    \end{aligned}
    \label{eq:CorrelationTimePaper}
\end{equation}
where the variance and covariance are with respect to the stationary distribution. Similar definitions can be found in, e.g., Ref. \cite{StochasticProcessesApplicationsPavliotis}. As long as the covariance decays according to a single exponential with scale $-\lambda_1$ it correctly yields $\tau^z = -1/\lambda_1$. Despite that, if a decay has two timescales $-1/\lambda_2 < -1 / \lambda_1$, the correlation time \Eqref{eq:CorrelationTimePaper} is larger than with a single time-scale, although they decorrelate equally fast with timescale $-1/\mathrm{Re}(\lambda_1)$. Therefore, we will refrain from using this definition of $\tau^z$ further, as it is somewhat inconsistent. Furthermore, this definition is only applicable for stationary processes in the limit $t\to \infty$, where the latter is hard to achieve experimentally. 

Regardless of the name, the "correlation time" $\tau^z$ can be bound from below by \cite{BoundsCorrelationTimes}
\begin{equation}
    \begin{aligned}
        \tau^z = \frac{\int_0^\infty\mathrm{d}\tau\mathrm{cov}_\mathrm{s}(z_\tau, z_0)}{\mathrm{var}_\mathrm{s}(z)} \geq \sup_\omega \left[\frac{\frac{\mathrm{cov}_\mathrm{s}(\omega, z)^2}{\mathrm{var}_\mathrm{s}(z)}}{\mathcal{D}^\omega + \Dot{S}_\mathrm{tot}\mathrm{var}_{\mathrm{ent}}(\omega)}\right]\,.
    \end{aligned}
    \label{eq:continuousCorrelationBound1}
\end{equation}
Here, $\mathcal{D}^\omega$ is the continuous fluctuation-scale function \Eqref{eq:continuousDiffusion} for $\omega$ in steady state. Additionally, $\mathrm{var}_{\mathrm{ent}}(\omega)$ is the variance of $\omega$ with the shifted NESS probability density $p_{\mathrm{ent}}(\boldsymbol{x}_\tau) = p^{\mathrm{s}}(\boldsymbol{x}_{\tau})\Dot{s}(\boldsymbol{x}_\tau)/\Dot{S}_\mathrm{tot}$, where $\Dot{s}(\boldsymbol{x}_\tau)$ is the stochastic entropy production rate and $\Dot{S}_\mathrm{tot} = \langle \Dot{s}(\boldsymbol{x}_\tau)\rangle$ is the total NESS entropy production rate. Leaving $\omega$ as a free function, i.e., removing the supremum, and rewriting \Eqref{eq:continuousCorrelationBound1}, we get a bound for the total entropy production rate in NESS
\begin{equation}
    \begin{aligned}
        \Dot{S}_\mathrm{tot}\mathrm{var}_{\mathrm{ent}}(\omega)\geq \frac{\mathrm{cov}_\mathrm{s}(\omega, z)^2}{\int_0^\infty\mathrm{d}\tau\mathrm{cov}_\mathrm{s}(z_\tau, z_0)} - \mathcal{D}^\omega\,.
    \end{aligned}
    \label{eq:LiteratureCorrelationBound}
\end{equation}
This bound is only valid in the limit $t\to\infty$, hence it is reasonable to bound the total entropy production rate, as the total entropy production $\Delta S_\mathrm{tot} = t\Dot{S}_\mathrm{tot}\xrightarrow{t\to\infty}\infty$ diverges while $\Dot{S}_\mathrm{tot}$ is constant. 

The requirement of infinite times does, however, limit the use significantly. To apply this inequality in practice, one would require a significant amount of very long trajectories to be able to, at least, estimate $\int_0^\infty\mathrm{d}\tau\mathrm{cov}_\mathrm{s}(z_\tau, z_0)$. Additionally, it does not apply to transient systems. The continuous space correlation bound can be extended to transient dynamics \cite{DieballCorrelationBound}. Hence, we aim to generalise this result to MJP, which can be found in Sec.~\ref{sec:Correlationproof}.

\section{Advantages of Direct Proofs\label{sec:AdvantagesDirectProof}}


As mentioned in the previous section, thermodynamic bounds are useful to estimate entropy production without requiring knowledge of the entire system. However, with every bound comes the question of how to saturate it, i.e., which observable quantities have to be considered to predict the precise cost of a process. This is one of the main motivations of this thesis: finding a method to \textit{directly} prove a wide range of thermodynamic bounds. Before going into this further, we discuss two proofs found in the literature. 

\subsection{A Selection of Proofs}

One method often seen in the proofs of the TUR uses the Fisher information \cite{InformationTheory, Shiraishi2021}
\begin{equation}
    \begin{aligned}
        F_t(\theta) = -\left\langle\partial_\theta^2\ln{P}_\theta\left[(n_\tau)_{0\leq\tau\leq t}\right]\right\rangle_\theta\,,
    \end{aligned}
    \label{eq:FisherInformation}
\end{equation}
which requires a path measure ${P}_\theta\left[(n_\tau)_{0\leq\tau\leq t}\right]$ of a tilted dynamics to be evaluated. For MJP, the chosen tilting is in the rates $r^\theta_{ij}(\tau) = r_{ij}\mathrm{e}^{\theta Z_{ij}(\tau)}$ that enter ${P}_\theta\left[(n_\tau)_{0\leq\tau\leq t}\right]$, where \cite{Shiraishi2021}
\begin{equation}
    \begin{aligned}
        Z_{ij}(\tau) = \frac{r_{ij}p_i(\tau) - r_{ji}p_j(\tau)}{r_{ij}p_i(\tau) + r_{ji}p_j(\tau)}\,.
    \end{aligned}
    \label{eq:Zdefinition}
\end{equation}
The average is with respect to the tilted path measure. With this and the generalised Cram\'er-Rao-inequality, one can prove the transient TUR, \Eqref{eq:continuousTURtransient}, without the modified current \Eqref{eq:continuousCurrentModified} for MJP \cite{Shiraishi2021, TURArbitraryInit}. As shown in the supplementary material of Ref. \cite{DirectTUR}, the modified current is necessary to guarantee that the inequality holds.

Another method is using the scaled cumulant generating function (SCGF), e.g., in Ref. \cite{TURTimeDependentDriving} in the context of time-dependent driving, which also makes use of the path measure by considering the SCGF
\begin{equation}
    \begin{aligned}
        \lambda(z) = \frac{1}{t}\ln{\left\langle\mathrm{e}^{ztX_t}\right\rangle}\,,
    \end{aligned}
\end{equation}
where $X_t=X\left[({n}_\tau)_{0\leq\tau\leq t}\right]$ is a placeholder for a fluctuating path-observable (for instance, $X_t$ may denote a current) and the average is taken over all paths. Jensen's inequality is used together with an auxiliary path weight to get a lower bound on $\lambda(z)$. With the appropriate identifications, one finds a version of the TUR applicable to systems with time-dependent driving with parameter $v$ that looks similar to \Eqref{eq:continuousTURtransient}. However, from the proof provided, it is unclear how to choose $X_t$ to saturate the bound.

Other proof methods include, e.g., large deviation theory \cite{Gingrich2016, Proesmans_2017}, variational approaches \cite{BoundsCorrelationTimes}, and Hilbert space approach \cite{Falasco_2020}. A comprehensive list can also be found in Ref.~\cite{DirectTUR}. What should become clear from the selection of proofs presented is that they all involve some additional steps. To add to that, in many cases they depend on (path-)ensemble-level quantities, such as the SCGF, and auxiliary path weights. Additionally, we generally cannot easily extend these to the most general description of systems as well as gain insight into the saturation of the bounds. Finally, these methods do not easily generalise to other bounds. 

\subsection{Direct Proofs as a General Approach}
Before going into the advantages of a \textit{direct} proof, we first have to specify what we mean by that. To do so, it is easier to classify what we mean by \textit{indirect} proof, namely all proofs that involve some further concepts and objects in addition to the underlying dynamics. Examples of this are presented in the previous section, where the additional methods are the Fisher information together with the generalised Cram\'er-Rao inequality and the SCGF. Hence, a direct proof follows directly from the underlying dynamics, e.g., \Eqref{eq:Langevin} or \Eqref{eq:OverdampedLangevin}, together with the Cauchy-Schwarz inequality for expectation values of observables $\hat{X}_t, \hat{Y}_t$ \cite{Rogers_Williams_2000_1}
\begin{align}
    \langle \hat{X}_t\hat{Y}_t\rangle^2\leq \langle \hat{X}_t^2\rangle\langle \hat{Y}_t^2\rangle\,.
    \label{eq:CSI}
\end{align}
In other words, calculating the (auto)correlations of observables using the underlying dynamics and, using \Eqref{eq:CSI}, we get the desired bounds. As this approach requires only methods or mathematical tools close to path level descriptions, we refer to it as a direct method.

\section{On the Necessity of Stochastic Calculus for Observables}
So far, we have discussed both continuous and discrete systems, although we so far only have discussed observables in the former. The current for continuous dynamics we define as a Stratonovich integral using the stochastic differential $\mathrm{d}\boldsymbol{x}_\tau$. The question is whether we can make similar definitions of path-based observables in discrete systems using stochastic differentials. In other words, we want a stochastic equation of motion for MJP which corresponds to the Langevin equation Eq.~\ref{eq:OverdampedLangevin} in continuous space. We then want to express observables in terms of the stochastic differentials entering the equation of motion. In the literature, such stochastic differentials for MJP are not well-established, and we will thus focus this section on the need for a path-based approach in the form of stochastic integrals using currents and densities as examples. The resulting stochastic equations of motion that are required are then presented in the next section. 

First, we  consider the following form of a current often found in literature \cite{SEIFERT2018176, Seifert_2012, CurrentCharacteristicsZia, PietzonkaCurrentFluctuations, SymmetryCurrentBarato, FluctuationTheoremHarris, BoundsCurrentFluctuationsBarato, DynamicalEquivalenceVerley}
\begin{equation}
    \begin{aligned}
        J_t = \sum_{i, j\in\Omega,i\neq j} d_{ij}n_{ij}(t)\,,
    \end{aligned}
    \label{eq:dumbCurrent}
\end{equation}
which requires an anti-symmetric generalised "metric" $d_{ij} = -d_{ji}$. Additionally, the $n_{ij}(t)$ is the path-dependent number of transitions $i\to j$ for a path of length $t$ already introduced in Sec.~\ref{sec:Path}. This definition of a current, however, leaves much to be desired. 


The definition does not allow for the metric $d_{ij}$ to be time-dependent. To give an example, consider the case where only transitions after some time $t_d$ are observed, i.e., that $d_{ij}(\tau) = d_{ij}\Theta(\tau - t_d)$ with some constant $d_{ij}$ and Heavi-side function $\Theta(\tau)$. With this metric, the observed current cannot be described by \Eqref{eq:dumbCurrent}, as there is no way of distinguishing whether the transitions $n_{ij}(t)$ all happen after $t_d$ or if there are some that do not "count".

Consider a corresponding definition for a density (see, e.g., Ref. \cite{BoundsCurrentFluctuationsBarato} for a similar definition)
\begin{equation}
    \begin{aligned}
        \rho_t = \sum_{i\in\Omega} V_i \tau_i(t)\,,
    \end{aligned}
\end{equation}
where $V_i$ is some state function and $\tau_i(t)$ is the total time a trajectory of length $t$ spends in state $i$, see Sec.~\ref{sec:Path}. Similar to above, this description breaks down if $V_i(\tau)$ explicitly depends on time. 


It should become clear at this point that the numbers $n_{ij}(t)$ and $\tau_i(t)$, although sufficient for the path measure \Eqref{eq:PathMeasure}, are not suitable for a general description of time-dependent path-observables. However, these are still the relevant quantities. In the next section, we will see how we can write $n_{ij}(t)$ and $\tau_i(t)$ in terms of stochastic integrals on which we build the remaining parts of the thesis.

\section{Stochastic Equations of Motion}\label{sec:StochasticCalculus}

The goal of this thesis is to find an approach to prove the bounds introduced in Sec.~\ref{sec:thermodynamicbounds} directly from the stochastic equations of motion of MJPs, as well as a mathematical foundation for describing observables. For this, we first have to introduce the equations of motion of MJPs (see Refs.~\cite{roldan2023martingales, Rogers_Williams_2000_martingales} for a more in-depth mathematical description). The path-based quantities $n_{ij}$ and $\tau_i$, see Sec.~\ref{sec:Path}, vary for different paths of length $t$. Hence, these are random quantities, which we from now on will write as $\hat{n}_{xy}([0, t])$ and $\hat{\tau}_x([0, t])$.\footnote{The hat will from this point on denote random quantities unless specified otherwise.} These may then be written as integrals of differentials
\begin{subequations}
    \begin{alignat}{2}
        \hat{n}_{xy}([0, t]) = \int_0^t\mathrm{d}\hat{n}_{xy}(\tau)\,,
    \label{eq:eom_n}\\
        \hat{\tau}_x([0, t]) = \int_0^t\mathrm{d}\hat{\tau}_x(\tau)\,.
    \label{eq:eom_tau} 
    \end{alignat}
\end{subequations}
The latter definition is sometimes found in literature as occupation time \cite{OccupationTimeMajumdar, ExactOccupationMajumdar, OccupationTimesDarling}, since one may write $\mathrm{d}\hat{\tau}_x(\tau) = \hat{\mathds{1}}_x(\tau)\mathrm{d}\tau$ with (random) window function $\hat{\mathds{1}}_x(\tau)$ of being in state $x$ at time $\tau$ \cite{DirectTUR}. Specifically, the differential for time spent in a state is simply
\begin{equation}
    \begin{aligned}[b]
         \mathrm{d}\hat{\tau}_{x}(\tau) = \begin{cases}
            \mathrm{d}\tau & \text{state is }x \text{ in } [\tau, \tau+\mathrm{d}\tau]\,,\\
            0 & \text{else}\,.
        \end{cases}
    \end{aligned}
    \label{eq:dtauDef}
\end{equation}

Assuming that $x_\tau = x$, the differential $\mathrm{d}\hat{n}_{xy}(\tau)\sim\mathrm{Poi}(r_{xy}\mathrm{d}\tau)$ is Poisson distributed with parameter $r_{xy}\mathrm{d}\tau$. Additionally, considering the limit of $\mathrm{d}\tau\to0$, which is assumed when writing integrals like \Eqref{eq:eom_n}, the differential jump count becomes
\begin{equation}
    \begin{aligned}[b]
         \mathrm{d}\hat{n}_{xy}(\tau) =
         \begin{cases}
            1 & \text{transition }x\to y \text{ in } [\tau, \tau+\mathrm{d}\tau]\,,\\
            0 & \text{else}\,,
        \end{cases}
    \end{aligned}
    \label{eq:dnDef}
\end{equation}
since all higher number of transition have probability scaling as $\mathcal{O}(\mathrm{d}\tau^b)$ with $b\geq 2$. If $x_\tau = x$ and $\mathrm{d}\tau\to 0$, the $\mathrm{d}\hat{n}_{xy}(\tau)$ are Bernoulli distributed with parameter $r_{xy}\mathrm{d}\tau$.

Considering how important the Wiener process is for continuous dynamics, it motivates us to find a corresponding formulation of noise for MJP. This should have the same statistical properties as the Wiener process \Eqref{eq:WienerStatistics}. Dieball and Godec propose the noise differential,
\begin{equation}
    \begin{aligned}
        \mathrm{d}\hat{\varepsilon}_{xy}(\tau) = \mathrm{d}\hat{n}_{xy}(\tau) - r_{xy}\mathrm{d}\hat{\tau}_x(\tau)\,,
    \end{aligned}
    \label{eq:EpsiloJumpIncrements}
\end{equation}
in Ref~\cite{DirectTUR} with the wanted statistical properties. The quantity $\mathrm{d}\hat{\varepsilon}_{xy}(\tau)$ quantifies how much a jump, $\mathrm{d}\hat{n}_{xy}(\tau)$, deviates from the expected jump, $r_{xy}\mathrm{d}\hat{\tau}_x(\tau)$. Rearranging Eq.~\eqref{eq:EpsiloJumpIncrements} for $\mathrm{d}\hat{n}_{xy}(\tau)$ yields the stochastic equation of motion
\begin{equation}
    \begin{aligned}
        \underbrace{\mathrm{d}\hat{n}_{xy}(\tau)}_{\text{"displacement"}} =  \underbrace{r_{xy}\hat{\mathds{1}}_x(\tau)}_{\text{"drift"}}\mathrm{d}{\tau} + \underbrace{\mathrm{d}\hat{\varepsilon}_{xy}(\tau)}_{\text{"noise"}}\,.
    \end{aligned}
    \label{eq:eomMJP}
\end{equation}
Structure-wise, this is similar to the overdamped Langevin equation Eq.~\eqref{eq:OverdampedLangevin}. The jump increments can be understood as a displacement, while the rates give the expected (hence "deterministic") drift.

Of interest are the statistical properties of these stochastic differentials, which follow from the definitions Eqs.~\eqref{eq:dtauDef}~and~\eqref{eq:dnDef}. The mean values are
\begin{equation}
    \begin{aligned}[b]
        \langle \mathrm{d}\hat{n}_{xy}(\tau)\rangle &= r_{xy}p_x(\tau)\mathrm{d}\tau\,,\\
        \langle \mathrm{d}\hat{\tau}_{x}(\tau)\rangle &=p_x(\tau)\mathrm{d}\tau\,,
    \end{aligned}
    \label{eq:n_mean}
\end{equation}
while the variances are more complicated and are shown in Sec.~\ref{sec:EoMcorrelations}. This also allows us to compute the averages of Eqs.~\eqref{eq:eom_n} and \eqref{eq:eom_tau}
\begin{equation}
    \begin{aligned}[b]
        \langle \hat{n}_{xy}([0, t])\rangle &= \int_0^t\mathrm{d}\tau r_{xy}p_x(\tau)\,,\\
        \langle \hat{\tau}_x([0, t])\rangle &=\int_0^t\mathrm{d}\tau p_x(\tau)\,.
    \end{aligned}
\end{equation}
Using Eqs.~\eqref{eq:n_mean}, it follows that $\langle\mathrm{d}\hat{\varepsilon}_{xy}(\tau)\rangle=0$.

\section{Correlations\label{sec:EoMcorrelations}}

Central to proving the various bounds directly from the equation of motion Eq.~\eqref{eq:eomMJP} are correlators of the stochastic differentials. In this section, we calculate these explicitly. 

\subsection{Noise-Noise Correlation}

Since we consider Markovian systems, a transition from a state $x$ to a state $y$ at some time $\tau$ is independent of another transition from a state $i$ to a state $j$ at a later time $\tau^\prime$. Therefore, we only get contributions for $\tau=\tau^\prime$. Also, if states $i$ and state $x$ are different, then $\langle \mathrm{d}\hat{\varepsilon}_{xy}(\tau)\mathrm{d}\hat{\varepsilon}_{ij}(\tau^\prime)|x\neq i\rangle = 0$. Lastly, since we consider a Poisson distributed $\mathrm{d}\hat{n}_{xy}(\tau)$ with mean and variance $r_{xy}\mathrm{d}\tau$  if $x_\tau=x$, we end up with a covariance 
\begin{equation}
    \begin{aligned}[b]
        \langle \mathrm{d}\hat{\varepsilon}_{xy}(\tau)\mathrm{d}\hat{\varepsilon}_{ij}(\tau^\prime)\rangle = \delta(\tau - \tau^\prime) \delta_{ix}\delta_{jy}r_{xy}p_x(\tau)\mathrm{d}\tau\mathrm{d}\tau^\prime\,.
    \end{aligned}
    \label{eq:jump_covariance}
\end{equation}
Hence, $\mathrm{d}\hat{\varepsilon}_{xy}(\tau)$ has the required statistical.

\subsection{ Noise-Time Correlation}

The first correlation we consider is between the random noise increment and the time spent at a state, i.e., 
$\langle \mathrm{d}\hat{\varepsilon}_{xy}(\tau)\mathrm{d}\hat{\tau}_i(\tau^\prime)\rangle$.
This can be split into three different regimes:  the noise increment from $x$ to $y$ happens (i) before $\tau^\prime$, (ii) at $\tau^\prime$, or (iii) after $\tau^\prime$, corresponding to $\tau < \tau^\prime$, $\tau = \tau^\prime$, and $\tau > \tau^\prime$, respectively. In case (iii), the Markovian nature of the system makes the expectation value vanish. In other words, being in state $i$ at time $\tau^\prime$ and a noise increment from $x$ to $y$ at time $\tau >\tau^\prime$ are two independent events. Continuing to the second case, the only contribution to the expectation value is when $i=x$, which then yields $\langle \mathrm{d}\hat{\varepsilon}_{xy}(\tau)\rangle = 0$. Case (i), however, needs more consideration; there is a non-trivial correlation between the initial noise increment and being in a specific case at a later time. For instance, imagine a jump happening at $\tau$ and the state at $\tau+\mathrm{d}\tau$ being observed, then there certainly is no independence between the jump (where it starts and ends) and the observed state. Thus, the expectation value can be written as
\begin{equation}
    \begin{aligned}
        \langle \mathrm{d}\hat{\varepsilon}_{xy}(\tau)\mathrm{d}\hat{\tau}_i(\tau^\prime)\rangle = \mathds{1}_{\tau < \tau^\prime}\langle \mathrm{d}\hat{\varepsilon}_{xy}(\tau)\mathrm{d}\hat{\tau}_i(\tau^\prime)\rangle\,.
    \end{aligned}
    \label{eq:HardExpectation}
\end{equation}
Recalling the definition of $\mathrm{d}\hat{\varepsilon}_{xy}(\tau)$ in \Eqref{eq:EpsiloJumpIncrements}, we may write\footnote{Note that this is exact only for $\mathrm{d}\tau\to0$. In general, $\mathrm{d}\hat{n}_{xy}(\tau)$ is distributed according to a Poisson distribution with parameter $r_{xy}\mathrm{d}\tau$. Including higher-order terms will therefore only lead to higher-order corrections, which we can neglect in the limit mentioned above.}
\begin{equation}
    \begin{aligned}[b]
        \mathrm{d}\hat{\varepsilon}_{xy}(\tau) \xrightarrow{\mathrm{d}\tau\to 0} \hat{\mathds{1}}_x(\tau)
        \begin{cases}
        - r_{xy}\mathrm{d}\tau &\text{ if there is no transition from $x$ to $y$}\,,\\
        1 - r_{xy}\mathrm{d}\tau &\text{ if there is a transition from $x$ to $y$}\,.
        \end{cases}
    \end{aligned}
\end{equation}
Given $x_\tau=x$, the probabilities for a jump and no jump to occur are $r_{xy}\mathrm{d}\tau$ and $1 - r_{xy}\mathrm{d}\tau$, respectively. To evaluate \Eqref{eq:HardExpectation}, we need the probability of being in state $i$ at time $\tau^\prime$ given we are in state $x$ at time $\tau$, which is given by the propagator \Eqref{eq:Propagator}. With this in place, we use the same method as in Ref.~\cite{DieballCurrentVariance}, namely using an intermediate step to calculate the expectation value. To clarify how the calculation works in discrete space, we introduce the state-function
\begin{align}
    \gamma(\mathrm{d}\hat{\varepsilon}_{xy}|x_\tau=x) = 
    \begin{cases}
        y & \text{ if }\mathrm{d}\hat{\varepsilon}_{xy} = 1 - r_{xy}\mathrm{d}\tau\,,\\
        x & \text{ if }\mathrm{d}\hat{\varepsilon}_{xy} = - r_{xy}\mathrm{d}\tau\,,
    \end{cases}
\end{align}
which is conditioned on $x_\tau=x$. 

\begin{figure}
    \centering
    \includegraphics[width=0.5\textwidth]{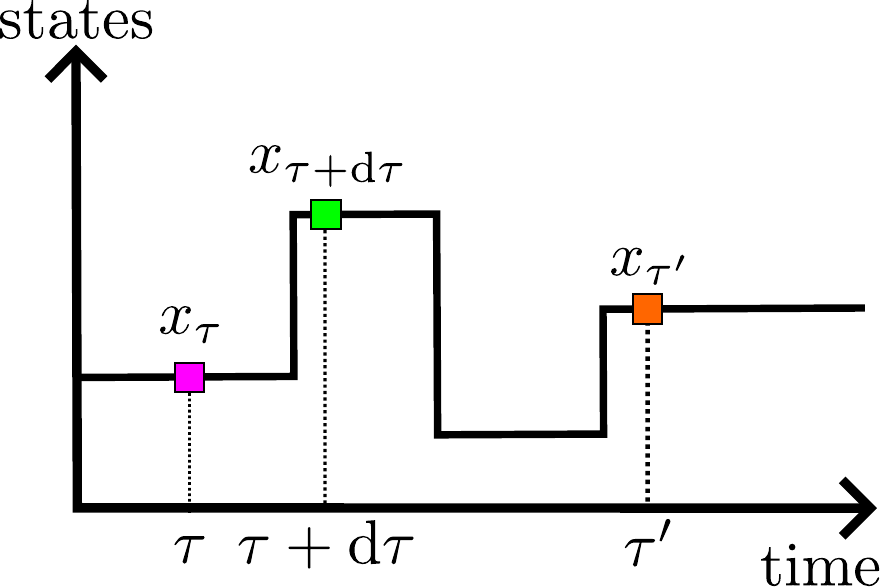}
    \caption[Sketch for proof of noise-time correlation lemma]{Sketch of the proof method to calculate the correlation \Eqref{eq:HardExpectation}. The general idea is to include the intermediate state $x_{\tau+\mathrm{d}\tau}$ and average over all possible noise increments while taking into account which state $x_{\tau+\mathrm{d}\tau}$ occurs for the various increments.}
    \label{fig:Proofsketch}
\end{figure}

The sketch in Fig.~\ref{fig:Proofsketch} helps to visualise the following calculation. The goal is to calculate the correlation between increment at $x_\tau$ and $x_{\tau'}$. To do so, the intermediate $x_{\tau+\mathrm{d}\tau}$ is included, so that \Eqref{eq:HardExpectation} becomes the combined average of the noise from $x_\tau$ to $x_{\tau+\mathrm{d}\tau}$ and of being in states $x_{\tau+\mathrm{d}\tau}$ and $x_{\tau'}$ at time $\tau+\mathrm{d}\tau$ and $\tau'$, respectively. With this, \Eqref{eq:HardExpectation} can be explicitly calculated for $\mathrm{d}\tau\to 0$
\begin{equation}
    \begin{aligned}[b]
        \langle \mathrm{d}\hat{\varepsilon}_{xy}(\tau)\mathrm{d}\hat{\tau}_i(\tau^\prime)\rangle &= \mathds{1}_{\tau < \tau^\prime}\mathrm{d}\tau^\prime\sum_{\mathrm{d}\hat{\varepsilon}_{xy}(\tau)} p(\mathrm{d}\hat{\varepsilon}_{xy}(\tau)) \mathrm{d}\hat{\varepsilon}_{xy}(\tau)P(i, \tau^\prime|\gamma(\mathrm{d}\hat{\varepsilon}_{xy}(\tau)|x_\tau=x), \tau+\mathrm{d}\tau) p_x(\tau)\\
        &=\mathds{1}_{\tau < \tau^\prime}\mathrm{d}\tau^\prime p_x(\tau) r_{xy}\mathrm{d}\tau\left(1-r_{xy}\mathrm{d}\tau\right)\left[P(i, \tau^\prime|y, \tau + \mathrm{d}\tau) - P(i, \tau^\prime|x, \tau + \mathrm{d}\tau)\right]\\
        &= \mathds{1}_{\tau < \tau^\prime}\mathrm{d}\tau^\prime \left[P(i, \tau^\prime|y, \tau) - P(i, \tau^\prime|x, \tau)\right]p_x(\tau) r_{xy}\mathrm{d}\tau\,.
    \end{aligned}
    \label{eq:EpsilonCorrelator}
\end{equation}

\subsection{Time-Time Correlation}

Next we will tackle the time-time correlator $\langle \mathrm{d}\hat{\tau}_i(\tau)\mathrm{d}\hat{\tau}_x(\tau^\prime)\rangle$. Similarly to the previous section, it can again be split into three regimes: (i) $\tau<\tau^\prime$, (ii) $\tau=\tau^\prime$, and (iii) $\tau>\tau^\prime$. These terms each then give the respective parts in the expectation value
\begin{equation}
    \begin{aligned}[b]
        \langle \mathrm{d}\hat{\tau}_i(\tau)\mathrm{d}\hat{\tau}_x(\tau^\prime)\rangle =& \mathds{1}_{\tau<\tau^\prime}\mathrm{d}\tau\mathrm{d}\tau^\prime P(x, \tau^\prime|i, \tau)p_i(\tau) + \mathds{1}_{\tau=\tau'}\mathrm{d}\tau\mathrm{d}\tau^\prime\delta_{ix}p_i(\tau) \\&+ \mathds{1}_{\tau>\tau^\prime}\mathrm{d}\tau\mathrm{d}\tau^\prime P(i, \tau|x, \tau^\prime)p_x(\tau^\prime)\,.
    \end{aligned}
    \label{eq:taucorr1}
\end{equation}
We can use the fact that $P(i, \tau|x, \tau) = \delta_{ix} = P(x, \tau| i, \tau)$ to rewrite \Eqref{eq:taucorr1} as 
\begin{equation}
    \begin{aligned}[b]
        \langle \mathrm{d}\hat{\tau}_i(\tau)\mathrm{d}\hat{\tau}_x(\tau^\prime)\rangle &= \mathds{1}_{\tau\leq\tau^\prime}\mathrm{d}\tau\mathrm{d}\tau^\prime P(x, \tau^\prime|i, \tau)p_i(\tau) + \mathds{1}_{\tau>\tau^\prime}\mathrm{d}\tau\mathrm{d}\tau^\prime P(i, \tau|x, \tau^\prime)p_x(\tau^\prime)\\
        &=\mathds{1}_{\tau<\tau^\prime}\mathrm{d}\tau\mathrm{d}\tau^\prime P(x, \tau^\prime|i, \tau)p_i(\tau) + \mathds{1}_{\tau\geq\tau^\prime}\mathrm{d}\tau\mathrm{d}\tau^\prime P(i, \tau|x, \tau^\prime)p_x(\tau^\prime)\,.
    \end{aligned}
    \label{eq:taucorr2}
\end{equation}

\subsection{Jump-Time Correlation}

We can use the previous two correlators Eqs.~\eqref{eq:EpsilonCorrelator} and \eqref{eq:taucorr2} to calculate the expectation value $\langle \mathrm{d}\hat{n}_{xy}(\tau)\mathrm{d}\hat{\tau}_i(\tau^\prime)\rangle$ by using the definition of $\mathrm{d}\hat{\varepsilon}_{xy}(\tau)$ and the linearity of the expectation value
\begin{equation}
    \begin{aligned}[b]
        \langle \mathrm{d}\hat{n}_{xy}(\tau)\mathrm{d}\hat{\tau}_{i}(\tau^\prime)\rangle =& \langle \mathrm{d}\hat{\varepsilon}_{xy}(\tau)\mathrm{d}\hat{\tau}_{i}(\tau^\prime)\rangle +r_{xy}\langle \mathrm{d}\hat{\tau}_{x}(\tau)\mathrm{d}\hat{\tau}_{i}(\tau^\prime)\rangle\\
        =& \mathds{1}_{\tau < \tau^\prime}\mathrm{d}\tau\mathrm{d}\tau^\prime r_{xy}\left[P(i, \tau^\prime|y, \tau) - P(i, \tau^\prime|x, \tau)\right]p_x(\tau)  \\
        &+ \mathrm{d}\tau\mathrm{d}\tau^\prime r_{xy} \left(\mathds{1}_{\tau^\prime\leq\tau}P(x, \tau|i, \tau^\prime)p_i(\tau^\prime) + \mathds{1}_{\tau^\prime>\tau}P(i, \tau^\prime|x, \tau)p_x(\tau)\right)\\
        =& \mathrm{d}\tau\mathrm{d}\tau^\prime r_{xy}\left[\mathds{1}_{\tau < \tau^\prime}P(i, \tau^\prime|y, \tau)p_x(\tau) +\mathds{1}_{\tau^\prime\leq\tau}P(x, \tau|i, \tau^\prime)p_i(\tau^\prime)\right]\,.
    \end{aligned}
    \label{eq:ntaucorrelator}
\end{equation}
The first term comes from a transition from $x$ to $y$ taking place at time $\tau$ and then being at $i$ at some later time $\tau^\prime>\tau$, while the second term is the reverse path, i.e., being at $i$ at some time $\tau^\prime$ and then transitioning from $x$ to $y$ at some later time $\tau > \tau^\prime$. 


\subsection{Jump-Jump Correlation}

To calculate the jump-jump correlation, we use that
\begin{equation}
    \begin{aligned}[b]
        \langle \mathrm{d}\hat{\varepsilon}_{xy}(\tau)\mathrm{d}\hat{\varepsilon}_{ij}(\tau^\prime)\rangle =& \langle \left(\mathrm{d}\hat{n}_{xy}(\tau) -r_{xy}\mathrm{d}\hat{\tau}_{x}(\tau)\right)\left(\mathrm{d}\hat{n}_{ij}(\tau^\prime) -r_{ij}\mathrm{d}\hat{\tau}_{i}(\tau^\prime)\right) \rangle \\
        =&\langle \mathrm{d}\hat{n}_{xy}(\tau)\mathrm{d}\hat{n}_{ij}(\tau^\prime)\rangle + r_{xy}r_{ij}\langle \mathrm{d}\hat{\tau}_{x}(\tau)\mathrm{d}\hat{\tau}_{i}(\tau^\prime)\rangle - r_{ij}\langle \mathrm{d}\hat{n}_{xy}(\tau)\mathrm{d}\hat{\tau}_{i}(\tau^\prime)\rangle \\
        &- r_{xy}\langle \mathrm{d}\hat{\tau}_{x}(\tau)\mathrm{d}\hat{n}_{ij}(\tau^\prime)\rangle\,.
    \end{aligned}
    \label{eq: correlator_epsilon_n}
\end{equation}
Rearranging \Eqref{eq: correlator_epsilon_n} and using \Eqref{eq:ntaucorrelator}, we get that
\begin{equation}
    \begin{aligned}[b]
        \langle \mathrm{d}\hat{n}_{xy}(\tau)\mathrm{d}\hat{n}_{ij}(\tau^\prime)\rangle =&\langle \mathrm{d}\hat{\varepsilon}_{xy}(\tau)\mathrm{d}\hat{\varepsilon}_{ij}(\tau^\prime)\rangle +r_{ij}\langle \mathrm{d}\hat{\varepsilon}_{xy}(\tau)\mathrm{d}\hat{\tau}_{i}(\tau^\prime)\rangle + r_{xy}\langle \mathrm{d}\hat{\tau}_{x}(\tau)\mathrm{d}\hat{\varepsilon}_{ij}(\tau^\prime)\rangle \\&+ r_{ij}r_{xy}\langle \mathrm{d}\hat{\tau}_{x}(\tau)\mathrm{d}\hat{\tau}_{i}(\tau^\prime)\rangle\,.
    \end{aligned}
\end{equation}
Using Eqs.~\eqref{eq:jump_covariance}, \eqref{eq:EpsilonCorrelator}, and \eqref{eq:taucorr1} we finally find
\begin{equation}
    \begin{aligned}[b]
        \langle \mathrm{d}\hat{n}_{xy}(\tau)\mathrm{d}\hat{n}_{ij}(\tau^\prime)\rangle =&\delta(\tau - \tau^\prime) \delta_{ix}\delta_{jy}r_{xy}p_x(\tau)\mathrm{d}\tau\mathrm{d}\tau^\prime\\
        &+ \mathds{1}_{\tau < \tau^\prime}\mathrm{d}\tau\mathrm{d}\tau^\prime r_{xy}r_{ij}\left[P(i, \tau^\prime|y, \tau) - P(i, \tau^\prime|x, \tau)\right]p_x(\tau)\\
        &+ \mathds{1}_{\tau^\prime < \tau}\mathrm{d}\tau\mathrm{d}\tau^\prime r_{xy}r_{ij}\left[P(x, \tau|j, \tau^\prime) - P(x, \tau|i, \tau^\prime)\right]p_i(\tau^\prime)\\
        &+ \mathds{1}_{\tau>\tau^\prime}\mathrm{d}\tau\mathrm{d}\tau^\prime r_{xy}r_{ij}P(x, \tau|i, \tau^\prime)p_i(\tau^\prime) + \mathds{1}_{\tau=\tau'}\mathrm{d}\tau\mathrm{d}\tau^\prime\delta_{ix} r_{xy}r_{ij}p_i(\tau) \\
        &+ r_{xy}r_{ij}\mathds{1}_{\tau<\tau^\prime}\mathrm{d}\tau\mathrm{d}\tau^\prime P(i, \tau^\prime|x, \tau)p_x(\tau)\\
        =& 
        \delta(\tau - \tau^\prime) \delta_{ix}\delta_{jy}r_{xy}p_x(\tau)\mathrm{d}\tau\mathrm{d}\tau^\prime+ \mathds{1}_{\tau < \tau^\prime}\mathrm{d}\tau\mathrm{d}\tau^\prime r_{xy}r_{ij}P(i, \tau^\prime|y, \tau)p_x(\tau)\\
        &+ \mathds{1}_{\tau^\prime < \tau}\mathrm{d}\tau\mathrm{d}\tau^\prime r_{xy}r_{ij}P(x, \tau|j, \tau^\prime)p_i(\tau^\prime) + \mathds{1}_{\tau=\tau'}\mathrm{d}\tau\mathrm{d}\tau^\prime\delta_{ix} r_{xy}r_{ij}p_i(\tau)\,,
    \end{aligned}
\end{equation}
and thus the covariance is
\begin{equation}
    \begin{aligned}[b]
        \mathrm{cov}(\mathrm{d}\hat{n}_{xy}(\tau),\mathrm{d}\hat{n}_{ij}(\tau')) =& \delta(\tau - \tau^\prime) \delta_{ix}\delta_{jy}r_{xy}p_x(\tau)\mathrm{d}\tau\mathrm{d}\tau^\prime+ \mathds{1}_{\tau < \tau^\prime}\mathrm{d}\tau\mathrm{d}\tau^\prime r_{xy}r_{ij}P(i, \tau^\prime|y, \tau)p_x(\tau)\\
        &+ \mathds{1}_{\tau^\prime < \tau}\mathrm{d}\tau\mathrm{d}\tau^\prime r_{xy}r_{ij}P(x, \tau|j, \tau^\prime)p_i(\tau^\prime)  + \mathds{1}_{\tau=\tau'}\mathrm{d}\tau\mathrm{d}\tau^\prime\delta_{ix} r_{xy}r_{ij}p_i(\tau)
        \\&- \mathrm{d}\tau\mathrm{d}\tau^\prime r_{xy}r_{ij}p_x(\tau)p_i(\tau^\prime)\,.
    \end{aligned}
    \label{eq:n_cov}
\end{equation}
\textcolor{black}{The first term is what we expect from uncorrelated jumps, while the last term simply is the product of means. The second term, which consists of being in state $x$ and jumping to $y$ in the time $[\tau, \tau+\mathrm{d}\tau]$ and being at state $i$ and jumping to $j$ at some later time in $[\tau^\prime, \tau^\prime + \mathrm{d}\tau^\prime]$. It reflects how jump result in later jumps. An analogous identification is made for the third term.} 

\section{Driving Systems with Invariant Stationary Density and Cycle Perturbations\label{sec:DrivingSystems}}

\begin{figure}
    \centering
    \includegraphics[width=0.6\textwidth]{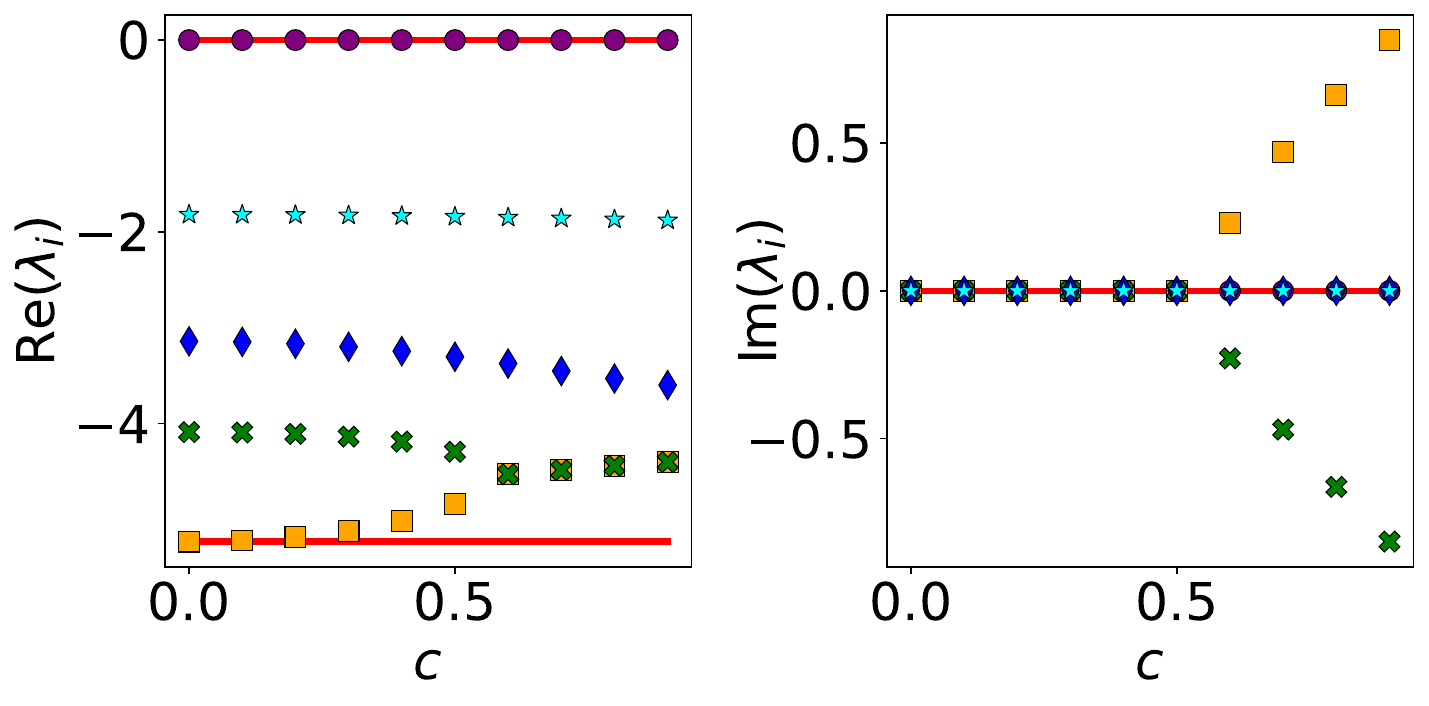}
    \caption[Eigenvalues of systematically driven system]{Real and imaginary parts of the eigenvalues $\lambda_i$ of $\mathbf{L}$ for various driving strengths $c$ of the cycle $\mathcal{C}=\{1\to2,2\to3,3\to1\}$ in the toy model. The colour and symbols distinguish between distinct eigenvalues. The red lines are the smallest and largest eigenvalues of $\mathbf{L}_s$. The energies used can be found in Tab. \ref{tab:ToyModelEnergies}.}
    \label{fig:DrivingEW}
\end{figure}

There are mainly two reasons why we are interested in driving systems while keeping the steady-state fixed. Firstly, as shown in Ref. \cite{Kaiser2017}, breaking DB\footnote{Breaking DB refers to the breaking of DB for the stationary probability.} leads to the system relaxing faster to the stationary distribution for long times and mixed initial conditions. This can, e.g., seen by the real part of the eigenvalues of the generator $\mathbf{L}$ in Fig.~\ref{fig:DrivingEW} using the method we show below. This comparison requires to keep the same steady state with and without DB, which brings us to the second reason; it allows us to better compare results in and out of equilibrium, and between non-equilibrium systems. 

\subsection{General}
The goal of this section is to establish a method with which we can perturb a system by changing the transition rates while leaving the resulting stationary probability density unchanged. This will prove useful for numerical investigations of the toy model. We emphasise that this is not a general method, i.e., it does not allow us to get every possible NESS. An alternative method for driving a system is shown in App. \ref{sec:AlternativeDriving}. While we do not use this alternative approach for any numerical methods presented in this thesis, it is nonetheless interesting from a mathematical perspective. 

As discussed in Sec.~\ref{sec:SystemClassification}, a system with time-reversal symmetry, i.e., a system in equilibrium, satisfies the detailed balance condition \Eqref{eq:DetailedBalance}. Let $\mathbf{L}_\mathrm{s}$ be the master operator of such a system, and let $\mathbf{P}^\mathrm{eq}=\mathrm{diag}(p_1^\mathrm{eq},\dots,p_N^\mathrm{eq})$ be the diagonal matrix with steady-state probabilities. The master operator can be symmetrised using a similarity transform
\begin{align}
    \mathcal{L}_\mathrm{s} = \sqrt{\mathbf{P}^\mathrm{eq}}^{-1} \mathbf{L}_\mathrm{s}\sqrt{\mathbf{P}^\mathrm{eq}}\,,
\end{align}
with entries $(\mathcal{L}_\mathrm{s})_{ij} = r_{ji}\sqrt{p^\mathrm{eq}_j}/\sqrt{p^\mathrm{eq}_i} \stackrel{\text{DB}}{=} r_{ij}\sqrt{p^\mathrm{eq}_i}/\sqrt{p^\mathrm{eq}_j} = (\mathcal{L}_\mathrm{s})_{ji}$. Adding an anti-symmetric matrix $\mathcal{L}_a=-\mathcal{L}_a^T$ to the symmetric part then gives a master operator,
\begin{align}
    \mathbf{L} = \sqrt{\mathbf{P}^\mathrm{eq}} \left(\mathcal{L}_\mathrm{s} + \mathcal{L}_\mathrm{a}\right) \sqrt{\mathbf{P}^\mathrm{eq}}^{-1}= \mathbf{L}_\mathrm{s} + \mathbf{L}_\mathrm{a}\,,
\end{align}
which now describes a non-reversible system. Note that $\mathbf{L}_\mathrm{s} \neq \mathbf{L}_\mathrm{s}^T$ and $\mathbf{L}_\mathrm{a} \neq -\mathbf{L}_\mathrm{a}^T$. There is, however, the need for $\mathbf{L}$ to have positive off-diagonal elements, leading to conditions on the entries of $\mathcal{L}_\mathrm{a}$. Denoting the entries of $\mathbf{L}_\mathrm{s}$ and $\mathbf{L}_\mathrm{a}$ as $r_{ij}^\mathrm{s}$ and $r_{ij}^\mathrm{a}$, respectively, this condition leads to $0 < r_{ij}^\mathrm{s} + r_{ij}^\mathrm{a}$ for all edges $i\to j\in\mathcal{V}$. If $i\to j\notin\mathcal{V}$, then $r_{ij}^\mathrm{s} = 0$ which implies $r_{ij}^\mathrm{a}=0$.

We want the steady-state vector to remain invariant under the driving, hence we enforce
\begin{align}
    \mathbf{L}\mathbf{p}^\mathrm{eq} = \left(\mathbf{L}_\mathrm{s} + \mathbf{L}_\mathrm{a}\right)\mathbf{p}^\mathrm{eq} = \mathbf{L}_a\mathbf{p}^\mathrm{eq} = 0\,,
\end{align}
as $\mathbf{L}_\mathrm{s}$ is the reversible part of the master operator. Thus, for every $i$, we require
\begin{align}
    \sum_{j\in\Omega} r_{ji}^\mathrm{a} p_j = 0\,.
    \label{eq: invariant_measure_condition}
\end{align}
The goal of this section is to identify the entries $r_{ij}^\mathrm{a}$ satisfying the conditions.

Identifying the reversible part with the (free) energies $E_i$, so that the equilibrium probability of being in some state $i$ is $p_i^\mathrm{eq} = \mathrm{e}^{-E_i}/Z$ with normalisation $Z$, still leaves the rates satisfying the DB condition undetermined. We will therefore keep to one form of rates that obey this manifestly, namely
\begin{align}
    r_{ij}^\mathrm{s} = \mathrm{e}^{(E_i-E_j)/2}\,.
\end{align}
For the anti-symmetric part, we follow the example presented in Ref. \cite{Kaiser2017}
\begin{align}
    r_{ij}^\mathrm{a} = j_{ij}\mathrm{e}^{E_i}
    \label{eq:AntisymmetricRates}
\end{align}
with anti-symmetric coefficients $j_{ij} = -j_{ji}$. These coefficients are constrained by the two conditions discussed above: (i) off-diagonal entries of $\mathbf{L}$ have to be positive and (ii) the steady state density should remain invariant. The first condition of this construction gives
\begin{align}
    |j_{ij}| < \mathrm{e}^{-(E_i + E_j)/2}\,.
    \label{eq:DrivingCoefficientBound}
\end{align}
The second condition simplifies significantly here. Writing out \Eqref{eq: invariant_measure_condition} yields
\begin{align}
    \sum_{j\in\Omega} j_{ji}\mathrm{e}^{E_j} \mathrm{e}^{-E_j}/Z = \sum_{j\in\Omega} j_{ji} = 0\,,
    \label{eq:j_zero_sum}
\end{align}
which physically means that the probability currents arising from the driving conserve the probability. A consequence of this is that one cannot perturb a single transition rate. Additionally, the rates into a node cannot only increase, as the coefficients would be positive and, thus, would not add up to $0$. 

\subsection{Application to Individual Cycles\label{sec:IndividualCycle}}

In this section, we explicitly show how to perturb a single cycle in a network. To begin, consider a cycle $\mathcal{C} = \{m_1\to m_2,m_2\to m_3, \dots, m_M\to m_1\}$ of size $|\mathcal{C}|=M\geq 3$. Due to the structure of a cycle, each of its states has exactly two neighbours that are also part of the cycle, e.g., state $m_2$ has $m_1$ and $m_3$ as neighbours on the cycle. Recalling the condition \Eqref{eq:j_zero_sum} for the anti-symmetric coefficients, it becomes obvious that
\begin{equation}
    \begin{aligned}[b]
        j_{m_Mm_1} + j_{m_2m_1} &= 0\,,\\
        j_{m_1m_2} + j_{m_3m_2} &= 0\,,\\
        \vdots\\
        j_{m_1m_M} + j_{m_{M-1}m_M} &= 0\,,\\
    \end{aligned}
\end{equation}
and using that they are anti-symmetric eventually leads to $j_{ij}=\mathrm{const}\,\,\forall i\to j\in\mathcal{C}$. Additionally, each of these coefficients is bounded by $\mathrm{e}^{-(E_i+E_j)/2}$, respectively. Putting all of this together, the coefficients take the form
\begin{equation}
    \begin{aligned}[b]
        j_{ij} = \pm  c \min_{l\to z\in\mathcal{C}}\left(\mathrm{e}^{-(E_l+E_z)/2}\right)
        \begin{cases}
            1 & i\to j \in\mathcal{C}\,,\\
            -1 & i\to j \in \Tilde{\mathcal{C}}\,,\\
            0 & \mathrm{else}\,,
        \end{cases}
        \label{eq:solution_j}
    \end{aligned}
\end{equation}
where $0<c<1$ is a constant modulating the strength of the cycle perturbation and $\Tilde{\mathcal{C}}$ is the reverse cycle of $\mathcal{C}$. The sign of the coefficients is not fixed, since one can either drive the cycle such that the current goes parallel or anti-parallel to it. Choosing "$+$", the rates along the cycle increase while the rates in the reverse direction decrease. Therefore, the current $0 = r^\mathrm{s}_{ij}p_i - r^\mathrm{s}_{ji}p_j < r_{ij}p_i - r_{ji}p_j = \mathcal{J}_{ij}$ is positive along the cycle. 

The sign of the coefficients can also be related to the sign of the cycle affinity \Eqref{eq:CycleAffinity}. By the same logic as with the currents, one can show that for $i\to j\in\mathcal{C}$
\begin{align}
    \mathrm{sgn}\left(\mathcal{A}_{\mathcal{C}}\right)= \mathrm{sgn}\left(\mathcal{A}_{ij}\right) = \mathrm{sgn}\left(j_{ij}\right) = \mathrm{sgn}\left(\mathcal{J}_{ij}\right)\,.
\end{align}


\chapter{Direct Proof of Thermodynamic Bounds\label{ch:directProofs}}

\section{Thermodynamic Uncertainty Relation \label{sec:TURproof}}

To prove the TUR we need to specify currents and densities. Especially their variances and covariances are of interest; for continuous space see Ref. \cite{DieballCurrentVariance}. Following Ref. \cite{DieballCurrentVariance, DirectTUR}, we introduce an auxiliary integral with a type of "pseudo entropy", which is central to all proofs of bounds in this thesis. Subsequently, we prove the TUR and discuss the optimisation the inference via the TUR.

\subsection{Currents and Densities \label{sec:currentsdensities}}

The proof of the TUR is based on three quantities: currents, pseudo entropy, and entropy. Additionally, to saturate the bound one can include densities. Here, we introduce these observables and their variances and covariances. We start with the time-integrated current \cite{DirectTUR}
\begin{equation}
    \begin{aligned}
        \hat{J}_t &=  \sum_{\substack{i, j\in\Omega\\i\neq j}}\int_{\tau=0}^{\tau=t} d_{ij}(\tau)\mathrm{d}\hat{n}_{ij}(\tau)\,,
    \end{aligned}
    \label{eq:Current}
\end{equation}
which is a central observable for stochastic systems \cite{DirectTUR, DieballCoarseGraining, DieballCurrentVariance, TURArbitraryInit, TURBiomolecular, TURTimeDependentDriving, KineticTUR, VuUnderdampedTUR, KwonUnderdampedTUR, LeeUnderdanmpedTUR, EspositoUnderdampedTUR, CrutchfieldUnderdampedTUR, PietzonkaCurrentFluctuations, TighteningUncertaintyPolettini, FluctuationTheoremHarris, SymmetryCurrentBarato, BoundsCurrentFluctuationsBarato}.
Here, $d_{ij}(\tau)=-d_{ji}(\tau)$ is a, possibly time-dependent, anti-symmetric metric which characterises the current. The mean and variance of $\hat{J}_t$ follow directly from Eqs.~\eqref{eq:n_mean} and \eqref{eq:n_cov} and read
\begin{subequations}
    \begin{alignat}{2}
        \langle \hat{J}_t\rangle =& \int_0^t\mathrm{d}\tau \sum_{\substack{i, j\in\Omega\\i\neq j}}d_{ij}(\tau)r_{ij}p_i(\tau)\,,\label{eq:J_mean}\\
        \mathrm{var}(\hat{J}_t) =& \int_0^t\mathrm{d}\tau \sum_{\substack{x, y\in\Omega\\x\neq y}}d_{xy}^2(\tau)r_{xy}p_x(\tau) - \langle \hat{J}_t\rangle^2\nonumber\\
        &+ 2\int_0^t\mathrm{d}\tau^\prime \sum_{\substack{i, j\in\Omega\\i\neq j}}d_{ij}(\tau^\prime)r_{ij} \int_0^{\tau^\prime}\mathrm{d}\tau\sum_{\substack{x, y\in\Omega\\x\neq y}}d_{xy}(\tau)r_{xy}P(i, \tau^\prime|y, \tau)p_x(\tau)\label{eq:J_var}\,,
    \end{alignat}
\end{subequations}
where we used the fact that $d_{xy}d_{ij}$ is symmetric under exchange of index pairs $(i, j) \leftrightarrow (x, y)$. 
Interestingly, the last term in the variance in \Eqref{eq:J_var} includes the average current at time $\tau$ from state $x$ to $y$ weighted by the probability to be in state $i$ at some later time $\tau^\prime> \tau$ and the resulting average current there.

In a generalisation one may consider two currents with different metrics and their covariance. If $\hat{J}_t^\alpha$ and $\hat{J}_t^\beta$ are two currents corresponding to $d_{ij}^\alpha$ and $d_{xy}^\beta$, respectively, the covariance takes the form
\begin{equation}
    \begin{aligned}[b]
        \mathrm{cov}\left(\hat{J}_t^\alpha,\hat{J}_t^\beta\right)=&
        \int_0^t\mathrm{d}\tau \sum_{\substack{x, y\in\Omega\\x\neq y}}d_{xy}^\alpha(\tau) d_{xy}^\beta(\tau)r_{xy}p_x(\tau) - \langle \hat{J}_t^\alpha\rangle\langle \hat{J}_t^\beta\rangle\\
        &+\int_0^t\mathrm{d}\tau^\prime \sum_{\substack{i, j\in\Omega\\i\neq j}}d_{ij}^\alpha(\tau^\prime)r_{ij} \int_0^{\tau^\prime}\mathrm{d}\tau\sum_{\substack{x, y\in\Omega\\x\neq y}}d_{xy}^\beta(\tau)r_{xy}P(i, \tau^\prime|y, \tau)p_x(\tau)\\
        &+\int_0^t\mathrm{d}\tau \sum_{\substack{x, y\in\Omega\\x\neq y}}d_{xy}^\beta(\tau)r_{xy} \int_0^{\tau}\mathrm{d}\tau^\prime\sum_{\substack{i, j\in\Omega\\i\neq j}}d_{ij}^\alpha(\tau^\prime)r_{ij}P(x, \tau|j, \tau^\prime)p_i(\tau^\prime)\,.\\
    \end{aligned}
    \label{eq:J_cov}
\end{equation}
Since there are different metrics, the weighted jumps observed in the variance are split such that we get contributions from observing the $J_t^\alpha$ before $J_t^\beta$ and vice versa.

\begin{figure}[ht]
    \centering
    \includegraphics[width=1\textwidth]{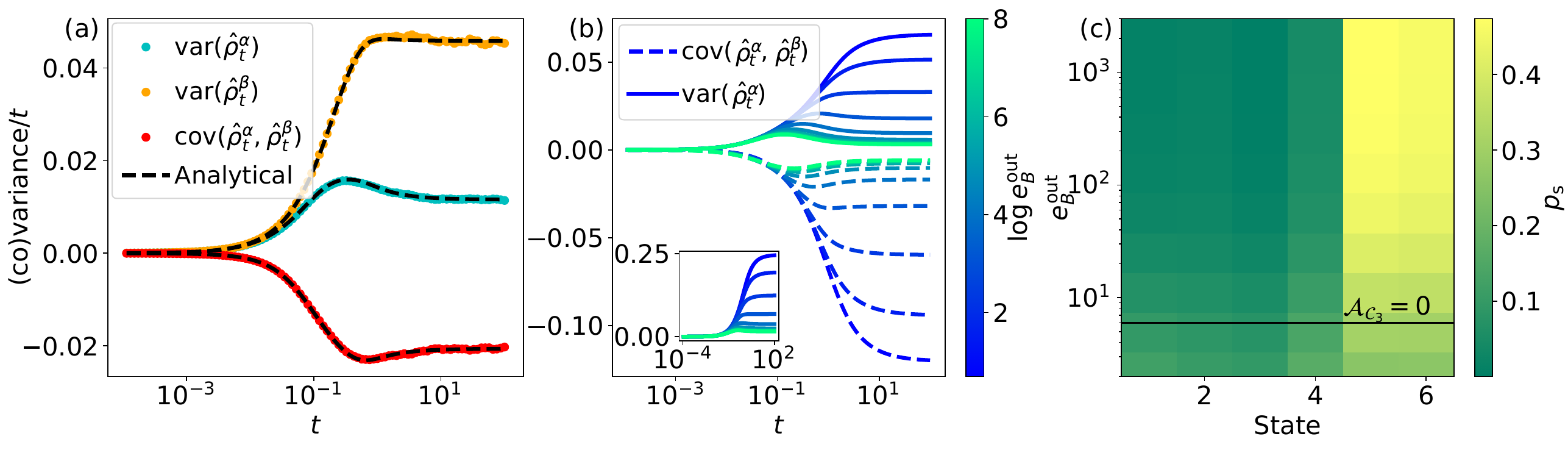}
    \caption[Density variance and covariance in SAT]{Density (co)variances in the SAT model derived in Sec.~\ref{sec:SecondaryTransport}. Two densities are considered, $\hat{\rho}_t^\alpha$ and $\hat{\rho}_t^\beta$, corresponding to the state functions $V_i^\alpha = \delta_{i1} + \delta_{i2}$ and $V_i^\beta = \delta_{i5} + \delta_{i6}$ measuring if molecules of type $A$ and $B$ are in the channel, respectively. In (a), the variances $\mathrm{var}(\hat{\rho}^{\alpha/\beta}_t)/t$ are shown together with the covariance between them $\mathrm{cov}(\hat{\rho}^{\alpha}_t,\hat{\rho}^{\beta}_t)/t$ from numerical simulations (symbols) and analytical solutions Eqs.~\eqref{eq:density_var} and \eqref{eq:density_cov} (dashed lines). The analytical variance of $\hat{\rho}_t^\alpha$ (solid lines) and covariance (dashed lines), both modulated by $1/t$, are shown in (b) for various values of $e^\mathrm{out}_B$. The inset shows $\mathrm{var}(\hat{\rho}_t^\beta)/t$. The steady state is presented in (c) for each state and the same values of $e^\mathrm{out}_B$ as in (b). The black line corresponds the values of $e^\mathrm{out}_B$ where $\mathcal{A}_{\mathcal{C}_3}=0$. The numerics in (a) are evaluated using $N=10^4$ trajectories sampled using the celebrated Gillespie algorithm. The initial condition is $p_i=(\delta_{i1}+\delta_{i3}+\delta_{i5})/3$ and the values of the parameters can be found in Tab. \ref{tab:Parameter SAT}. In (a), $e^\mathrm{out}_B = 40$ is used.}
    \label{fig:SecondaryTransportDensity}
\end{figure}

Next, we will look at densities which we define as
\begin{equation}
    \begin{aligned}
        \hat{\rho}_t = \int_{\tau=0}^{\tau=t}\sum_{i\in\Omega} V_i(\tau) \mathrm{d}\hat{\tau}_i(\tau)\,,
    \end{aligned}
    \label{eq:rho}
\end{equation}
with some state function $V_i(\tau)$. 
The mean of $\hat{\rho}_t$ is
\begin{align}
    \langle \hat{\rho}_t\rangle = \int_0^t\mathrm{d}\tau\sum_{i\in\Omega} V_i(\tau) p_i(\tau)\,.
\end{align}
Recalling \Eqref{eq:taucorr1}, the variance becomes
\begin{equation}
    \begin{aligned}[b]
        \mathrm{var}(\hat{\rho}_t) =& \int_0^t\mathrm{d}\tau^\prime \int_0^{\tau^\prime}\mathrm{d}\tau \sum_{i,x\in\Omega} V_x(\tau')V_i(\tau)P(x, \tau^\prime|i, \tau)p_i(\tau) - \langle \hat{\rho}_t\rangle ^2\,,
    \end{aligned}
    \label{eq:density_var}
\end{equation}
while the covariance for densities $\hat{\rho}_t^\gamma$ and $\hat{\rho}_t^\sigma$, corresponding to state functions $V_i^\gamma(\tau)$ and $V_j^\sigma(\tau)$, respectively, is
\begin{equation}
    \begin{aligned}[b]
        \mathrm{cov}(\hat{\rho}_t^\gamma, \hat{\rho}_t^\sigma) =& \int_0^t\mathrm{d}\tau^\prime \int_0^{\tau^\prime}\mathrm{d}\tau \sum_{i,x\in\Omega} V_x^\gamma(\tau')V_i^\sigma(\tau)P(x, \tau^\prime|i, \tau)p_i(\tau)\\
        &+  \int_0^t\mathrm{d}\tau \int_0^{\tau}\mathrm{d}\tau' \sum_{i,x\in\Omega} V_i^\sigma(\tau)V_x^\gamma(\tau')P(i, \tau|x, \tau')p_x(\tau') - \langle \hat{\rho}_t^\gamma\rangle\langle \hat{\rho}_t^\sigma\rangle\,.
    \end{aligned}
    \label{eq:density_cov}
\end{equation}
A comparison of numerical and analytical results for the variances and covariances is shown in Fig.~\hyperref[fig:SecondaryTransportDensity]{\ref*{fig:SecondaryTransportDensity}a} for the SAT model with $V_i^\alpha = \delta_{i1} + \delta_{i2}$ and $V_i^\beta = \delta_{i5} + \delta_{i6}$. A good agreement is found between the numerical and analytical results. Additionally, how the variances and covariance change when increasing $e^\mathrm{out}_B$ is presented in Figs.\hyperref[fig:SecondaryTransportDensity]{\ref*{fig:SecondaryTransportDensity}b} and~\hyperref[fig:SecondaryTransportDensity]{\ref*{fig:SecondaryTransportDensity}c} shows variations of the steady state $p_i^\mathrm{s}$ upon increasing $e^\mathrm{out}_B$. The latter seems to have a well-defined limit for $e^\mathrm{out}_B\to\infty$. In fact, we can prove this in general for finite systems with a single diverging rate, see Prop.~\ref{prop:ConvergenceEdgeDrivingSteadyState}. This leads to the variances and covariance modulated by $1/t$, e.g., $\mathrm{var}(\hat{\rho}_t^\alpha)/t$, in Fig.~\hyperref[fig:SecondaryTransportDensity]{\ref*{fig:SecondaryTransportDensity}b} converging for large $t$ in the limit $e^\mathrm{out}_B\to\infty$.

\begin{figure}[ht]
    \centering
    \includegraphics[width=\textwidth]{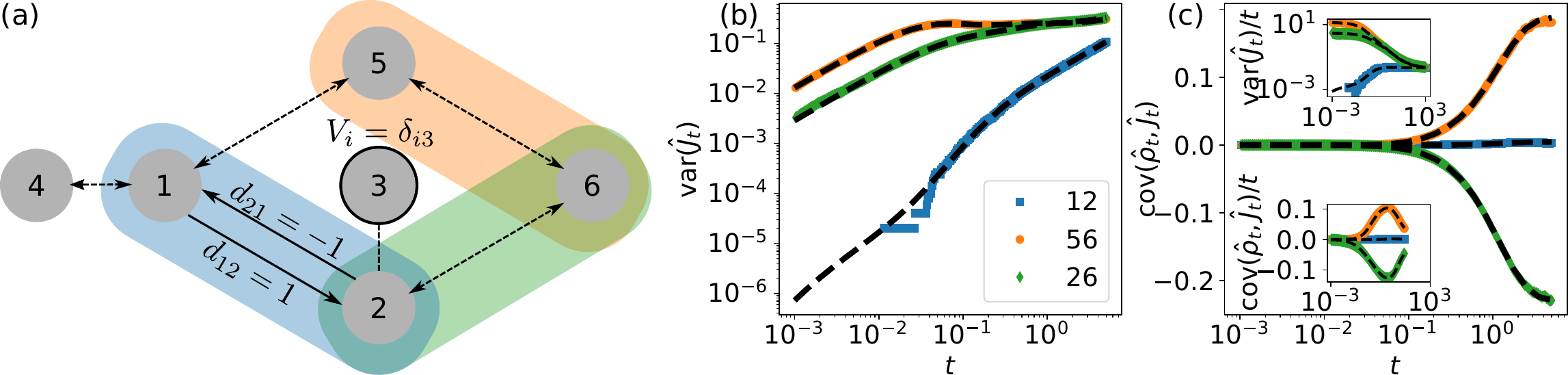}
    \caption[Current and density in calmodulin]{Comparison of numerical and analytical current variances and current-density covariances. The model is shown in (a), where the shading of the pairs of states indicates the currents observed. The respective metrics are $d_{ij}^{(12)}=\delta_{i1}\delta_{j2} - \delta_{i2}\delta_{j1}$, $d_{ij}^{(56)}=\delta_{i5}\delta_{j6} - \delta_{i6}\delta_{j5}$, and $d_{ij}^{(26)}=\delta_{i2}\delta_{j6} - \delta_{i6}\delta_{j2}$. Additionally, the density used is $V_i(\tau) = \delta_{i3}$, which is marked in the figure as a black edge on the respective state. In (b), the variances of the before mentioned currents $\mathrm{var}(\hat{J}_t)$ are shown for numerical (markers) and analytical (lines) results. The colours correspond to the shadings in (a) and the analytical solution \Eqref{eq:J_var} is used. Lastly, in (c) the covariances between the currents and density $\mathrm{cov}(\hat{\rho}_t,\hat{J}_t)$ are compared for numerical (markers) and analytical (lines) values using \Eqref{eq:density_J_cov}. The colours again correspond to the shadings in (a). The insets show the long-time behaviour of the rescaled current variance $\mathrm{var}(\hat{J}_t)/t$ and current-density covariance $\mathrm{cov}(\hat{\rho}_t,\hat{J}_t)/t$, which reveals the expected linearity in $t$ for large $t$. These figures show that the respective numerical and analytical solutions agree nicely. The initial probability density is $p_i(0) = \delta_{i6}$. The numerical values are evaluated using $N=10^5$ trajectories using the celebrated Gillespie algorithm and the transition rates are found in Tab. \ref{tab:Calmodulin_Rates}.}
    \label{fig:CalmodulinJvarJrhocov}
\end{figure}

Using \Eqref{eq:ntaucorrelator}, the current-density covariance can be written as
\begin{equation}
    \begin{aligned}[b]
        \mathrm{cov}(\hat{J}_t, \hat{\rho}_t) =&  \int_0^t\mathrm{d}\tau^\prime \int_0^{\tau^\prime}\mathrm{d}\tau \sum_{i\in\Omega}\sum_{\substack{x, y\in\Omega\\x\neq y}} V_i(\tau')d_{xy}(\tau)r_{xy}P(i, \tau^\prime|y, \tau)p_x(\tau)\\
        &+ \int_0^t\mathrm{d}\tau \int_0^{\tau}\mathrm{d}\tau^\prime \sum_{i\in\Omega}\sum_{\substack{x, y\in\Omega\\x\neq y}} V_i(\tau')d_{xy}(\tau)r_{xy}P(x, \tau|i, \tau')p_i(\tau')\\
        &- \langle \hat{\rho}_t\rangle \langle \hat{J}_t\rangle\,.
    \end{aligned}
    \label{eq:density_J_cov}
\end{equation}
The first double integral is for a transition from $x$ to $y$ before the system is in state $i$, while the second double integral accounts for the transition taking place after being in state $i$ sometime before.  In Fig.~\hyperref[fig:CalmodulinJvarJrhocov]{\ref*{fig:CalmodulinJvarJrhocov}b}, the variances of three different currents are shown in the calmodulin example, while Fig.~\hyperref[fig:CalmodulinJvarJrhocov]{\ref*{fig:CalmodulinJvarJrhocov}c} shows the covariance of these currents with a density. The numerical values, which are generated using the celebrated Gillespie algorithm, agree well with the theoretical predictions in Eqs.~\eqref{eq:J_var} and \eqref{eq:density_J_cov}. The initial condition is $p_i(0) = \delta_{i6}$, and since the current corresponding to red in Figs.~\hyperref[fig:CalmodulinJvarJrhocov]{\ref*{fig:CalmodulinJvarJrhocov}a} and \hyperref[fig:CalmodulinJvarJrhocov]{\ref*{fig:CalmodulinJvarJrhocov}b} measures the net transitions between states $1$ and $2$, there is a finite time where no such transitions are observed for the finite number ($N=10^5$) of trajectories considered.

\subsubsection{Green-Kubo Relation}

We now show how the results above, i.e., Eqs.~\eqref{eq:J_cov}, \eqref{eq:density_cov}, and \eqref{eq:density_J_cov}, can be written as Green-Kubo relations for steady state dynamics \cite{DieballCoarseGraining}. Starting by defining an integration operator,
\begin{equation}
    \begin{aligned}
        {\mathcal{I}}^{t}_{U^\alpha, U^\beta}[\cdot] \equiv \int_0^t\mathrm{d}\tau \int_0^t\mathrm{d}\tau' \sum_{x, y\in\Omega}U^\alpha_{xy}\sum_{i, j\in\Omega}U^\beta_{ij}[\cdot]\,,
    \end{aligned}
\end{equation}
so that
\begin{equation}
    \begin{aligned}[b]
        \mathrm{cov}(\hat{\rho}_t^\alpha, \hat{\rho}_t^\beta) &= {\mathcal{I}}^{t}_{V^\alpha\delta, V^\beta\delta}\left[\boldsymbol{\Xi}^1_{ijxy} - p_x^\mathrm{s}p_i^\mathrm{s}\right]\,,\\
        \mathrm{cov}(\hat{J}_t^\alpha, \hat{\rho}_t^\beta) &= {\mathcal{I}}^{t}_{d^\alpha, V^\beta\delta}\left[\boldsymbol{\Xi}^2_{ijxy} - J^\mathrm{s}_{xy}p_{i}^\mathrm{s}\right]\,,\\
        \mathrm{cov}(\hat{J}_t^\alpha, \hat{J}_t^\beta) &= {\mathcal{I}}^{t}_{d^\alpha, d^\beta}\left[\boldsymbol{\Xi}^3_{ijxy} - J^\mathrm{s}_{xy}J^\mathrm{s}_{ij}\right]\,,
    \end{aligned}
    \label{eq:IntegralDefinitionCovariance}
\end{equation}
where $J_{xy}^\mathrm{s}$ is the steady state probability flux from $x$ to $y$, see \Eqref{eq:flux}, and 
\begin{equation}
    \begin{aligned}[b]
        \boldsymbol{\Xi}^1_{ijxy} &= \frac{\langle \mathrm{d}\hat{\tau}_{x}(\tau)\mathrm{d}\hat{\tau}_{i}(\tau') \rangle}{\mathrm{d}\tau\mathrm{d}\tau'}\,,\\
        \boldsymbol{\Xi}^2_{ijxy} &= \frac{\langle \mathrm{d}\hat{n}_{xy}(\tau)\mathrm{d}\hat{\tau}_{i}(\tau') \rangle}{\mathrm{d}\tau\mathrm{d}\tau'}\,,\\
        \boldsymbol{\Xi}^3_{ijxy} &= \frac{\langle \mathrm{d}\hat{n}_{xy}(\tau)\mathrm{d}\hat{n}_{ij}(\tau') \rangle}{\mathrm{d}\tau\mathrm{d}\tau'}\,.
    \end{aligned}
\end{equation}  
To bring this into a Green-Kubo-like relation, we need the dual dynamics $r_{ij}^\dagger = r_{ji}p_j^\mathrm{s}/p_{i}^\mathrm{s}$ \cite{Seifert_2012}. The steady state probability in normal and dual dynamics is the same, however, the net probability flux changes sign $\mathcal{J}_{xy}^\dagger = - \mathcal{J}_{xy}$, i.e., the net probability flux has opposite direction for dual dynamics. We can define the current $\hat{j}_{xy}$ and dual current $\hat{j}^\dagger_{xy}$ operators\footnote{These are not random quantities.} acting on the probability density as
\begin{align}
    \hat{j}_{xy} p^\mathrm{s}_{x} &\equiv r_{xy}p_x^\mathrm{s}\,,\nonumber\\
    \hat{j}_{xy}^{\dagger} p^\mathrm{s}_{x} &\equiv r_{xy}^\dagger p_x^\mathrm{s} = r_{yx}\frac{p_y^\mathrm{s}}{p_x^\mathrm{s}}p_x^\mathrm{s} = r_{yx}p_y^\mathrm{s}\,.
\end{align}
The steady-state covariances in \Eqref{eq:IntegralDefinitionCovariance} can therefore be written as
\begin{align}
    \mathrm{cov}(\hat{\rho}_t^\alpha, \hat{\rho}_t^\beta) &= {\mathcal{I}}^{t}_{V^\alpha\delta, V^\beta\delta}\left[P(x,\tau;i,\tau') - p_x^\mathrm{s}p_i^\mathrm{s}\right]\nonumber\,,
    \\
    \mathrm{cov}(\hat{J}_t^\alpha, \hat{\rho}_t^\beta) &= {\mathcal{I}}^{t}_{d^\alpha, V^\beta\delta}\left[\hat{j}_{xy}P(x, \tau| i, \tau')p_{i}^\mathrm{s} + \hat{j}_{yx}^\dagger P(i, \tau'| y, \tau)p_y^\mathrm{s}- J^\mathrm{s}_{xy}p_{i}^\mathrm{s}\right]\,,
    \\
    \mathrm{cov}(\hat{J}_t^\alpha, \hat{J}_t^\beta) &=  t\sum_{\substack{x, y\in\Omega\\x\neq y}}d_{xy}^\alpha d_{xy}^\beta r_{xy}p_x^\mathrm{s} + {\mathcal{I}}^{t}_{d^\alpha, d^\beta}\left[\hat{j}_{xy}\hat{j}_{ji}^\dagger P(x, \tau| j, \tau')p_{j}^\mathrm{s} + \hat{j}_{ij}\hat{j}_{yx}^\dagger P(i, \tau'| y, \tau)p_y^\mathrm{s} - J^\mathrm{s}_{xy}J^\mathrm{s}_{ij}\right]\nonumber\,,
\end{align}
assuming that $d_{ij}^{\alpha/\beta}$ do not depend on time.

\subsubsection{Operationally Accessibility}

Using a current $\hat{J}_t$ as an example, we show how to evaluate path observables integrated w.r.t. $\mathrm{d}\hat{n}_{xy}(\tau)$. Let $\hat{t}^{xy}_i$ be the (random) time of the $i$th transition from $x$ to $y$ of an individual path with $\hat{t}^{xy}_1 < \hat{t}^{xy}_2 <\dots$. Recall that the total number of transitions from $x$ to $y$ in the time $[0,t]$ is given by $\hat{n}_{xy}(t) = \hat{n}_{xy}([0,t])$. Hence, \Eqref{eq:Current} can be written as
\begin{equation}
    \begin{aligned}
        \hat{{J}}_t = \sum_{\substack{x, y\in\Omega\\x\neq y}}\sum_{i=1}^{\hat{n}_{xy}(t)}d_{xy}(\hat{t}^{xy}_i)\,.
    \end{aligned}
    \label{eq:TrajectoryCurrentIntegral}
\end{equation}
Note that Eq.~\eqref{eq:TrajectoryCurrentIntegral} does not require any sort of binning in time, only the times of transition between different states are required.

\subsection{Pseudo Entropy}
Next, we turn to the so-called pseudo entropy and show how it relates to the entropy production of the system. We start from the auxiliary integral \cite{ DirectTUR}
\begin{align}
    \hat{A}_t = \sum_{\substack{x, y\in\Omega\\x\neq y}}\int_0^t Z_{xy}(\tau)\mathrm{d}\hat{\varepsilon}_{xy}(\tau)\,,
    \label{eq:A_observable}
\end{align}
with $Z_{xy}(\tau) = -Z_{yx}(\tau)$ defined in \Eqref{eq:Zdefinition}. $\hat{A}_t$ defined in Eq.~\eqref{eq:A_observable} has zero mean, see \Eqref{eq:EpsiloJumpIncrements}, while its variance reads
\begin{equation}
    \begin{aligned}
        \langle \hat{A}_t^2\rangle = \sum_{\substack{x, y\in\Omega\\x\neq y}}\int_0^t\mathrm{d}\tau Z_{xy}^2(\tau)r_{xy}p_x(\tau)
        =\frac{1}{2}\sum_{\substack{x, y\in\Omega\\x\neq y}}\int_0^t\mathrm{d}\tau \frac{\left(p_x(\tau)r_{xy}-p_y(\tau)r_{yx}\right)^2}{p_x(\tau)r_{xy}+p_y(\tau)r_{yx}}\,,
    \end{aligned}
    \label{eq:PseudoEntropy}
\end{equation}
and only vanishes in equilibrium. For systems out-of-equilibrium $\langle \hat{A}_t^2\rangle > 0$, which motivates calling $\langle \hat{A}_t^2\rangle$ the pseudo entropy production in the system. The farther the system is from equilibrium, the larger $\langle \hat{A}_t^2\rangle$ becomes. In other words, it maps the system to some number in $\mathds{R}^+$, where it only becomes zero if the system is in equilibrium. In that sense, it behaves similarly to the entropy production in the system. However, it does not have the functional form of entropy which is needed for relating it to other physical quantities, such as heat flow. 
This form is, however, very useful as it can be related to the total entropy production \cite{Dechant_2019, Shiraishi2021, DirectTUR}. Using the well known inequality for the logarithm \cite{Dechant_2019}
\begin{align}
    \frac{(a-b)^2}{a+b}\leq \frac{1}{2}(a-b)\log{\frac{a}{b}}\,,
    \label{eq:LogIdentity}
\end{align}
for $a, b>0$, the pseudo entropy production gives a lower bound on the total entropy production
\begin{align}
    \langle \hat{A}_t^2\rangle \leq \frac{\Delta S_\mathrm{tot}(t)}{2}\,,
    \label{eq:ASigmaBound}
\end{align}
which saturates when approaching equilibrium with $\Delta S_\mathrm{tot}=0$.

To evaluate the covariance between $\hat{A}_t$ and $\hat{J}_t$, one has to calculate $\langle\mathrm{d}\hat{\varepsilon}_{xy}(\tau)\mathrm{d}\hat{n}_{ij}(\tau')\rangle$. Since the covariances Eqs.~\eqref{eq:jump_covariance} and \eqref{eq:EpsilonCorrelator} are already known, it is useful to split the expectation value as
\begin{align}
    \langle\mathrm{d}\hat{\varepsilon}_{xy}(\tau)\mathrm{d}\hat{n}_{ij}(\tau')\rangle = r_{ij}\langle\mathrm{d}\hat{\varepsilon}_{xy}(\tau)\mathrm{d}\hat{\tau}_{i}(\tau')\rangle + \langle\mathrm{d}\hat{\varepsilon}_{xy}(\tau)\mathrm{d}\hat{\varepsilon}_{ij}(\tau')\rangle\,,
\end{align}
and, therefore,
\begin{equation}
    \begin{aligned}[b]
        \langle \hat{A}_t\hat{J}_t\rangle = \int_0^t\int_0^t \sum_{\substack{i, j\in\Omega\\i\neq j}}\sum_{\substack{x, y\in\Omega\\x\neq y}}Z_{xy}(\tau)d_{ij}(\tau')\left(r_{ij}\langle\mathrm{d}\hat{\varepsilon}_{xy}(\tau)\mathrm{d}\hat{\tau}_{i}(\tau')\rangle + \langle\mathrm{d}\hat{\varepsilon}_{xy}(\tau)\mathrm{d}\hat{\varepsilon}_{ij}(\tau')\rangle\right)\,,
    \end{aligned}
    \label{eq:AJ_correlator}
\end{equation}
which again motivates splitting the current into two sub-currents \cite{DirectTUR}
\begin{equation}
    \begin{aligned}
        \hat{J}_t &= \hat{J}_t^\mathrm{I}+\hat{J}_t^\mathrm{II} = \underbrace{\sum_{\substack{i, j\in\Omega\\i\neq j}}\int_0^t d_{ij}(\tau) \mathrm{d}\hat{\varepsilon}_{ij}(\tau)}_{\hat{J}_t^\mathrm{I}} + \underbrace{\sum_{\substack{i, j\in\Omega\\i\neq j}}\int_0^t d_{ij}(\tau) r_{ij}\mathrm{d}\hat{\tau}_{i}(\tau)}_{\hat{J}_t^\mathrm{II}}\,.
    \end{aligned}
    \label{eq:SplitCurrent}
\end{equation}
We can understand $\hat{J}_t^\mathrm{II}$ as the expected current from being in various states along the trajectory, while $\hat{J}_t^\mathrm{I}$ is how much the trajectory deviates from the expected $\hat{J}_t^\mathrm{II}$. This is also reflected in the average, as $\langle \hat{J}_t \rangle = \langle \hat{J}_t^\mathrm{II}\rangle$ and $\langle \hat{J}_t^\mathrm{I}\rangle = 0$, showing that we will not deviate from the expected current on average. To compute \Eqref{eq:AJ_correlator}, we simply have to calculate the expectation values with the two sub-currents, respectively, which can be done with Eqs.~\eqref{eq:jump_covariance} and \eqref{eq:EpsilonCorrelator}. For $\hat{J}_t^\mathrm{I}$ this is a simple calculation yielding
\begin{equation}
    \begin{aligned}
        \langle \hat{A}_t\hat{J}_t^\mathrm{I}\rangle &= \int_0^t\mathrm{d}\tau\sum_{\substack{x, y\in\Omega\\x\neq y}} Z_{xy}(\tau)d_{xy}(\tau)r_{xy}p_x(\tau) = \int_0^t\mathrm{d}\tau\sum_{\substack{x, y\in\Omega\\x\neq y}} d_{xy}(\tau)r_{xy}p_x(\tau) = \langle \hat{J}_t\rangle\,.
    \end{aligned}
    \label{eq:AJI_correlator}
\end{equation}
Plugging \Eqref{eq:EpsilonCorrelator} into the expectation value of $\hat{A}_t\hat{J}_t^\mathrm{II}$,
\begin{equation}
    \begin{aligned}[b]
        \langle \hat{A}_t\hat{J}_t^\mathrm{II}\rangle = \int_0^t\mathrm{d}\tau^\prime \int_0^t \mathrm{d}\tau \sum_{\substack{i, j\in\Omega\\i\neq j}}d_{ij}(\tau^\prime)r_{ij}\mathds{1}_{\tau < \tau^\prime} \sum_{\substack{x, y\in\Omega\\x\neq y}}\left[P(i, \tau^\prime|y, \tau) - P(i, \tau^\prime|x, \tau)\right]p_x(\tau) r_{xy}Z_{xy}(\tau)\,.
    \end{aligned}
\end{equation}
Simplifying it with Eqs.~\eqref{eq:MasterEquation} and \eqref{eq:NetFlux} gives
\begin{equation}
    \begin{aligned}
        \langle \hat{A}_t\hat{J}_t^\mathrm{II}\rangle =\int_0^t\mathrm{d}\tau^\prime \int_0^t \mathrm{d}\tau \sum_{\substack{i, j\in\Omega\\i\neq j}}d_{ij}(\tau^\prime)r_{ij}\mathds{1}_{\tau < \tau^\prime} \sum_{x\in\Omega} P(i, \tau'|x, \tau)\partial_{\tau}p_x(\tau)\,.
    \end{aligned}
\end{equation}
By performing an integration by parts and recalling that the propagator only depends on the time difference, such that $\partial_\tau P(i, \tau'|x, \tau) = -\partial_{\tau'} P(i, \tau'|x, \tau)$, and we can explicitly evaluate the inner integral
\begin{equation}
    \begin{aligned}[b]
        \langle \hat{A}_t\hat{J}_t^\mathrm{II}\rangle &=-\int_0^t\mathrm{d}\tau^\prime \sum_{\substack{i, j\in\Omega\\i\neq j}}d_{ij}(\tau^\prime)r_{ij}\left[p_i(\tau') + \int_0^t\mathrm{d}\tau  \sum_{x\in\Omega} p_x(\tau)\partial_{\tau}\left\{\mathds{1}_{\tau < \tau^\prime} P(i, \tau'|x, \tau)\right\}\right]\\
        &=-\langle \hat{J}_t\rangle + \int_0^t\mathrm{d}\tau^\prime \sum_{\substack{i, j\in\Omega\\i\neq j}}d_{ij}(\tau^\prime)r_{ij}\partial_{\tau'}\left[\tau'p_i(\tau')\right]\,,
    \end{aligned}
    \label{eq:AJ2covariance_partial_int}
\end{equation}
where we use \Eqref{eq:forwardKolmogorov} and $\int_0^t\mathrm{d}\tau\mathds{1}_{\tau < \tau^\prime} = \tau'$. Upon another integration by parts and by defining
\begin{align}
    \hat{\Tilde{J}}_t = \int_0^t \sum_{\substack{i, j\in\Omega\\i\neq j}}\tau\mathrm{d}\hat{n}_{ij}(\tau)\partial_\tau d_{ij}(\tau)\,,
    \label{eq:modifiedCurrent}
\end{align}
we finally recover
\begin{equation}
    \begin{aligned}
        \langle \hat{A}_t\hat{J}_t^\mathrm{II}\rangle = -\langle \hat{J}_t\rangle + t\partial_t\langle \hat{J}_t\rangle - \langle \hat{\Tilde{J}}_t\rangle\,.
    \end{aligned}
    \label{eq:AJII_correlator}
\end{equation}
Adding Eqs.~\eqref{eq:AJI_correlator} and \eqref{eq:AJII_correlator} we find that \Eqref{eq:AJ_correlator} takes the simple form
\begin{align}
    \langle \hat{A}_t\hat{J}_t\rangle = t\partial_t\langle \hat{J}_t\rangle - \langle \hat{\Tilde{J}}_t\rangle\,,
    \label{eq:AJ_final_correlator}
\end{align}
which agrees with the corresponding calculation in continuous systems \cite{DirectTUR}.


\subsection{Proof: From Cauchy-Schwarz to the TUR}
The proof is based on using $\hat{A}_t$ and $\Delta \hat{J}_t = \hat{J}_t - \langle \hat{J}_t\rangle$ in the Cauchy-Schwarz inequality, so that \Eqref{eq:CSI} becomes
\begin{align}
    \langle \hat{A}_t\Delta\hat{J}_t\rangle \leq \langle \hat{A}_t^2\rangle \mathrm{var}(\hat{J}_t)\,.
    \label{eq:CSI_AJ}
\end{align}
Using the expressions in the previous section, namely Eqs.~\eqref{eq:ASigmaBound} and \eqref{eq:AJ_final_correlator}, this in turn becomes
\begin{align}
    \left(t\partial_t\langle \hat{J}_t\rangle - \langle \hat{\Tilde{J}}_t\rangle\right)^2 \leq \langle \hat{A}_t^2\rangle \mathrm{var}(\hat{J}_t)\leq \frac{\Delta S_\mathrm{tot}(t)}{2}\mathrm{var}(\hat{J}_t)\,.
    \label{eq:TURtransientversion}
\end{align}
This proof is analogous to the proof in Ref. \cite{DirectTUR} for dynamics in continuous space. 

In many cases it is more useful to consider the quality factor,
\begin{align}
    \mathcal{Q}(t) = \frac{\text{Inferred entropy production}}{\text{Actual entropy production}}=\frac{2\left(t\partial_t\langle \hat{J}_t\rangle - \langle \hat{\Tilde{J}}_t\rangle\right)^2}{\Delta S_\mathrm{tot}(t) \mathrm{var}(\hat{J}_t)}\,,
    \label{eq:transient_quality_factor}
\end{align}
to compare the inferred to the actual entropy produced.  
It should be clear that $0\leq \mathcal{Q}(t)\leq 1$. For $\mathcal{Q}(t)=1$, one has a perfect inference of the total entropy production from the mean and variance of the current $\hat{J}_t$. 

For numerical comparison, we also introduce the quality factor for the pseudo entropy production
\begin{align}
    \mathcal{Q}_A(t) = \frac{\left(t\partial_t\langle \hat{J}_t\rangle - \langle \hat{\Tilde{J}}_t\rangle\right)^2}{\langle \hat{A}_t^2\rangle\mathrm{var}(\hat{J}_t)}\,,
    \label{eq:transient_A_quality_factor}
\end{align}
which we refer to as a pseudo-quality factor. By looking at \Eqref{eq:TURtransientversion}, one can immediately identify that $\mathcal{Q}(t)\leq\mathcal{Q}_A(t)$. In fact, $\mathcal{Q}(t) = \Delta S_\mathrm{tot}(t)\mathcal{Q}_A(t)/2\langle \hat{A}_t^2\rangle$. As mentioned earlier, the pseudo entropy production has no relation to physical quantities. It is a mathematical construct which helps with the proof. We will, however, consider it further to understand the saturation of quality factors. 

\begin{figure}[ht]
    \centering
    \includegraphics[width=.6\textwidth]{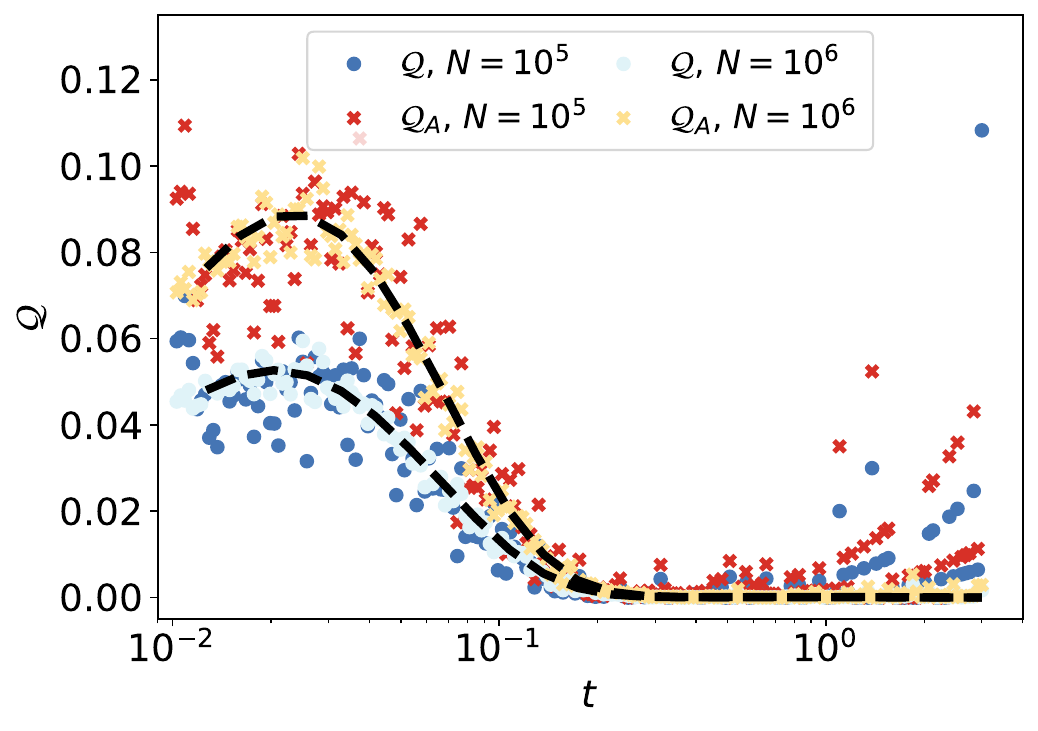}
    \caption[Effect of sample size]{Showcase of data precision for transient dynamics in the calmodulin example, see Fig.~\ref{fig:CalmodulinJvarJrhocov}. The quality factors Eqs.~\eqref{eq:transient_quality_factor} and \eqref{eq:transient_A_quality_factor} are shown as a function of $t$. The circle and cross markers correspond to $\mathcal{Q}$ and $\mathcal{Q}_A$, respectively, where the inferred entropy production is obtained using the celebrated Gillespie algorithm for $N=10^5$ (dark orange and blue) and $N_\mathrm{sample}=10^6$ (light orange and blue) trajectories. The respective analytical prediction is marked as the black dashed line. The metric is $\delta_{ij}^{(12)} = \delta_{i1}\delta_{j2} - \delta_{i2}\delta_{j1}$ with an initial condition $p_i(0) = 1/6$. To calculate the derivative $\partial_t\langle\hat{J}_t\rangle$, we approximate it by $(\langle\hat{J}_{t+\mathrm{d}t}\rangle - \langle\hat{J}_t\rangle)/\mathrm{d}t$ with $\mathrm{d}t=10^{-3}$.}
    \label{fig:Derivative_sample}
\end{figure}

To achieve reasonable accuracy, transient systems generally need a large number of trajectories. An example of this can be seen in Fig.~\ref{fig:Derivative_sample} for a transient calmodulin system, where the spread in the inferred entropy production\footnote{We limit the discussion to entropy production in this paragraph, as every statement applies to the pseudo-entropy production due to $\mathcal{Q}\propto\mathcal{Q}_A$.} is visibly greater for $N=10^5$ trajectories compared to $N=10^6$ trajectories. The figure also shows how the sample size compares to the analytic solution using Eqs.~\eqref{eq:J_mean}, \eqref{eq:J_var}, \eqref{eq:transient_quality_factor}, and \eqref{eq:transient_A_quality_factor}. Especially on timescales greater than the relaxation time of the system, i.e., where the system is close to equilibrium in this example, we see a clear deviation between the expected $\mathcal{Q}_\mathrm{an}$ and the numerically calculated $\mathcal{Q}_\mathrm{num}$ quality factors. There,  $\mathcal{Q}_\mathrm{an}$ goes to zero as seen in Fig.~\ref{fig:Derivative_sample} by the black dashed line. However, since the numerical sampling does not warrant perfect statistics, it results in a large $\mathcal{Q}_\mathrm{num}$ due to fluctuations in $\partial_t\langle\hat{J}_t\rangle$. These fluctuations get amplified by $t$, which becomes especially apparent for large $t$. 

\subsection{Saturation and Stationary Systems \label{sec:TURsaturation}}

As bounds on the total entropy production play an important role in this thesis, it is natural to wonder if and when these bounds saturate. Two inequalities are used in the proof: the logarithm identity \Eqref{eq:LogIdentity} and the Cauchy-Schwarz inequality \Eqref{eq:CSI}. We will in the following consider stationary systems so that \Eqref{eq:AJ_final_correlator} becomes $\langle \hat{A}_t\hat{J}_t\rangle = \langle \hat{J}_t\rangle$. Since \Eqref{eq:CSI} holds for any observable, we can shift the current by a deterministic factor multiplied by the density in \Eqref{eq:rho}. In steady state, the expectation $\langle \hat{A}_t\hat{\rho}_t\rangle = 0$, and, thus, the stationary TUR is
\begin{equation}
    \begin{aligned}[b]
        \langle \hat{J}_t\rangle ^2 \leq \frac{\Delta S_\mathrm{tot}(t)}{2}\mathrm{var}(\hat{J}_t - c(t) \hat{\rho}_t)\,,
    \end{aligned}
    \label{eq:SSTURshifted}
\end{equation}
where $c(t)\in\mathds{R}$ is a deterministic weighting function for the density contribution. In fact, optimising the variance with respect to $c(t)$ by using $\mathrm{var}(\hat{J}_t - c(t) \hat{\rho}_t) = \mathrm{var}(\hat{J}_t) + c^2 \mathrm{var}(\hat{\rho}_t) - 2c\mathrm{cov}(\hat{J}_t,\hat{\rho}_t)$, yields the optimal weighting function
\begin{align}
    c(t) = \frac{\mathrm{cov}(\hat{J}_t,\hat{\rho}_t)}{\mathrm{var}(\hat{\rho_t})}\,.
    \label{eq:SS_optimal_c}
\end{align}
Using Eqs.~\eqref{eq:density_var} and \eqref{eq:density_J_cov}, $c(t)$ can be determined explicitly. Choosing this optimal $c(t)$, the r.h.s. of \Eqref{eq:SSTURshifted} can be minimised. 

\begin{figure}
    \centering
    \includegraphics[width=\textwidth]{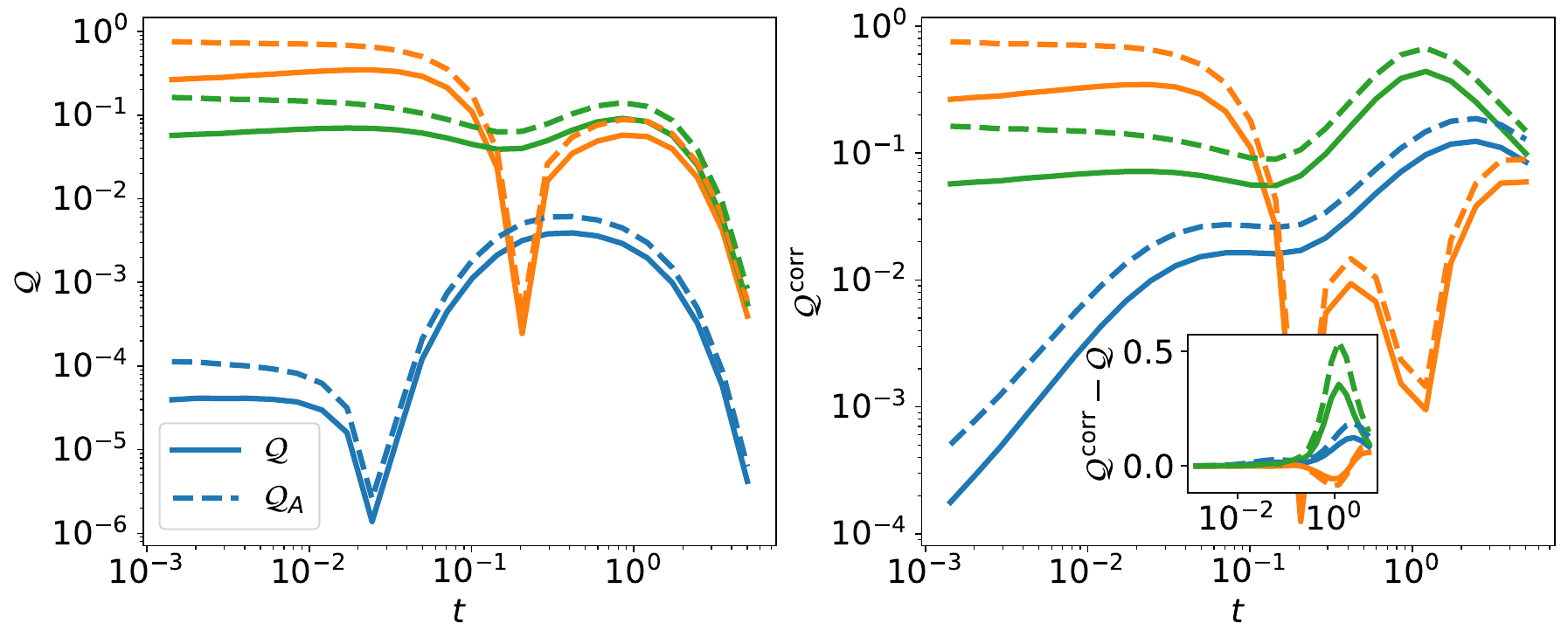}
    \caption[Comparison of quality factors]{Quality factors of currents $\mathcal{Q}(t)$ \Eqref{eq:TURtransientversion} (left) and currents+densities $\mathcal{Q}^\mathrm{corr}(t)$ \Eqref{eq:transient_shifted_TUR} (right) from Fig.~\ref{fig:CalmodulinJvarJrhocov} shown as a function of time with the corresponding colour coding. For the right plot, the coefficient $c(t) = 1$ is chosen. The inset shows the difference of the quality factors with and without including densities.}
    \label{fig:TUR_Q1}
\end{figure}

\begin{figure}
    \centering
    \includegraphics[width=0.5\textwidth]{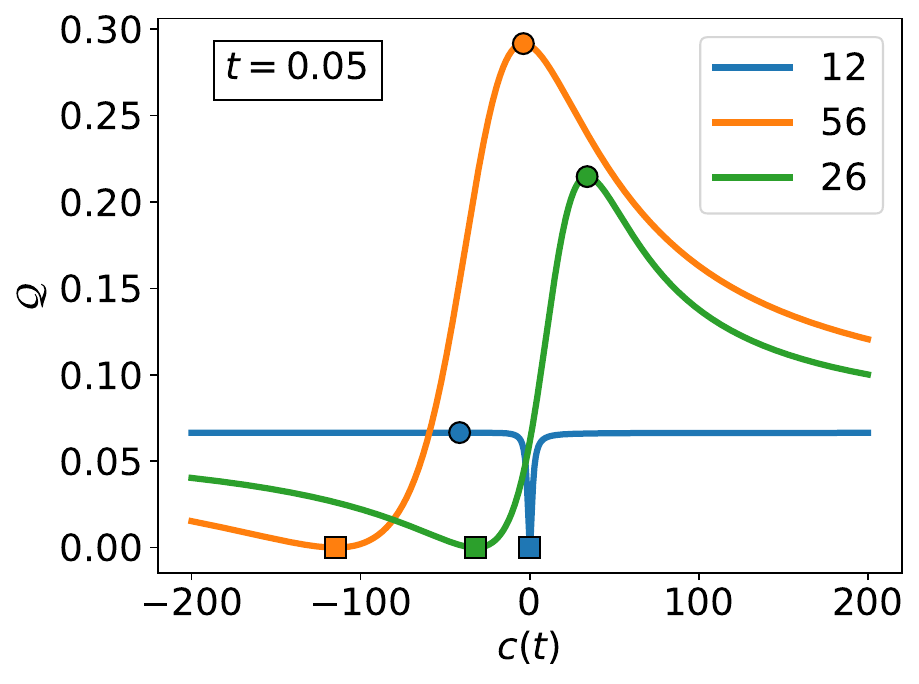}
    \caption[Optimal shifting at fixed $t$]{Correlation quality factor $\mathcal{Q}(t)$ as a function of $c(t)$ at a fixed length $t=0.05$. The colours correspond to Figs.~\ref{fig:CalmodulinJvarJrhocov} and \ref{fig:TUR_Q1}. The value of $\mathcal{Q}(t)$ from the values of $c(t)$ in \Eqref{eq:optimal_c} are shown as squares and circles for $c^\dagger(t)$ and $c^*(t)$, respectively. As predicted, choosing $c^\dagger(t)$ results in $\mathcal{Q}=0$.}
    \label{fig:TUR_optimisation_tfixed}
\end{figure}

If we consider transient dynamics, the TUR \Eqref{eq:SSTURshifted} generalises to correlation TUR (cTUR)
\begin{equation}
    \begin{aligned}
        \frac{2\left(t\partial_t\langle \hat{J}_t\rangle - \langle\hat{\Tilde{J}}_t\rangle - c(t)\left[(t\partial_t-1)\langle \hat{\rho}_t\rangle -\langle\hat{\Tilde{\rho}}_t\rangle \right]\right)^2}{\mathrm{var}(\hat{J}_t - c(t)\hat{\rho}_t)} \leq \Delta S_\mathrm{tot}(t)\,,
    \end{aligned}
    \label{eq:transient_shifted_TUR}
\end{equation}
where 
\begin{equation}
    \begin{aligned}
        \langle\hat{\Tilde{\rho}}_t\rangle = \frac{1}{t}\int_0^t\mathrm{d}\tau \sum_{i\in\Omega} p_i(\tau) \tau\partial_\tau V_i(\tau)\,.
    \end{aligned}
\end{equation}

Figure~\ref{fig:TUR_Q1} shows the quality factor of the current TUR $\mathcal{Q}$ in (left) and correlation TUR $\mathcal{Q}^\mathrm{corr}$ in (right) in the calmodulin example. The inset shows the difference $\mathcal{Q}^\mathrm{corr}-\mathcal{Q}$, demonstrating the non-trivial effect of $c(t)$. Additionally, the impact of various values of $c(t)$ on correlation $\mathcal{Q}(t)$ calculated from \Eqref{eq:transient_shifted_TUR} at a fixed $t$ is shown in Fig.~\ref{fig:TUR_optimisation_tfixed} for the calmodulin example in Fig.~\ref{fig:CalmodulinJvarJrhocov}. Taking the derivative of the l.h.s. in \Eqref{eq:transient_shifted_TUR}, which we denote by $h(c, t)$, with respect to $c(t)$ to find the extrema by solving $\partial_c h(c,t) = 0$ for $c(t)$, yields two solutions
\begin{equation}
    \begin{aligned}[b]
        c^\dagger(t) &= \frac{t\partial_t\langle \hat{J}_t\rangle - \langle\hat{\Tilde{J}}_t\rangle}{(t\partial_t-1)\langle \hat{\rho}_t\rangle - \langle\hat{\Tilde{\rho}}_t\rangle}\,,\\
        c^*(t) &= \frac{(t\partial_t\langle \hat{J}_t\rangle - \langle\hat{\Tilde{J}}_t\rangle)\mathrm{cov}(\hat{\rho}_t, \hat{J}_t) - ((t\partial_t-1)\langle \hat{\rho}_t\rangle - \langle\hat{\Tilde{\rho}}_t\rangle)\mathrm{var}(\hat{J}_t)}{(t\partial_t\langle \hat{J}_t\rangle - \langle\hat{\Tilde{J}}_t\rangle)\mathrm{var}(\hat{\rho}_t) - ((t\partial_t-1)\langle \hat{\rho}_t\rangle - \langle\hat{\Tilde{\rho}}_t\rangle)\mathrm{cov}(\hat{\rho}_t, \hat{J}_t)}\,.
    \end{aligned}
    \label{eq:optimal_c}
\end{equation}
A quick calculation shows that
\begin{equation}
    \begin{aligned}[b]
        \frac{\partial^2}{\partial c^2}h(c, t)|_{c(t) = c^\dagger(t)} &= \frac{4b^2}{\mathrm{var}\left(\hat{J}_t - \frac{a}{b}\hat{\rho}_t\right)}\geq 0\,,\\
        \frac{\partial^2}{\partial c^2}h(c, t)|_{c(t) = c^*(t)} &= \frac{-4b^2\mathrm{var}\left(\hat{J}_t - \frac{a}{b}\hat{\rho}_t\right)}{\mathrm{var}\left(\hat{J}_t - c^*(t)\hat{\rho}_t\right)^2}\leq 0\,,
    \end{aligned}
\end{equation}
where $a = t\partial_t\langle \hat{J}_t\rangle - \langle\hat{\Tilde{J}}_t\rangle$ and $b = (t\partial_t-1)\langle \hat{\rho}_t\rangle  - \langle\hat{\Tilde{\rho}}_t\rangle$. Hence, $c^\dagger(t)$ is the least optimal value for saturating \Eqref{eq:transient_shifted_TUR} regardless of $t$, while $c^*(t)$ is the optimal value. Furthermore, plugging in $c^\dagger(t)$ into \Eqref{eq:transient_shifted_TUR} the numerator vanishes, so that one recovers \textcolor{black}{the second law of thermodynamics} \Eqref{eq:secondlaw}. This is, however, the weakest possible non-trivial bound. Consequently, choosing any other $c$ will give an improved bound compared to the second law. Additionally, the coefficients \Eqref{eq:optimal_c} agree with the optimal $c$ from steady state dynamics \Eqref{eq:SS_optimal_c} in the limit where the dynamics are close to a (non-)equilibrium steady state, denoted by "$\xrightarrow{\text{sd}}$" in the following. 
\footnote{There are two ways this is realised: (i) taking the long time limit $t\to \infty$ or (ii) taking the limit where the initial probability measure goes to the steady state measure $p_x(0) \to p_x^\mathrm{s}$. Since we are interested in the time-dependency of the optimal $c(t)$, the limit (ii) is generally considered in this thesis.}
There, $(t\partial_t-1)\langle \hat{\rho}_t\rangle - \langle\hat{\Tilde{\rho}}_t\rangle\xrightarrow{\text{sd}} 0$ and $t\partial_t\langle \hat{J}_t\rangle - \langle\hat{\Tilde{J}}_t\rangle \xrightarrow{\text{sd}} \langle \hat{J}_t\rangle$, so $c^\dagger(t)\xrightarrow{\text{sd}} \pm \infty$ and $c^*(t)\xrightarrow{\text{sd}} \mathrm{cov}(\hat{\rho}_t, \hat{J}_t) / \mathrm{var}(\hat{\rho}_t)$. Since the r.h.s. of \Eqref{eq:SSTURshifted} consists of the entropy production times a convex parabola in $c$, the bound can be made arbitrarily bad by increasing the absolute value of $c(t)$ away from $c^*(t)$. Taking the limit $|c|\to \infty$ is therefore equivalent to using the second law to bound the entropy production. 

\begin{figure}
    \centering
    \includegraphics[width=\textwidth]{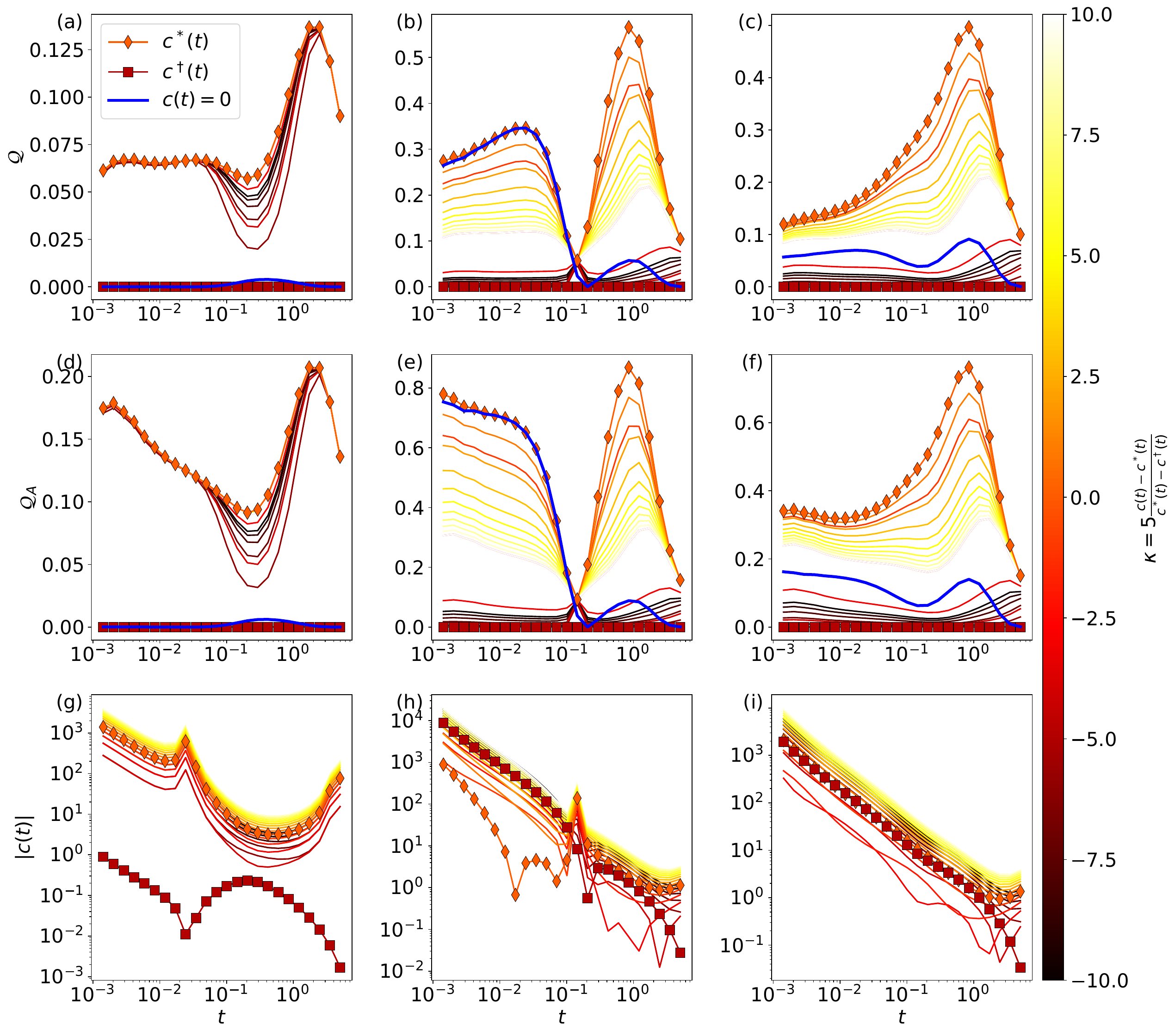}
    \caption[Time dependent TUR optimisation]{Impact of parameter $c(t)$ on the quality and pseudo quality factor as a function of time in (a)-(c) and (d)-(f), respectively. The last row, (g)-(i), shows the absolute value of parameter $|c(t)|$ used in the rows above. The left column, i.e., (a), (d), and (g), corresponds to the $d^{(12)}$ curve in Figs.~\ref{fig:CalmodulinJvarJrhocov} and \ref{fig:TUR_Q1}. Likewise, the middle and right columns correspond to $d^{(56)}$ and $d^{(26)}$ in the previously mentioned figures, respectively. In all subplots, the optimal curve with $c(t) = c^*(t)$ is marked with diamonds, while the least optimal $c(t)=c^\dagger(t)$ is marked with squares. The colour bar $\kappa$ indicates the "distance" of $c(t) = c^*(t) + \kappa(c^*(t) - c^\dagger(t))/5$ from the optimal value $c^*(t)$. The blue line corresponds to the value of $\mathcal{Q}$ and $\mathcal{Q}_A$ for $c(t)=0$, which are the curves in the left part of Fig.~\ref{fig:TUR_Q1}. We can see that the choice of $c(t)$ is non-trivial as seen in, e.g., (b), where there is a switch-over from $c^*(t)<c^\dagger(t)$ to $c^*(t) > c^\dagger(t)$. 
    }
    \label{fig:optc_TUR_comp}
\end{figure}

Figure \ref{fig:optc_TUR_comp} shows the impact which various $c(t)$, including $c^*(t)$ and $c^\dagger(t)$, have on $\mathcal{Q}$ and $\mathcal{Q}_A$, as well as the time-dependence of $|c(t)|$, for three example currents in the calmodulin example, see Fig.~\ref{fig:CalmodulinJvarJrhocov}. This figure shows that the choice of $c(t)$ is non-trivial. For instance, one may naively think that choosing any $c(t)\neq0$ may improve the inference. However, as can be seen in Figs.~\hyperref[fig:optc_TUR_comp]{\ref*{fig:optc_TUR_comp}b} and \hyperref[fig:optc_TUR_comp]{\ref*{fig:optc_TUR_comp}e}, there are cases where using $c(t)=0$ yields inference close to the optimal value using $c^*(t)$ for some ranges of $t$, while in other cases, e.g., Fig.~\hyperref[fig:optc_TUR_comp]{\ref*{fig:optc_TUR_comp}a}, one almost always improves when choosing a non-zero $c(t)$. The coefficients $c(t)$ have complicated dependencies on $t$. For example, Fig.~\hyperref[fig:optc_TUR_comp]{\ref*{fig:optc_TUR_comp}h} shows that there is a crossover at some $t$ in $0.1\leq t\leq 0.2$, where $c^*(t)<c^\dagger(t)$ reverts to $c^*(t)>c^\dagger(t)$, showing that it initially is a good choice to use $c(t)=0$, but in the long run is close to the least optimal choice. The impact of this observation is readily seen in Figs.~\hyperref[fig:optc_TUR_comp]{\ref*{fig:optc_TUR_comp}b} and \hyperref[fig:optc_TUR_comp]{\ref*{fig:optc_TUR_comp}e}, as the crossover of the blue line corresponding to $c(t)=0$ from the most to the least optimal inference becomes visible.

However, optimising $c(t)$ does in most cases not saturate the bound. Recalling that the inequality in \Eqref{eq:CSI} saturates, i.e., becomes an equality, for $\hat{X}_t\propto\hat{Y}_t$, choosing $c(t)$ and $V_i$ such that $\hat{J}_t - c(t)\hat{\rho}_t\propto \hat{A}_t$ the inequality \Eqref{eq:CSI_AJ} will saturate. This can be done by choosing $c(t)\hat{\rho}_t= \hat{J}_t^\mathrm{II}$, i.e., $V_i(\tau) = \sum_{j\in\Omega, j\neq i} d_{ij}(\tau)r_{ij}/c(t)$, and
\begin{align}
    d_{ij}(\tau) = c^\prime \frac{r_{ij}p_i(\tau) - r_{ji}p_j(\tau)}{r_{ij}p_i(\tau) + r_{ji}p_j(\tau)} = -d_{ji}(\tau)\,,
    \label{eq:optimalMetric}
\end{align}
with $c^\prime \in \mathds{R}$. The shifted current then becomes
\begin{equation}
    \begin{aligned}
        \hat{J}_t - c(t) \hat{\rho}_t = \hat{J}_t - \hat{J}_t^\mathrm{II} = \hat{J}_t^\mathrm{I} = \int_0^t\sum_{\substack{i, j\in\Omega\\i\neq j}}c^\prime \frac{r_{ij}p_i(\tau) - r_{ji}p_j(\tau)}{r_{ij}p_i(\tau) + r_{ji}p_j(\tau)}\mathrm{d}\hat{\varepsilon}_{ij}(\tau) = c'\hat{A}_t\,,
    \end{aligned}
\end{equation}
and the steady state TUR \Eqref{eq:SSTURshifted} saturates $\langle \hat{A}_t(\hat{J}_t - c(t)\hat{\rho}_t) \rangle ^2= \langle \hat{J}_t \rangle ^2/c' = {c'}\langle \hat{A}_t^2\rangle^2$. While this saturates the Cauchy-Schwarz inequality, it physically only tells us something about the pseudo-entropy production. The second inequality, i.e., \Eqref{eq:LogIdentity}, is only saturated for $a=b$, which in this case corresponds to DB where neither pseudo- nor proper entropy is produced. In other words,  we are unable to saturate \Eqref{eq:LogIdentity} for transient or out-of-equilibrium systems
\begin{align}
    \frac{1}{{c'}}\langle \hat{J}_t\rangle^2 = \langle \hat{A}_t^2\rangle^2 \leq \frac{\Delta S_\mathrm{tot}(t)^2}{4}\,,
\end{align}
which means that $\mathcal{Q}_A=1$ and $\mathcal{Q}<1$. 
However, we can still saturate the first inequality for transient systems (with $c'=1$)
\begin{equation}
    \begin{aligned}
        \langle \hat{A}_t (\hat{J}_t - c(t)\hat{\rho}_t)\rangle^2 = \langle \hat{A}_t \hat{J}_t^\mathrm{I}\rangle^2 = \langle \hat{J}_t\rangle^2 = \langle \hat{A}_t^2\rangle^2\,,
    \end{aligned}
\end{equation}
which leads to the optimal TUR for arbitrary systems.

While, in theory, it warrants the best inference of $\Delta S_\mathrm{tot}$, choosing the metric as \Eqref{eq:optimalMetric} is not applicable in practice. It requires both the (transient) probabilities and, more challenging, the transition rates. Knowing the latter, there would be no need to estimate the entropy production, as it could be calculated explicitly. 

\section{Thermodynamic Bound on Transport of Observables \label{sec:Transportproof}}

\subsection{Proof of Transport Bound}
Next, we prove the transport bounds for MJP derived in Ref. \cite{dieball2024thermodynamic} for continuous processes.
We again start from \Eqref{eq:CSI} and \Eqref{eq:A_observable} and introduce
\begin{equation}
    \begin{aligned}[b]
        \hat{B}_t &= \sum_{\substack{x, y\in\Omega\\x\neq y}}\int_{\tau=0}^{\tau=t}\Delta z_{yx}(\tau)\mathrm{d}\hat{\varepsilon}_{xy}(\tau)\,,
    \end{aligned}
\end{equation}
where $\Delta z_{yx}(\tau) = z_y(\tau) - z_x(\tau) = -\Delta z_{xy}(\tau)$ is the difference in state functions of the observable $z_x(\tau)$. 
Recalling that $\langle \hat{A}_t^2\rangle \leq \Delta S_\mathrm{tot}(t) /2$, the remaining quantities to be calculated are $\langle \hat{A}_t\hat{B}_t\rangle$ and $\langle \hat{B}_t^2\rangle$. Starting with the latter and using \Eqref{eq:jump_covariance} one gets
\begin{equation}
    \begin{aligned}[b]
        \langle \hat{B}_t^2\rangle &= \sum_{\substack{x, y\in\Omega\\x\neq y}}\int_0^t\mathrm{d}\tau \Delta z_{yx}^2p_x(\tau)r_{xy} = \sum_{\substack{x, y\in\Omega\\x\neq y}}\int_0^t\mathrm{d}\tau \left( z_{y} -z_x\right)^2p_x(\tau)r_{xy}\,,\\
        \langle \hat{A}_t\hat{B}_t\rangle &= \sum_{\substack{x, y\in\Omega\\x\neq y}}\int_0^t\mathrm{d}\tau Z_{xy}(\tau)\Delta z_{yx}(\tau)p_{x}(\tau)r_{xy}\,.
    \end{aligned}
\end{equation}
Since $Z_{xy}(\tau)\Delta z_{yx} = Z_{yx}(\tau)\Delta z_{xy}$, the double sum simplifies as follows
\begin{equation}
    \begin{aligned}[b]
        \sum_{\substack{x, y\in\Omega\\x\neq y}} Z_{xy}(\tau)\Delta z_{yx}(\tau)p_{x}(\tau)r_{xy} &= \frac{1}{2}\sum_{\substack{x, y\in\Omega\\x\neq y}} Z_{xy}(\tau)\Delta z_{yx}(\tau)(p_{x}(\tau)r_{xy}+p_y(\tau)r_{yx})\\
        &=\sum_{\substack{x, y\in\Omega\\x\neq y}} z_x(\tau) (p_y(\tau)r_{yx}-p_{x}(\tau)r_{xy})\,.
    \end{aligned}
\end{equation}
Identifying the net flux in \Eqref{eq:NetFlux} yields
\begin{equation}
    \begin{aligned}[b]
        \langle \hat{A}_t\hat{B}_t\rangle &= \int_0^t\mathrm{d}\tau\sum_{x\in\Omega} z_x(\tau) \sum_{y\neq x, x} (-\mathcal{J}_{xy}(\tau)) =  \int_0^t\mathrm{d}\tau\sum_{x\in\Omega} z_x(\tau) \partial_\tau p_x(\tau)\\
        &= \sum_x \left[z_x(t)p_x(t) - z_x(0)p_x(0)\right] - \int_0^t\mathrm{d}\tau \sum_{x\in\Omega} p_x(\tau)\partial_\tau z_x(\tau)\,,
    \end{aligned}
\end{equation}
where we performed a partial integration in the last line. Putting all together, we find
\begin{equation}
    \begin{aligned}[b]
        \frac{\Delta S_\mathrm{tot}(t)}{2}&\geq \frac{\left(\sum_{x\in\Omega} \left[z_x(t)p_x(t) - z_x(0)p_x(0)\right] - \int_0^t\mathrm{d}\tau \sum_{x\in\Omega} p_x(\tau)\partial_\tau z_x(\tau)\right)^2}{\sum_{\substack{x, y\in\Omega\\x\neq y}}\int_0^t\mathrm{d}\tau \left( z_{x} -z_y\right)^2p_x(\tau)r_{xy}}\\
        &=\frac{\langle z_x(t) - z_x(0)  - \int_0^t\mathrm{d}\tau\partial_\tau z_x(\tau)\rangle^2}{t\mathcal{D}_t}\,,
    \end{aligned}
    \label{eq:TransportBound}
\end{equation}
where we introduce the fluctuation-scale function 
\begin{align}
    \mathcal{D}_t = \frac{1}{t}\int_0^t\mathrm{d}\tau \sum_{\substack{x, y\in\Omega\\x\neq y}} (z_x-z_y)^2r_{xy}p_x(\tau) = \frac{1}{2t}\int_0^t\mathrm{d}\tau \sum_{\substack{x, y\in\Omega\\x\neq y}} (z_x-z_y)^2\left(r_{xy}p_x(\tau) + r_{yx}p_y(\tau)\right)\,.
\end{align}
The quantity $\mathcal{D}_t$ can be seen as a weighted time-averaged activity. The time-averaged activity has already been introduced in the literature \cite{ThermodynamicCorrelationInequality, Ohga2023}. Additionally, $\mathcal{D}_t$ can also be seen as a (time-dependent) diffusion coefficient of $z_x(\tau)$ on the network, since it contains the second moment of change in the state function weighted by the activity.

\subsection{Connection Between Diffusivity and Short Time Fluctuations\label{sec:DiffusionFluctuation}}
We now ask whether \Eqref{eq:TransportBound} is operationally accessible. The problem lies in the \textcolor{black}{diffusion coefficient $\mathcal{D}_t$}. For continuous overdamped dynamics, see Sec.~\ref{sec:continuous}, \textcolor{black}{$\mathcal{D}_t$} quantifies the short-time fluctuations \cite{BoundsCorrelationTimes}
\begin{equation}
    \begin{aligned}
        \mathcal{D}_t = \lim_{\mathrm{d}\tau\to0}\frac{\mathrm{var}(\mathrm{d}z_\tau)}{\mathrm{d}\tau}\,,
    \end{aligned}
    \label{eq:DiffusionConstantEmpirical}
\end{equation}
where $\mathrm{d}z_\tau = z({\tau + \mathrm{d}\tau}) - z(\tau)$. 

It remains to be investigated whether this connection can be made for discrete state spaces. We first show it in equilibrium and then generalise the statement to general dynamics. The mean squared can be neglected in equilibrium, as 
\begin{equation}
    \begin{aligned}
        \langle z({\tau + \mathrm{d}\tau}) - z(\tau) \rangle ^ 2 &= \left(\sum_{x,y\in\Omega} (z_y - z_x)P(y, \tau + \mathrm{d}\tau|x, \tau) p_x^\mathrm{eq}\right)^2\\
        &= \left(\sum_{\substack{x, y\in\Omega\\x\neq y}} (z_y - z_x)\mathrm{d}\tau r_{xy} p_x^\mathrm{eq} + \mathcal{O}(\mathrm{d}\tau^2)\right)^2\\
        &\stackrel{\mathrm{DB}}{=} \mathcal{O}(\mathrm{d}\tau^4)\,,
    \end{aligned}
\end{equation}
which vanishes in the limit \Eqref{eq:DiffusionConstantEmpirical}. The second moment, however, needs some more consideration, as the variance can be written as
\begin{equation}
    \begin{aligned}[b]
        \mathrm{var}(\mathrm{d}z_\tau) &= \langle (z({\tau + \mathrm{d}\tau}) - z(\tau))^2\rangle\\
        &= \sum_{x,y\in\Omega} (z_y - z_x)^2P(y, \tau + \mathrm{d}\tau|x, \tau) p_x^\mathrm{eq}\\
        &= \sum_{x,y\in\Omega} (z_y - z_x)^2\left(\delta_{xy} + \mathrm{d}\tau L_{yx}\right)p_x^\mathrm{eq} + \mathcal{O}(\mathrm{d}\tau^2)\\
        &= \sum_{\substack{x, y\in\Omega\\x\neq y}} (z_y - z_x)^2\mathrm{d}\tau r_{xy} p_x^\mathrm{eq} + \mathcal{O}(\mathrm{d}\tau^2)\,.
    \end{aligned}
\end{equation}
Hence,
\begin{equation}
    \begin{aligned}[b]
        \mathcal{D}_t &= \lim_{\mathrm{d}\tau\to0}\frac{\mathrm{var}(\mathrm{d}z_\tau)}{\mathrm{d}\tau} = \lim_{\mathrm{d}\tau\to0}\left(\sum_{\substack{x, y\in\Omega\\x\neq y}} (z_y - z_x)^2 r_{xy} p_x^\mathrm{eq} + \mathcal{O}(\mathrm{d}\tau)\right)\\
        &=\sum_{\substack{x, y\in\Omega\\x\neq y}} (z_y - z_x)^2 r_{xy} p_x^\mathrm{eq}\,,
    \end{aligned}
\end{equation}
which shows that $\mathcal{D}_t$ in fact can be written as \Eqref{eq:DiffusionConstantEmpirical} in equilibrium. This also means that \textcolor{black}{$\mathcal{D}_t$} is accessible experimentally, as long as $z$ can be measured with sufficient precision in time, \textcolor{black}{i.e., as long as one can perform the limit $\mathrm{d}\tau\to 0$ reasonably well and the assumption of a MJP is not violated.}

Note that the statements above generalise to transient dynamics as well. The second moment, and thus the variance, look the same, only that the density is no longer for the equilibrium density. For the mean squared, we must proceed differently than for DB before. However, the leading term is $\mathcal{O}(\mathrm{d}\tau^2)$, which therefore can be neglected as well. Lastly, since the density and $z$ may depend on time in general, the diffusion coefficient $\mathcal{D}_t$ in some time-interval $[0,t]$ can be written as
\begin{equation}
    \begin{aligned}
        \mathcal{D}_t=\frac{1}{t}\int_0^t\mathrm{var}(\mathrm{d}z_\tau)\,,
    \end{aligned}
    \label{eq:DiscreteDiffusion}
\end{equation}
which is the same identification as in \Eqref{eq:continuousDiffusion}.

While accessible in principle, it is difficult to handle, seeing that it requires sampling the variance of $\mathrm{d}z_\tau$ at each $\tau$ in $[0,t]$. This can, of course, be approximated by choosing a small $\mathrm{d}\tau$ and partitioning the integral into bins. Experiments may have a time resolution which leads to an {a priori} partitioning of the data. However, as we are required to properly evaluate the limit $\mathrm{d}\tau\to0$ is required to properly calculate it, such a partitioning may be sub-optimal for achieving a reasonable accuracy. 

Considering a path-wise diffusion coefficient,
\begin{equation}
    \begin{aligned}
        \hat{\mathcal{D}}_t = \frac{1}{t}\int_0^t\sum_{\substack{x, y\in\Omega\\x\neq y}}(z_x(\tau) - z_y(\tau))^2\mathrm{d}\hat{n}_{xy}(\tau)\,,
    \end{aligned}
    \label{eq:TrajectoryDiffusion}
\end{equation}
yields a more efficient way of calculating $\mathcal{D}_t$.  One can immediately recognise that $\langle \hat{\mathcal{D}}_t\rangle = \mathcal{D}_t$ and \Eqref{eq:TrajectoryDiffusion} is easy to evaluate for individual trajectories for certain conditions, as the integral reduces to a sum, see Sec.~\ref{sec:currentsdensities}. This is discussed in the next section.

\subsection{Generalisation of Transport Bound: From Transport to Currents}
In generalising the TUR to underdamped dynamics in Ref. \cite{CrutchfieldUnderdampedTUR}, the authors concomitantly proposed a generalisation of the transport bound. However, they have not made this connection. This generalisation comes from using the observable $\hat{J}_t^\mathrm{I}$, see  \Eqref{eq:SplitCurrent}, together with the auxiliary integral $\hat{A}_t$. Using Eqs.~\eqref{eq:AJI_correlator} and \eqref{eq:CSI}, we get
\begin{equation}
    \begin{aligned}
        \frac{\langle \hat{J}_t\rangle^2}{\mathrm{var}(\hat{J}_t^\mathrm{I})}\leq \langle \hat{A}_t^2\rangle \leq \frac{\Delta S_\mathrm{tot}(t)}{2}\,.
    \end{aligned}
    \label{eq:GeneralisedTransport}
\end{equation}
It should be noted that $\hat{J}_t^\mathrm{I}$ cannot be measured, as it depends on the noise $\mathrm{d}\hat{\varepsilon}_{xy}(\tau)$ which is not operationally accessible. While it may seem counter intuitive to include the variance of $\hat{J}_t^\mathrm{I}$, we can see that this can be measured as
\begin{equation}
    \begin{aligned}
        \left\langle \left(\hat{J}_t^\mathrm{I}\right)^2\right\rangle = \int_0^t\mathrm{d}\tau\sum_{\substack{x, y\in\Omega\\x\neq y}} d_{xy}(\tau)^2r_{xy}p_x(\tau) = \left\langle\int_0^t\sum_{\substack{x, y\in\Omega\\x\neq y}} d_{xy}(\tau)^2\mathrm{d}\hat{n}_{xy}(\tau) \right\rangle\,.
    \end{aligned}
    \label{eq:EmpiricalCurrentVariance}
\end{equation}
This has been elaborately discussed in Sec.~\ref{sec:currentsdensities}.

But is this operationally accessible? As long as the state functions in Eq.~\eqref{eq:TrajectoryDiffusion} and metric in Eq.~\eqref{eq:EmpiricalCurrentVariance} are known, the respective quantities are accessible. If this is not the case, the evaluation must be carried out, e.g., using \Eqref{eq:DiscreteDiffusion}. As an example, consider measuring a lumped trajectory. Then, the metric and state function can be chosen to optimise the (c)TUR and transport bound. However, if the measured quantity is a current, i.e., without explicit knowledge about the metric, one cannot use \Eqref{eq:EmpiricalCurrentVariance}.

The reason why this is a generalisation of the transport bound becomes obvious when choosing $d_{xy}(\tau)=z_y(\tau)-z_x(\tau)$. With this choice, the transport bound is recovered. Furthermore, \Eqref{eq:GeneralisedTransport} is also connected to the cTUR, as one gets it when $\hat{J}_t^\mathrm{II}$ is cancelled by a shift with some density, see Sec.~\ref{sec:TURsaturation}. In contrast to the TUR, this bound only requires calculating averages and is thus less prone to sampling errors.

\subsection{Saturation of Transport Bound}\label{sec:TransportSaturation}

Similar to the TUR, it is of great interest to identify when, and how, the transport bound in \Eqref{eq:TransportBound} saturates. The conditions in this case are simpler to identify, as both $\hat{B}_t$ and $\hat{A}_t$ are proportional to $\mathrm{d}\hat{\varepsilon}_{xy}(\tau)$. Hence, \Eqref{eq:TransportBound} saturates when the proportionality, up to a constant factor, is the same, i.e., $\Delta z_{yx}(\tau) = z_y - z_x\propto Z_{xy}(\tau)$. This means that for every pair of states $x,y$, the condition reads
\begin{equation}
    \begin{aligned}
        z_y(\tau) - z_x(\tau) = c(\tau)\frac{r_{xy}p_x(\tau) - r_{yx}p_y(\tau)}{r_{xy}p_x(\tau) + r_{yx}p_y(\tau)}\,,
    \end{aligned}
\end{equation}
where $c(\tau)$ is some, possibly time-dependent, proportionality factor. However, this is generally a condition that cannot be satisfied. A counter-example for this can be found in App. \ref{sec:TBSaturationCounterExample}.

\section{Correlation Bound \label{sec:Correlationproof}}
The correlation bound is introduced in terms of continuous dynamics for stationary systems in the limit of infinite trajectory length $t\to\infty$ \cite{BoundsCorrelationTimes}. We will prove the correlation bound in Secs.~\ref{sec:EQcoorbound} and \ref{sec:TRANSIENTcoorbound} for equilibrium and transient dynamics for finite $t$, respectively. Splitting it this way is useful from a practical perspective. For instance, looking at equilibrium systems allows us to use spectral decomposition to investigate the saturation, see App.~\ref{sec:CorrelationSpectral}. The work we present here is complementary to the continuous space generalisation \cite{DieballCorrelationBound}.

\subsection{Proof of Equilibrium Correlation Bound\label{sec:EQcoorbound}}
We start by proving the correlation bound in equilibrium using the \Eqref{eq:CSI}. Hence, we need two observables. Motivated by Ref. \cite{BoundsCorrelationTimes}, the diffusion coefficient $\mathcal{D}_t$ plays a role. Thus, we use the following observables
\begin{equation}
    \begin{aligned}[b]
        \hat{B}_t &= \frac{1}{\sqrt{t}}\sum_{\substack{x, y\in\Omega\\x\neq y}}\int_{\tau=0}^{\tau=t}\Delta z_{yx}(\tau)\mathrm{d}\hat{\varepsilon}_{xy}(\tau)\,,\\\
        \hat{C}_t &= \frac{1}{\sqrt{t}}\sum_{x\in\Omega}\int_{\tau=0}^{\tau=t}(\omega_x(\tau) - \langle \omega_y(\tau) \rangle_\mathrm{s})\mathrm{d}\hat{\tau}_x(\tau)\,.
    \end{aligned}
\end{equation}
We also require
\begin{align}
    \overline{z}_t = \frac{1}{t}\int_0^t\sum_{x\in\Omega}\mathrm{d}\hat{\tau}_x(\tau) z_x(\tau)\,,
\end{align}
which simply adds up the contributions of $z\mathrm{d}\tau$ along the trajectory. In a steady state, a straightforward calculation then yields 
\begin{equation}
    \begin{aligned}
        \mathrm{var}_\mathrm{s}(\overline{z}_t) = \frac{2}{t^2}\int_0^t\mathrm{d}t'(t - t')\mathrm{cov}_\mathrm{s}(z(t'), z(0))\,,
    \end{aligned}
\end{equation}
and
\begin{equation}
    \begin{aligned}
        t\mathrm{var}_\mathrm{s}(\overline{z}_t) = 2\int_0^t\mathrm{d}t'(1 - \frac{t'}{t})\mathrm{cov}_\mathrm{s}(z(t'), z(0))\xrightarrow{t\to\infty}2\int_0^\infty\mathrm{d}t'\mathrm{cov}_\mathrm{s}(z(t'), z(0))\,.
    \end{aligned}
\end{equation}

The second moment of $\hat{C}_t$ in steady state is
\begin{equation}
    \begin{aligned}[b]
        \langle \hat{C}_t^2 \rangle = \frac{1}{t}\sum_{x, i\in\Omega}\int_{\tau=0}^{\tau=t}\int_{\tau=0}^{\tau=t}(\omega_x(\tau^\prime) - \langle \omega_y (\tau^\prime)\rangle_\mathrm{s})(\omega_i(\tau) - \langle \omega_j(\tau) \rangle_\mathrm{s})\langle\mathrm{d}\hat{\tau}_i(\tau)\mathrm{d}\hat{\tau}_x(\tau^\prime)\rangle\,,
    \end{aligned}
\end{equation}
so that, with $\Omega_x(\tau^\prime)= \omega_x(\tau^\prime) - \langle \omega_y (\tau^\prime)\rangle_\mathrm{s}$,
\begin{equation}
    \begin{aligned}[b]
        \langle \hat{C}_t^2 \rangle 
        &=\frac{2}{t}\sum_{x, i\in\Omega}\int_{0}^{t}\mathrm{d}\tau^\prime\int_{\tau^\prime}^{t}\mathrm {d}\tau\Omega_x(\tau^\prime)\Omega_i(\tau)P(i, \tau|x, \tau^\prime)p_x(\tau^\prime)\,.
    \end{aligned}
\end{equation}
As $P(i, \tau|x, \tau^\prime)p_x(\tau^\prime)$ is the joint probability, the integrand is simply the two-time covariance of $\omega$, i.e.,
\begin{equation}
    \begin{aligned}[b]
        \langle \hat{C}_t^2 \rangle
        &=\frac{2}{t}\int_{0}^{t}\mathrm{d}\tau^\prime\int_{\tau^\prime}^{t}\mathrm {d}\tau\mathrm{cov}(\omega(\tau^\prime), \omega(\tau))\,.
    \end{aligned}
\end{equation}
In a steady state, the covariance depends on the time difference only, so
\begin{equation}
    \begin{aligned}[b]
        \langle \hat{C}_t^2 \rangle&=\frac{2}{t}\int_{0}^{t}\mathrm{d}t^\prime (t - t^\prime) \mathrm{cov}_\mathrm{s}(\omega(t^\prime), \omega(0))\,.
    \end{aligned}
\end{equation}
What remains is calculating the correlation between $\hat{C}_t$ and $\hat{B}_t$ using \Eqref{eq:EpsilonCorrelator}
\begin{equation}
    \begin{aligned}[b]
        \langle \hat{C}_t\hat{B}_t \rangle
        &=\frac{1}{t}\sum_{\substack{i,x, y\in\Omega\\x\neq y}}\int_{\tau=0}^{\tau=t}\int_{\tau=0}^{\tau=t}\Delta z_{yx}(\tau)\left(\omega_i(\tau^\prime) - \langle \omega_j(\tau^\prime)\rangle\right)\langle\mathrm{d}\hat{\varepsilon}_{xy}(\tau)\mathrm{d}\hat{\tau}_i(\tau^\prime)\rangle\\
        &=\frac{1}{t}\sum_{\substack{i,x, y\in\Omega\\x\neq y}}\int_{0}^{t}\mathrm{d}\tau^\prime\int_{0}^{t}\mathrm{d}\tau \Delta z_{yx}(\tau) \Omega_i(\tau^\prime)\mathds{1}_{\tau < \tau^\prime}p_xr_{xy}\left[P(i, \tau^\prime|y, \tau) - P(i, \tau^\prime|x, \tau)\right]\,.
    \end{aligned}
    \label{eq:CorrectionCBintermediate}
\end{equation}
We will look at the sum over $x$ and $y$, and for notational simplicity omit the time argument as the dependencies should be clear from the context. Rearranging the inner sum in Eq.~\eqref{eq:CorrectionCBintermediate}, one recovers\footnote{Since the propagator also evolves according to the master equation, we get that $\partial_\tau \mathrm{e}^{\mathbf{L}\tau} = \mathbf{L}\mathrm{e}^{\mathbf{L}\tau} = \mathrm{e}^{\mathbf{L}\tau}\mathbf{L}$, as $\mathbf{L}$ commutes with the propagator \cite{MarkovJumpProcessesBook}. Hence, we can write $\partial_\tau P(i, \tau|y) = \sum_{x\in\Omega} r_{xi}P(x, \tau|y) = \sum_{x\in\Omega} P(i, \tau|x)r_{yx}$ identifying that $L_{xy} = r_{yx}$.}
\begin{equation}
    \begin{aligned}[b]
        &\sum_{\substack{x, y\in\Omega\\x\neq y}} z_yp_xr_{xy}(P(i|y)- P(i|x)) - \sum_{\substack{x, y\in\Omega\\x\neq y}} z_xp_xr_{xy}(P(i|y)- P(i|x))\\ 
        &\stackrel{\text{DB}}{=}2\sum_{\substack{x, y\in\Omega\\x\neq y}} z_yp_yr_{yx}P(i|y) - 2\sum_{\substack{x, y\in\Omega\\x\neq y}} z_xp_xr_{xy}P(i|y)\\
        &=-2 \sum_{x\in\Omega}z_xp_x\underbrace{\sum_{y\in\Omega}r_{xy}P(i|y)}_{= \partial_\tau P(i, \tau'|x, \tau)} = 2 \sum_{x\in\Omega}z_xp_x\partial_{\tau'} P(i, \tau'|x, \tau)\,.\\
    \end{aligned}
\end{equation}
Here, we again  shift the time derivative of the propagator and change the labelling of the indices in the last line to find
\begin{equation}
    \begin{aligned}[b]
        \langle \hat{C}_t\hat{B}_t \rangle
        &=\frac{2}{t}\sum_{i, x\in\Omega}\int_{0}^{t}\mathrm{d}\tau^\prime\int_{0}^{t}\mathrm{d}\tau  z_{x}(\tau) \Omega_i(\tau^\prime)\mathds{1}_{\tau < \tau^\prime}p_x\partial_{\tau'} P(i, \tau'|x, \tau)\,.
    \end{aligned}
    \label{eq:CorrectionCBintermediate2}
\end{equation}
Assuming $z_x(\tau)$ and $\Omega_i(\tau')$ to be time-independent, Eq.~\eqref{eq:CorrectionCBintermediate2} can be written as
\begin{align}
    \langle \hat{C}_t\hat{B}_t \rangle
        &=2\sum_{i, x\in\Omega}\int_{0}^{t}\mathrm{d}t^\prime\left(1-\frac{t'}{t}\right)  z_{x} \Omega_ip_x\partial_{t'} P(i, t'|x)\,.
\end{align}
An integration by parts yields
\begin{equation}
    \begin{aligned}[b]
        \langle \hat{C}_t\hat{B}_t \rangle
        &=2\sum_{i, x\in\Omega}z_{x} \Omega_ip_x \left[-\underbrace{P(i, 0|x)}_{=\delta_{ix}} - \int_{0}^{t}\mathrm{d}t^\prime  P(i, t'|x)\underbrace{\partial_{t'}\left(1-\frac{t'}{t}\right)}_{=-\frac{1}{t}}\right]\\
        &= - 2\sum_{i, x\in\Omega}z_{x} \Omega_x p_x + \frac{2}{t}\int_{0}^{t}\mathrm{d}t^\prime \sum_{i, x\in\Omega}z_{x} \Omega_ip_x P(i, t'|x)\,.\\
    \end{aligned}
\end{equation}
One can immediately recognise, that this also holds for a shifted $\Tilde{z}_x = z_x - \langle z \rangle$. Note that this does not change the value of $\hat{B}_t$, as the mean values cancel. With this, we recover the desired result
\begin{equation}
    \begin{aligned}[b]
        \langle \hat{C}_t\hat{B}_t \rangle
        &=- 2\mathrm{cov}_\mathrm{eq}(z, \omega) + \frac{2}{t}\int_{0}^{t}\mathrm{d}t^\prime \mathrm{cov}_\mathrm{eq}(z(0), \omega(t'))\,,
    \end{aligned}
\end{equation}
and
\begin{equation}
    \begin{aligned}[b]
        \mathcal{D}_t \frac{2}{t}\int_{0}^{t}\mathrm{d}t^\prime (t - t^\prime) \mathrm{cov}_\mathrm{eq}(\omega(t^\prime), \omega(0)) \geq \left(2\mathrm{cov}_\mathrm{eq}(z, \omega - \frac{2}{t}\int_{0}^{t}\mathrm{d}t^\prime \mathrm{cov}_\mathrm{eq}(z(0), \omega(t'))\right)^2\,,
    \end{aligned}
\end{equation}
resulting in
\begin{equation}
    \begin{aligned}[b]
        \mathcal{D}_t t\mathrm{var}_\mathrm{eq}(\overline{\omega}_t) \geq \left(2\mathrm{cov}_\mathrm{eq}(z, \omega) - \frac{2}{t}\int_{0}^{t}\mathrm{d}t^\prime \mathrm{cov}_\mathrm{eq}(z(0), \omega(t'))\right)^2\,.
    \end{aligned}
    \label{eq:eqcorrbound}
\end{equation}
This is the generalisation for finite $t$ of the result in Ref. \cite{BoundsCorrelationTimes}.

\subsection{Proof for Transient Dynamics\label{sec:TRANSIENTcoorbound}}
It is shown in Ref. \cite{BoundsCorrelationTimes} that the bound also can be applied to out-of-equilibrium systems for $t\to\infty$ to estimate the entropy production rate. We want to extend this further to transient systems with finite $t$. To get a bound which is operationally accessible, we have to restrict the state functions $z$ and $\omega$ to have no explicit time dependence. However, we write, e.g., $\langle z(\tau)\rangle$ and $\mathrm{var}(z(\tau))$, to emphasise the implicit time dependence through the probability density. We start by labelling the mean pseudo-entropy production rate
\begin{equation}
    \begin{aligned}[b]
        \Sigma^\mathrm{ps}(\tau) &= \sum_{\substack{x, y\in\Omega\\x\neq y}} p_x(\tau)r_{xy}Z_{xy}^2(\tau) = \sum_{x\in\Omega} p_x(\tau)\sum_{\substack{y\in\Omega\\y\neq x}}r_{xy}Z_{xy}^2(\tau) \\&= \left\langle\sum_{\substack{y\in\Omega\\y\neq x}}r_{xy}Z_{xy}^2(\tau) \right\rangle = \langle {\Sigma}_x^\mathrm{p}(\tau)\rangle\,,
    \end{aligned}
\end{equation}
where ${\Sigma}_x^\mathrm{p}(\tau)=\sum_{\substack{y\in\Omega\\y\neq x}}r_{xy}Z_{xy}^2(\tau)$.
We can rescale the probability to get 
\begin{equation}
    \begin{aligned}
        \Tilde{p}_{x,y}^\mathrm{ps}(\tau) = \frac{p_x(\tau)r_{xy}Z_{xy}^2(\tau)}{\Sigma^\mathrm{ps}(\tau)}\,.
    \end{aligned}
    \label{eq:ScaleProbabilityDensity}
\end{equation}
This is now a two-point probability
and normalised when summing over $x,y\neq x$. 
Denoting expectation values w.r.t. this probability as $\langle\cdot \rangle_\mathrm{ps}$, we can generalise the observables to include an additional part which vanishes in equilibrium
\begin{equation}
    \begin{aligned}[b]
        \hat{B}_t &= \frac{1}{\sqrt{t}}\sum_{\substack{x, y\in\Omega\\x\neq y}}\int_{\tau=0}^{\tau=t}\left[\Delta z_{yx} - Z_{xy}(\tau)(z_x+z_y-2C(\tau))\right]\mathrm{d}\hat{\varepsilon}_{xy}(\tau)\,,\\\
        \hat{C}_t &= \frac{1}{\sqrt{t}}\sum_{x\in\Omega}\int_{\tau=0}^{\tau=t}(\omega_x - \langle \omega_y (\tau)\rangle)\mathrm{d}\hat{\tau}_x(\tau)\,.
    \end{aligned}
    \label{eq:ObsCorrelationBound}
\end{equation}
Here, $C(\tau)$ is a time-dependent, but state-independent, shifting function. For example, $C(\tau) = \langle z_x(\tau)\rangle$ will be a choice which is operationally accessible. As we will see later, this is one choice allowing further simplifications. We can rewrite the second moment of $\hat{B}_t$ as
\begin{equation}
    \begin{aligned}[b]
        \langle \hat{B}_t^2\rangle 
        &= \mathcal{D}_t + \frac{1}{t}\int_0^t\mathrm{d}\tau \Sigma^\mathrm{ps}(\tau)\langle (z_x(\tau) + z_y(\tau)-2C(\tau))^2 \rangle_\mathrm{ps} \\
        -& \frac{2}{t}\int_0^t\mathrm{d}\tau\sum_{\substack{x, y\in\Omega\\x\neq y}}\left[(z_y(\tau)-C(\tau))^2 - (z_x(\tau)-C(\tau))^2\right]Z_{xy}(\tau)r_{xy}p_x(\tau)\,.
    \end{aligned}
    \label{eq:B_corrBound_second_moment}
\end{equation}
The last term will vanish in NESS. However, for transient dynamics, it will yield a generally non-zero contribution, as
\begin{equation}
    \begin{aligned}[b]
        &\sum_{\substack{x, y\in\Omega\\x\neq y}}((z_y(\tau)-C(\tau))^2 - (z_x(\tau)-C(\tau))^2)Z_{xy}(\tau)r_{xy}p_x(\tau) \\
        &= \frac{1}{2}\sum_{\substack{x, y\in\Omega\\x\neq y}}((z_y(\tau)-C(\tau))^2 - (z_x(\tau)-C(\tau))^2)(r_{xy}p_x(\tau) - r_{yx}p_y(\tau))\\
        &= \sum_{y\in\Omega} (z_y(\tau)-C(\tau))^2\partial_\tau p_y(\tau)\,.
    \end{aligned}
\end{equation}
We now introduce the semi-variance $\mathrm{var}^C(z_y(\tau))\equiv\sum_{y\in\Omega}(z_y-C(\tau))^2p_y(\tau)$, that becomes the variance for $C(\tau) = \langle z_y(\tau)\rangle$. Consider $\partial_\tau \mathrm{var}^C(z_y(\tau))$ with the choice $C(\tau) = \langle z_y(\tau)\rangle$, then $\partial_\tau \mathrm{var}^C(z_y(\tau)) = \partial_\tau \mathrm{var}(z_y(\tau)) = \sum_{y\in\Omega} (z_y-\langle z_y(\tau)\rangle)^2\partial_\tau p_y(\tau)$, since
\begin{align}
    \sum_{y\in\Omega} \partial_\tau(z_y(\tau)-\langle z_y(\tau)\rangle)^2 p_y(\tau) \propto \langle z_y(\tau)\rangle - \langle z_y(\tau)\rangle = 0\,,
\end{align}
as $\partial_\tau z_y= 0$. Then, the last integral in \Eqref{eq:B_corrBound_second_moment} becomes
\begin{equation}
    \begin{aligned}[b]
        \int_0^t\mathrm{d}\tau \sum_{y\in\Omega} (z_y-\langle z_y(\tau)\rangle)^2\partial_\tau p_y(\tau)&=\int_0^t\mathrm{d}\tau \partial_\tau\mathrm{var}(z_y(\tau))= \mathrm{var}(z_y(t)) - \mathrm{var}(z_y(0))\,.
    \end{aligned}
\end{equation}
We stress that this result is only valid for the particular choice $C(\tau) = \langle z_x(\tau)\rangle$.


Next, we again look at the cross term between $\hat{B}_t$ and $\hat{C}_t$, where we use the notation $z_x^\Delta(\tau) = z_x - \langle z_y(\tau)\rangle$
\begin{equation}
    \begin{aligned}[b]
        t\langle \hat{B}_t\hat{C}_t\rangle =& \int_{0}^{t}\mathrm{d}\tau^\prime\int_{0}^{t}\mathrm {d}\tau \sum_{\substack{x, y\in\Omega\\x\neq y}}\sum_{i\in\Omega}\mathds{1}_{\tau < \tau'}\Omega_i(\tau')\\
         &\times\left[\Delta z_{yx} - Z_{xy}(\tau)(z_x^\Delta(\tau) + z_y^\Delta(\tau))r_{xy}p_x(\tau)\right]\left(P(i, \tau'|y, \tau) - P(i, \tau'|x, \tau)\right)
    \end{aligned}
    \label{eq:CorrectionLabel3}
\end{equation}
Simplifying the inner sums first, i.e., the sums over $x,y$, and omitting the time arguments where clear from context, the first term can be written as
\begin{equation}
    \begin{aligned}[b]
        \sum_{\substack{x, y\in\Omega\\x\neq y}}\Delta z_{yx}r_{xy}p_x( P(i|y)-P(i|x)) =& \sum_{\substack{x, y\in\Omega\\x\neq y}}z_y^\Delta r_{xy}p_x( P(i|y)-P(i|x)) \\&- \sum_{\substack{x, y\in\Omega\\x\neq y}}z_x^\Delta r_{xy}p_x( P(i|y)-P(i|x))\\
        =&-\sum_{\substack{x, y\in\Omega\\x\neq y}}z_x^\Delta(r_{xy}p_x + r_{yx}p_y)( P(i|y)-P(i|x))\,,
    \end{aligned}
    \label{eq:Sum1}
\end{equation}
while the second term becomes
\begin{equation}
    \begin{aligned}[b]
        &\sum_{\substack{x, y\in\Omega\\x\neq y}} Z_{xy}(z_x^\Delta + z_y^\Delta)r_{xy}p_x\left(P(i|y) - P(i|x)\right)
         = \sum_{\substack{x, y\in\Omega\\x\neq y}} z_x^\Delta(r_{xy}p_x- r_{yx}p_y)\left(P(i|y) - P(i|x)\right)\,.
    \end{aligned}
    \label{eq:Sum2}
\end{equation}
Plugging Eqs.~\eqref{eq:Sum1} and \eqref{eq:Sum2} into Eq.~\eqref{eq:CorrectionLabel3}, we get
\begin{equation}
    \begin{aligned}[b]
        t\langle \hat{B}_t\hat{C}_t\rangle =& -2\int_{0}^{t}\mathrm{d}\tau^\prime\int_{0}^{t}\mathrm {d}\tau \sum_{\substack{x, y\in\Omega\\x\neq y}}\sum_{i\in\Omega}\mathds{1}_{\tau < \tau'}\Omega_i(\tau')z_x^\Delta(\tau)r_{xy}p_x(\tau)\left(P(i, \tau'|y, \tau) - P(i, \tau'|x, \tau)\right)\\
        =& -2\int_{0}^{t}\mathrm{d}\tau^\prime\int_{0}^{t}\mathrm {d}\tau \sum_{x,i\in\Omega}\mathds{1}_{\tau < \tau'}\Omega_i(\tau')z_x^\Delta(\tau)p_x(\tau)\underbrace{\sum_{y\in\Omega, y\neq x}r_{xy}\left(P(i, \tau'|y, \tau) - P(i, \tau'|x, \tau)\right)}_{=\partial_\tau P(i, \tau'|x, \tau) = -\partial_{\tau'}P(i, \tau'|x, \tau)}\\
        =& 2\int_{0}^{t}\mathrm{d}\tau^\prime\int_{0}^{t}\mathrm {d}\tau \mathds{1}_{\tau < \tau'}\partial_{\tau'}\mathrm{cov}(\omega_i(\tau'), z_x(\tau))\,,
    \end{aligned}
\end{equation}
where in the last line we used
\begin{equation}
    \begin{aligned}[b]
        \partial_{\tau'}\mathrm{cov}(\omega_i(\tau'), z_x(\tau)) =& \sum_{x,i\in\Omega}\Omega_i(\tau')z_x^\Delta(\tau)p_x(\tau)\partial_{\tau'}P(i, \tau'|x, \tau) \\&+ \sum_{x,i\in\Omega}\partial_{\tau'}\Omega_i(\tau')z_x^\Delta(\tau)p_x(\tau)P(i, \tau'|x, \tau)\,,
    \end{aligned}
\end{equation}
and where the last term vanishes since
\begin{equation}
    \begin{aligned}
        \sum_{x,i\in\Omega}\partial_{\tau'}\Omega_i(\tau')z_x^\Delta(\tau)p_x(\tau)P(i, \tau'|x, \tau) &= -\partial_{\tau'}\langle \omega(\tau')\rangle \underbrace{\sum_{x\in\Omega}z_x^\Delta(\tau)p_x(\tau)\underbrace{\sum_{i\in\Omega}P(i, \tau'|x, \tau)}_{=1}}_{=0}\,.\\
    \end{aligned}
\end{equation}
Performing an integration by parts in $\tau'$ we find
\begin{equation}
    \begin{aligned}[b]
        t\langle \hat{B}_t\hat{C}_t\rangle =& 2\int_0^t\mathrm{d}\tau\left[\mathds{1}_{\tau < \tau'}\mathrm{cov}(\omega_i(\tau'), z_x(\tau))|_{\tau'=0}^{\tau'=t} - \int_0^t\mathrm{d}\tau'\mathrm{cov}(\omega_i(\tau'), z_x(\tau))\partial_{\tau'}\mathds{1}_{\tau < \tau'}\right]\\
        &= 2\int_0^t\mathrm{d}\tau\left[\mathrm{cov}(\omega_i(t), z_x(\tau))- \int_0^t\mathrm{d}\tau'\mathrm{cov}(\omega_i(\tau'), z_x(\tau))\textcolor{black}{\delta(\tau'-\tau)}\right]\\
        &= 2t\mathrm{cov}(\omega_i(t), \overline{z}_t) - \textcolor{black}{2\int_0^t\mathrm{d}\tau \mathrm{cov}(\omega_i(\tau), z_x(\tau))}\,.
    \end{aligned}
\end{equation}
Hence, we get the correlation bound
\begin{equation}
    \begin{aligned}[b]
        &\frac{2\left[\mathrm{cov}(\omega_i(t), \overline{z}_t) - {\frac{1}{t}\int_0^t\mathrm{d}\tau \mathrm{cov}(\omega_i(\tau), z_x(\tau))}\right]^2}{\frac{1}{t}\int_{0}^{t}\mathrm{d}\tau^\prime\int_{\tau^\prime}^{t}\mathrm {d}\tau\mathrm{cov}(\omega(\tau^\prime), \omega(\tau))}\\&\leq \left(\mathcal{D}_t + \frac{1}{t}\int_0^t\mathrm{d}\tau \Sigma^\mathrm{ps}(\tau)\mathrm{var}_\mathrm{ps}^C(z_x(\tau) + z_y(\tau)) - \frac{2}{t}\int_0^t\mathrm{d}\tau \sum_{y\in\Omega} (z_y(\tau)-C(\tau))^2\partial_\tau p_y(\tau)\right)\\
        &\stackrel{C(\tau)=\langle z_x(\tau)\rangle}{=}\left(\mathcal{D}_t + \frac{1}{t}\int_0^t\mathrm{d}\tau \Sigma^\mathrm{ps}(\tau)\mathrm{var}^C_\mathrm{ps}(z_x(\tau) + z_y(\tau)) - \frac{2}{t}\left[\mathrm{var}(z_y(t)) - \mathrm{var}(z_y(0))\right]\right)\,.
    \end{aligned}
    \label{eq:CorrelationBound}
\end{equation}
In steady state, Eq.~\eqref{eq:CorrelationBound} simplifies to 
\begin{equation}
    \begin{aligned}
        \frac{4\left[\mathrm{cov}_\mathrm{s}(\omega_i(t), \overline{z}_t) - \textcolor{black}{\mathrm{cov}_\mathrm{s}(\omega_i, z_x})\right]^2}{t\mathrm{var}_\mathrm{s}(\overline{\omega}_t)}\leq \left(\mathcal{D}_t + \Sigma^\mathrm{ps}\mathrm{var}_\mathrm{ps}^C(z_x + z_y)\right)\,.
    \end{aligned}
    \label{eq:StationaryCorrelationBound}
\end{equation}
If $|z_x-C|$ is bounded by some value $a/2$, then $\mathrm{var}_\mathrm{ps}^C(z_x + z_y)\leq a^2$, so that we get a bound for the NESS pseudo entropy production rate, and thus the total entropy production
\begin{equation}
    \begin{aligned}
        \frac{4\left[\mathrm{cov}_\mathrm{s}(\omega_i(t), \overline{z}_t) - \textcolor{black}{\mathrm{cov}_\mathrm{s}(\omega_i(t), z_x(t))}\right]^2}{t\mathrm{var}_\mathrm{s}(\overline{\omega}_t)} - \mathcal{D}_t \leq a^2\Sigma^\mathrm{ps}\leq \frac{a^2}{2t}\Delta S_\mathrm{tot}(t)\,,
    \end{aligned}
\end{equation}
with the more general transient form
\begin{equation}
    \begin{aligned}
        \frac{2\left[\mathrm{cov}(\omega_i(t), \overline{z}_t) - {\frac{1}{t}\int_0^t\mathrm{d}\tau \mathrm{cov}(\omega_i(\tau), z_x(\tau))}\right]^2}{\frac{1}{t}\int_{0}^{t}\mathrm{d}\tau^\prime\int_{\tau^\prime}^{t}\mathrm {d}\tau\mathrm{cov}(\omega(\tau^\prime), \omega(\tau))} - \mathcal{D}_t + \frac{2}{t}\left[\mathrm{var}(z_y(t)) - \mathrm{var}(z_y(0))\right] \leq \frac{a^2}{t}\Delta S_\mathrm{tot}(t)\,.
    \end{aligned}
    \label{eq:TransientCorrelationBoundBounded}
\end{equation}

\subsection{Discussion of Transient Correlation Bound}

How useful is the bound in Eq.~\eqref{eq:CorrelationBound}? In comparison to Eqs.~\eqref{eq:TURtransientversion} and \eqref{eq:TransportBound}, the correlation bound in either transient (\Eqref{eq:CorrelationBound}) or stationary (\Eqref{eq:StationaryCorrelationBound}) form can yield a negative lower bound for the $\Dot{S}_\mathrm{tot}(t)$. In this case, the bound is less sharp than the second law \Eqref{eq:secondlaw}. Furthermore, to evaluate the correlation bound, one either needs to know $\Tilde{p}^\mathrm{ps}_{x,y}(\tau)$, which again requires knowing the rates $r_{xy}$, or using that $|z_x-C|\leq a/2$ is bounded by some $a\in\mathds{R}_{>0}$. The former makes using bounds redundant and the latter suffers from the fact that $\mathrm{var}_\mathrm{ps}^C(z_x + z_y)\leq a^2$ may be a very loose bound. 

\begin{figure}
    \centering
    \includegraphics[width=.9\textwidth]{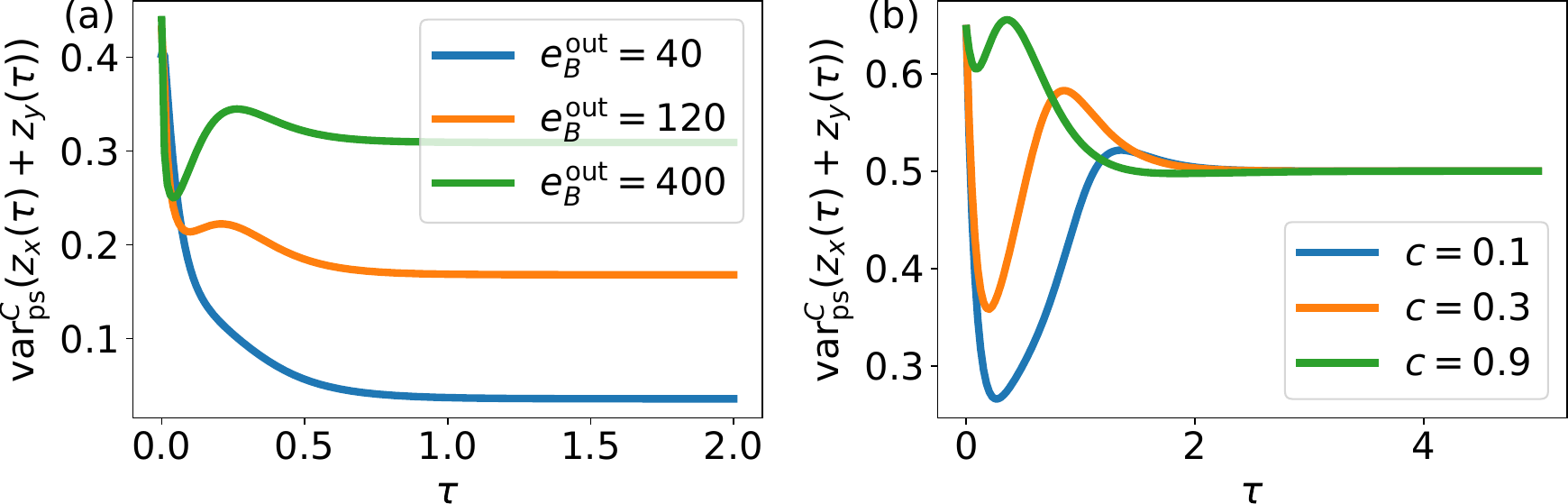}
    \caption[Driving dependent semi-variance]{Transient semi-variance in the SAT model (a) and toy model (b) as a function of time $\tau$ for various driving strengths. In (a), the driving is controlled by the value of $e^\mathrm{out}_B$, where a uniform initial distribution is used. The state function is $z^\mathrm{SAT}_x = \delta_{i1} + \delta_{i2}$, corresponding to molecule $A$ in the funnel on either side of the cell membrane, and $C^\mathrm{SAT}(\tau) = \langle z^\mathrm{SAT}_x(\tau) \rangle = \sum_{x\in\Omega} z^\mathrm{SAT}_xp_x(\tau)$. The driving in (b) is by the method in Sec.~\ref{sec:DrivingSystems}, where $c$ refers to the driving strength of the cycle $\mathcal{C}=\{1\to 2, 2\to 3, 3\to 1\}$. The state function here is $z^\mathrm{TM}_x = \delta_{i2} + \delta_{i3}$ and $C^\mathrm{TM}(\tau) = \langle z^\mathrm{TM}_x(\tau)\rangle$. The corresponding parameters are found in Tabs.~\ref{tab:ToyModelEnergies}~and~\ref{tab:Parameter SAT}. }
    \label{fig:SemiVariance}
\end{figure}

To make this point clear, consider the example shown in Fig.~\ref{fig:SemiVariance} of semi-variances as a function of time for various driving strengths in transient SAT and toy models. In both cases the state function $z$ is a linear combination of two Kronecker deltas: $z^\mathrm{SAT}_x = \delta_{i1} + \delta_{i2}$ and $z^\mathrm{TM}_x = \delta_{i2} + \delta_{i3}$. The $C(\tau) = \langle z(\tau)\rangle$ is chosen to be the respective mean value of state functions w.r.t. the original transient probability measure. Hence, we can identify $\max_{x,y\in\Omega}|z_x + z_y - 2\langle z(\tau)\rangle| = 2 - 2(z_kp_k(\tau) + z_lp_l(\tau))\leq 2$, where $(k,l) = (1,2)$ for the SAT and $(k,l) = (2,3)$ for toy model. If the mean is not accessible, the natural bound to choose is $a=2$. Coming back to Fig.~\ref{fig:SemiVariance}, we see that this choice of upper bound yields a large overestimation $a^2=4$ of the actual value of the semi-variance. When systematically driving systems according to the method presented in Sec. \ref{sec:DrivingSystems}, the driving strength $c$ cancel in the steady state semi-variance. Hence, when choosing a $C(\tau)$ independent of the rates, the semi-variances for different driving strengths converge to constant values for $t\to\infty$, as can be seen in Fig.~\hyperref[fig:SemiVariance]{\ref*{fig:SemiVariance}b}. The details of this can be found in App.~\ref{sec:InvariantSemiVariance}.

\begin{figure}
    \centering
    \includegraphics[width=.8\textwidth]{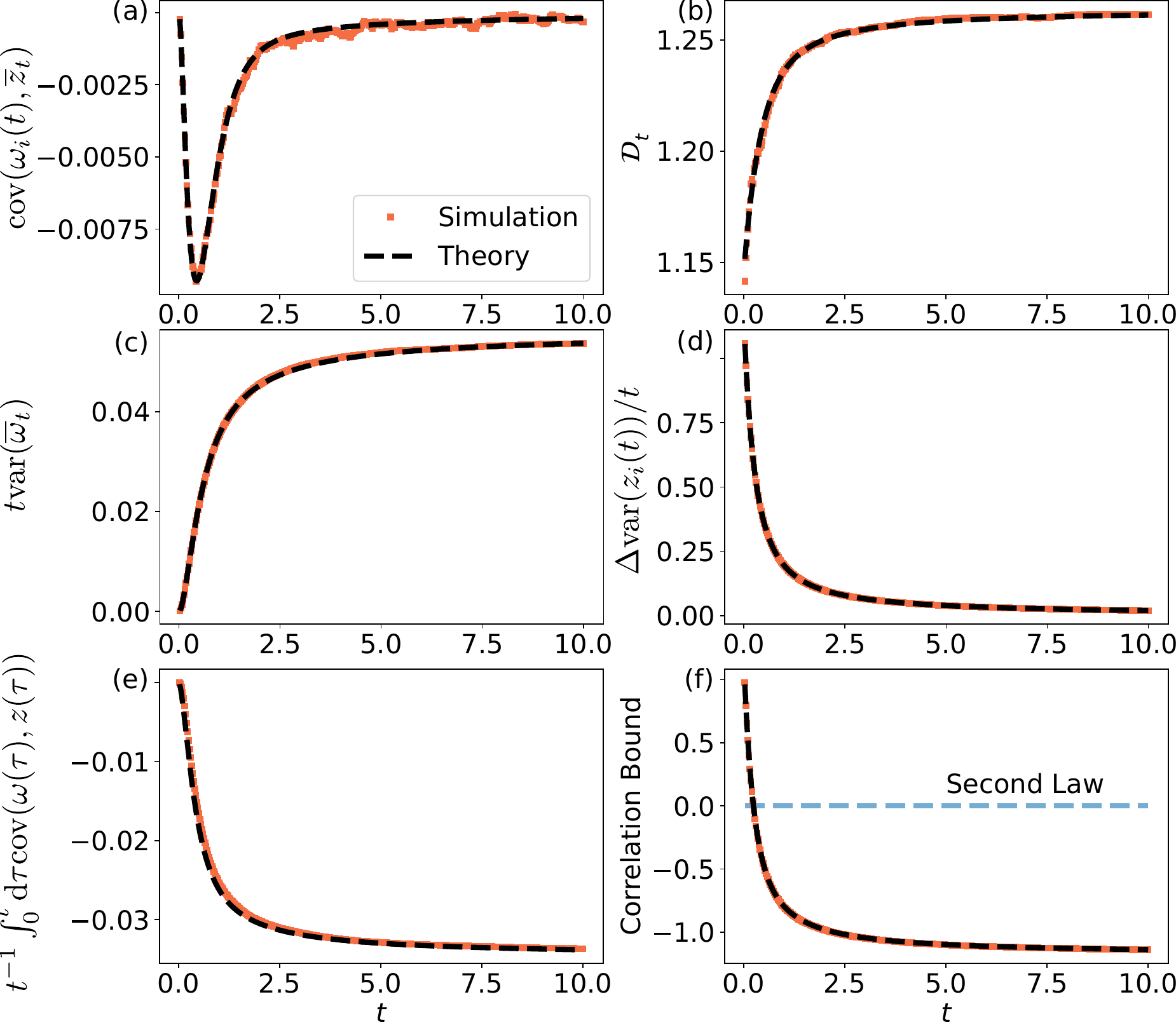}
    \caption[Correlation Bound Quantities]{Comparison of various quantities entering the transient correlation bound in \Eqref{eq:CorrelationBound} in the toy model. The initial condition is $p_i(0) = \delta_{i0}$ with state functions $z_i=\delta_{i3}$ and $\omega_i = \delta_{i2}$. The model is driven out of equilibrium on the cycle $\mathcal{C}=\{1\to 2, 2\to 3, 3\to 1\}$ with strength $c=0.1$, see Sec.~\ref{sec:DrivingSystems}. In (a)-(e), the $\mathrm{cov}(\omega_i(t), \overline{z}_t)$, $\mathcal{D}_t$, $t\mathrm{var}(\overline{\omega}_t)$, $\Delta \mathrm{var}(z_i(t))/t = \mathrm{var}(z_i(t))/t - \mathrm{var}(z_i(0))/t$, and $\int_0^t\mathrm{d}\tau \mathrm{cov}(\omega(\tau), z(\tau))/t$ are shown for numerical simulations (orange markers) and theoretical prediction (black dashed lines), respectively. The theoretical lower bound on the total entropy production, i.e., the l.h.s. of \Eqref{eq:TransientCorrelationBoundBounded}, is shown in (f) together with the numerical bound using the quantities in (a)-(e). Additionally, the dashed blue line shows the bound using the second law \Eqref{eq:secondlaw}. For the simulation, a total of $N=10^5$ trajectories are used and the parameters are given in Tab.~\ref{tab:ToyModelEnergies}. }
    \label{fig:CBquantities}
\end{figure}

In Fig.~\hyperref[fig:CBquantities]{\ref*{fig:CBquantities}a-e}, the various quantities entering the transient correlation bound in \Eqref{eq:TransientCorrelationBoundBounded} are shown as a function of $t$ in the driven toy model. There, the theoretical predictions are shown together with the respective numerical results from $N=10^5$ trajectories. A good agreement is observed. However, combining all these into the l.h.s. of \Eqref{eq:TransientCorrelationBoundBounded}, which is shown in Fig.~\hyperref[fig:CBquantities]{\ref*{fig:CBquantities}f}, ultimately results in a negative lower bound on the total entropy production for $t\gtrsim 0.5$, i.e., a bound \textit{less} sharp than the second law of thermodynamics \Eqref{eq:secondlaw}. In other words, the correlation bound fails to provide a sharper lower bound for all $t$. It should be noted that there is an initial time interval where the correlation bound yields a positive bound on the total entropy production, so that it is sharper than the second law. This example shows what one can see in  \Eqref{eq:TransientCorrelationBoundBounded}, namely that the case of a negative lower bound on the entropy production can and does occur.

\begin{figure}
    \centering
    \includegraphics[width=.8\textwidth]{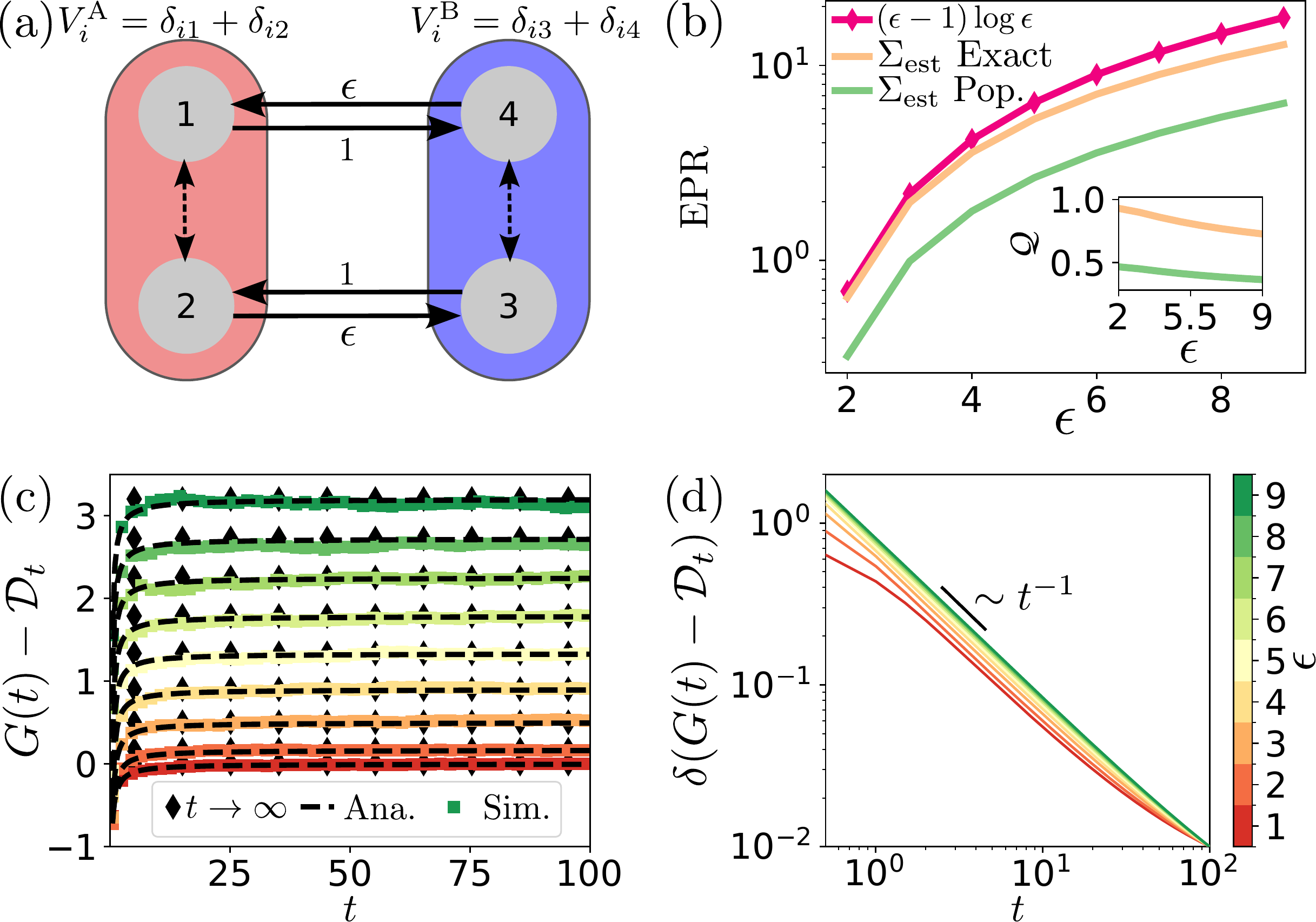}
    \caption{ Four-state example model with correlation bounds on entropy production and long-time limit quality factor. In (a), the microscopic four-state model is shown together with the lumping into two states $\mathrm{A}$ and $\mathrm{B}$. The system is uniformly driven in one direction with rates $\epsilon$. The exact entropy production (EPR) is shown in (b) compared to the estimated EPR using the correlation bound with exact semi-variance (orange) and Popoviciu's inequality (green). The inset of figure (b) then shows the long-time limit of the correlation bound quality factor as a function of $\epsilon$. In (c), the l.h.s. in Eq.~\eqref{eq:StationaryCorrelationBound} which we denote $G(t)-\mathcal{D}_t$ here, is shown for different values of driving rate $\epsilon$ as a function of time. This is done using simulations (coloured squares) and theory (dashed lines), together with the long-time limit (black diamonds). The leading order correction $\delta(G(t)-\mathcal{D}_t) =G(t)-\mathcal{D}_t - \lim_{t\to\infty}[G(t)-\mathcal{D}_t]\sim\mathcal{O}(t^{-1})$ of the the l.h.s. in Eq.~\eqref{eq:StationaryCorrelationBound} is shown in (d). The colourbar denotes the $\epsilon$ for the squares and lines in (c) and (d), respectively. The simulations are performed with $N=50000$ trajectories.} 
    \label{fig:PRX_CB}
\end{figure}

Consider a stationary four-state system on a ring driven uniformly in one direction with rate $\epsilon$ while the rates in the opposite direction are unity, see Fig.~\hyperref[fig:PRX_CB]{\ref*{fig:PRX_CB}a}. If states $1$ and $2$ are lumped into $\mathrm{A}$ and likewise $3$ and $4$ lumped into $\mathrm{B}$, then only the correlation bound yields a positive bound on the entropy production (for sufficiently large $t$) out of the bounds presented in this thesis, see Fig.~\hyperref[fig:PRX_CB]{\ref*{fig:PRX_CB}c}. Specifically, the transport bound and TUR yields $0\leq \Delta S_\mathrm{tot}$, the latter is since the metric $d_{ij}(\tau) = \mathds{1}_{i\in \mathrm{A}}\mathds{1}_{j\in \mathrm{B}} - \mathds{1}_{j\in \mathrm{A}}\mathds{1}_{i\in \mathrm{B}}$ yields a steady-state current with zero mean. Adding a density, i.e., using the cTUR also gives $0\leq \Delta S_\mathrm{tot}$, as the density only contributes for transient dynamics. The estimated entropy production rate $\Sigma_\mathrm{est}$ and exact $\Dot{S}_\mathrm{tot}^\mathrm{s}=(\epsilon-1)\log\epsilon$ are shown as functions of $\epsilon$ in Fig.~\hyperref[fig:PRX_CB]{\ref*{fig:PRX_CB}b}. The inset of Fig.~\hyperref[fig:PRX_CB]{\ref*{fig:PRX_CB}b} shows the exact long-time limit of the correlation bound quality factor is compared to the estimated version using Popoviciu's inequality $\mathrm{var}(z)\leq (z_\mathrm{max}-z_\mathrm{min})^2/4$ for a bounded $z_\mathrm{min}\leq z\leq z_\mathrm{max}$ applied to the semi-variance. Up to roughly $50\%$ of the entropy production can be inferred using Popoviciu's inequality and this occurs as $\epsilon$ becomes small. Increasing $\epsilon$ decreases the fraction of inferred entropy produced. Lastly, in Fig.~\hyperref[fig:PRX_CB]{\ref*{fig:PRX_CB}d}, we show that the leading correction of the long-time limit in Fig.~\hyperref[fig:PRX_CB]{\ref*{fig:PRX_CB}c} is of order $\mathcal{O}(t^{-1})$.

\chapter{Comparison of Asymptotic and Steady-State Behaviour of Thermodynamic Bounds\label{ch:ComparisonBounds}}

In the following, we compare the various thermodynamic bounds proven in Ch.~\ref{ch:directProofs} for transient and NESS systems in Secs.~\ref{sec:comparisonTransient} and \ref{sec:comparisonNESS}, respectively. For this, the quality factors of the bounds are compared. The TUR can be either in terms of currents or densities leading to the corresponding quality factors $\mathcal{Q}_J^\mathrm{TUR}(t)$ and $\mathcal{Q}_\rho^\mathrm{TUR}(t)$, see Sec.~\ref{sec:TURsaturation}. Note that the cTUR is not shown here, as the saturation is already discussed in Sec.~\ref{sec:TURsaturation}. Moreover, the asymptotic behaviour of $\mathcal{Q}^\mathrm{corr}(t)$ can easily be recovered from the asymptotic behaviour of quantities entering $\mathcal{Q}_J^\mathrm{TUR}(t)$ and $\mathcal{Q}_\rho^\mathrm{TUR}(t)$. Furthermore, the quality factor of transport bound is denoted $\mathcal{Q}^\mathrm{TB}(t)$.

\section{Transient Dynamics\label{sec:comparisonTransient}}

We want to show how our results compare in various situations. As the extension and derivation of bounds for transient dynamics is a central part of this thesis, we use this section to compare the behaviour in transient systems. Two cases have to be considered for this: does the system relax towards equilibrium or a NESS? The need for this becomes obvious when considering the current TUR for time-independent metrics, which contains the factor $\partial_t\langle \hat{J}_t\rangle$. This derivative vanishes in equilibrium for $t\to\infty$, while it remains non-zero in NESS. In other words, even though entropy is produced during the relaxation towards equilibrium, the quality factor $\mathcal{Q}_J^\mathrm{TUR}(t)$ will become zero. This is not the case if the system relaxes towards a NESS, which will always give a finite inferred estimation of the entropy production.

\subsection{Relaxation Towards Equilibrium\label{sec:ComparisonRelaxationEQ}}

\begin{figure}[ht]
    \centering
    \includegraphics[width=\textwidth]{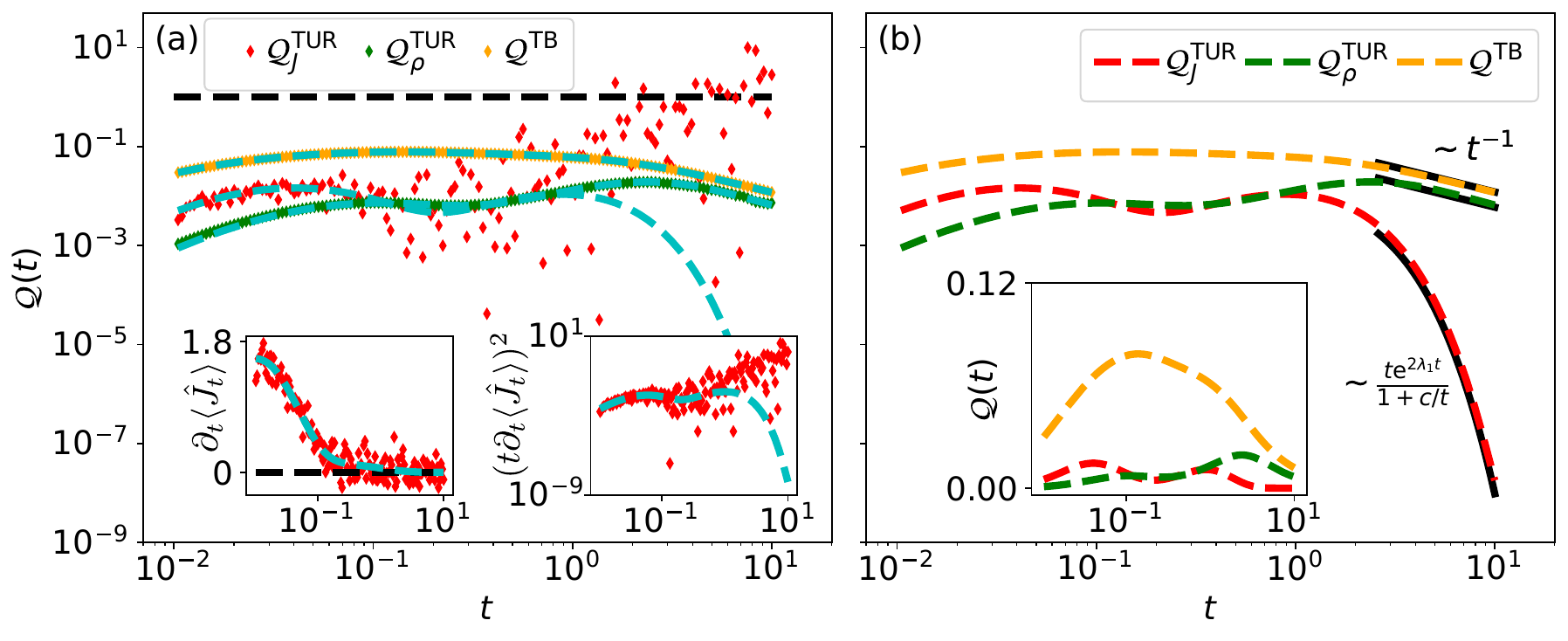}
    \caption[Long time behaviour for relaxation toward equilibrium]{TUR quality factors for a current $\mathcal{Q}_J^\mathrm{TUR}(t)$ (red) and a density $\mathcal{Q}_\rho^\mathrm{TUR}(t)$ (green) together with the transport bound quality factor $\mathcal{Q}^\mathrm{TB}(t)$ (orange) as a function of time $t$ for the Calmodulin system. For each quality factor, the initial distribution is set to $p_i(0) = 1/6$. The metric is $d_{ij}^{(23)}(\tau) = \delta_{i2}\delta_{j3} - \delta_{i3}\delta_{j2}$ and the state function is $V_i(\tau) = z_i(\tau) = \delta_{i2} + \delta_{i3}$.
    Numerical quality factors (diamonds) sampled using $N=1.3\cdot10^6$ trajectories are compared to the analytical values (dashed cyan lines) in (a). Additionally, the solid black line is the upper bound, $\mathcal{Q}(t) = 1$, for the quality factor. The insets in (a) show $\partial_t\langle \hat{J}_t\rangle$ (left) and $(t\partial_t\langle \hat{J}_t\rangle)^2$ (right) versus $t$, which enter in $\mathcal{Q}_J^\mathrm{TUR}(t)$, recovered from sampling trajectories (red diamonds) compared to the expected result (dashed cyan lines). In the left inset in (a), the long-time limit $\partial_t\langle \hat{J}_t\rangle\xrightarrow{t\to\infty} 0$ is shown as the dashed black line. The respective large-time behaviour of the quality factors (red, green, and orange dashed lines) is shown in (b) as the solid black lines and the explicit expression for these limits is found in Eqs.~\eqref{eq:TURJQualityScaling}, \eqref{eq:TURrhoQualityScaling}, and \eqref{eq:TBQualityScaling} for the respective quality factors. To better visualise the quality factors, a log-linear plot is shown in the inset of (b).  The largest non-zero eigenvalue of the generator for the Calmodulin system is $\lambda_1\approx -1.00$ and $c=\kappa/\gamma=1.68\cdot10^{4}$. Hence, we are in the intermediate regime of $\mathcal{Q}_J^\mathrm{TUR}(t)$, see \Eqref{eq:TURJQualityScaling}.}
    \label{fig:TUR_TB_comp1}
\end{figure}

As mentioned at the beginning of this section, one can immediately see from, e.g., \Eqref{eq:TURtransientversion}, that the TUR quality factors vanish if the system relaxes towards equilibrium. In fact, as can be seen in Fig.~\ref{fig:TUR_TB_comp1}, $\mathcal{Q}_J^\mathrm{TUR}(t)$ decays significantly faster than $\mathcal{Q}_\rho^\mathrm{TUR}(t)$ and $\mathcal{Q}^\mathrm{TB}(t)$. This is shown for the Calmodulin system with $d_{ij}^{(23)}$, $V_i = z_i = \delta_{i2}+\delta_{i3}$, and flat initial distribution. To further quantify this, we will look at the asymptotic behaviour of each quality factor. 

Before going into the individual quality factors, we need to mention how we handle the entropy production of the system. For sufficiently large times, the total produced entropy can be regarded as constant $\Delta S_\mathrm{tot}^\infty = \lim_{t\to\infty}\Delta S_\mathrm{tot}(t)$, as all time-dependencies are decaying exponentials. This assumption is only valid for systems relaxing to equilibrium; for NESS systems the long time limit of entropy production rate becomes constant, such that the total produced entropy for long times is linear in $t$ (see Sec.~\ref{sec:Entropy}).

Recalling the mean current \Eqref{eq:J_mean} and assuming the metric to be time-independent, the derivative is
\begin{equation}
    \begin{aligned}
        \partial_t\langle \hat{J}_t\rangle = \sum_{\substack{x, y\in\Omega\\x\neq y}} d_{xy}r_{xy} p_x(t)\,.
    \end{aligned}
\end{equation}
We can use \Eqref{eq:Propagator} to write this in terms of decaying exponentials (see Sec.~\ref{sec:Spec})
\begin{equation}
    \begin{aligned}
        \partial_t\langle \hat{J}_t\rangle = \sum_{\substack{x, y,j\in\Omega\\x\neq y}} d_{xy}r_{xy}\sum_{i\geq 0} \psi^R_{i,x} \mathrm{e}^{\lambda_i t}\psi^L_{i,j}p_j(0)\,.
    \end{aligned}
\end{equation}
Hence, we can see that
\begin{equation}
    \begin{aligned}
        t\partial_t\langle \hat{J}_t\rangle = t\sum_{\substack{x, y,j\in\Omega\\x\neq y}} d_{xy}r_{xy}\left(p_x^\mathrm{eq} + \sum_{i>0} \psi^R_{i,x}\mathrm{e}^{\lambda_i t}\psi^L_{i,j}p_j(0)\right)\,.
    \end{aligned}
\end{equation}
The first term vanishes due to DB. For large times $t\gg -1/\lambda_1$, the second term also vanishes exponentially, hence $\mathcal{Q}_J^\mathrm{TUR}(t)\xrightarrow{t\to\infty}0$ if the system relaxes to equilibrium. We emphasise that this is valid regardless of how much entropy is produced in the relaxation process. To be more precise, the last expression scales as $t\partial_t\langle \hat{J}_t\rangle\sim t\mathrm{e}^{\lambda_1t}$. Additionally, the current variance \Eqref{eq:J_var} can be written as (see App.~\ref{sec:VarianceScaling})
\begin{equation}
    \begin{aligned}
        \mathrm{var}(\hat{J}_t) = \kappa t + \gamma + \mathcal{O}(\mathrm{e}^{\lambda_1t})\,,
    \end{aligned}
    \label{eq:currentvariancecomparison}
\end{equation}
where $\kappa$ and $\gamma$ are the scaling coefficients of the linear and constant order in $t$, respectively. The exact expressions can be found in Eq.~\eqref{eq:varJSPEC}.
Thus, the quality factor $\mathcal{Q}_J^\mathrm{TUR}(t)$ can be written as
\begin{equation}
    \begin{aligned}
        \mathcal{Q}_J^\mathrm{TUR}(t) \stackrel{c\gg t\gg -1/\lambda_1}{\sim}\frac{t\mathrm{e}^{2\lambda_1t}}{\left(1 + \frac{c}{t}\right)}\stackrel{t\gg c\gg -1/\lambda_1}{\sim}t\mathrm{e}^{2\lambda_1t}\xrightarrow{t\to\infty}0\,,
    \end{aligned}
    \label{eq:TURJQualityScaling}
\end{equation}
with $c=\gamma/\kappa$. If $c \ll -1/\lambda_1$, the intermediate regime will not take place. The reason why the $\mathcal{O}(1)$ term of the variance is included here, is that systems where the rates $r_{ij}$ are very large need significantly longer for this term to become negligible. In the Calmodulin example, the rates go between $0.121$ and $1514.820$ \cite{FirstPassageRick} (see Tab. \ref{tab:Calmodulin_Rates}). In Fig.~\hyperref[fig:TUR_TB_comp1]{\ref*{fig:TUR_TB_comp1}b}, the time-dependency of $\mathcal{Q}_J^\mathrm{TUR}$ is shown together with $\mathcal{Q}_\rho^\mathrm{TUR}$ and $\mathcal{Q}^\mathrm{TB}$ (see below), where the metric $d_{ij}^{(23)}$ is used\footnote{This was chosen specifically such that the largest rates enter, to showcase the effect of it, as these will enter the timescale set by $\gamma/\kappa$.}. For these values, $c=\gamma/\kappa\approx 1.68\cdot 10^4$, hence the constant term cannot be neglected in the time-interval shown in Fig.~\hyperref[fig:TUR_TB_comp1]{\ref*{fig:TUR_TB_comp1}b}.

A similar analysis can be done for $\mathcal{Q}_\rho^\mathrm{TUR}(t)$. Starting from 
\begin{equation}
    \begin{aligned}
        \partial_t\langle \hat{\rho}_t\rangle = \sum_{i\in\Omega} V_i p_i(t)\,.
    \end{aligned}
\end{equation}
Hence, we can identify
\begin{equation}
    \begin{aligned}
        t\partial_t\langle \hat{\rho}_t\rangle = t\sum_{i\in\Omega} V_i \left(p_i^\mathrm{eq} + \sum_{j>0, n\in\Omega} \psi^R_{j,i} \psi^L_{j,n} p_n(0)\mathrm{e}^{\lambda_j t} \right)\,,
    \end{aligned}
\end{equation}
which to leading order increases as $\mathcal{O}(t^{1})$. As $(t\partial_t - 1)\langle \hat{\rho}_t\rangle = \mathcal{O}(1)$ and $\mathrm{var}(\hat{\rho}_t) = \mathcal{O}(t)$, the fraction entering $\mathcal{Q}_\rho^\mathrm{TUR}$ scales as $( (t\partial_t-1)\langle \hat{\rho}_t\rangle)^2/\mathrm{var}(\hat{\rho}_t) = \mathcal{O}(t^{-1})$. In other words, $\mathcal{Q}_\rho^\mathrm{TUR}(t)$ is expected to decay slower than $\mathcal{Q}_J^\mathrm{TUR}(t)$ for large times. To be precise, for large times $t\gg -1/\lambda_1$ the quality factor is
\begin{equation}
    \begin{aligned}
        \mathcal{Q}_\rho^\mathrm{TUR}(t) \xrightarrow{t\gg -1/\lambda_1}\frac{1}{t}\frac{\left(\sum_{i,n\in\Omega}\sum_{j>0}V_i\psi^R_{j,i} \psi^L_{j,n} p_n(0)\frac{1}{\lambda_j}\right)^2}{\Delta S_\mathrm{tot}^\infty\left(-\sum_{i,n\in\Omega}\sum_{j>0}V_nV_i \psi^R_{j,i} \psi^L_{j,n} \frac{1}{\lambda_j}\right)}\,.
    \end{aligned}
    \label{eq:TURrhoQualityScaling}
\end{equation}

Next, we consider the transport bound \Eqref{eq:TransportBound}. As above, we will consider $z_i(\tau)$ not to depend on time for simplicity. Thus, we can immediately see that
\begin{equation}
    \begin{aligned}
        \langle z_x(t) - z_x(0)\rangle &= \sum_{x\in\Omega} z_x \left(\sum_{j\in\Omega, i\geq 0} \psi^R_{i,x}  \mathrm{e}^{\lambda_i t}\psi^L_{i,j} p_j(0) - p_x(0)\right) \\
        &= \underbrace{\sum_{x\in\Omega} z_x(p_x^\mathrm{eq} - p_x(0))}_{=\langle z_x\rangle_\mathrm{eq} - \langle z_x(0)\rangle} + \mathcal{O}(\mathrm{e}^{t\lambda_1})\,.
    \end{aligned}
\end{equation}
By integrating over the spectral decomposition of the propagator Eq.~\eqref{eq:SpectralPropagator} the diffusion coefficient decomposes as
\begin{equation}
    \begin{aligned}[b]
        \mathcal{D}_t &= \frac{1}{t}\sum_{\substack{x, y\in\Omega\\x\neq y}}(z_y-z_x)^2r_{xy}\left(tp_x^\mathrm{eq} + \sum_{j>0, n\in\Omega} \psi^R_{j,x} \psi^L_{j,n} p_n(0)\frac{1}{\lambda_j}(\mathrm{e}^{\lambda_j t}-1)\right)\\
        &= \sum_{\substack{x, y\in\Omega\\x\neq y}}(z_y-z_x)^2r_{xy}p_x^\mathrm{eq} + \mathcal{O}(t^{-1})\,.
    \end{aligned}
\end{equation}
Thus,
\begin{equation}
    \begin{aligned}
        \mathcal{Q}^\mathrm{TB}(t) \xrightarrow{t\gg -1/\lambda_1} \frac{1}{t}\frac{2(\langle z_x\rangle_\mathrm{eq} - \langle z_x(0)\rangle)^2}{\Delta S_\mathrm{tot}^\infty \sum_{\substack{x, y\in\Omega\\x\neq y}}(z_y-z_x)^2r_{xy}p_x^\mathrm{eq}}\,.
    \end{aligned}
    \label{eq:TBQualityScaling}
\end{equation}

Coming back to the example introduced at the beginning of this section, i.e., Fig.~\ref{fig:TUR_TB_comp1}, we can see that the expected large $t$ behaviour from Eqs.~\eqref{eq:TURJQualityScaling}, \eqref{eq:TURrhoQualityScaling}, and \eqref{eq:TBQualityScaling} is observed. Since $-1/\lambda_1\approx 1\ll \max(t)\ll c\approx 1.68\cdot 10^4$, the intermediate regime in \Eqref{eq:TURJQualityScaling} is observed. 

It should become obvious from Fig.~\ref{fig:TUR_TB_comp1} that $\mathcal{Q}_J^\mathrm{TUR}(t)$ is not useful for long trajectories that relax towards equilibrium. Moreover, as mentioned in Sec.~\ref{sec:TURproof}, specifically in regards to Fig.~\ref{fig:Derivative_sample}, the number of trajectories required to be able to make any reasonable statement about the system is very large, especially for large times. There, it is not observed that the numerical quality factor exceeds the upper bound $\mathcal{Q}_J^\mathrm{TUR}(t)\leq 1$. In Fig.~\hyperref[fig:TUR_TB_comp1]{\ref*{fig:TUR_TB_comp1}a}, exactly this violation is observed, due to the fluctuations in $\partial_t\langle \hat{J}_t\rangle$ and $(t\partial_t\langle \hat{J}_t\rangle)^2$, see insets in Fig.~\hyperref[fig:TUR_TB_comp1]{\ref*{fig:TUR_TB_comp1}a}. Somewhat surprisingly, these fluctuations are not observed in $\mathcal{Q}_\rho^\mathrm{TUR}(t)$, which also requires the derivative in $t$. The numerical quality factors of $\mathcal{Q}_\rho^\mathrm{TUR}(t)$ and $\mathcal{Q}^\mathrm{TB}(t)$ are in good agreement with the theoretical prediction, further supporting their choice over $\mathcal{Q}_J^\mathrm{TUR}(t)$. 

\subsection{Relaxation Towards NESS\label{sec:ComparisonRelaxationNESS}}

Now we will see what can be said about the aforementioned quality factors for a system which relaxes towards a NESS. We again start with the total entropy production. In NESS, the total produced entropy $\Delta S_\mathrm{tot}(t) = t \Dot{S}_\mathrm{tot}^\mathrm{s}$ is simply given by the NESS entropy production rate $\Dot{S}_\mathrm{tot}^\mathrm{s}$. In general, i.e., transient, the rate at which entropy is produced $\Dot{S}_\mathrm{tot}(\tau)$ is time-dependent and differs from the NESS rate. However, in the long time limit, we may say that $\Delta S_\mathrm{tot}/t\xrightarrow{t\gg-1/\mathrm{Re}(\lambda_1)}\Dot{S}_\mathrm{tot}^\mathrm{s}$. 

While the long-time behaviour of the quality factors is of interest, in general, the master operator $\mathbf{L}$ for out-of-equilibrium systems is not diagonalisable (see App. \ref{sec:NonDiagonalisable} for a simple counter-example). This means that the approach of using spectral decomposition does not generally apply to all driven systems. However, as the examples we consider are diagonalisable, we will show the calculation using spectral decomposition, i.e., we consider only diagonalisable systems in the following. 

For systems relaxing towards a NESS, the mean current scales linearly in $t$ for $t\gg-1/\mathrm{Re}(\lambda_1)$ so that
\begin{equation}
    \begin{aligned}
        t\partial_t\langle \hat{J}_t\rangle \xrightarrow{t\gg-1/\mathrm{Re}(\lambda_1)} t\sum_{\substack{x, y\in\Omega\\x\neq y}}d_{xy}r_{xy}p_x^\mathrm{s}\,.
    \end{aligned}
\end{equation}
Similar to the previous section, the variance still scales like \Eqref{eq:currentvariancecomparison}.
Therefore, the current TUR quality factor scales as
\begin{equation}
    \begin{aligned}
        \mathcal{Q}_J^\mathrm{TUR}(t) \stackrel{c\gg t\gg -1/\mathrm{Re}(\lambda_1)}{\sim}\frac{1}{\left(1 + \frac{c}{t}\right)}\stackrel{t\gg {c}}{\sim}1\,,
    \end{aligned}
    \label{eq:TURJNESSQualityScaling}
\end{equation}
and in the limit of infinitely long trajectories, it becomes
\begin{equation}
    \begin{aligned}
         \mathcal{Q}_J^\mathrm{TUR}(t)\xrightarrow{t\to\infty} \frac{2\left(\sum_{\substack{x, y\in\Omega\\x\neq y}}d_{xy}r_{xy}p_x^\mathrm{s}\right)^2}{\kappa\Dot{S}_\mathrm{tot}^\mathrm{s}}\,.
    \end{aligned}
    \label{eq:TURJNESSQualityScalingExact}
\end{equation}

\begin{figure}
    \centering
    \includegraphics[width=\textwidth]{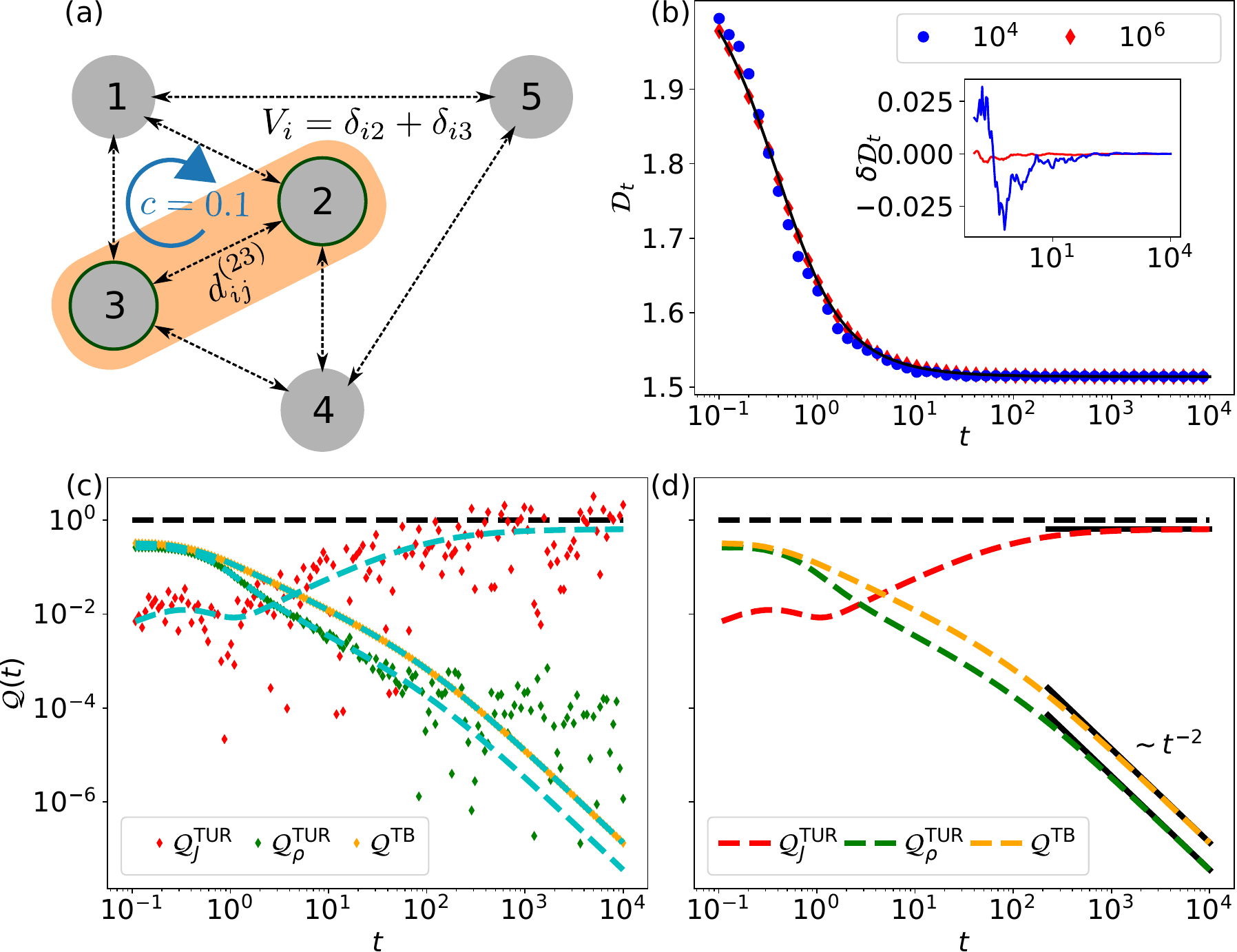}
    \caption[Long time behaviour for relaxation toward NESS]{Driven toy model with resulting diffusion coefficient $\mathcal{D}_t$ and quality factors. In (a), the model is shown with the direction of driving. The system is driven with scaling coefficient $c=0.1$ along the cycle $\mathcal{C}=\{1\to2,2\to3,3\to1\}$, as marked by the blue arrow. See Sec.~\ref{sec:DrivingSystems} for details on the driving. The energies entering in the construction of the rates are drawn from a uniform distribution between zero and one and are given in Tab. \ref{tab:ToyModelEnergies}. The metric $d_{ij}^{(23)}$ and state functions $V_i = \delta_{i2} +\delta_{i3} = z_i$ are used, which are marked with orange shading and thick edge, respectively. As initial distribution we use $p_i(0) = \delta_{i1}$. Figure (b) shows the diffusion coefficient $\mathcal{D}_t$ as a function of $t$ numerically sampled from $N=10^4$ (blue) and $N=10^6$ (red) trajectories using \Eqref{eq:TrajectoryDiffusion}. The black curve is the expected result $\mathcal{D}_t^\mathrm{ana}$ and the inset shows how much the numerical value deviates $\delta \mathcal{D}_t = \mathcal{D}_t^\mathrm{num} - \mathcal{D}_t^\mathrm{ana}$ from the analytical result. In (c) and (d), the TUR quality factors for a current $\mathcal{Q}_J^\mathrm{TUR}$ (red) and a density $\mathcal{Q}_\rho^\mathrm{TUR}$ (green) together with the transport bound quality factor $\mathcal{Q}^\mathrm{TB}$ (orange) are shown as a function of time $t$. The numerical quality factors in (c), which are represented by the diamonds, are evaluated using $N=10^6$ trajectories. In (d), the solid black lines are the long time limits shown in Eqs.~\eqref{eq:TURJNESSQualityScalingExact}, \eqref{eq:TURrhoNESSQualityScaling}, and \eqref{eq:TBNESSQualityScaling}. In both (c) and (d), the dashed black line visualises the theoretical upper bound $\mathcal{Q}<1$.
    }
    \label{fig:DrivenComparison}
\end{figure}

For the density quality factor, we can immediately see that the scaling of the mean and variance are the same as when relaxing towards equilibrium. Thus, we get that
\begin{equation}
    \begin{aligned}[b]
        \mathcal{Q}_\rho^\mathrm{TUR}(t) \xrightarrow{t\gg -1/\mathrm{Re}(\lambda_1)}\frac{1}{t^2}\frac{\left(\sum_{i,n\in\Omega}\sum_{j>0}V_i\psi^R_{j,i}\psi^L_{j,n}p_n(0)\frac{1}{\lambda_j}\right)^2}{\Dot{S}_\mathrm{tot}^\mathrm{s}\left(-\sum_{i,n\in\Omega}\sum_{j>0}V_nV_i \psi^R_{j,i}\psi^L_{j,n}\frac{1}{\lambda_j}\right)}\,.
    \end{aligned}
    \label{eq:TURrhoNESSQualityScaling}
\end{equation}
Similarly, the transport bound quality factor behaves asymptotically as
\begin{equation}
    \begin{aligned}
        \mathcal{Q}^\mathrm{TB}(t) \xrightarrow{t\gg -1/\mathrm{Re}(\lambda_1)} \frac{1}{t^2}\frac{2(\langle z_x\rangle_\mathrm{s} - \langle z_x(0)\rangle)^2}{\Dot{S}_\mathrm{tot}^\mathrm{s} \sum_{\substack{x, y\in\Omega\\x\neq y}}(z_y-z_x)^2r_{xy}p_x^\mathrm{s}}\,.
    \end{aligned}
    \label{eq:TBNESSQualityScaling}
\end{equation}
In Fig.~\ref{fig:DrivenComparison}, these asymptotic behaviours are shown in the context of the toy model driven along the cycle $\mathcal{C}=\{1\to2,2\to3,3\to1\}$ with strength $c=0.1$, see Sec.~\ref{sec:DrivingSystems} and Fig.~\hyperref[fig:DrivenComparison]{\ref*{fig:DrivenComparison}a}. This system is initially transient, as $p_i(0) = \delta_{i1}$, and the functions for the observables are $V_i =z_i= \delta_{i2} +\delta_{i3} $ and $d_{ij}^{(23)}$. Figure \hyperref[fig:DrivenComparison]{\ref*{fig:DrivenComparison}b} shows the diffusion coefficient as a function of time evaluated numerically for $N=10^4$ and $N=10^6$ trajectories, and the analytical solution is added for comparison. In addition, the inset shows to what degree the numerical values deviate from the theoretical predictions. The asymptotic behaviour in Eqs.~\eqref{eq:TURJNESSQualityScaling}, \eqref{eq:TURrhoNESSQualityScaling}, and \eqref{eq:TBNESSQualityScaling} can be seen in Figs.~\hyperref[fig:DrivenComparison]{\ref*{fig:DrivenComparison}c} and \hyperref[fig:DrivenComparison]{\ref*{fig:DrivenComparison}d}. Similar to the previous section, the numerical values in Fig.~\hyperref[fig:DrivenComparison]{\ref*{fig:DrivenComparison}c} for $\mathcal{Q}_J^\mathrm{TUR}(t)$ and $\mathcal{Q}_\rho^\mathrm{TUR}(t)$ spread around the analytical prediction. While the density TUR seems to be stable, i.e., does not spread much, when the system relaxes towards equilibrium, see Fig.~\hyperref[fig:TUR_TB_comp1]{\ref*{fig:TUR_TB_comp1}a}, we find that if the system relaxes to a NESS the numerical $\mathcal{Q}_\rho^\mathrm{TUR}(t)$ spreads significantly more for larger $t$, see Fig.~\hyperref[fig:DrivenComparison]{\ref*{fig:DrivenComparison}c}. On the other hand, we again see that the numerical current TUR quality factor violates the upper bound, stemming from sampling errors. While the transport bound does not yield the best possible inference for all $t$, it is more stable than the TUR counterparts, as can be seen in Figs.~\hyperref[fig:TUR_TB_comp1]{\ref*{fig:TUR_TB_comp1}a} and \hyperref[fig:DrivenComparison]{\ref*{fig:DrivenComparison}c}.

We make one final remark on Eqs.~\eqref{eq:TURJNESSQualityScaling} and \eqref{eq:TURJNESSQualityScalingExact}. The scaling in Eq~\eqref{eq:TURJNESSQualityScaling} is only observed if the $d_{xy}$ allows driven transitions to be observed. For example, if the transition between states $1$ and $5$ in Fig.~\hyperref[fig:DrivenComparison]{\ref*{fig:DrivenComparison}a} is observed, then the mean current and resulting quality factor also decay exponentially, since $\mathcal{A}_{15}=0$. 

\section{Comparison of Bounds for Periodic SAT in NESS\label{sec:comparisonNESS}}

\begin{figure}
    \centering
    \includegraphics[width=\textwidth]{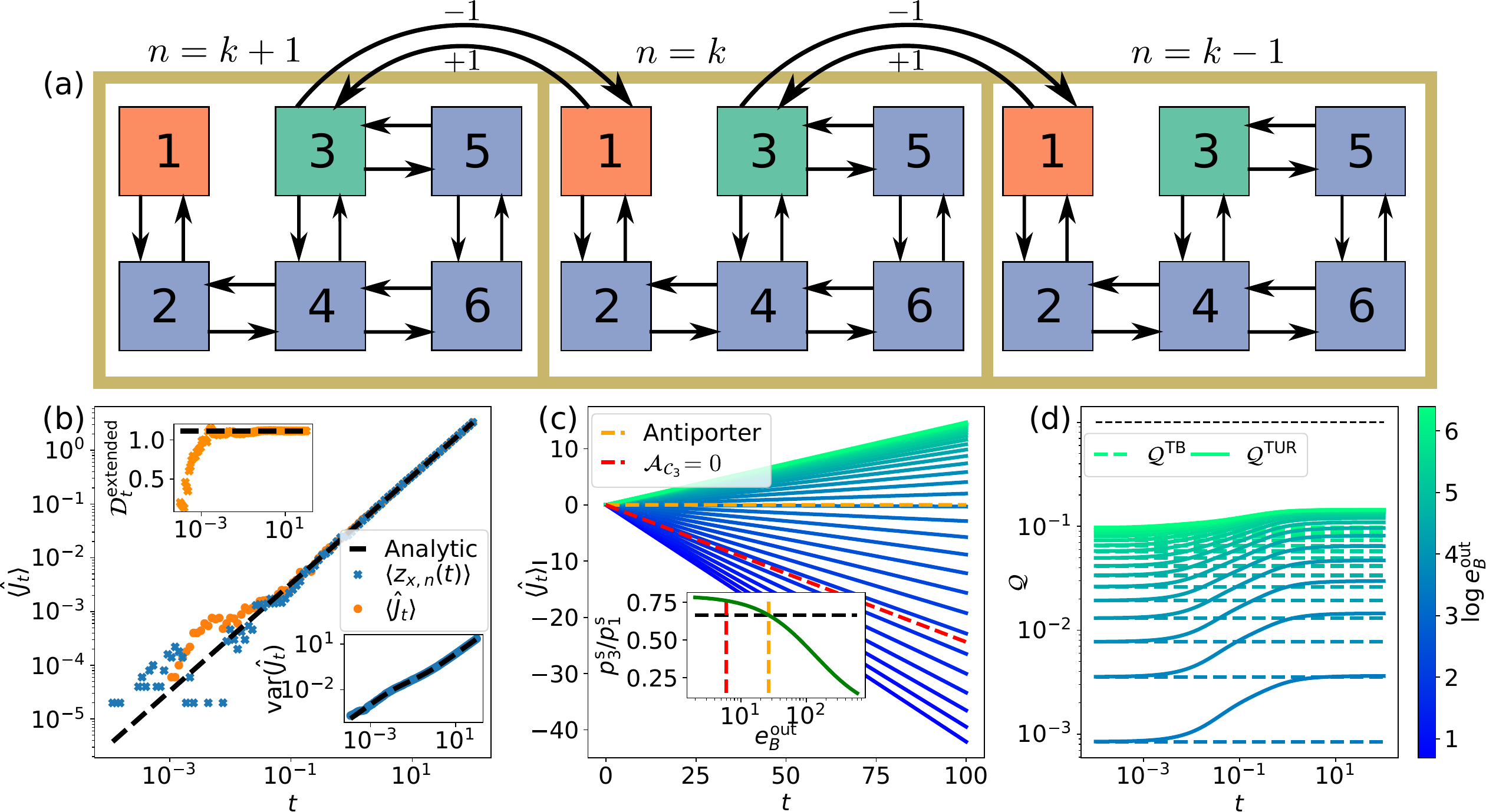}
    \caption[Periodic SAT bound comparison]{Extension of SAT with a comparison of TUR and transport bound. In (a), a schematic of a part of a periodic SAT model is shown. The $k\in\mathds{Z}$ are the number of A molecules transported through the membrane. The mean of $\hat{J}_t$ (orange) and $z_{x,n}$ (blue) from $N=50000$ trajectories is shown as a function of $t$, compared to the analytic $\langle \hat{J}_t\rangle$ (see \Eqref{eq:J_mean}). The insets show numerical values of $\mathcal{D}_t^\mathrm{ext}$ and $\mathrm{var}(\hat{J}_t)$ compared to the analytical values $l_A p^\mathrm{s}_1 + e^\mathrm{out}_A p^\mathrm{s}_3$ and \Eqref{eq:J_var}, respectively. That the numerical $\mathcal{D}_t^\mathrm{ext}$ deviates from the theoretical value for small $t$ is due to the small sample size. The mean current is shown in (c) for various values of $e^\mathrm{out}_B$. The dashed orange and red lines indicate the values of $e^\mathrm{out}_B$ where the antiporter regime begins and where $\mathcal{A}_{\mathcal{C}_3}=0$, respectively. The inset shows the fraction $p_3^\mathrm{s}/p_1^\mathrm{s}$ for different values of $e^\mathrm{out}_B$. The orange and red lines correspond to the values of $e^\mathrm{out}_B$ from the main plot, while the black dashed line is the upper limit $p_3^\mathrm{s}/p_1^\mathrm{s}$ can have so that the system is an antiporter. The steady-state quality factors for TUR (solid) and TB (dashed) are shown in (c) as a function of $t$ for various values of $e^\mathrm{out}_B$ in the antiporter regime. The parameters are listed in Tab. \ref{tab:Parameter SAT}.}
    \label{fig:SATperiodic}
\end{figure}

It may be more interesting and relevant to infer the entropy production in a NESS system, since it is commonly done in the literature to investigate thermodynamic bounds \cite{BoundsCorrelationTimes, ImprovingBoundsCorrelations, SeifertPartiallyAccessibleNetworks}. We use the SAT model (Sec.~\ref{sec:SecondaryTransport}) to compare the current TUR and transport bound. For this, we change the frame of reference so that the transport bound can be applied. Specifically, consider the SAT in the antiporter regime, i.e., $e^\mathrm{out}_B>e^\mathrm{out}_Ae^\mathrm{in}_B/e^\mathrm{in}_A$, so that $\mathcal{A}_{\mathcal{C}_3}<1$. The quantity of interest is the number of molecules $A$ exported through the membrane. Hence, we consider a chain of connected SAT systems as shown in the schematic in Fig.~\hyperref[fig:SATperiodic]{\ref*{fig:SATperiodic}a}, so that the state 1 in one system is connected to state 3 in the next (reading right to left). Physically, this setting can be motivated by the fact that each of these systems corresponds to a specific number of $A$ being transported through the membrane. So the six-state model is extended to a six-state model coupled to an additional degree of freedom representing the number of transported $A$ molecules. This means that the state space becomes $\Tilde{\Omega}=\Omega\cross\mathds{Z}$, where $\Omega$ is the six-state SAT state space. For simplicity, the transition rates are kept constant along the chain.\footnote{As mentioned in Sec.~\ref{sec:SecondaryTransport}, the transition rates of $A$ and $B$ entering the protein channel depend on the concentration of the species. Transport changes the concentrations, thus complicating the problem further. This is neglected here.}

For the TUR, the current metric to calculate the number of transitioned $A$ molecules in the six-state model is $d^A_{ij} = \delta_{i1}\delta_{j3} - \delta_{i3}\delta_{j1}$, and it corresponds to the difference in left and right SAT system transitions in the extended model. 

We can enumerate the SAT systems with an index $n\in\mathds{Z}$, indicating the net number of exported $A$ molecules, i.e., 
\begin{align}
    n = \# A \text{ exported} - \# A \text{ imported}.
\end{align}
The probability of being in state $x$ in system $n$ at time $\tau$ is denoted by $p_{x,n}(\tau)$.\footnote{Note that the projected six-state model can be recovered by marginalising over $n$.} The state function $z_{x,n}$ is then a function of $x$ and $n$ as well, so that the average is $\langle z_{x,n}(\tau)\rangle = \sum_{x\in\Omega,n\in\mathds{Z}} z_{x,n}p_{x,n}(\tau)$. Similarly, the extended diffusion constant  becomes
\begin{equation}
    \begin{aligned}[b]
        \mathcal{D}_t^\mathrm{ext} &= \frac{1}{t}\sum_{n\in\mathds{Z}} \sum_{\substack{x, y\in\Omega\\x\neq y}}\int_0^t\mathrm{d}\tau\left[ (z_{y,n} - z_{x,n})^2\delta_{x\to y ,n}+(z_{y,n+1} - z_{x,n})^2\delta_{x\to y, n\to n+1}\right.\\
        &+\left.(z_{y,n-1} - z_{x,n})^2\delta_{x\to y,n \to n-1}\right]r_{xy}p_{x,n}(\tau)\,.
    \end{aligned}
\end{equation}
The first sum in the integral accounts for the diffusion within each subsystem, while the latter two accounts for the diffusion across boundaries between systems. If $z_{x,n}=z_x$ is independent of $n$, this simplifies to the diffusion coefficient in \Eqref{eq:DiscreteDiffusion}. Coming back to the example, since we are interested in the number of transported $A$ molecules, it is natural to choose $z_{x,n} = n$ as a state function. Hence, it is independent of the specific state in each SAT system. A consequence of this is that the internal diffusion of each system vanishes, i.e., $\mathcal{D}_t^\mathrm{ext}$ becomes
\begin{equation}
    \begin{aligned}[b]
        \mathcal{D}_t^\mathrm{ext} =& \frac{1}{t}\sum_{n\in\mathds{Z}} \sum_{\substack{x, y\in\Omega\\x\neq y}}\int_0^t\mathrm{d}\tau r_{xy}p_{x,n}(\tau)\\
        &\times\left[ (z_{y,n+1} - z_{x,n})^2\delta_{x\to y ,n\to n+1} + (z_{y,n-1} - z_{x,n})^2\delta_{x\to y, n\to n-1}\right]\,.
    \end{aligned}
    \label{eq:corrdiff1}
\end{equation}
Since we know which transitions lead to change in the SAT system, we can simplify Eq.~\eqref{eq:corrdiff1} more as $z_{y,n\pm1} - z_{x,n} = \pm1$ to get
\begin{equation}
    \begin{aligned}[b]
        \mathcal{D}_t^\mathrm{ext} &= \frac{1}{t}\sum_{n\in\mathds{Z}} \sum_{\substack{x, y\in\Omega\\x\neq y}}\int_0^t\mathrm{d}\tau r_{xy}p_{x,n}(\tau)\left[ \delta_{x\to y, 1\to3} + \delta_{x\to y, 3\to1}\right]\,\\
        &=\frac{1}{t} \int_0^t\mathrm{d}\tau \left[r_{13}p_{1}(\tau)+ r_{31}p_{3}(\tau)\right]\,.
    \end{aligned}
\end{equation}
Thus, the transport bound for this specific example can be written as
\begin{equation}
    \begin{aligned}
        \frac{\Delta S_\mathrm{tot}(t)}{2}\geq \frac{(\sum_{n\in\mathds{Z}} n\sum_{x\in\Omega} p_{x,n}(t))^2}{t\mathrm{D}_t^\mathrm{ext}}\,,
    \end{aligned}
\end{equation}
where we assume that we start in the SAT system with zero $A$ molecules, i.e., $\sum_{n\in\mathds{Z},x\in\Omega}np_{x,n}(0) = \sum_{n\in\mathds{Z},x\in\Omega}np_{x}(0)\delta_{n,0}=0$. Since $\sum_{n\in\mathds{Z}} n\sum_{x\in\Omega} p_{x,n}(t)$ corresponds to the average number of transitions of boundaries, it is the same as the average current $\langle \hat{J}_t\rangle$ with the metric mentioned above. The agreement between $\sum_{n\in\mathds{Z}} n\sum_{x\in\Omega} p_{x,n}(t)$ and $\langle \hat{J}_t\rangle$ can be seen in Fig.~\hyperref[fig:SATperiodic]{\ref*{fig:SATperiodic}b}. The transport bound can therefore be written as
\begin{equation}
    \begin{aligned}
        \frac{\Delta S_\mathrm{tot}(t)}{2}\geq \frac{\langle\hat{J}_t\rangle^2}{t\mathrm{D}_t^\mathrm{ext}}\,,
    \end{aligned}
    \label{eq:corrgTB}
\end{equation}
which allows us to compare it with the current TUR if the transitions between state $1$ and $3$ are the only accessible part of the system. Equation~\eqref{eq:corrgTB} has the form of the generalised transport bound \Eqref{eq:GeneralisedTransport}, as $\mathcal{D}^\mathrm{ext}_t = \mathrm{var}(\hat{J}_t^\mathrm{I})/t$.

Using the mean current, we can also identify which values of $e^\mathrm{out}_B$ lead to antiporter behaviour. Of interest is a positive mean current, i.e., on average more $A$ molecules being transported out of the cell, see Fig.~\hyperref[fig:SATperiodic]{\ref*{fig:SATperiodic}c}. Since, in steady state, the mean current is $\langle \hat{J}_t\rangle = l_A p^\mathrm{s}_1 - e^\mathrm{out}_A p^\mathrm{s}_3$ the antiporter regime occurs when $p_3^\mathrm{s}/p_1^\mathrm{s} < l_A/e^\mathrm{out}_A$. As the probabilities depend on $e^\mathrm{out}_B$, the threshold value can be varied until the condition is satisfied. The inset in Fig.~\hyperref[fig:SATperiodic]{\ref*{fig:SATperiodic}c} shows how $p_3^\mathrm{s}/p_1^\mathrm{s}$ depends on the values of $e^\mathrm{out}_B$. The black dashed line is the value, $l_A/e^\mathrm{out}_A$, below which the antiporter regime takes place. In both the main part and inset of Fig.~\hyperref[fig:SATperiodic]{\ref*{fig:SATperiodic}c} the orange and red dashed lines mark the value for $e^\mathrm{out}_B$ where the anitporter regime emerges and $\mathcal{A}_{\mathcal{C}_3}=0$, respectively. Of particular note is that $\mathcal{A}_{\mathcal{C}_3}<0$ is not sufficient a condition for the system to be an antiporter.

A similar observation can be made for $\mathcal{D}_t^\mathrm{ext}$ and $\mathrm{var}(\hat{J}_t)/t$ in a steady state. As $\mathcal{D}_t^\mathrm{ext} = r_{13}p_{1}^\mathrm{s}+ r_{31}p_{3}^\mathrm{s}$ for all $t$ if $p_i(0) = p_i^\mathrm{s}$, the current variance can be identified as 
\begin{equation}
    \begin{aligned}[b]
        \mathrm{var}(\hat{J}_t) =& t\mathcal{D}_t^\mathrm{ext} - \langle \hat{J}_t\rangle^2\\
        &+ 2\int_0^t\mathrm{d}\tau^\prime \sum_{\substack{i, j\in\Omega\\i\neq j}}d_{ij}(\tau^\prime)r_{ij} \int_0^{\tau^\prime}\mathrm{d}\tau\sum_{\substack{x, y\in\Omega\\x\neq y}}d_{xy}(\tau)r_{xy}P(i, \tau^\prime|y, \tau)p_x(\tau)\,.
    \end{aligned}
\end{equation}
Hence, considering the limit $t\to 0$, one can identify $\mathrm{var}(\hat{J}_t)/t\to \mathcal{D}_t^\mathrm{ext}$, since $\langle \hat{J}_t\rangle\sim t$ and the double integral vanishes (even when divided by $t$) in this limit. Therefore, we expect $\mathcal{Q}^\mathrm{TUR}\xrightarrow{t\to 0} \mathcal{Q}^\mathrm{TB}$, which can be observed in Fig.~\hyperref[fig:SATperiodic]{\ref*{fig:SATperiodic}d}, where the quality factors are shown for various $e^\mathrm{out}_B$ in the antiporter regime. For $t\to \infty$, there are contributions to $\mathrm{var}(\hat{J}_t)/t$ that may increase or decrease it relative to $\mathcal{D}_t^\mathrm{ext}$. In the example considered here, the corrections decrease the value s.t. $\mathrm{var}(\hat{J}_t)/t \stackrel{t\to \infty}{<}\mathcal{D}_t^\mathrm{ext}$. As a consequence, the TUR quality factor becomes larger than the TB quality factor, as can also be seen in Fig.~\hyperref[fig:SATperiodic]{\ref*{fig:SATperiodic}d}. Hence, the precision of the antiporter process allows for a better inference of the total entropy production than the transport of $A$ molecules independent on $e^\mathrm{out}_B$. 

Since the steady state converges for $e^\mathrm{out}_B\to \infty$ (see Prop. \ref{prop:ConvergenceEdgeDrivingSteadyState}), there is an upper bound $U<1$ for the quality factor in this particular example, so that $\mathcal{Q}^\mathrm{TB}\leq \mathcal{Q}^\mathrm{TUR}<U$. The latter can also be observed in Fig.~\hyperref[fig:SATperiodic]{\ref*{fig:SATperiodic}d}. Physically, this means that increasing $e^\mathrm{out}_B$ does not necessarily allow for an improved inference of the entropy produced. Considering that additional energy has to be put into the system by increasing the concentration of $B$ molecules outside the cell, there is an energy/concentration-independent part of the entropy that cannot be inferred. It should be mentioned that the latter statement is only valid in the antiporter regime; decreasing $e^\mathrm{out}_B$, so that the system becomes a symporter, can allow for improved estimations compared to $U$. However, we specifically construct the example to investigate the antiporter behaviour and thus choose not to show the symporter case. 

\chapter{Conclusion \label{ch:conclusion}}

\section{Summary}

To start, in this thesis, we presented a stochastic calculus formulation of MJP. For this, the stochastic differentials $\mathrm{d}\hat{n}_{xy}(\tau)$ and $\mathrm{d}\hat{\tau}_x(\tau)$ for jumps between states and waiting in states were used. By additionally introducing a noise increment $\mathrm{d}\hat{\varepsilon}_{xy}(\tau)$, we were able to describe not only the mean values but also (co)variances of observables integrated w.r.t. the stochastic differentials. For example, the variance of an integrated current and covariance between an integrated current and density is required to evaluate the correlation TUR (and the saturation thereof); see Sec.~\ref{sec:TURsaturation}. 

In addition, the stochastic calculus approach provides a simple method, or even "scheme", to prove and find thermodynamic bounds. We proved the TUR, transport bound, and correlation bound in Secs.~\ref{sec:TURproof}, \ref{sec:Transportproof}, and \ref{sec:Correlationproof}, respectively, to showcase this. These proofs fundamentally require two steps: choosing observables/auxiliary integrals and using the Cauchy-Schwarz inequality when evaluating expectation values. Central to the first is how entropy production enters the bounds. For example, in the TUR and transport bound, the total entropy production $\Delta S_\mathrm{tot}(t)$ enters as a factor independent of the remaining quantities, while the entropy production rate enters the correlation bound by tilting the probability distribution. However, as long as the state functions entering are bounded, another inequality can be used to again "isolate" $\Delta S_\mathrm{tot}(t)$. Furthermore, in contrast to continuous systems, there exists no auxiliary integral whose second moment gives $\Delta S_\mathrm{tot}(t)$. Instead, we need to proceed via a pseudo entropy production, a lower bound on the total entropy production. How to saturate this bound, as well as the Cauchy-Schwarz inequality, is known, hence we only needed to show how to optimise the respective bounds in Secs.~\ref{sec:TURsaturation}, \ref{sec:TransportSaturation}, and \ref{sec:CorrelationSpectral}. 

Returning to the bounds themselves, we saw that these are similar to the continuous counterparts presented in Sec.~\ref{sec:thermodynamicbounds}. In particular, the quantities entering the current TUR are the same as in the continuous case, i.e. we found that the current TUR consist of $\mathrm{var}(\hat{J}_t)$, $\langle \hat{J}_t\rangle$, and $\langle \hat{\Tilde{J}}_t\rangle$. For the transport bound, the only apparent difference lies in the diffusion coefficient $\mathcal{D}_t$; in the continuous case, the latter depends on the gradient of the observed state functions. The discrete diffusion coefficient, similar to the continuous one, also describes the spread of the state function. However, instead of looking at the gradient, which is not defined in the discrete case, it consists of the variance of small increments of the state function over time, which was presented in Sec.~\ref{sec:DiffusionFluctuation}. Furthermore, we showed that $\mathcal{D}_t$ is accessible from data by only keeping track of transitions in the system. Moreover, the correlation bound we found in this thesis is more general than the one proposed in literature \cite{BoundsCorrelationTimes}, as it is valid even for finite time and transient, i.e. for non-stationary, processes. In the appropriate limits, i.e. the system starting in a stationary state $p_i(0)\to p_i^\mathrm{s}$ and of infinite trajectory length $t\to \infty$, we recover the bound derived in Ref. \cite{BoundsCorrelationTimes} for discrete systems. 

While making quantitative statements about which bound is sharper, i.e., which one provides better inference of $\Delta S_\mathrm{tot}(t)$, is in general not possible, we could still make observations about long-time behaviour and compare the bounds in this limit. The reason we claim that a general comparison is not possible, is because depending on the system considered and what part of the system is observable will greatly impact how the various bounds perform. However, various statements can be made about the asymptotic behaviour in the limit $t\to\infty$. The quality factors of the density TUR and transport bound decay as $t^{-1}$ and $t^{-2}$, respectively, for large $t$ when applied to a transient system which decays towards equilibrium or NESS, respectively. On the other hand, $\mathcal{Q}^\mathrm{TUR}_{J}$ decays as $t\mathrm{e}^{2\lambda_1t}$ for large $t$ if the system relaxes towards equilibrium. Additionally, there may be an intermediate regime yielding a possible correction even for $c\gg t\gg -1/\lambda_1$ depending on $c=\gamma/\kappa$, consisting of the leading and next-to-leading order terms in the current variance $\mathrm{var}(\hat{J}_t) = \kappa t + \gamma + \mathcal{O}(\mathrm{e}^{\lambda_1 t})$. If the system instead relaxes towards a NESS, the quality factor $\mathcal{Q}^\mathrm{TUR}_{J}\to \mathrm{const.}$ for $t\to\infty$. 

The methods presented in this thesis can be used to investigate a broad selection of biological examples. The systems we chose are the calmodulin folding dynamics for a detailed balance system (see Sec.~\ref{sec:calmodulin}) and the secondary active transport for a driven system (see Sec.~\ref{sec:SecondaryTransport}). Especially the latter is of interest, as the choice of $e^\mathrm{out}_B$ impacts the biological function of the system. The impact of $e^\mathrm{out}_B$ on various quantities, such as $\mathrm{var}(\hat{\rho}_t)$ and the transport of $A$ molecules, were investigated. One key finding we presented is that increasing $e^\mathrm{out}_B$ will lead the quality factors, both current TUR and transport bound, converging to some $U<1$. In other words, we cannot infer a larger fraction of $\Delta S_\mathrm{tot}(t)$ by simply increasing $e^\mathrm{out}_B$. We also found that, for the parameters, metrics, and state functions used, the current TUR infers more of the total entropy production than the transport bound. 

By investigating these examples numerically it was shown that the numerical TUR quality factors deviate from the expected values in case of limited sample sizes. Specifically, as the derivative of mean current and density, i.e., $\partial_t\langle \hat{J}_t\rangle$ and $\partial_t\langle \hat{\rho}_t\rangle$, are one-time quantities, they require significantly more data to get a good agreement with the expected results. This is important for applications to experiments. 

Lastly, some mathematical aspects surrounding the systematic driving of systems were investigated at in this thesis. Specifically, two complementary methods to drive systems out of equilibrium while keeping the steady state density fixed were derived. The first, presented in Sec.~\ref{sec:DrivingSystems}, is an explicit application of the method in Ref.~\cite{Kaiser2017}, where we showed how to drive up to multiple cycles even when they are not topologically independent. The second method, found in Apx. \ref{sec:AlternativeDriving}, is a more mathematical approach that unlike the first method, where transition rates are directly adjusted, divides a given cycle affinity $\mathcal{A}_\mathcal{C}$ onto the individual edges. We could prove that there is always a unique way to do this. 

To summarise, this thesis provides a method based on stochastic calculus for MJP to evaluate statistical properties of pathwise observables and, thus, proves to be a powerful candidate for proving thermodynamic inequalities directly. The only inequalities we used are the Cauchy-Schwarz inequality and a fundamental logarithm identity. We provided insight into how to saturate the bounds. The bounds were also compared for different dynamics using (biological) examples, e.g., for asymptotic behaviour.

\section{Outlook}



As the stochastic calculus approach and the resulting bounds are valid for all MJPs, they can be applied to a broad family of  systems. For instance, one could investigate the folding by looking at the scattering of polymers by using a two-state model \cite{Dieball_2022}, signal transduction \cite{Ohga2023}, and phosphorylation in KaiC molecules \cite{CoherenceBiochemicalOscillation}. Using the bounds we proved in this thesis may give valuable insight into the workings of such, and many more, systems. 

The Langevin equation in the language of stochastic calculus has been extended to account for memory in the system, which is relevant for biological systems \cite{Banos2019}. Hence, it motivates how this can be implemented in the framework of jump processes we presented in this thesis. This is an interesting question which goes beyond the scope of this work and, thus, is left for future study.


Finally, constraining the diagonal elements to be constant may not be the only and best physical motivated way to drive a system while keeping the steady state density fixed. It remains to be investigated whether, e.g., a fixed total activity expressed through $-\Tr \mathbf{L}=\mathrm{const.}$ is a more appropriate constraint. While physically relevant, it is unclear at this point if this will yield a solution, and, if there are solutions, whether it is unique.

\cleardoublepage

\newpage
\appendix

\counterwithin{figure}{section}

\chapter{\label{Ch:Apx_details} Examples and Additional Information}



\section{Connecting Cycle Perturbation to Cycle Affinities and Entropy Production}
Returning to the case where a single cycle is perturbed, we want to show how the driving strength $c$ relates to the cycle affinity and entropy production. With $x = \pm c\min_{l\to z}(\mathrm{e}^{-(E_l+E_z)/2})$, the cycle affinity can be written as
\begin{equation}
    \begin{aligned}[b]
        \mathcal{A}_\mathcal{C} 
        &= \sum_{i\to j\in\mathcal{C}}\log{\left(\frac{\mathrm{e}^{-(E_i+E_j)/2} + x}{\mathrm{e}^{-(E_i+E_j)/2} - x}\right)}\\
        &= \sum_{i\to j\in\mathcal{C}}\log{\left(\frac{1 + x\mathrm{e}^{(E_i+E_j)/2}}{1 - x\mathrm{e}^{(E_i+E_j)/2}}\right)}\\
        &= 2\sum_{i\to j\in\mathcal{C}}\mathrm{arctanh}\left(x\mathrm{e}^{(E_i+E_j)/2}\right)\,.
    \end{aligned}
\end{equation}
Expanding this around $x=0$ yields
\begin{equation}
    \begin{aligned}[b]
        \mathcal{A}_\mathcal{C} &= 2\sum_{i\to j\in\mathcal{C}} \left(\pm c\min_{l\to z}(\mathrm{e}^{-(E_l+E_z)/2})\mathrm{e}^{(E_i+E_j)/2}\right) + \mathcal{O}(c^3)\\
        &= \pm 2c\Lambda_\mathcal{C} + \mathcal{O}(c^3)\,.
    \end{aligned}
    \label{eq:AffinityDrivingStrength}
\end{equation}
In the last line, we introduce the cycle coefficients 
\begin{align}
    \Lambda_\mathcal{C} = \min_{l\to z}(\mathrm{e}^{-(E_l+E_z)/2})\sum_{i\to j\in\mathcal{C}}\mathrm{e}^{(E_i+E_j)/2}\,.
\end{align}
The contributions can be split into the respective edge coefficients
\begin{align}
    \Lambda_{ij} = \min_{l\to z}(\mathrm{e}^{-(E_l+E_z)/2}) \mathrm{e}^{(E_i+E_j)/2}\,,
\end{align}
so that the edge affinities for small $c$ become
\begin{align}
    \mathcal{A}_{ij} = \pm 2c \Lambda_{ij} +\mathcal{O}(c^3)\,.
\end{align}

\begin{figure}
    \centering
    \includegraphics[width=\textwidth]{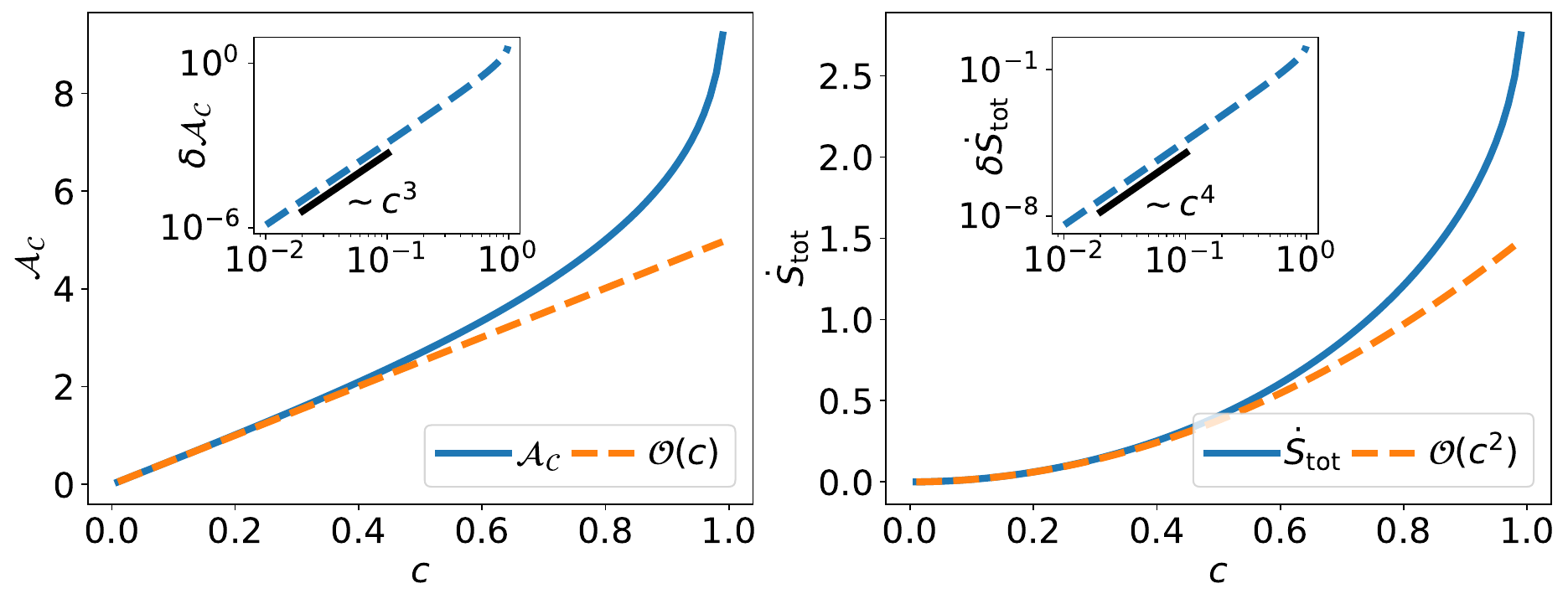}
    \caption[Affinity and entropy change with driving]{Cycle affinity $\mathcal{A}_\mathcal{C}$ (left) of a driven cycle and entropy production rate $\Dot{S}_\mathrm{tot}$ (right) in steady state for the toy model as functions of driving strength $c$. The cycle that is driven is $\mathcal{C}=\{1\to2,2\to3,3\to1\}$. The orange lines are the leading order terms in \Eqref{eq:AffinityDrivingStrength} and \Eqref{eq:EntropyRateDrivingStrength}, respectively. The insets show the difference of the exact and leading order of the cycle affinity $\delta \mathcal{A}_\mathcal{C} = \mathcal{A}_\mathcal{C} - 2c\Lambda_\mathcal{C}$ (left) and entropy production rate $\delta \Dot{S}_\mathrm{tot} = \Dot{S}_\mathrm{tot} - \frac{4c^2}{Z}\Lambda_\mathcal{C}\min_{l\to z\in \mathcal{C}}(\mathrm{e}^{-(E_l+E_z)/2})$ (right).}
    \label{fig:AffinityEntropyDriving}
\end{figure}

Considering the small $c$ expansion of a single cycle $\mathcal{C}$ in steady state, we get that the entropy production rate scales as $c^2$
\begin{align}
    \Dot{S}_\mathrm{tot} &= \frac{1}{Z}\sum_{i, j\in\Omega,i\neq j} j_{ij}\left(\pm 2c\Lambda_{ij} +\mathcal{O}(c^3)\right) = \frac{4c^2}{Z}\Lambda_\mathcal{C}\min_{l\to z\in \mathcal{C}}(\mathrm{e}^{-(E_l+E_z)/2}) +\mathcal{O}(c^4)\,.
    \label{eq:EntropyRateDrivingStrength}
\end{align}
The relation between $\Dot{S}_\mathrm{tot}$ and $c$, as well as $\mathcal{A}_\mathrm{C}$ and $c$, are shown in Fig.~\ref{fig:AffinityEntropyDriving} for the toy model. A good agreement is observed between the exact and leading order solutions, Eqs.~\eqref{eq:AffinityDrivingStrength} and \eqref{eq:EntropyRateDrivingStrength}, for small driving strengths $c$. The subset then shows that the expected next-to-leading order dominates for small $c$ when subtracting the leading order. For $c\to 1$ $\exists \mathcal{A}_{ij}$, i.e.,\footnote{This expression is valid for all $c\in[0,1)$}
\begin{align}
    \mathcal{A}_{ij} = \log\frac{\mathrm{e}^{-(E_i+E_j)/2} + x}{\mathrm{e}^{-(E_i+E_j)/2} - x}\,,
\end{align}
with $i\to j\in\mathcal{C}$, such that $\lim_{c\to1}\mathcal{A}_{ij}=\infty$ and, hence, 
\begin{equation}
    \begin{aligned}
        \lim_{c\to 1}\mathcal{A}_\mathcal{C} = \lim_{c\to 1}\Dot{S}_\mathrm{tot} = \infty\,.
    \end{aligned}
\end{equation}
The start of this divergence is also seen in Fig.~\ref{fig:AffinityEntropyDriving}.

\section{Invariance of Steady State Semi-Variance under Systematic Driving\label{sec:InvariantSemiVariance}}
Let $r_{xy}$ be the rates of a system driven along a single cycle $\mathcal{C}$ according to Sec.~\ref{sec:DrivingSystems} with some strength $0<c<1$ and $r_{xy}^\mathrm{s}$ the corresponding DB rates. Consider a $C$ independent of the rates and time, e.g., $C = \langle z\rangle_\mathrm{s}$, then the semi-variance in steady state can be identified as
\begin{equation}
    \begin{aligned}
        \mathrm{var}^C_\mathrm{ps}(z_x+z_y) = \frac{\sum_{x, y\in\Omega}(z_x+z_y-2C)^2\frac{(p_x^\mathrm{s}r_{xy} - p_y^\mathrm{s}r_{yx})^2}{p_x^\mathrm{s}r_{xy} + p_y^\mathrm{s}r_{yx}}}{\sum_{x, y\in\Omega}\frac{(p_x^\mathrm{s}r_{xy} - p_y^\mathrm{s}r_{yx})^2}{p_x^\mathrm{s}r_{xy} + p_y^\mathrm{s}r_{yx}}}\,.
    \end{aligned}
\end{equation}
As $\Sigma_\mathrm{ps}\propto c^2$, since
\begin{equation}
    p_x^\mathrm{s}r_{xy} - p_y^\mathrm{s}r_{yx} =
    \begin{cases}
        2c j_{xy} & x\to y \in \mathcal{C}\cup\Tilde{\mathcal{C}}\,,\\
        0 & \text{else}\,,
    \end{cases}
\end{equation}
we can write $\Sigma_\mathrm{ps}=\Xi c^2$ with $\Xi = 2\sum_{x, y\in\Omega}j_{xy}^2/r_{xy}^\mathrm{s}p_x^\mathrm{s}$. Hence, we get that
\begin{equation}
    \begin{aligned}
        \frac{\frac{(p_x^\mathrm{s}r_{xy} - p_y^\mathrm{s}r_{yx})^2}{p_x^\mathrm{s}r_{xy} + p_y^\mathrm{s}r_{yx}}}{\Sigma_{\mathrm{ps}}} = 
        \begin{cases}
            \frac{2j_{xy}^2}{r_{xy}^\mathrm{s}p_x^\mathrm{s}\Xi}& x\to y \in \mathcal{C}\cup\Tilde{\mathcal{C}}\,,\\
            0 & \text{else}\,.
        \end{cases}
    \end{aligned}
\end{equation}
In other words, if a single cycle is driven with this method, the semi-variance will for $t\to\infty$ converge to a constant value regardless of of driving strength $c$. 

\subsection{Spectral Representation for Long Time-Limit and Saturation\label{sec:CorrelationSpectral}}
In the long time limit, i.e., taking $t\to\infty$, the equilibrium correlation bound \Eqref{eq:eqcorrbound} simplifies to 
\begin{equation}
    \begin{aligned}[b]
        {\left(\sum_{\substack{x, y\in\Omega\\x\neq y}} (z_y-z_x)^2r_{xy}p_x^\mathrm{eq}\right)} \int_0^\infty\mathrm{d}\tau\mathrm{cov}_\mathrm{eq}({\omega}(\tau), \omega(0)) \geq 2\left(\mathrm{cov}_\mathrm{eq}(z, \omega) \right)^2\,.
    \end{aligned}
    \label{eq:eqcorrboundLongT}
\end{equation}
To better recognise the saturation and general behaviour of this bound, we use the spectral representation of the observables.

We start by defining a Hilbert space with scalar product $\bra{\boldsymbol{v}}\ket{\boldsymbol{u}} = \sum_{x\in\Omega}v_xp_x^\mathrm{eq}u_x$ and the diagonal matrix $\boldsymbol{P}^\mathrm{eq}=\mathrm{diag}(p_1^\mathrm{eq},p_2^\mathrm{eq},\dots)$ with equilibrium probabilities of $\mathbf{L}$ as entries. 
Next, we define the vectors $\boldsymbol{\psi}^{(i)}$ such that $\mathbf{L}\boldsymbol{P}^\mathrm{eq}\boldsymbol{\psi}^{(i)} = \lambda_i\boldsymbol{P}^\mathrm{eq}\boldsymbol{\psi}^{(i)}$, i.e., that $\boldsymbol{P}^\mathrm{eq}\boldsymbol{\psi}^{(i)}$ is the $i$-th eigenvector of $\mathbf{L}$ to the eigenvalue $\lambda_i$. The eigenvalues are ordered\footnote{Since we consider an equilibrium system, the eigenvalues are real and, thus, can be ordered.} $0=\lambda_0 > \lambda_1 \geq \lambda_2\geq \dots$ and the $\boldsymbol{\psi}^{(i)}$ are orthonormal in the induced norm
\begin{align}
    \bra{\boldsymbol{\psi}^{(i)}}\ket{\boldsymbol{\psi}^{(j)}} = \sum_{x\in\Omega} \boldsymbol{\psi}^{(i)}_xp_x^\mathrm{eq}\boldsymbol{\psi}^{(j)}_x = \delta_{ij}\,,
\end{align}
and one can easily identify $\boldsymbol{\psi}^{(0)}_x=1$ as the constant one vector. Thus, the operator $\mathbf{L}$ can be written as
\begin{align}
    \mathbf{L} = \sum_{i\geq 0}\lambda_i\ket{\boldsymbol{\psi}^{(i)}}\bra{\boldsymbol{\psi}^{(i)}}\,,
\end{align}
because
\begin{align}
    \bra{\boldsymbol{\psi}^{(i)}}\mathbf{L}\ket{\boldsymbol{\psi}^{(i)}} = \sum_{x,y\in\Omega} \boldsymbol{\psi}^{(i)}_xL_{xy}p_y^\mathrm{eq}\boldsymbol{\psi}^{(i)}_y = \lambda_i \sum_{x\in\Omega} \boldsymbol{\psi}^{(i)}_xp_x^\mathrm{eq}\boldsymbol{\psi}^{(i)}_x = \lambda_i\,.
\end{align}
Since the $\boldsymbol{\psi}^{(i)}$ form an orthonormal basis of the Hilbert space, we can write any state function $\boldsymbol{z} = (z_1, z_2, \dots)^T$ as a linear combination
\begin{align}
    z_x = \sum_{i\geq 0} c_i^z \boldsymbol{\psi}^{(i)}_x\,,
\end{align}
where the coefficients $c_i$ are
\begin{align}
    c_i^z = \bra{\boldsymbol{z}}\ket{\boldsymbol{\psi}^{(i)}} = \sum_{x\in\Omega}\boldsymbol{\psi}^{(i)}_xp_x^\mathrm{eq}z_x\,.
\end{align}
A consequence of this is that the equilibrium average value of $\boldsymbol{z}$ can be written as
\begin{align*}
    \langle \boldsymbol{z}\rangle_\mathrm{eq} = \sum_{x\in\Omega} z_xp_x^\mathrm{eq} = \sum_{x\in\Omega}\sum_{i\geq 0}c_i^z \boldsymbol{\psi}^{(i)}_xp_x^\mathrm{eq} = \sum_{i\geq 0} c_i^z\bra{\boldsymbol{\psi}^{(i)}}\ket{\boldsymbol{\psi}^{(0)}} = c_0^z\,.
\end{align*}

Turning to \Eqref{eq:eqcorrboundLongT}, we can write each part of the inequality in spectral representation. Specifically, we want to write the inequality in terms of coefficients $c_i$ and eigenvalues $\lambda_i$. The simplest is the r.h.s. of \Eqref{eq:eqcorrboundLongT}
\begin{equation}
    \begin{aligned}[b]
        \mathrm{cov}_\mathrm{eq}(\omega, z) &= \sum_{x\in\Omega} (\omega_x - \langle\omega\rangle _\mathrm{eq})(z_x - \langle z\rangle _\mathrm{eq}) p_x^\mathrm{eq}\\
        &= \sum_{x\in\Omega} \omega_xz_xp_x^\mathrm{eq} - c_0^zc_0^\omega\\
        &= \sum_{i,j\geq 0} c_i^zc_j^\omega \underbrace{\sum_x \boldsymbol{\psi}^{(i)}_x\boldsymbol{\psi}^{(j)}_xp_x^\mathrm{eq}}_{=\bra{\boldsymbol{\psi}^{(i)}}\ket{\boldsymbol{\psi}^{(j)}}=\delta_{ij}} - c_0^zc_0^\omega\\
        &=\sum_{i>0}c_i^zc_i^\omega\,.
    \end{aligned}
\end{equation}
The integrated covariance is
\begin{equation}
    \begin{aligned}[b]
        \int_0^\infty \mathrm{d}\tau \mathrm{cov}(\omega(\tau), \omega(0)) &= \int_0^\infty\mathrm{d}\tau \sum_{x,y\in\Omega} (\omega_x - \langle \omega\rangle_\mathrm{eq})(\omega_y - \langle \omega\rangle_\mathrm{eq})P(x, \tau|y) p_y^\mathrm{eq}\\
        &= \sum_{x,y\in\Omega} (\omega_x - \langle \omega\rangle_\mathrm{eq})(\omega_y - \langle \omega\rangle_\mathrm{eq})\sum_{i\geq 0}p_x^\mathrm{eq}\boldsymbol{\psi}^{(i)}_x \boldsymbol{\psi}^{(i)}_y p_y^\mathrm{eq}\int_0^\infty\mathrm{d}\tau \mathrm{e}^{\lambda_i\tau}\\
        &= \sum_{i\geq 0}\left(\int_0^\infty\mathrm{d}\tau \mathrm{e}^{\lambda_i\tau}\right)\left[\sum_{x\in\Omega}(\omega_x - c_0^\omega)\boldsymbol{\psi}^{(i)}_xp_x^\mathrm{eq}\right]^2
        = -\sum_{i > 0} \frac{\left(c_i^\omega\right)^2}{\lambda_i}\,,
    \end{aligned}
\end{equation}
and the fluctuation-scale function is
\begin{equation}
    \begin{aligned}[b]
        \mathcal{D}_t 
        &= \sum_{x,y\in\Omega} (z_y^2 + z_x^2-2z_xz_y) \sum_{i\geq0} p_y^\mathrm{eq}\boldsymbol{\psi}^{(i)}_y\lambda_i\boldsymbol{\psi}^{(i)}_xp_x^\mathrm{eq}
        \\
        &= \sum_{i\geq0} \lambda_i \left[2\delta_{i0}\sum_{y\in\Omega} z_y^2p_y^\mathrm{eq}\boldsymbol{\psi}^{(i)}_y - 2(c_i^z)^2\right]
        = -2\sum_{i>0} \lambda_i(c_i^z)^2\,.
    \end{aligned}
\end{equation}
Hence, the bound \Eqref{eq:eqcorrboundLongT} can be written as
\begin{equation}
    \begin{aligned}
        \left(\sum_{i>0} \lambda_i(c_i^z)^2\right)\left(\sum_{i > 0} \frac{\left(c_i^\omega\right)^2}{\lambda_i}\right) \geq \left(\sum_{i>0}c_i^zc_i^\omega\right)^2\,.
    \end{aligned}
\end{equation}
This is saturates for $c_i^z = c_i^\omega / \lambda_i\,\,\forall i>0$.

\section{Various Counterexamples}
\subsection{Non-Diagonalisable Master Operator\label{sec:NonDiagonalisable}}
Consider the following three-state generator of a possibly driven system
\begin{equation}
    \begin{aligned}
        \mathbf{L}=
        \begin{pmatrix}
            -1-a & b & 1\\
            1 & -1-b & b\\
            a & 1 & -1-b
        \end{pmatrix}\,.
    \end{aligned}
\end{equation}
If $a=b^{-2}$, this system will not be driven which can be seen from a cycle affinity of zero
\begin{equation}
    \begin{aligned}
        \mathcal{A}_\mathcal{C} = \log\left(ab^2\right) = 0\,.
    \end{aligned}
\end{equation}
However, if $a\neq 1/b^2$, this is non-zero and the system is thus driven. Let $a=5$ and $b=2$, then $\mathbf{L}$ has eigenvalues $\lambda_0=0$ and $\lambda_1=\lambda_2=-6$. The NESS probability distribution is then $\mathbf{p}^\mathrm{s}=(7, 13, 16)/36$, while the eigenvector to $\lambda_1$ and $\lambda_2$ is $\mathbf{v} = (-1, -1, 2)$. In other words, the span of the two times degenerate eigenvalue $-6$ is one-dimensional. Hence, the algebraic and geometric multiplicity does not match, which implies that $\mathbf{L}$ is not diagonalisable for this choice of $a$ and $b$. 

\subsection{Counterexample to Construction of Transport Bound Saturating State Functions \label{sec:TBSaturationCounterExample}}
    Consider a fully connected three-state system $\Omega=\{1,2,3\}$. Let $a,b,c>0$ and
\begin{equation}
    \begin{aligned}
        \mathbf{L}=
        \begin{pmatrix}
            -1-a & b & 1\\
            1 & -1-b & c\\
            a & 1 & -1-c
        \end{pmatrix}\,.
    \end{aligned}
\end{equation}
The stationary distribution is then
\begin{equation}
    \begin{aligned}
        \mathbf{p}^\mathrm{s} = 
        \frac{1}{Z}\begin{pmatrix}
            \frac{1+b+bc}{1+a+ab}\\
            \frac{1+c+ac}{1+a+ab}\\
            1
        \end{pmatrix}\,,
    \end{aligned}
\end{equation}
where $Z$ is the appropriate normalisation. We need to solve the following system of equations to saturate Cauchy-Schwarz
\begin{equation}
    \begin{aligned}
        z_2 - z_1 &= Z_{12}\,,\\
        z_3 - z_1 &= Z_{13}\,,\\
        z_3 - z_2 &= Z_{23}\,,
    \end{aligned}
    \label{eq:ExampleTBSaturation}
\end{equation}
where we assume $c(\tau) = 1$. This, of course, yields the conditions $z_2 = z_1 + Z_{12}$, $z_3 = z_1 + Z_{13}$, and $Z_{23} = Z_{13} - Z_{12}$. The last condition we can write out to be
\begin{equation}
    \begin{aligned}
        \frac{1-abc}{1+2c+2bc+abc} = \frac{abc-1}{1+2a+2ab+abc} - \frac{1-abc}{1+2b+2bc+abc}\,.
    \end{aligned}
\end{equation}
This only holds for $abc = 1$, e.g., in an equilibrium system with $a=b=c=1$. Considering $abc\neq 1$, then the condition becomes
\begin{equation}
    \begin{aligned}
        \frac{1}{1+2c+2bc+abc} \stackrel{?}{=} -\underbrace{\left(\frac{1}{1+2a+2ab+abc} + \frac{1}{1+2b+2bc+abc}\right)}_{>0}\,.
    \end{aligned}
\end{equation}
As $0<a,b,c<\infty$ by definition, this has no solution. Hence, we can only saturate the three-state example if $abc=1$, i.e., when the system is not driven. However, in this case, there is no entropy production. That this does not saturate for stationary driven systems is not a problem, as the transport bound yields a valid but trivial bound in this case. We finally look at transient systems, where $\mathbf{p}(\tau)$ is the probability vector and 
\begin{equation}
    \begin{aligned}
        Z_{12}(\tau) = \frac{p_1(\tau) - bp_2(\tau)}{p_1(\tau) + bp_2(\tau)}\,,\quad Z_{23}(\tau) = \frac{p_2(\tau) - cp_3(\tau)}{p_2(\tau) + cp_3(\tau)}\,, \quad Z_{13}(\tau) = \frac{ap_1(\tau) - p_3(\tau)}{ap_1(\tau) + p_3(\tau)}\,,
    \end{aligned}
\end{equation}
so that a solution of \Eqref{eq:ExampleTBSaturation} requires
\begin{equation}
    \begin{aligned}
        \frac{p_2(\tau) - cp_3(\tau)}{p_2(\tau) + cp_3(\tau)} = \frac{ap_1(\tau) - p_3(\tau)}{ap_1(\tau) + p_3(\tau)} - \frac{p_1(\tau) - bp_2(\tau)}{p_1(\tau) + bp_2(\tau)}\,.
    \end{aligned}
    \label{eq:ExampleTBSaturation2}
\end{equation}
For this to hold, the signs have to match, i.e., $\mathrm{sgn}(p_2(\tau) - cp_3(\tau)) = \mathrm{sgn}(abp_2(\tau) - p_3(\tau))$. One can immediately see that this is fulfilled for $abc=1$. On the other hand, one cannot satisfy this in general. For example, if $a=b=c=2$,  $p_1(0) = 0$, and $p_2(0) = p_3(0)=1/2$, then the l.h.s. of \Eqref{eq:ExampleTBSaturation2} is $-1/3$ while the r.h.s. is $0$. Hence, we generally cannot construct a state function $z_i(\tau)$ out of the $Z_{ij}(\tau)$ so that the transport bound saturates. 

\subsection{Indispensability of the Modified Current}

\begin{figure}
    \centering
    \includegraphics[width=0.7\textwidth]{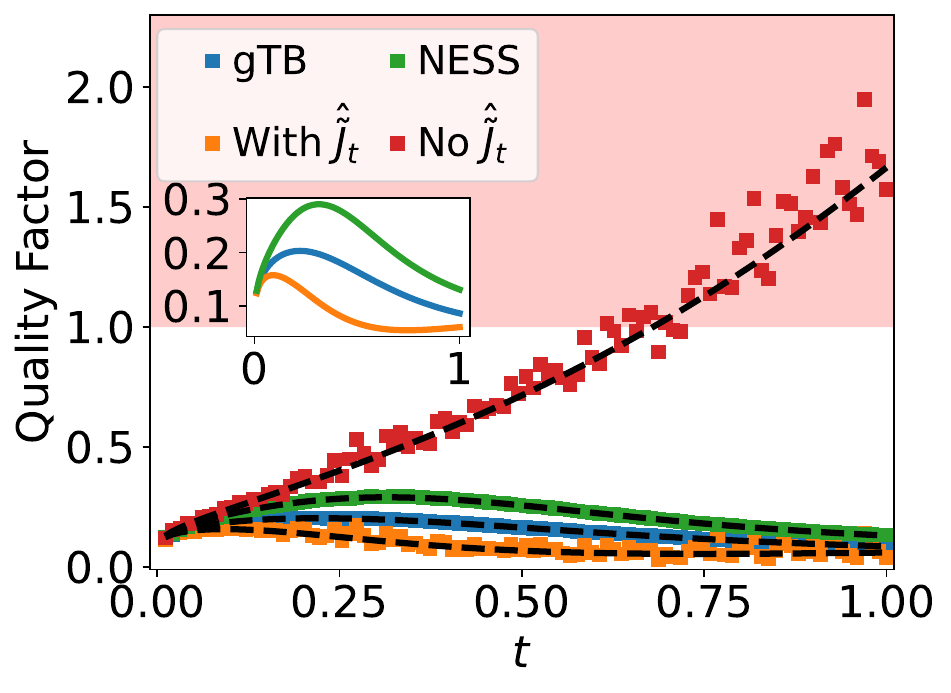}
    \caption[Necessity for modified current]{Validity and comparison of generalised transport bound and various TUR quality factors. A three-state system with generator \Eqref{eq:MoCurGenerator} and initial condition $p_i(0) = (\delta_{i1} + 2\delta_{i2})/3$. The metric is $d_{ij}(\tau) = \mathrm{e}^{4\tau}\left(\delta_{i1}\delta_{j2}-\delta_{i2}\delta_{j1}\right)$. The black dashed lines are the theoretical predictions to the corresponding numerical quality factors of generalised transport bound \Eqref{eq:GeneralisedTransport} (blue), NESS TUR (green), transient TUR with $ \hat{\Tilde{J}}_t$ (orange), and transient TUR without $ \hat{\Tilde{J}}_t$ (red). A clear violation (shaded red region) of $\mathcal{Q}\leq 1$ is seen for the transient TUR without $ \hat{\Tilde{J}}_t$. The inset shows the theory of the former three quality factors to better visualise these. The numerical values are extracted from $N=5\cdot 10^5$ trajectories using the celebrated Gillespie algorithm.}
    \label{fig:QualityFactorBreaking}
\end{figure}

Consider a three-state system with generator
\begin{equation}
    \begin{aligned}
        \mathbf{L} = 
        \begin{pmatrix}
        -4 & 2 & 1\\
        1 & -3 & 1\\
        3 & 1 & -2
        \end{pmatrix}\,,
    \end{aligned}
    \label{eq:MoCurGenerator}
\end{equation}
with eigenvalues $0, -4, -5$. Let the current defining metric be  $d_{ij}(\tau) = \mathrm{e}^{4\tau}\left(\delta_{i1}\delta_{j2}-\delta_{i2}\delta_{j1}\right)$. Figure~\ref{fig:QualityFactorBreaking} shows various quality factors of the transient system with initial distribution $p_i(0) = (\delta_{i1} + 2\delta_{i2})/3$. Explicitly, the generalised transport bound \Eqref{eq:GeneralisedTransport}, NESS TUR \Eqref{eq:SSTURshifted} (with $c(t)=0$), and the transient TUR \Eqref{eq:TURtransientversion} with and without modified current $\hat{\Tilde{J}}_t$, see \Eqref{eq:modifiedCurrent}. As can be seen in Fig.~\ref{fig:QualityFactorBreaking}, one needs $\hat{\Tilde{J}}_t$ so that the bound $\mathcal{Q}\leq 1$ is not violated. Mathematically, one can show that to leading order $t\partial_t\langle \hat{J}_t\rangle \sim t\mathrm{e}^{4t}$ for large $t$, while $t\partial_t\langle \hat{J}_t\rangle - \langle \hat{\Tilde{J}}_t\rangle \sim \mathrm{e}^{4t}$. Additionally, $\mathrm{var}(\hat{J}_t)\Delta S_\mathrm{tot}(t)\sim t\mathrm{e}^{8t}$ for large $t$, so that
\begin{equation}
    \begin{aligned}[b]
        \frac{2\left(t\partial_t\langle \hat{J}_t\rangle\right)^2}{\mathrm{var}(\hat{J}_t)\Delta S_\mathrm{tot}(t)}&\sim t\,,\\
        \frac{2\left(t\partial_t\langle \hat{J}_t\rangle - \langle \hat{\Tilde{J}}_t\rangle\right)^2}{\mathrm{var}(\hat{J}_t)\Delta S_\mathrm{tot}(t)}&\sim \frac{1}{t}\,.
    \end{aligned}
\end{equation}
Hence, the inclusion of the modified current is necessary for the quality factor to remain bounded, as it otherwise diverges for $t\to\infty$, which obviously is a violation of the upper bound of the quality factor. Note that $t$ in Fig.~\ref{fig:QualityFactorBreaking} is not large enough to observe the $t^{-1}$ decay.

\section{Limit of Steady State for Driving Single Edge}

\begin{prop}\label{prop:ConvergenceEdgeDrivingSteadyState}
    Let $\mathbf{L}(\lambda)\in M(N, \mathds{R})$ be a generator of a MJP on a graph $\mathcal{G}=(\Omega, \mathcal{V})$ with $L_{lk} =r_{kl}= \lambda>0$ for fixed $l,k\in\Omega$ with $l\neq k$. If $\mathbf{p}(\lambda)$ is the corresponding steady state, then $\exists \mathbf{p}^\infty\in\mathds{R}_{\geq 0}^N$ s.t. $\mathbf{p}^\infty=\lim_{\lambda\to\infty}\mathbf{p}(\lambda)$ element wise with ${p}^\infty_k=0$. This ${p}^\infty_n$, $n\in\Omega\setminus\{k\}$, is the steady state of a directed graph $\Tilde{\mathcal{G}}=(\Tilde{\Omega}, \Tilde{\mathcal{V}})$ where $\Tilde{\Omega}=\Omega\setminus\{k\}$ and 
    \begin{equation}
        \begin{aligned}
            \Tilde{\mathcal{V}} = \mathcal{V}\cup\left(\bigcup_{\substack{i\in\Omega\setminus\{l\}\\ i\to k\in\mathcal{V}}}\{i\to l\}\right)\setminus \left(\bigcup_{i\in\Omega}\{i\to k\}\cup\{k\to i\}\right)\,.
        \end{aligned}
        \label{eq:PropSteadyStateLimitSet}
    \end{equation}
\end{prop}
\begin{proof}
    For every $\lambda\in\mathds{R}_{>0}$, the steady state conditions require $\forall i\in\Omega$
    \begin{equation}
        \begin{aligned}
            \sum_{j\in\Omega} r_{ji}(\lambda) p_j(\lambda) = 0\,.
        \end{aligned}
        \label{eq:PropProofCurrentBalanceGeneral}
    \end{equation}
    In particular, considering $i=l$, then
    \begin{equation}
        \begin{aligned}
            \lambda p_k(\lambda) +\sum_{j\in\Omega\setminus\{k\}} r_{jl}p_j(\lambda)=0 \,,
        \end{aligned}
    \end{equation}
    and for $i=k$ it yields
    \begin{equation}
        \begin{aligned}
            0< \min_{\substack{j\in\Omega\setminus\{k\}}}r_{jk}\leq \underbrace{\left(\lambda + \sum_{j\in\Omega\setminus\{k,l\}}r_{kj}\right)}_{=-r_{kk}}p_k(\lambda) \stackrel{\text{Eq.~\eqref{eq:PropProofCurrentBalanceGeneral}}}{=} \sum_{j\in\Omega\setminus\{k\}}r_{jk}p_j(\lambda)\leq \sum_{j\in\Omega\setminus\{k\}}r_{jk}\,.
        \end{aligned}
        \label{eq:PropProofCurrentBalance}
    \end{equation}
    Because the right side of \Eqref{eq:PropProofCurrentBalance} is finite, the left side has to remain finite as well. Introducing the labeling $C_0 = \min_{\substack{j\in\Omega\setminus\{k\}}}r_{jk}$, $C_1 = \sum_{j\in\Omega\setminus\{k\}}r_{jk}$, and $C_2 = \sum_{\substack{j\in\Omega\setminus\{ k,l\}}}r_{kj}$ we get that
    \begin{equation}
        \begin{aligned}
            \frac{C_0}{\lambda + C_2} \leq p_k(\lambda) \leq \frac{C_1}{\lambda + C_2}\,.
        \end{aligned}
    \end{equation}
    Since $C_0$ and $C_1$ are independent of $\lambda$, the probability $p_k(\lambda)$ has to decay like $1/\lambda$ in the limit of large $\lambda$.
    
    This means that the probability fluxes from $k$ to all $j\neq l$ scale as $1/\lambda$. Using \Eqref{eq:PropProofCurrentBalanceGeneral} for $i\neq l,k$ yields that all remaining probabilities are $\mathcal{O}(1)$ in the limit of larger $\lambda$. Hence, the net flux $\mathcal{J}_{jk}(\lambda) = J_{jk}(\lambda) - J_{kj}(\lambda)$ (see \Eqref{eq:NetFlux}) into node $k$ from $j\neq l$ remains $\mathcal{O}(1)$ and, in particular, it becomes the net probability flux from $j$ to $k$ for $\lambda\to\infty$, i.e., $|\mathcal{J}_{jk}(\lambda)- J_{jk}(\lambda)|\sim 1/\lambda\xrightarrow{\lambda\to\infty}0$. 

    Consider now a graph $\Tilde{\mathcal{G}} = (\Tilde{\Omega}, \Tilde{\mathcal{V}})$, where $\Tilde{\Omega} = \Omega\setminus \{k\}$ and $\Tilde{\mathcal{V}}$ as in \Eqref{eq:PropSteadyStateLimitSet}. The master operator of this system can be written as the sum of two linear operators $\Tilde{\mathbf{T}}\in M(N-1, \mathds{R})$ and $\Tilde{\mathbf{S}}\in M(N-1, \mathds{R})$. The former is the master operator on the original graph with node $k$ and all connecting edges removed, i.e., on $\Tilde{\mathcal{G}}$ without the edges connecting $i\neq l$ to $k$. The latter operator then describes exactly these additional edges. That is
    \begin{equation}
        \begin{aligned}[b]
            (\Tilde{\mathbf{T}})_{ij} = 
            \begin{cases}
                r_{ji} &  i\neq k \neq j \neq i\,,\\
                -\sum_{\substack{n\in\Tilde{\Omega}\\ n\neq j}} r_{jn} & i = j \neq k\,,
            \end{cases}
        \end{aligned}
    \end{equation}
    and
    \begin{equation}
        \begin{aligned}[b]
            (\Tilde{\mathbf{S}})_{ij} = 
            \begin{cases}
                r_{jk} &  i=l\neq j \text{ and } j\to k\in\mathcal{V}\,,\\
                - r_{jk} & i = j \neq l \text{ and } j\to k\in\mathcal{V}\,.
            \end{cases}
        \end{aligned}
    \end{equation}
    Since the sum of these, $\Tilde{\mathbf{L}} = \Tilde{\mathbf{T}} + \Tilde{\mathbf{S}}$, is a generator for a directed MJP, the corresponding master equation has a unique steady state probability solution $\Tilde{\mathbf{p}}^\infty\in\ker(\Tilde{\mathbf{L}})$. A small four-state example is shown in Fig.~\ref{fig:proofSketchMapping}. 

    What remains is to show that $p_i(\lambda)\to \Tilde{p}^\infty_i$ $\forall i\in\Tilde{\Omega}$, as we already have established that $p_k(\lambda)\xrightarrow{\lambda\to\infty}0$. For this, we first recognise that $\partial_t p_k(\lambda) = -\lambda p_k(\lambda) + r_{lk}p_l(\lambda) + \sum_{\substack{j\in\Omega\\ j\neq k,l}}r_{jk} p_j(\lambda) + \mathcal{O}(1/\lambda) = 0$. The first two terms are (modulo minus sign) the net flux into site $l$ from $k$, while the sum is the flow into $k$ from the remaining sites. Furthermore, it follows immediately that 
    \begin{equation}
        \begin{aligned}
            \lambda p_k(\lambda) - r_{lk}p_l(\lambda)+ \mathcal{O}(1/\lambda) = \sum_{\substack{j\in\Omega\\ j\neq k,l}}r_{jk} p_j(\lambda)\,.
        \end{aligned}
        \label{eq:ProofPropSteadyStateBalance}
    \end{equation} 
    So redirecting the flow from $j\neq l$ to $k$ into $l$ recovers the structure of $\Tilde{\mathbf{L}}$. Let $\Tilde{\mathbf{L}}^+, \Tilde{\mathbf{T}}^+, \Tilde{\mathbf{S}}^+\in M(N, \mathds{R})$ be the extension of $\Tilde{\mathbf{L}},\Tilde{\mathbf{T}},\Tilde{\mathbf{S}}$ into the full graph $\mathcal{G}$ setting the $k$th row and column to zero, i.e.,
    \begin{equation}
        \begin{aligned}[b]
        \Tilde{{L}}^+_{ij} =
            \begin{cases}
                \Tilde{{L}}_{ij} & i,j\neq k\,,\\
                0 & i=k \lor j=k\,,
            \end{cases}
        \end{aligned}
    \end{equation}
    and analogously for $\Tilde{\mathbf{T}}$ and $\Tilde{\mathbf{S}}$. Since $\dim{\ker{\Tilde{\mathbf{L}}}} = 1$ (uniqueness of steady state, see Refs.~\cite{Schnakenberg, Horn_Johnson_2012}) and $\dim{\ker{\Tilde{\mathbf{L}}^+}} = \dim{\ker{\left(\text{span}(\mathbf{p}^\infty)\bigcup \text{span}(\mathbf{e}_k)\right)}} = 2$, with $\mathbf{p}^\infty=(\Tilde{p}^\infty_1, \dots, \Tilde{p}^\infty_{k-1}, 0, \Tilde{p}^\infty_{k+1}, \dots, \Tilde{p}^\infty_N)$ the extension of $\Tilde{\mathbf{p}}^\infty$ to the full space $\Omega$ and $\mathbf{e}_k$ the $k$th unit vector, we need to show that $\Tilde{\mathbf{L}}^+ (\mathbf{p}(\lambda) - \mathbf{p}^\infty)\xrightarrow{\lambda\to\infty} 0$. Since $\mathbf{p}^\infty\in\ker{\Tilde{\mathbf{L}}^+}$, what remains is to show that $\forall i\in\Omega$
    \begin{equation}
        \begin{aligned}
            (\Tilde{\mathbf{L}}^+ \mathbf{p}(\lambda))_i = \sum_{j\in\Omega}\Tilde{L}^+_{ij} {p}_j(\lambda)\xrightarrow{\lambda\to \infty} 0\,.
        \end{aligned}
        \label{eq:ProofPropConvergenceCondition}
    \end{equation}

    \begin{figure}
        \centering
        \includegraphics[width=.7\textwidth]{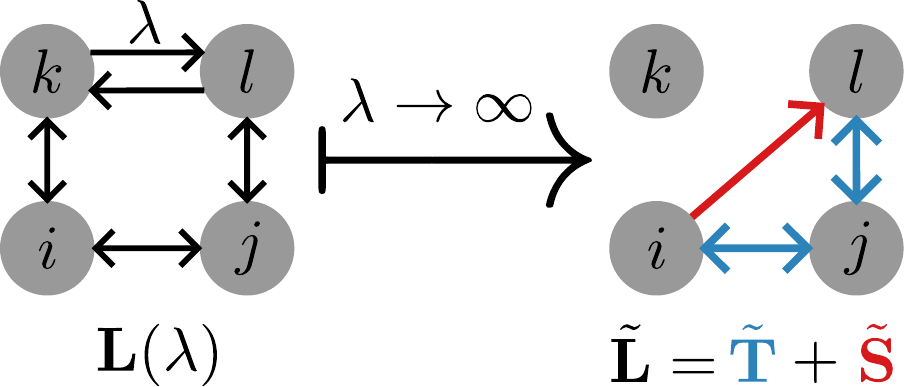}
        \caption{Sketch of mapping in the limit of single diverging transition rate $r_{kl}=\lambda\to\infty$. Here, the state spaces and edge sets are $\Omega = \{i,j,k,l\}$, $\Tilde{\Omega}=\{i, j, l\}$, ${\mathcal{V}=\bigcup_{\substack{n\in\{i,l\}\\m\in\{k,j\}}}(\{n\to m\}\cup \{m\to n\})}$, and ${\Tilde{\mathcal{V}}=\{i\to l\}\cup\bigcup_{n\in\{i,l\}}(\{n\to j\}\cup \{j\to n\})}$.}
        \label{fig:proofSketchMapping}
    \end{figure}
    
    For $i=k$ \Eqref{eq:ProofPropConvergenceCondition} is trivially satisfied. Let $i=l$, then
    \begin{equation}
        \begin{aligned}[b]
            (\Tilde{\mathbf{L}}^+ \mathbf{p}(\lambda))_l &= \sum_{j\in\Omega}(\Tilde{T}^+_{lj} + \Tilde{S}^+_{lj}) {p}_j(\lambda)\\
            &= \Tilde{T}_{ll}^+p_l(\lambda) + \underbrace{\Tilde{S}_{ll}^+}_{=0}p_l(\lambda) + \sum_{j\in\Omega\setminus\{l\}}(\Tilde{T}^+_{lj} + \Tilde{S}^+_{lj}) {p}_j(\lambda)\\
            &= -\sum_{n\in\Omega\setminus\{l,k\}}r_{ln}p_l(\lambda) + \sum_{j\in\Omega\setminus\{l,k\}}r_{jl}{p}_j(\lambda) + \sum_{j\in\Omega\setminus\{l,k\}}r_{jk} {p}_j(\lambda)\\
            &= \sum_{n\in\Omega\setminus\{l,k\}}\left[r_{nl}{p}_n(\lambda)-r_{ln}p_l(\lambda)\right] + \sum_{j\in\Omega\setminus\{l,k\}}r_{jk} {p}_j(\lambda)\\
            &\stackrel{\mathclap{\text{\Eqref{eq:ProofPropSteadyStateBalance}}}}{=}\sum_{n\in\Omega\setminus\{l,k\}}\left[r_{nl}{p}_n(\lambda)-r_{ln}p_l(\lambda)\right] +  \lambda p_k(\lambda) - r_{lk}p_l(\lambda)+ \mathcal{O}(1/\lambda)\\
            &=\underbrace{\sum_{n\in\Omega\setminus\{l\}}\left[r_{nl}{p}_n(\lambda)-r_{ln}p_l(\lambda)\right]}_{=\partial_\tau p_l(\lambda) = 0} + \mathcal{O}(1/\lambda)\xrightarrow{\lambda\to \infty} 0\,.
        \end{aligned}
    \end{equation}
    On the other hand, for $i\neq l,k$, we get
    \begin{equation}
        \begin{aligned}
            (\Tilde{\mathbf{L}}^+ \mathbf{p}(\lambda))_i &= \sum_{j\in\Omega}(\Tilde{T}^+_{ij} + \Tilde{S}^+_{ij}) {p}_j(\lambda)\\
            &= \Tilde{T}_{ii}^+p_i(\lambda) + \Tilde{S}_{ii}^+p_i(\lambda) + \sum_{j\in\Omega\setminus\{i\}}(\Tilde{T}^+_{ij} + \Tilde{S}^+_{ij}) {p}_j(\lambda)\\
            &= -\sum_{n\in\Omega\setminus\{i,k\}}r_{in}p_i(\lambda) - r_{ik}p_i(\lambda) + \sum_{n\in\Omega\setminus\{i,k\}}r_{ni}{p}_n(\lambda)\\
            &= -\sum_{n\in\Omega\setminus\{i\}}r_{in}p_i(\lambda) + \sum_{n\in\Omega\setminus\{i,k\}}r_{ni}{p}_n(\lambda) + r_{ki}p_{k}(\lambda) - r_{ki}p_{k}(\lambda)\\
            &= \underbrace{\sum_{n\in\Omega\setminus\{i\}}\left[r_{ni}{p}_n(\lambda) -r_{in}p_i(\lambda)\right]}_{=\partial_\tau p_i(\lambda) = 0} - \underbrace{r_{ki}p_{k}(\lambda)}_{=\mathcal{O}(1/\lambda)}\xrightarrow{\lambda\to\infty}0\,.
        \end{aligned}
    \end{equation}
    Hence, $\lim_{\lambda\to\infty}\mathbf{p}(\lambda)\in\ker \Tilde{\mathbf{L}}^+$ and, therefore, $\lim_{\lambda\to\infty}\mathbf{p}(\lambda) = \mathbf{p}^\infty$.
\end{proof}
\begin{example}
    Using the SAT Model in Sec.~\ref{sec:SecondaryTransport}, and choice of rates as in Figs. \ref{fig:SecondaryTransportModel} and \ref{fig:SATperiodic}, where we again vary the rate $r_{24} = e^\mathrm{out}_B=\lambda$. The master operator is then
    \begin{equation}
        \begin{aligned}
            \mathbf{L}(\lambda) = 
            \begin{pmatrix}
                -30 & 10 & 30 & 0 & 0 & 0\\
                10 & -30 & 0 & 10 & 0 & 0\\
                20 & 0 & -35-\lambda & 5 & 1 & 0\\
                0 & 20 & 5 & 17 & 0 & 1\\
                0 & 0 & \lambda & 0 & -11 & 10\\
                0 & 0 & 0 & 2 & 10 & -11
            \end{pmatrix}\,.
        \end{aligned}
    \end{equation}
    The steady state can be found to be
    \begin{equation}
        \begin{aligned}
            \mathbf{p}(\lambda) = 
            \frac{1}{1090 + 170\lambda}
            \begin{pmatrix}
                186 + \lambda\\
                132 + 3\lambda\\
                142\\
                210 + 8\lambda\\
                200 + 82\lambda\\
                220 + 76\lambda
            \end{pmatrix}\,.
        \end{aligned}
    \end{equation}
    The probability in $k=3$ can be seen to scale as $p_3(\lambda)\sim1/\lambda$ for large $\lambda$. In the limit $\lambda\to\infty$, the steady state becomes
    \begin{equation}
        \begin{aligned}
            \lim_{\lambda\to\infty}\mathbf{p}(\lambda)= 
            \frac{1}{170}
            \begin{pmatrix}
                1\\
                3\\
                0\\
                8\\
                82\\
                76
            \end{pmatrix}\,.
        \end{aligned}
    \end{equation}

    In this case, the master operator corresponding to $\Tilde{\mathcal{G}}$ is, in the numbering $(1, 2, 4, 5, 6)$,
    \begin{equation}
        \begin{aligned}[b]
            \Tilde{\mathbf{L}} &= 
            \begin{pmatrix}
                -x\gamma-l_A & x\gamma & 0 & 0 & 0\\
                x\gamma  & -x\gamma - l_A & e^\mathrm{in}_A & 0 & 0\\
                0 & l_A & -e^\mathrm{in}_A - e^\mathrm{in}_B - \gamma & 0 & l_B\\
                l_A & 0 & \gamma & -x\gamma & x\gamma\\
                0 & 0 & e^\mathrm{in}_B & x\gamma & -x\gamma-l_B
            \end{pmatrix}
            \\
            &=
            \begin{pmatrix}
                -30 & 10 & 0 & 0 & 0\\
                10  & -30 & 10 & 0 & 0\\
                0 & 20 & 17 & 0 & 1\\
                20 & 0 & 5 & -10 & 10\\
                0 & 0 & 2 & 10 & -11
            \end{pmatrix}\,,
        \end{aligned}
    \end{equation}
    where
    \begin{equation}
        \begin{aligned}[b]
            \Tilde{\mathbf{T}} &=
            \begin{pmatrix}
                -x\gamma & x\gamma & 0 & 0 & 0\\
                x\gamma  & -x\gamma - l_A & e^\mathrm{in}_A & 0 & 0\\
                0 & l_A & -e^\mathrm{in}_A - e^\mathrm{in}_B  & 0 & l_B\\
                0 & 0 & 0 & -x\gamma & x\gamma\\
                0 & 0 & e^\mathrm{in}_B & x\gamma & -x\gamma-l_B
            \end{pmatrix}\,,\\
            \Tilde{\mathbf{S}} &=
            \begin{pmatrix}
                -l_A & 0 & 0 & 0 & 0\\
                0  & 0 & 0 & 0 & 0\\
                0 & 0 & -\gamma  & 0 & 0\\
                l_A & 0 &\gamma & 0 & 0\\
                0 & 0 & 0 &0 & 0
            \end{pmatrix}\,,
        \end{aligned}
    \end{equation}
    with corresponding steady state
    \begin{equation}
        \begin{aligned}
            \Tilde{\mathbf{p}}^\infty= 
            \frac{1}{170}
            \begin{pmatrix}
                1\\
                3\\
                8\\
                82\\
                76
            \end{pmatrix}\,.
        \end{aligned}
        \label{eq:LimitSteadyState}
    \end{equation}
    Hence, $p_i(\lambda)\xrightarrow{\lambda\to\infty}\Tilde{p}_i^\infty$ for all $i\in\Tilde{\Omega}=\{1, 2, 4, 5, 6\}$. This can be seen in Fig.~\ref{fig:SteadyStateConvergence}, together with the expected $1/\lambda$ scaling of $p_3(\lambda)$.

    \begin{figure}[ht]
        \centering
        \includegraphics[width=.6\textwidth]{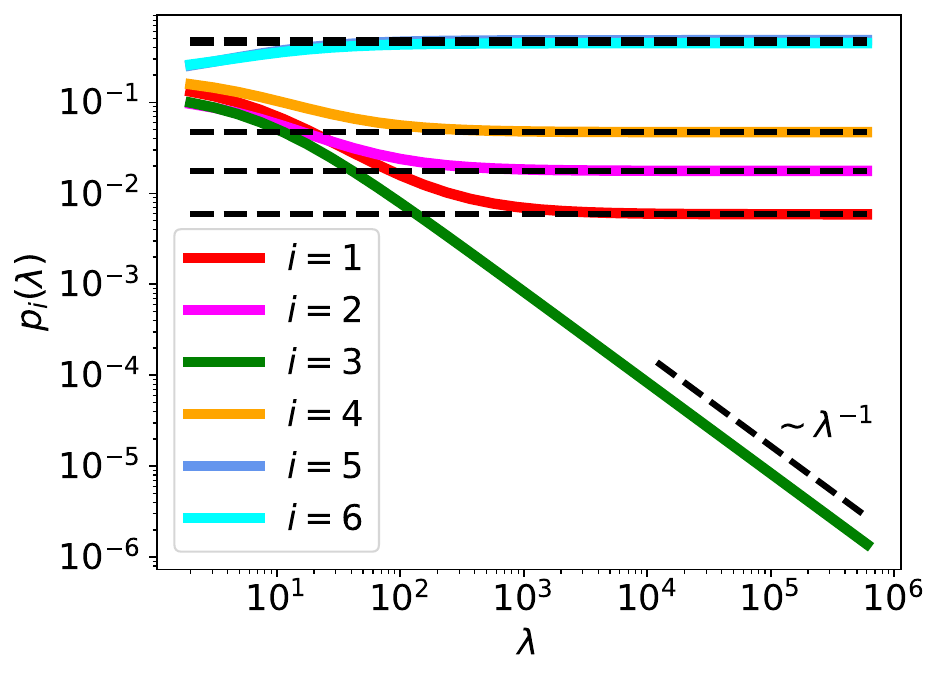}
        \caption[Example of convergence of NESS]{Steady state distribution in each state as a function of transition rate $\lambda = e^\mathrm{out}_B$ along the edge $3\to 5$ in the SAT model. The black dotted lines are the probability densities \Eqref{eq:LimitSteadyState} in the graph $\Tilde{\mathcal{G}}$, see Prop. \ref{prop:ConvergenceEdgeDrivingSteadyState}. }
        \label{fig:SteadyStateConvergence}
    \end{figure}
\end{example}

\section{Multi-Cycle Driving}
For completness, we present the extension of Sec.~\ref{sec:DrivingSystems} to multiple cycles. Suppose we want to drive $L$ cycles $\mathcal{C}_m$, $1\leq m\leq L$. The anti-symmetric coefficients entering in the rate \Eqref{eq:AntisymmetricRates} are the sum $j_{ij} = \sum_{m=1}^kj_{ij}^m$ of the respective coefficients from individual cycles $j^m_{ij}$. We can distinguish between three different cases, namely (i) independent, (ii) single-overlap, and (iii) multiple-overlap cycles. By defining overlap, we ignore the direction. In other words, if $i\to j\in\mathcal{C}_m$ and $j\to i\in\mathcal{C}_n$, we say that $i\to j$ is in the overlap of $\mathcal{C}_n$ and $\mathcal{C}_m$. For (i), the implementation follows trivially with \Eqref{eq:solution_j} for each cycle. If, however, the cycles share at least one edge, i.e., $\exists i\neq j \in \{1, \dots, L\}$ so that $(\mathcal{C}_i\cup \Tilde{\mathcal{C}}_i)\cap \mathcal{C}_j\neq \emptyset$, the cycles \emph{cannot} be driven independently. 

Consider that two cycles $\mathcal{C}_n$ and $\mathcal{C}_m$ sharing a single edge $k\to l$ overlap, which can be written as$(\mathcal{C}_n\cup \Tilde{\mathcal{C}}_n)\cap \mathcal{C}_m = \{k\to l\}$, where we assume that $k\to l\in\mathcal{C}_m$ without loss of generality. We denote the anti-symmetric coefficients \Eqref{eq:solution_j} of the cycles as $j_{ij}^n$ and $j_{ij}^m$, respectively, which are bounded as
\begin{equation}
    \begin{aligned}
        |j_{ij}^m| < \min_{x\to y\in\mathcal{C}_m}\left(\mathrm{e}^{-(E_x+E_y)/2}\right) = a_m\,, \quad |j_{ij}^n| < \min_{x\to y\in\mathcal{C}_n}\left(\mathrm{e}^{-(E_x+E_y)/2}\right) = a_n\,,
    \end{aligned}
\end{equation}
if the minimum is not on $k\to l$ for the respective cycles. Using the condition \Eqref{eq:DrivingCoefficientBound}, we require that $|j_{kl}^m + j_{kl}^n|<\mathrm{e}^{-(E_k+E_l)/2}$. Depending on whether the cycles are driven in the same or opposite direction along $k\to l$, this condition can be written as
\begin{equation}
    \begin{aligned}
        \mathrm{e}^{-(E_k+E_l)/2}>|j_{kl}^m + j_{kl}^n| = 
    \begin{cases}
        c_m a_m + c_n a_n   & \text{if }\mathrm{sgn}(j_{kl}^m) =  \mathrm{sgn}(j_{kl}^n)\,,\\
        |c_m a_m - c_n a_n| & \text{if }\mathrm{sgn}(j_{kl}^m) = -\mathrm{sgn}(j_{kl}^n)\,,
    \end{cases}
    \end{aligned}
    \label{eq:TwoCycleCondition}
\end{equation}
where $c_n$ and $c_m$ are the driving strength factors in \Eqref{eq:solution_j} for the individual cycles.  As $a_m$ and $a_n$ are determined by the cycles, \Eqref{eq:TwoCycleCondition} yields a condition on $c_n$ and $c_m$. Note that this condition may allow for either $c_n$ or $c_m$ to be $\geq 1$ if the driving is anti-parallel on $k\to l$ and the minimum of the corresponding cycle is on that overlap edge. This statement also holds for the remaining part of the section. If $L$ cycles $\mathcal{C}_{m}$, $\forall m\in\{1, \dots, L\}$, have a single edge $k\to l\in\mathcal{C}_{1}$ (again without loss of generality) overlap, then the condition on the coefficients $c_{m}$ for case (ii) is given as
\begin{equation}
    \begin{aligned}
        \left|\sum_{m=1}^L j_{kl}^{m}\right| < \mathrm{e}^{-(E_k+E_l)/2}\,.
    \end{aligned}
    \label{eq:GeneralSingleOverlapCycleCondition}
\end{equation}

\begin{figure}[H]
    \centering
    \includegraphics[width=0.5\textwidth]{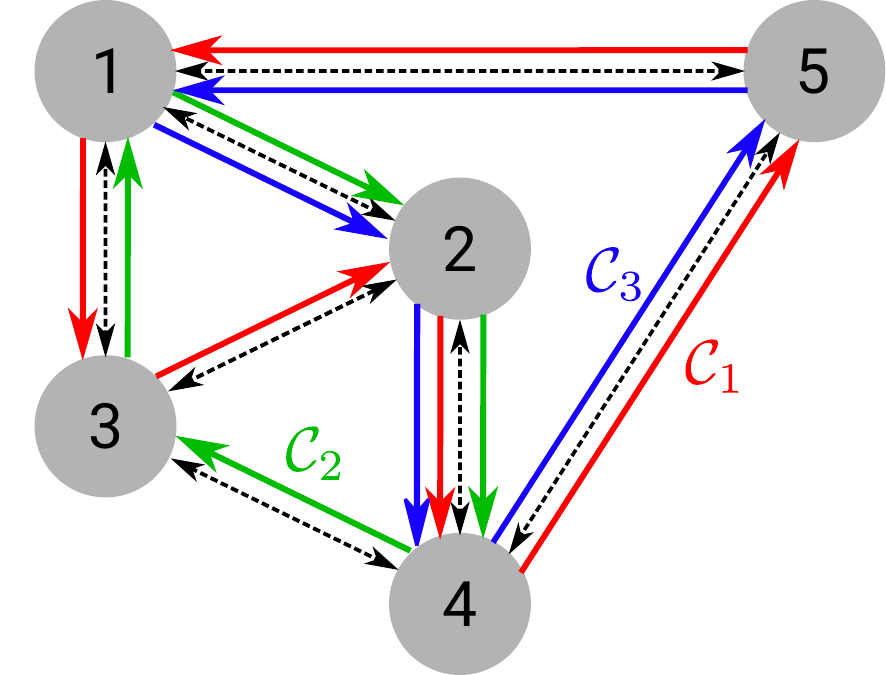}
    \caption[Cycle overlap example]{Toy model with three cycles $\mathcal{C}_1$ (red), $\mathcal{C}_2$ (green), and $\mathcal{C}_3$ (blue). The cycles have multiple overlaps, e.g., $\mathcal{C}_1$ and $\mathcal{C}_2$ have overlap $\{1\to3\}=\mathcal{C}_1\cap\Tilde{\mathcal{C}}_2$ and $\{2\to4\}= \mathcal{C}_1\cap\mathcal{C}_2$.}
    \label{fig:CycleOverlapExample}
\end{figure}

In the last case, i.e., case (iii), some more consideration has to go into finding the appropriate condition. For instance, the relative orientation of the overlap of cycles does not have to be the same for all edges in the overlap. An example is shown in Fig.~\ref{fig:CycleOverlapExample}, where the cycles $\mathcal{C}_1=\{1\to3,3\to2,2\to4,4\to5,5\to1\}$ and $\mathcal{C}_2=\{1\to2, 2\to4, 4\to3, 3\to1\}$ are shown for the toy model. Here, the overlap $\{1\to3\}=\mathcal{C}_1\cap\Tilde{\mathcal{C}}_2$ and $\{2\to4\}= \mathcal{C}_1\cap\mathcal{C}_2$ have opposing relative orientation, that is, the former overlap belongs to $\Tilde{\mathcal{C}}_2$ while the latter belongs to $\mathcal{C}_2$. Taking this into consideration, \Eqref{eq:TwoCycleCondition} generalises for two cycles to
\begin{equation}
    \begin{aligned}
        \left|j_{ij}^m + j_{ij}^n\right| < \min_{l\to k\in\mathcal{C}_n\cap\left(\mathcal{C}_m\cup\Tilde{\mathcal{C}}_m\right)}\left(\mathrm{e}^{-(E_l+E_k)/2}\right)\,.
    \end{aligned}
\end{equation}

For $L\geq3$ cycles, the generalisation does not follow similarly, since the perturbation on every edge $k\to l$ has to satisfy \Eqref{eq:GeneralSingleOverlapCycleCondition}. If two cycles have an overlap, it does not necessarily mean that this overlap is part of the remaining cycles. To write all conditions, we therefore need to consider all edges with respective overlapping cycles. For this, let $A_{ij}=\{n|i\to j\in\mathcal{C}_n\cup\Tilde{\mathcal{C}}_n\}\subseteq\{1, \dots, L\}$ be the set of cycle indices that overlap on $i\to j$. Thus, \Eqref{eq:GeneralSingleOverlapCycleCondition} generalises to
\begin{equation}
    \begin{aligned}
        |\sum_{m\in A_{ij}}j_{ij}^{m}| < \min_{k\to l\in B_{ij} }e^{-(E_k+E_l)/2}\,,
    \end{aligned}
    \label{eq:MultiCycleGeneral}
\end{equation}
where 
\begin{equation}
    \begin{aligned}
        B_{ij}=\left(\bigcap_{m\in A_{ij}} \mathcal{C}_m\cup\Tilde{\mathcal{C}}_m\right) \setminus \left(\bigcup_{n\in \{1,\dots,L\}\setminus A_{ij}} \mathcal{C}_n\cup\Tilde{\mathcal{C}}_n \right)\,.
    \end{aligned}
\end{equation}
This is the set of all edges $k\to l$ which have the same cycles $A_{kl}$ as the edge $i\to j$. The reason we need to subtract the latter part becomes clear when considering the example in Fig.~\ref{fig:CycleOverlapExample} with $L=3$ cycles. Considering the edge $5\to 1$, then $A_{51}=\{1,3\}$ and $\bigcap_{m\in A_{51}} \mathcal{C}_m\cup\Tilde{\mathcal{C}}_m = \{5\to1, 4\to5, 2\to4, 1\to5, 5\to4, 4\to2\}$. However, if we also drive $\mathcal{C}_2$ we cannot say that $|j_{51}^1 + j_{51}^3|<\mathrm{e}^{-(E_2+E_4)/2}$, since \Eqref{eq:GeneralSingleOverlapCycleCondition} has to hold. Hence, we need to remove the edges $2\to4$ and $4\to 2$. 

As can be immediately identified, the number of constraints reduces from the number of cycles and the number of edges that have at least two cycles overlap to the number of different $A_{ij}$ or, equivalently, the number of distinct $B_{ij}$. Coming back to the previously mentioned example, every edge is driven, which gives seven conditions using \Eqref{eq:GeneralSingleOverlapCycleCondition}. However, we can reduce this to six conditions using \Eqref{eq:MultiCycleGeneral} since $A_{45}=A_{15}$.

\section{Determining Driven Transition Rates from Cycle 
Affinity\label{sec:AlternativeDriving}}

In Sec.~\ref{sec:DrivingSystems}, we provide a method to drive a system out of equilibrium while keeping the steady state distribution fixed. The method presented there is based on adding anti-symmetric parts to every edge in a cycle determined by the underlying (free) energy landscape. Here, we provide a method to drive a cycle in a system out of equilibrium by specifying the cycle affinity. In a sense, this is the counterpart to the method in Sec.~\ref{sec:DrivingSystems}, as the rates are adjusted according to the cycle affinity. 

We start by proving a couple of useful relations between DB transition rates, driven transition rates, and edge affinities. 


\begin{theorem}[Connection Between Transition Rates and Edge Affinity]\label{th:RateAffinityConnection}
    Let $r_{ij}$ be the transition rates of a finite Markov jump process with detailed balance and $p_i$ be the corresponding steady-state probability distribution. The transition rates $k_{ij}$ of a driven system with the same structure and steady state as the detailed balance system can be expressed as
    \begin{align}
        k_{ij} = \frac{2r_{ij}}{1 + \mathrm{e}^{-\mathcal{A}_{ij}}}\,,
    \end{align}
    where $\mathcal{A}_{ij}$ is the driven edge affinity. 
\end{theorem}
\begin{proof}
    From the definition of the edge affinities \Eqref{eq:affinities}, the probabilities are
    \begin{align}
        \frac{p_i}{p_j} = \frac{k_{ji}}{k_{ij}}\mathrm{e}^{\mathcal{A}_{ij}}\,.
        \label{eq:DrivingDetailedBalance}
    \end{align}
    Let $\mathbf{L}_\mathrm{s}$ be the detailed balance generator with entries $(\mathbf{L}_\mathrm{s})_{ji}=r_{ij}$. With $\mathbf{P} = \mathrm{diag}(p_1, p_2, \dots, p_N)$, the matrix $\mathcal{L}_\mathrm{s} = \sqrt{\mathbf{P}}^{-1}\mathbf{L}_\mathrm{s}\sqrt{\mathbf{P}}$ is symmetric. If $(\mathbf{L})_{ji}=k_{ij}$ is the generator of the driven system, the matrix $\mathcal{L}_\mathrm{a} = \sqrt{\mathbf{P}}^{-1}\mathbf{L}\sqrt{\mathbf{P}} - \mathcal{L}_\mathrm{s}$ is anti-symmetric
    \begin{equation}
        \begin{aligned}
            \left(k_{ji} - r_{ji}\right)\sqrt{\frac{p_j}{p_i}} = -\left(k_{ij} - r_{ij}\right)\sqrt{\frac{p_i}{p_j}}\,.
        \end{aligned}
        \label{eq:DrivingAntiSymmetricCondition}
    \end{equation}
    Rearranging this and using the DB condition \Eqref{eq:DetailedBalance} yields
    \begin{equation}
        \begin{aligned}
            k_{ij} = 2r_{ij} - k_{ji}\frac{p_j}{p_i}
        \end{aligned}
    \end{equation}
    Using \Eqref{eq:DrivingDetailedBalance} and $\mathcal{A}_{ij} = -\mathcal{A}_{ji}$ yields the wanted expression. 
\end{proof}

\begin{corollary}[Iterative Formula for Edge Affinities in Cycles]
    Consider the same assumptions as in Th. \ref{th:RateAffinityConnection}. Let the system be driven only on the cycle $\mathcal{C}=\{i_1\to i_2, i_2\to i_3, \dots, i_{N-1}\to i_N, i_N \to i_1 \}$ of length $|\mathcal{C}|=N$. Then $\forall k \in\mathds{N}_{>1}$
    \begin{equation}
        \begin{aligned}[b]
           \mathrm{e}^{\mathcal{A}_{i_k i_{k+1}}} =\frac{(r_{i_k i_{k+1}} +   r_{i_k i_{k-1}})\mathrm{e}^{\mathcal{A}_{i_{k-1} i_{k}}} + r_{i_k i_{k+1}}- r_{i_k i_{k-1}}}{(r_{i_k i_{k+1}} -   r_{i_k i_{k-1}})\mathrm{e}^{\mathcal{A}_{i_{k-1} i_{k}}} + r_{i_k i_{k+1}}+ r_{i_k i_{k-1}}}\,,
        \end{aligned}
        \label{eq:InitialAffinityEquation}
    \end{equation}
    with periodicity on the cycle $i_{N+l} = i_l\,\,\forall l\in\mathds{N}_{>0}$.
\end{corollary}
\begin{proof}
    Using the condition that $p_i$ is stationary together with Th. \ref{th:RateAffinityConnection} yields, for $i\in\Omega$,
    \begin{equation}
        \begin{aligned}
            \sum_{\substack{j\in\Omega\\j\neq i}} \frac{r_{ij}}{1 + \mathrm{e}^{-\mathcal{A}_{ij}}} p_i = \sum_{\substack{j\in\Omega\\j\neq i}}\frac{r_{ji}}{1 + \mathrm{e}^{-\mathcal{A}_{ji}}}p_j \stackrel{\mathrm{DB}}{=} \sum_{\substack{j\in\Omega\\j\neq i}} \frac{r_{ij}}{1 + \mathrm{e}^{\mathcal{A}_{ij}}} p_i\,.
        \end{aligned}
    \end{equation}
    Let now $i = i_k$, $k=2, \dots, N-1$, be on the cycle. The sums go over all neighbours. However, since only the cycle is driven, all non-driven contributions vanish due to DB. Hence,
    \begin{equation}
        \begin{aligned}
            \frac{r_{i_k i_{k+1}}}{1 + \mathrm{e}^{-\mathcal{A}_{i_k i_{k+1}}}} + \frac{r_{i_{k} i_{k-1}}}{1 + \mathrm{e}^{-\mathcal{A}_{i_{k} i_{k-1}}}}= \frac{r_{i_k i_{k+1}}}{1 + \mathrm{e}^{\mathcal{A}_{i_k i_{k+1}}}} + \frac{r_{i_{k} i_{k-1}}}{1 + \mathrm{e}^{\mathcal{A}_{i_{k} i_{k-1}}}}\,.
        \end{aligned}
    \end{equation}
    Rearranging for $\mathrm{e}^{\mathcal{A}_{i_k i_{k+1}}}$ yields the wanted result. With mapping $i_{N+l} = i_l\,\,\forall l\in\mathds{N}_{>0}$, this extends to $k\in\mathds{N}_{>1}$.
\end{proof}

\begin{theorem}[Induced Edge Affinities]\label{th:InducedEdgeAffinities}
    Let $r_{ij}$ be the transition rates of a MJP system with DB and $\mathcal{C} = \{i_1\to i_2, i_2\to i_3, \dots, i_{N-1}\to i_{N}, i_N\to i_1\}$ be a cycle in that system. If the system is driven out of equilibrium while keeping the probability distribution unchanged, the edge affinity $\mathcal{A}_{i_ki_{k+1}}$ for any $k\in\mathds{N}_{>0}$ is given by
    \begin{equation}
        \begin{aligned}[b]
            \mathrm{e}^{\mathcal{A}_{i_k i_{k+1}}} &= \frac{(\prod_{n=2}^k r_{i_n i_{n+1}} + \prod_{n=2}^k r_{i_n i_{n-1}})\mathrm{e}^{\mathcal{A}_{i_{1} i_{2}}} + \prod_{n=2}^k r_{i_n i_{n+1}} - \prod_{n=2}^k r_{i_n i_{n-1}}}{(\prod_{n=2}^k r_{i_n i_{n+1}} - \prod_{n=2}^k r_{i_n i_{n-1}})\mathrm{e}^{\mathcal{A}_{i_{1} i_{2}}} + \prod_{n=2}^k r_{i_n i_{n+1}} + \prod_{n=2}^k r_{i_n i_{n-1}}}\\
            &\equiv  \frac{\Gamma_{k}\mathrm{e}^{\mathcal{A}_{i_{1} i_{2}}} +\Theta_k}{\Theta_k\mathrm{e}^{\mathcal{A}_{i_{1} i_{2}}} + \Gamma_{k}}\,,
        \end{aligned}
        \label{eq:InducedAffinities}
    \end{equation}
    where the periodicity of the cycles is induced by mapping $i_{N+l} = i_l\,\,\forall l\in\mathds{N}_{>0}$. Here, $\Gamma_k$ and $\Theta_k$ are the prefactors in \Eqref{eq:InducedAffinities}
    \begin{equation}
        \begin{aligned}[b]
            \Gamma_k &\equiv 
            \begin{cases}
                1 & k = 1\,,\\
                \prod_{n=2}^k r_{i_n i_{n+1}} + \prod_{n=2}^k r_{i_n i_{n-1}} & k>1\,,
            \end{cases}
            \\
            \Theta_k &\equiv 
            \begin{cases}
                0 & k = 1\,,\\
                \prod_{n=2}^k r_{i_n i_{n+1}} - \prod_{n=2}^k r_{i_n i_{n-1}} & k>1\,.
            \end{cases}
        \end{aligned}
    \end{equation}
\end{theorem}

\begin{proof}
For $k=2$ in \Eqref{eq:InitialAffinityEquation}, we get
\begin{equation}
    \begin{aligned}[b]
        \mathrm{e}^{\mathcal{A}_{i_2 i_{3}}} 
       &=\frac{(r_{i_2 i_{3}} +   r_{i_2 i_{1}})\mathrm{e}^{\mathcal{A}_{i_{1} i_{2}}} + r_{i_2 i_{3}}- r_{i_2 i_{1}}}{(r_{i_2 i_{3}} -   r_{i_2 i_{1}})\mathrm{e}^{\mathcal{A}_{i_{1} i_{2}}} + r_{i_2 i_{3}}+ r_{i_2 i_{1}}}\,,
    \end{aligned}
\end{equation}
which agrees with what is from \Eqref{eq:InducedAffinities}. Now, for affinity $\mathcal{A}_{i_{k+1}i_{k+2}}$ Eqs.~\eqref{eq:InitialAffinityEquation} and \eqref{eq:InducedAffinities} yields
\begin{equation}
    \begin{aligned}[b]
        \mathrm{e}^{\mathcal{A}_{i_{k+1} i_{k+2}}}
       &=\frac{(r_{i_{k+1} i_{k+2}} + r_{i_{k+1} i_{k}})\mathrm{e}^{\mathcal{A}_{i_{k} i_{k+1}}} + r_{i_{k+1} i_{k+2}}- r_{i_{k+1} i_{k}}}{(r_{i_{k+1} i_{k+2}} -   r_{i_{k+1} i_{k}})\mathrm{e}^{\mathcal{A}_{i_{k} i_{k+1}}} + r_{i_{k+1} i_{k+2}}+ r_{i_{k+1} i_{k}}}\\
       &= \frac{(r_{i_{k+1} i_{k+2}} + r_{i_{k+1} i_{k}})\left(\Gamma_{k}\mathrm{e}^{\mathcal{A}_{i_{1} i_{2}}} +\Theta_k\right) + (r_{i_{k+1} i_{k+2}}- r_{i_{k+1} i_{k}})\left(\Theta_{k}\mathrm{e}^{\mathcal{A}_{i_{1} i_{2}}} +\Gamma_k\right)}{(r_{i_{k+1} i_{k+2}} - r_{i_{k+1} i_{k}})\left(\Gamma_{k}\mathrm{e}^{\mathcal{A}_{i_{1} i_{2}}} +\Theta_k\right) + (r_{i_{k+1} i_{k+2}}+ r_{i_{k+1} i_{k}})\left(\Theta_{k}\mathrm{e}^{\mathcal{A}_{i_{1} i_{2}}} +\Gamma_k\right)}\,.
    \end{aligned}
    \label{eq:InductionCalculation}
\end{equation}
In the numerator and denominator, the term proportional to $\mathrm{e}^{\mathcal{A}_{i_{1} i_{2}}}$ has prefactor
\begin{equation}
    \begin{aligned}[b]
        (&r_{i_{k+1} i_{k+2}} \pm r_{i_{k+1} i_{k}})\Gamma_{k} + (r_{i_{k+1} i_{k+2}} \mp r_{i_{k+1} i_{k}})\Theta_{k}\\
        =& (r_{i_{k+1} i_{k+2}} \pm r_{i_{k+1} i_{k}})\left(\prod_{n=2}^k r_{i_n i_{n+1}} + \prod_{n=2}^k r_{i_n i_{n-1}}\right) \\
        &+ (r_{i_{k+1} i_{k+2}} \mp r_{i_{k+1} i_{k}})\left(\prod_{n=2}^k r_{i_n i_{n+1}} - \prod_{n=2}^k r_{i_n i_{n-1}}\right)\\
        =& \prod_{n=2}^{k+1} r_{i_n i_{n+1}} + r_{i_{k+1} i_{k+2}}\prod_{n=2}^k r_{i_n i_{n-1}} \pm r_{i_{k+1} i_{k}}\prod_{n=2}^k r_{i_n i_{n+1}} \pm \prod_{n=2}^{k+1} r_{i_n i_{n-1}}\\
        &+\prod_{n=2}^{k+1} r_{i_n i_{n+1}} - r_{i_{k+1} i_{k+2}}\prod_{n=2}^k r_{i_n i_{n-1}} \mp r_{i_{k+1} i_{k}}\prod_{n=2}^k r_{i_n i_{n+1}} \pm \prod_{n=2}^{k+1} r_{i_n i_{n-1}}\\
        =& 2\prod_{n=2}^{k+1} r_{i_n i_{n+1}}\pm 2\prod_{n=2}^{k+1} r_{i_n i_{n-1}}\,,
    \end{aligned}
\end{equation}
which therefore is $2\Gamma_{k+1}$ and $2\Theta_{k+1}$, respectively. The remaining terms can be simplified in an analogous manner
, which yield $2\Theta_{k+1}$ and $2\Gamma_{k+1}$, respectively. Hence, \Eqref{eq:InductionCalculation} can be written as 
\begin{equation}
    \begin{aligned}[b]
        \mathrm{e}^{\mathcal{A}_{i_{k+1} i_{k+2}}} = \frac{\Gamma_{k+1}\mathrm{e}^{\mathcal{A}_{i_{1} i_{2}}} +\Theta_{k+1}}{\Theta_{k+1}\mathrm{e}^{\mathcal{A}_{i_{1} i_{2}}} + \Gamma_{k+1}}\,,
    \end{aligned}
\end{equation}
which concludes the proof. Using the periodicity of the cycle, this proof also holds for any $k\geq N+1$, since $\prod_{n=2}^{N+1} r_{i_n i_{n+1}} = r_{i_1 i_{2}}r_{i_2 i_{3}}\dots r_{i_{N-1} i_{N}}r_{i_{N} i_{1}} = \prod_{n=2}^{N+1} r_{i_n i_{n-1}}$. 
\end{proof}

\begin{lemma}\label{prop:PhysicalSolution}
    Let $z>0$ and $x_k = \Theta_k/\Gamma_k$. The fraction $(\Gamma_k z + \Theta_k)/(\Theta_k z + \Gamma_k)$ is positive for $x_k\in\left(-\min\left(z, \frac{1}{z}\right), 1\right)$.
\end{lemma}
\begin{proof}
    Values of $x_k$ are bounded $-1\leq x_k\leq 1$. The fraction is positive if numerator and denominator are simultaneously (i) positive or (ii) negative. In the case of (i), the conditions are $x_k>-z$ and $x_k>-1/z$, which can be fulfilled $\forall z>0$. For (ii), the conditions are $x_k<-z$ and $x_k<-1/z$, which can never be satisfied.
\end{proof}

\begin{definition}
    We denote the set of physical possible values of $z$ as
    \begin{align}
        \Omega = \left\{z\big|x_k\in\left(-\min\left(z, \frac{1}{z}\right), 1\right) \forall k\geq 1\right\}\,.
    \end{align}
\end{definition}

\begin{corollary}\label{cor:AffinitiesSign}
    Let $\mathcal{A}_{i_1i_2}$ be a physical affinity, i.e., $\mathrm{e}^{\mathcal{A}_{i_1i_2}}\in\Omega$. Then, for a system with assumptions as in Th. \ref{th:InducedEdgeAffinities}, the edge affinities $\mathcal{A}_{i_ki_{k+1}}$ induced by $\mathcal{A}_{i_1i_2}$  satisfy
    \begin{equation}
        \begin{aligned}
            \mathrm{sgn}(\mathcal{A}_{i_ki_{k+1}}) = \mathrm{sgn}(\mathcal{A}_{i_1i_2})\,,
        \end{aligned}
    \end{equation}
    and the sign of the cycle affinity \Eqref{eq:CycleAffinity} is 
    \begin{equation}
        \begin{aligned}
            \mathrm{sgn}(\mathcal{A}_\mathcal{C}) = \mathrm{sgn}(\mathcal{A}_{i_1i_2})\,.
        \end{aligned}
    \end{equation}
\end{corollary}
\begin{proof}
    Using \Eqref{eq:InducedAffinities}
    \begin{equation}
        \begin{aligned}[b]
            \mathrm{sgn}(\mathcal{A}_{i_ki_{k+1}}) &= \mathrm{sgn}\left((\Gamma_k - \Theta_k)\mathrm{e}^{\mathcal{A}_{i_1i_2}} + \Theta_k - \Gamma_k\right)\\
            &= \mathrm{sgn}\left(2(\mathrm{e}^{\mathcal{A}_{i_1i_2}}-1)\prod_{n=2}^kr_{i_ki_{k-1}}\right)\\
            &=\mathrm{sgn}\left(\mathrm{e}^{\mathcal{A}_{i_1i_2}}-1\right)\,,
        \end{aligned}
    \end{equation}
    since $r_{ij}>0$ $\forall i\to j\in \mathcal{C}$ by definition. Using that $\mathcal{A}_{i_1i_2}\in\mathds{R}$ and, thus, $\mathrm{sgn}\left(\mathcal{A}_{i_1i_2}\right) = \mathrm{sgn}\left(\mathrm{e}^{\mathcal{A}_{i_1i_2}}-1\right)$ proves the first part. Next, since $\mathcal{A}_\mathcal{C} = \sum_{i\to j\in\mathcal{C}}\mathcal{A}_{ij}$, and each term has the same sign, the cycle affinity has the same sign as each term.
\end{proof}

With the shorthand $y = \mathrm{e}^{\mathcal{A}_{i_1i_2}}$ the fixed cycle affinity condition can then be written as
\begin{equation}
    \begin{aligned}[b]
        \mathrm{e}^{\mathcal{A}_\mathcal{C}} &= \prod_{k=1}^N \mathrm{e}^{\mathcal{A}_{i_ki_{k+1}}}= \prod_{k=1}^N\frac{\Gamma_k y + \Theta_k}{\Theta_k y + \Gamma_k}\,.
    \end{aligned}
\end{equation}
Hence, the edge affinity of the edge $i_1\to i_2$ is given by the logarithm of physical solutions of
\begin{equation}
    \mathrm{e}^{\mathcal{A}_\mathcal{C}}\prod_{k=2}^N \left(\Theta_k y + \Gamma_k\right) = y \prod_{k=2}^N\left(\Gamma_k y + \Theta_k\right)\,,
    \label{eq:CycleAffinityToEdge}
\end{equation}
which is a polynomial of degree $N$ in $y$. Therefore, the roots generally need to be calculated numerically. Moreover, the existence of such physical solutions is {a priori} unclear. In the remaining part of this section, we prove that there exists one physical solution of \Eqref{eq:CycleAffinityToEdge} for all possible values of $\mathcal{A}_\mathcal{C}$. 

Let $a_k$ be the coefficient of $y$, so that the polynomial equation \Eqref{eq:CycleAffinityToEdge} can be written as
\begin{equation}
    \begin{aligned}[b]
        \sum_{k=0}^N a_k y^k = 0\,.
    \end{aligned}
    \label{eq:AffinityPolynomial}
\end{equation}
Let $A_k=\{A|A\subseteq\{1, 2, \dots, N\}, |A| = k\}$ be the set of all subsets of $\{1, \dots, N\}$ of size $k$. Recalling that $\Gamma_1=1$ and $\Theta_1=0$, we get that the coefficients of the polynomial are given by
\begin{equation}
    \begin{aligned}[b]
        a_k &=\sum_{A\in A_{k}} \left[\left(\prod_{n\in A} \Gamma_n\right)\left(\prod_{n\in \{1, \dots, N\}\setminus A} \Theta_n\right) -  \mathrm{e}^{\mathcal{A}_\mathcal{C}}\left(\prod_{n\in A} \Theta_n\right)\left(\prod_{n\in \{1, \dots, N\}\setminus A} \Gamma_n\right)\right]\,.
    \end{aligned}
    \label{eq:AffinityPolynomialCoefficients}
\end{equation}

\begin{lemma}
    The roots of the polynomial $\sum_{k=0}^N a_ky^k$ are also roots of $\sum_{k=0}^N d_ky^k$, with
    \begin{align}
        d_k = \sum_{A\in A_k}\prod_{n\in A^c}x_n - \mathrm{e}^{\mathcal{A}_\mathcal{C}}\sum_{A\in A_k}\prod_{n\in A}x_n\,.
        \label{eq:NewPolynomialCoefficients}
    \end{align}
\end{lemma}
\begin{proof}
    Since $\Gamma_k>0$ $\forall k\geq 1$, we can write
    \begin{align}
        \sum_{k=0}^N a_ky^k = a_N\sum_{k=0}^N d_ky^k = 0\,.
    \end{align}
    Hence, the rescaled coefficients are
    \begin{equation}
        \begin{aligned}
            d_k &= \frac{a_k}{a_N}\\
            &= \frac{\sum_{A\in A_{k}} \left[\left(\prod_{n\in A} \Gamma_n\right)\left(\prod_{n\in \{1, \dots, N\}\setminus A} \Theta_n\right) -  \mathrm{e}^{\mathcal{A}_\mathcal{C}}\left(\prod_{n\in A} \Theta_n\right)\left(\prod_{n\in \{1, \dots, N\}\setminus A} \Gamma_n\right)\right]}{\prod_{n=1}^N\Gamma_n}\\
            &= \sum_{A\in A_{k}} \left[\prod_{n\in \{1, \dots, N\}\setminus A} \frac{\Theta_n}{\Gamma_n}-  \mathrm{e}^{\mathcal{A}_\mathcal{C}}\prod_{n\in A} \frac{\Theta_n}{\Gamma_n}\right]\,.
        \end{aligned}
    \end{equation}
    With $x_n = \Theta_n/\Gamma_n$ and $A^c=\{1, \dots, N\}\setminus A $ the complement, we get the coefficients \Eqref{eq:NewPolynomialCoefficients}.
\end{proof}


\begin{prop}\label{cor:OddNumberRoots}
    The polynomial $\sum_{k=0}^N a_k y^k$, and thus $\sum_{k=0}^N d_k y^k$, has an odd number of, possibly degenerate, positive roots.
\end{prop}
\begin{proof}
    The leading order coefficient and constant term of the polynomial are $a_N = \prod_{n=1}^N \Gamma_n>0$ and $a_0 = - \mathrm{e}^{\mathcal{A}_\mathcal{C}}\prod_{n=1}^N \Gamma_n<0$, respectively. Hence, $a_N / a_0 = - \mathrm{e}^{\mathcal{A}_\mathcal{C}} <0$. Since the polynomial is continuous in $x$, it has to have an odd number of positive roots. 
\end{proof}

\begin{theorem}[Bound on Number of Physical Solutions]\label{th:NumberPhysicalSol}
    Let $n$ be odd and $y_i$, $i=1, \dots, n$, be positive roots of the polynomial $\sum_{k=0}^N d_k y^k$
    with coefficients \Eqref{eq:NewPolynomialCoefficients}. There exists at most one, possibly degenerate, physical root. 
\end{theorem}

\begin{proof}
    Let $z_1$, $z_2$ be two physical solutions of the polynomial with $z_1\neq z_2$. Without loss of generality, we may assume that $z_1>z_2$. By definition, we know that the affinities ${\mathcal{A}_{i_ki_{k+1}}^1},\,{\mathcal{A}_{i_ki_{k+1}}^2}$ induced by $z_1, z_2$ obey $\mathrm{e}^{\mathcal{A}_{i_ki_{k+1}}^1}, \mathrm{e}^{\mathcal{A}_{i_ki_{k+1}}^2}>0$ for any $k\in\mathds{N}_{>0}$. Consider
    \begin{equation}
        \begin{aligned}[b]
            \mathrm{e}^{\mathcal{A}_{i_ki_{k+1}}^1} - \mathrm{e}^{\mathcal{A}_{i_ki_{k+1}}^2} &= \frac{\Gamma_{k}z_1 +\Theta_k}{\Theta_kz_1 + \Gamma_{k}} - \frac{\Gamma_{k}z_2 +\Theta_k}{\Theta_kz_2 + \Gamma_{k}}\\
            &= \frac{(\Gamma_k^2 - \Theta_k^2)\left(z_1 - z_2\right)}{\left(\Theta_kz_1 + \Gamma_{k}\right)\left(\Theta_kz_2 + \Gamma_{k}\right)}\,.
        \end{aligned}
    \end{equation}
    Using that $\Gamma_k > |\Theta_k|$ $\forall k\in\mathds{N}_{>0}$, the numerator is positive for every $k\in\mathds{N}_{>0}$. 
    Since the solutions are physical, this leads to $\Gamma_{k}z_\kappa +\Theta_k>0$, and thus $\Theta_{k}z_\kappa +\Gamma_k>0$. 
    Hence, $\mathrm{e}^{\mathcal{A}_{i_ki_{k+1}}^1} - \mathrm{e}^{\mathcal{A}_{i_ki_{k+1}}^2}>0$. However, since both are solutions of the polynomial, they need to satisfy
    \begin{equation}
        \begin{aligned}
            \mathrm{e}^{\mathcal{A}_\mathcal{C}}=\prod_{k=1}^N\mathrm{e}^{\mathcal{A}_{i_ki_{k+1}}^1} = \prod_{k=1}^N\mathrm{e}^{\mathcal{A}_{i_ki_{k+1}}^2}\,,
        \end{aligned}
    \end{equation}
    which is not possible. Therefore, there cannot be more than one, possibly degenerate, physical solution of the polynomial. 
\end{proof}

\begin{corollary}\label{cor:NoNegativeX}
    If $\Theta_k\geq0$ $\forall k\geq1$, then there exists exactly one positive root of the polynomial in Th.~\ref{th:NumberPhysicalSol}. This root is also physical.
\end{corollary}
\begin{proof}
    If $\Theta_k\geq0$, then $0\leq x_k\leq 1$. If $z$ denotes a positive root of the polynomial, then $x_k\in [0, 1)\subset\left(-\min\left(z, \frac{1}{z}\right), 1\right)$ $\forall k\geq 1$. With Prop.~\ref{cor:OddNumberRoots}, there has to be an odd number of positive roots. Therefore, with Th.~\ref{th:NumberPhysicalSol}, there can only be one positive root which also is physical. 
\end{proof}
\begin{corollary}\label{cor:OneNegativeX}
    If there is a single $l\in\{2, \dots, N\}$ with $\Theta_l<0$, then there exists exactly one positive root of the polynomial in Th.~\ref{th:NumberPhysicalSol}. This root is also physical.
\end{corollary}
\begin{proof}
    Since $\mathcal{A}_\mathcal{C}\in\mathds{R}$, only an even number of $\mathcal{A}_{i_ki_{k+1}}$ can be complex. Hence, if $z$ is a positive root and there is a single $l\in\{2, \dots, N\}$ for which $\Theta_l$, and thus $x_l$, is negative, the fraction $(\Gamma_l z + \Theta_l)/(\Theta_l z + \Gamma_l)$ remains positive.
\end{proof}
\begin{remark}
The logic in this proof is not generalisable to any odd number $>1$, as there may exist pairs of $i,j$ s.t. $x_i, x_j<-\min(z, 1/z)$, where $z$ again is a positive root.
\end{remark}

\begin{prop}\label{prop:SymmetrySolution}
    Let $z$ be a positive root of the polynomial $\sum_{k=0}^N d_ky^k$ with $z\in\Omega$ and $\mathcal{A}_\mathcal{C}>0$. Then $1/z$ is a positive and physical root of the polynomial $\sum_{k=0}^N \Tilde{d}_ky^k$, where
    \begin{align}
        \Tilde{d}_k =  \sum_{A\in A_k}\prod_{n\in A^c}x_n - \mathrm{e}^{-\mathcal{A}_\mathcal{C}}\sum_{A\in A_k}\prod_{n\in A}x_n\,.
    \end{align}
\end{prop}
\begin{proof}
    Since $z>0$ is a root, we can write
    \begin{equation}
        \sum_{k=0}^N d_kz^k=z^N\sum_{k=0}^N d_kz^{k-N}=z^N\sum_{k=0}^N d_k\left(\frac{1}{z}\right)^{N-k} = 0\,.
        \label{eq:SymmetrySolutionProofStart}
    \end{equation}
    The coefficients can be written as
    \begin{equation}
        \begin{aligned}[b]
            d_k &= \sum_{A\in A_k}\prod_{n\in A^c}x_n - \mathrm{e}^{\mathcal{A}_\mathcal{C}}\sum_{A\in A_k}\prod_{n\in A}x_n\\
            &= - \mathrm{e}^{\mathcal{A}_\mathcal{C}}\left(\sum_{A\in A_k}\prod_{n\in A}x_n - \mathrm{e}^{-\mathcal{A}_\mathcal{C}}\sum_{A\in A_k}\prod_{n\in A^c}x_n\right)\,.
        \end{aligned}
    \end{equation}
    Next, we use that
    \begin{equation}
        \begin{aligned}
            \sum_{A\in A_k}\prod_{n\in A}x_n = \sum_{A\in A_{N-k}}\prod_{n\in A^c}x_n\,,
        \end{aligned}
    \end{equation}
    and therefore
    \begin{equation}
        \begin{aligned}
            d_k &= - \mathrm{e}^{\mathcal{A}_\mathcal{C}}\left(\sum_{A\in A_{N-k}}\prod_{n\in A^c}x_n - \mathrm{e}^{-\mathcal{A}_\mathcal{C}}\sum_{A\in A_{N-k}}\prod_{n\in A}x_n\right)\\
            &= - \mathrm{e}^{\mathcal{A}_\mathcal{C}}\Tilde{d}_{N-k}\,.
        \end{aligned}
    \end{equation}
    Thus, \Eqref{eq:SymmetrySolutionProofStart} can be written as
    \begin{equation}
        \begin{aligned}
            z^N\sum_{k=0}^N d_k\left(\frac{1}{z}\right)^{N-k} = -\mathrm{e}^{\mathcal{A}_\mathcal{C}}z^N\sum_{k=0}^N\Tilde{d}_{N-k}\left(\frac{1}{z}\right)^{N-k} = -\mathrm{e}^{\mathcal{A}_\mathcal{C}}z^N\sum_{k=0}^N\Tilde{d}_{k}\left(\frac{1}{z}\right)^{k} = 0\,.
        \end{aligned}
    \end{equation}
    Lastly, since $z\in\Omega$ it follows that $1/z\in\Omega$. 
\end{proof}

A consequence of Prop. \ref{prop:SymmetrySolution} is that one only needs to check whether a system has a physical solution for either positive or negative cycle affinities. 

\begin{prop}\label{prop:ZeroAffinityRoot}
    If $\mathcal{A}_\mathcal{C}=0$, then $z=1$ is a physical root of the polynomial $\sum_{k=0}^N d_ky^k$.
\end{prop}
\begin{proof}
    Let $\mathcal{A}_\mathcal{C}=0$, it follows that $d_k = -d_{N-k}$. If $N$ is even, then
    \begin{equation}
        \begin{aligned}
            d_{N/2}=\sum_{A\in A_{N/2}} \left[\prod_{n\in A^c} x_n-  \prod_{n\in A} x_n\right]\,.
        \end{aligned}
    \end{equation}
    As $\forall A\in A_{N/2} \exists A^c\in A_{N/2}$, it implies that $d_{N/2}=0$. Thus, $\sum_k d_{k} = 0$, i.e., $z=1$ is a positive root of the polynomial. The root $z$ is physical since $|x_n|<1 \forall n\geq 1$.
\end{proof}

\begin{theorem}[Existence of Unique Physical Root]
    The polynomial $\sum_{k=0}^N d_kx^k$ has a unique physical root $z>0$ for all $\mathcal{A}_\mathcal{C}\in\mathds{R}$.
\end{theorem}
\begin{proof}
    We only need to consider finite systems where at least two $x_k$ are negative, as the existence of a physical root in the remaining cases follows with Cors.~\ref{cor:NoNegativeX} and \ref{cor:OneNegativeX}. From Prop.~\ref{prop:ZeroAffinityRoot}, we know that if $\mathcal{A}_\mathcal{C}=0$ then there exist a unique physical root $z=1$.  Furthermore, roots of a polynomial are known to be continuous functions of the polynomial coefficients \cite{BookPolynomialRootsContinuous} and the function $-\min(z, 1/z)$ is continuous in $z$. Moreover, $-1\leq-\min(z, 1/z)\leq 0$. Let $M\subset\{2, \dots, N\}$ such that $x_m=\min_k x_k \forall m\in M$. Suppose $\exists R\subset\mathds{R}_{<0}$ such that $\forall \mathcal{A}_\mathcal{C}\in R$ the roots are $\min_{k\notin M}x_k>-\min(z, 1/z) > x_m$, where $m\in M$. Because of the continuity of $z$ when changing $\mathcal{A}_\mathcal{C}$, there exists a $\mathcal{A}^*_\mathcal{C}\in \partial R\subset\mathds{R}$ where $-\min(z^*, 1/z^*) = x_m$, and thus
    \begin{equation}
        \begin{aligned}
            \frac{z + x_m}{x_m z + 1} \xrightarrow{\mathcal{A}_\mathcal{C}\to\mathcal{A}^*_\mathcal{C}} 
            \begin{cases}
                \infty & z^*>1\,,\\
                0 & z^*<1\,.
            \end{cases}
        \end{aligned}
    \end{equation}
    Hence, in the limit of approaching $\mathcal{A}^*_\mathcal{C}$, the induced edge affinity becomes
    \begin{equation}
        \begin{aligned}
            \mathcal{A}_{i_mi_{m+1}} \xrightarrow{\mathcal{A}_\mathcal{C}\to\mathcal{A}^*_\mathcal{C}} 
            \begin{cases}
                \infty & z^*>1\,,\\
                -\infty & z^*<1\,.
            \end{cases}
        \end{aligned}
    \end{equation}
    Note that the sign of all $\mathcal{A}_{i_mi_{m+1}}$, $m\in M$, is the same. All other induced edge affinities remain finite, as $\min_{k\notin M}x_k>-\min(z, 1/z)$. We can write the cycle affinity as $\mathcal{A}_\mathcal{C} = \sum_{m\in M}\mathcal{A}_{i_mi_{m+1}} + \sum_{n\notin M}\mathcal{A}_{i_ni_{n+1}}$. The first sum diverges to $\pm \infty$ when approaching $\mathcal{A}^*_\mathcal{C}$, while the second sum remains finite. As $\mathcal{A}_\mathcal{C}$ is finite by assumption, this forms a contradiction. Hence, no such crossing points $\mathcal{A}^*_\mathcal{C}$ exist. In other words, this means that $-\min(z, 1/z)< x_m$ for all $m\in\{2, \dots, N\}$, i.e., $z\in\Omega$ is physical for all $\mathcal{A}_\mathcal{C}\in\mathds{R}$.
\end{proof}

\begin{example}
    Consider a four-state model on a ring, i.e., where the DB generator is
    \begin{equation}
        \begin{aligned}
            \mathbf{L} = 
            \begin{pmatrix}
                -r_{14}-r_{12} & r_{21} & 0 & r_{41}\\
                r_{12} & -r_{21}-r_{23} & r_{32} & 0\\
                0 & r_{23} & -r_{32}-r_{34} & r_{43}\\
                r_{14} & 0 & r_{34} & -r_{41} - r_{43}
            \end{pmatrix}\,.
        \end{aligned}
    \end{equation}
    The only cycles are along the ring in either direction. Choosing $\mathcal{C}=\{1\to 2, 2\to 3, 3\to 4, 4\to 1\}$ yields
    \begin{equation}
        \begin{aligned}[b]
            x_1 &= \frac{\Theta_1}{\Gamma_1} = 0\,,\\
            x_2 &= \frac{\Theta_2}{\Gamma_2} = \frac{r_{23}-r_{21}}{r_{23}+r_{21}}\,,\\
            x_3 &= \frac{\Theta_3}{\Gamma_3} = \frac{r_{23}r_{34}-r_{21}r_{32}}{r_{23}r_{34}+r_{21}r_{32}}\,,\\
            x_4 &= \frac{\Theta_4}{\Gamma_4} = \frac{r_{23}r_{34}r_{41}-r_{21}r_{32}r_{43}}{r_{23}r_{34}r_{41}+r_{21}r_{32}r_{43}}\,.
        \end{aligned}
    \end{equation}
    The coefficients of the corresponding polynomial are therefore
    \begin{equation}
        \begin{aligned}[b]
            d_0 &= -\mathrm{e}^{\mathcal{A}_\mathcal{C}}\,,\\
            d_1 &= x_2x_3x_4 - \mathrm{e}^{\mathcal{A}_\mathcal{C}}(x_2+x_3+x_4)\,,\\
            d_2 &= (x_2x_3 + x_2x_4 + x_3x_4)\left(1- \mathrm{e}^{\mathcal{A}_\mathcal{C}}\right)\,,\\
            d_3 &= x_2+x_3+x_4 - \mathrm{e}^{\mathcal{A}_\mathcal{C}}x_2x_3x_4\,,\\
            d_4 &= 1\,.
        \end{aligned}
    \end{equation}
    Here, we can nicely see the symmetry from Prop. \ref{prop:SymmetrySolution} which also is visualised in Fig.~\ref{fig:PolynomialRootBound1}.
    \begin{figure}
        \centering
        \includegraphics[width=.6\textwidth]{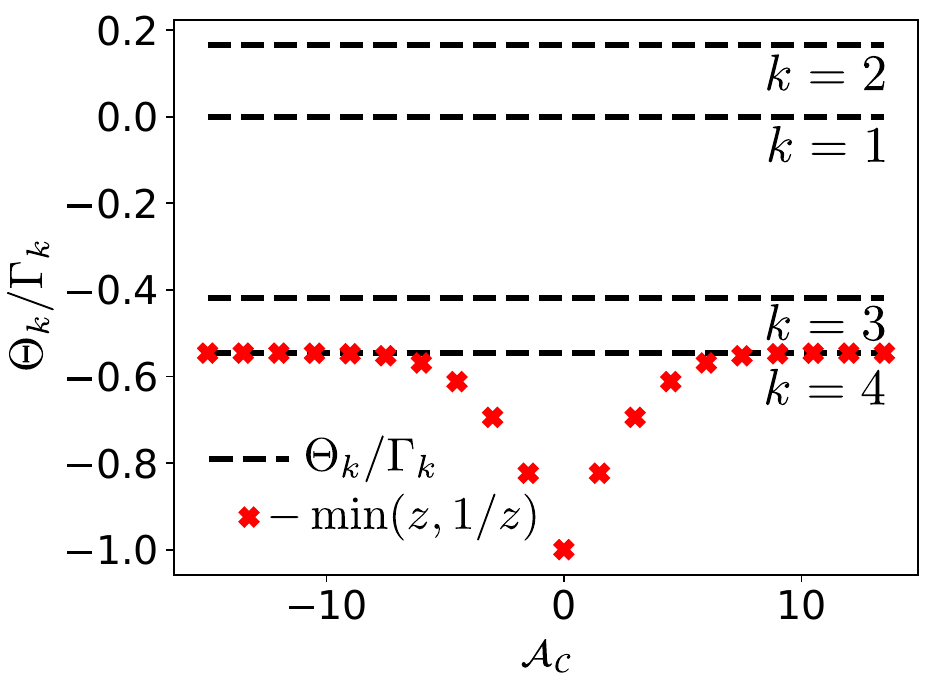}
        \caption[Visualisation of physical solution]{Visualisation of lower bound on $x_k=\Theta_k/\Gamma_k$ for various values of $\mathcal{A}_\mathcal{C}$ for a 4 state ring example. The lower bound $-\min(z, 1/z)$ is generated by the physical root $z$ of the polynomial \Eqref{eq:AffinityPolynomial}.}
        \label{fig:PolynomialRootBound1}
    \end{figure}
    
\end{example}

\chapter{Numerical Parameters and Spectral Decomposition\label{Ch:NumericalMethods}}

\section{Transition Rates}

\begin{table}[H]
    \centering
    \begin{tabular}{cc}
         \hline 
         Transition $i\to j$ & Rate $r_{ij}$\\
         \hline
         $1\to2$ &  5.997\\
         $2\to1$ &  0.774\\
         
         $1\to4$ &  13.439\\
         $4\to1$ &  127.968\\
         
         $1\to5$ &  15.330\\
         $5\to1$ &  0.121\\
         
         $5\to6$ &  3.749\\
         $6\to5$ &  13.326\\
         
         $2\to3$ &  1514.820\\
         $3\to2$ &  53.0661\\
         
         $2\to6$ &  13.441\\
         $6\to2$ &  2.922\\
         \hline
    \end{tabular}
    \caption{Transition rates used in the calmodulin example from Ref.~\cite{FirstPassageRick} which are adapted from Ref.~\cite{CalmodulinStigler}.}
    \label{tab:Calmodulin_Rates}
\end{table}

\begin{table}[H]
    \centering
    \begin{tabular}{cc}
         \hline 
         Site $i$ & Energies $E_i$\\
         \hline
         $1$ & $4.17022005\cdot 10^{-1}$ \\
         $2$ & $7.20324493\cdot 10^{-1}$ \\
         $3$ & $1.14374817\cdot 10^{-4}$ \\
         $4$ & $3.02332573\cdot 10^{-1}$ \\
         $5$ & $1.46755891\cdot 10^{-1}$ \\
         \hline
    \end{tabular}
    \caption{Energies in units of $k_\mathrm{B}T$ used for Toy Model simulations.}
    \label{tab:ToyModelEnergies}
\end{table}

\begin{table}[H]
    \centering
    \begin{tabular}{cc}
         \hline 
        Symbol &  Value\\
         \hline
         $\gamma$ & 5\\
         $x$ & 2\\
         $l_A$ & 20\\
         $l_B$ & 1\\
         $e^\mathrm{out}_A$ & 30\\
         $e^\mathrm{in}_A$ & 10\\
         $e^\mathrm{in}_B$ & 2\\
         \hline
    \end{tabular}
    \caption{Values of fixed parameters in SAT simulations which are adapted from Ref. \cite{Berlaga2022}.}
    \label{tab:Parameter SAT}
\end{table}

\section{Spectral Decomposition \label{Sec:SpectralDecomposition}}
In this section, we provide a lookup table for spectral decompositions of various observables and other quantities. For the details on the spectral decomposition see Sec.~\ref{sec:Spec}. As some expressions are quite lengthy, abbreviations are used. These can be found in Tab. \ref{tab:AbbreviationSpectral}. 

\begin{table}[ht]
    \centering
    \begin{tabular}{c|c}
         \hline
         Abbreviation & Full Expression\\
         \hline\\
         $G_{x, a}$ & $\psi^R_{a,x}\sum_n \psi^L_{a,n}p_n(0)$\\
         &\\
         $H_{i, x, y, a, b}$ & $\psi^R_{a,i}\psi^R_{b,x}\psi^L_{a,y}\sum_n \psi^L_{b,n}p_n(0)$\\
         &\\
         $D_{i,j,x,y}$ & $d_{ij}r_{ij}d_{xy}r_{xy}$\\
         &\\
         $D^{\alpha, \beta}_{i,j,x,y}$ & $d^{\alpha}_{ij}r_{ij}d^{\beta}_{xy}r_{xy}$\\
         \\
         \hline
    \end{tabular}
    \caption{Lookup table for abbreviated terms}
    \label{tab:AbbreviationSpectral}
\end{table}

\subsection{Propagator and Probabilities}

\begin{equation}
    \begin{aligned}[b]
        P(i, t'| y) =& \sum_a \psi^R_{a,i} \mathrm{e}^{\lambda_a t'}\psi^L_{a,y}\\
        p_i(\tau) =& \sum_{y} P(i, \tau| y)p_y(0) = \sum_{a} \mathrm{e}^{\lambda_a t'}G_{i, a}\\
        \int_0^t\mathrm{d}\tau^\prime \int_0^{\tau^\prime}\mathrm{d}\tau P(l, \tau^\prime|k, \tau)p_m(\tau)
        =&\sum_{a, b} H_{l, m, k, a, b}\left[\delta_{ab}\delta_{a0}\frac{t^2}{2} + \delta_{ab}(1-\delta_{a0})\frac{\mathrm{e}^{\lambda_a t}(\lambda_a t - 1) + 1}{\lambda_a^2} \right.\\
        &+ (1-\delta_{ab})\frac{1}{\lambda_b-\lambda_a}\left[t\delta_{b0} + (1 - \delta_{b0})\frac{\mathrm{e}^{\lambda_b t} - 1}{\lambda_b} - t\delta_{a0}\right.\\
        &-\left.\left. (1-\delta_{a0})\frac{\mathrm{e}^{\lambda_a t} - 1}{\lambda_a}\right]\right]\\
        Z_{xy}(\tau) =& \frac{p_x(\tau) r_{xy} - p_y(\tau)r_{yx}}{p_x(\tau) r_{xy} + p_y(\tau)r_{yx}}\\
        =& \frac{\sum_{i,j}\left(r_{xy} \psi^R_{i,x} - r_{yx} \psi^R_{i,y}\right) \mathrm{e}^{\lambda_i t'}\psi^L_{i,j} p_j(0)}{\sum_{i,j}\left(r_{xy} \psi^R_{i,x} + r_{yx} \psi^R_{i,y}\right) \mathrm{e}^{\lambda_i t'}\psi^L_{i,j} p_j(0)}\\
        =& \frac{\sum_i\left(r_{xy} G_{x,i} - r_{yx}G{y, i}\right)\mathrm{e}^{\lambda_i t}}{\sum_i\left(r_{xy} G_{x,i} + r_{yx}G{y, i}\right)\mathrm{e}^{\lambda_i t}}
    \end{aligned}
\end{equation}

\subsection{Mean and Derivatives}

\begin{equation}
    \begin{aligned}[b]
        \langle \hat{J}_t\rangle 
        =& \sum_{\stackrel{x\neq y}{a}}d_{xy}r_{xy}G_{x,a}\left(\delta_{a0}t + (1-\delta_{a0})\frac{\mathrm{e}^{\lambda_a t} - 1}{\lambda_a}\right)\\
        \langle \hat{\rho}_t\rangle 
        =& \sum_{x, a}V_xG_{x,a}\left(\delta_{a0}t + (1-\delta_{a0})\frac{\mathrm{e}^{\lambda_a t} - 1}{\lambda_a}\right)\\
        \partial_t\langle \hat{J}_t\rangle =& \sum_{\stackrel{x\neq y}{a}}d_{xy}r_{xy}G_{x,a}\mathrm{e}^{\lambda_at}\\
        =& \sum_{\stackrel{x\neq y}{a}}d_{xy}r_{xy}G_{x,a}\left(\delta_{a0} + (1-\delta_{a0})\mathrm{e}^{\lambda_a t}\right)\\
        \partial_t\langle \hat{\rho}_t\rangle =& \sum_{x, a}V_xG_{x,a}\left(\delta_{a0} + (1-\delta_{a0})\mathrm{e}^{\lambda_a t}\right)\\
        \mathcal{D}_t =& \sum_{x\neq y, a}(z_y-z_x)^2r_{xy}G_{x,a}\left(\delta_{a0} + (1-\delta_{a0})\frac{\mathrm{e}^{\lambda_a t} - 1}{\lambda_a t}\right)\\
        \langle \overline{\omega}_t\rangle =& \sum_{x, a}V_xG_{x,a}\left(\delta_{a0} + (1-\delta_{a0})\frac{\mathrm{e}^{\lambda_a t} - 1}{t\lambda_a}\right)
    \end{aligned}
    \label{eq:Spectral_rho_mean_der}
\end{equation}

\subsection{Covariances}

\begin{equation}
    \begin{aligned}[b]
        \mathrm{cov}(\hat{J}^\alpha_t, \hat{J}^\beta_t) =&  \sum_{\stackrel{x\neq y}{a}}d^\alpha_{xy}d^\beta_{xy}r_{xy}G_{x,a}\left(\delta_{a0}t + (1-\delta_{a0})\frac{\mathrm{e}^{\lambda_a t} - 1}{\lambda_a}\right)\\
        &+ \sum_{\stackrel{x\neq y}{i\neq j}}\sum_{a,b}D^{\alpha, \beta}_{i,j,x,y}\left(H_{i,x,y,a,b} + H_{x,i,j,a,b}\right)\left[\delta_{ab}\delta_{a0}\frac{t^2}{2} \right.\\
        &+ \left.\delta_{ab}(1-\delta_{a0})\frac{\mathrm{e}^{\lambda_a t}(\lambda_a t - 1) + 1}{\lambda_a^2} + (1-\delta_{ab})\frac{1}{\lambda_b-\lambda_a}\left[t\delta_{b0} \right.\right.\\
        &\left.\left. + (1 - \delta_{b0})\frac{\mathrm{e}^{\lambda_b t} - 1}{\lambda_b}- t\delta_{a0} - (1-\delta_{a0})\frac{\mathrm{e}^{\lambda_a t} - 1}{\lambda_a}\right]\right]-\langle\hat{J}_t\rangle^2\\
        \mathrm{cov}(\hat{\rho}^\alpha_t, \hat{\rho}^\beta_t) =&\sum_{x, i, a, b}V_{i}^\alpha V_x^\beta\left(H_{i,x,x,a,b} +H_{x,i,i,a,b}\right)\left[\delta_{ab}\delta_{a0}\frac{t^2}{2} \right.\\
        &+ \left.\delta_{ab}(1-\delta_{a0})\frac{\mathrm{e}^{\lambda_a t}(\lambda_a t - 1) + 1}{\lambda_a^2} + (1-\delta_{ab})\frac{1}{\lambda_b-\lambda_a}\left[t\delta_{b0} \right.\right.\\
        &\left.\left.+ (1 - \delta_{b0})\frac{\mathrm{e}^{\lambda_b t} - 1}{\lambda_b} - t\delta_{a0} - (1-\delta_{a0})\frac{\mathrm{e}^{\lambda_a t} - 1}{\lambda_a}\right]\right]-\langle\hat{\rho}_t\rangle^2\\
        \mathrm{cov}(\hat{\rho}_t, \hat{J}_t) =& \sum_{x\stackrel{x\neq y}{i, a, b}}V_{i}d_{xy}r_{xy}\left(H_{i,x,y,a,b} +H_{x,i,i,a,b}\right)\left[\delta_{ab}\delta_{a0}\frac{t^2}{2} \right.\\
        &+ \left.\delta_{ab}(1-\delta_{a0})\frac{\mathrm{e}^{\lambda_a t}(\lambda_a t - 1) + 1}{\lambda_a^2} + (1-\delta_{ab})\frac{1}{\lambda_b-\lambda_a}\left[t\delta_{b0} \right.\right.\\
        &\left.\left. + (1 - \delta_{b0})\frac{\mathrm{e}^{\lambda_b t} - 1}{\lambda_b}- t\delta_{a0} - (1-\delta_{a0})\frac{\mathrm{e}^{\lambda_a t} - 1}{\lambda_a}\right]\right]-\langle\hat{\rho}_t\rangle\langle\hat{J}_t\rangle\\
        \mathrm{cov}(\omega(t), \overline{z}_t) =& \sum_{x,y,a,b} (\omega_x - \langle \overline{\omega}_t\rangle) H_{x,y,y,a,b}\left[\delta_{ab}z_y \frac{\mathrm{e}^{\lambda_a t}}{t} + (1-\delta_{ab})z_y\frac{\mathrm{e}^{\lambda_b t} - \mathrm{e}^{\lambda_a t}}{t(\lambda_b-\lambda_a)}\right.\\
        &-\left.\sum_{g,k}z_kG_{k,g}\left(\delta_{\lambda_a-\lambda_b-\lambda_g, 0}\mathrm{e}^{\lambda_a t} + (1-\delta_{\lambda_a-\lambda_b-\lambda_g, 0})\frac{\mathrm{e}^{(\lambda_b +\lambda_g)t} - \mathrm{e}^{\lambda_a t}}{t(\lambda_b+\lambda_g-\lambda_a)})\right)\right]
        \\
        t\mathrm{var}(\overline{\omega}_t) =&2\sum_{x,y,a,b}\omega_x\omega_y H_{x,y,y,a,b}\left[\delta_{ab}\delta_{a0}\frac{t}{2} + \delta_{ab}(1-\delta_{a0})\frac{\mathrm{e}^{\lambda_a t}(\lambda_a t - 1) + 1}{t\lambda_a^2} \right.\\
        &+ \left.(1-\delta_{ab})\frac{1}{\lambda_b-\lambda_a}\left[\delta_{b0} + (1 - \delta_{b0})\frac{\mathrm{e}^{\lambda_b t} - 1}{t\lambda_b} - \delta_{a0} - (1-\delta_{a0})\frac{\mathrm{e}^{\lambda_a t} - 1}{t\lambda_a}\right]\right] \\
        &- t\langle \overline{\omega}_t\rangle^2\\
        \int_0^t\mathrm{d}\tau \mathrm{cov}(\omega(\tau), z(\tau)) =&\sum_{x, y,a,b}G_{x,a}\omega_x\left[\delta_{ab}\delta_{xy}z_x\left(\delta_{a0}t + (1-\delta_{a0})\frac{\mathrm{e}^{\lambda_a t} - 1}{\lambda_a}\right)\right. \\
        &- \left.z_yG_{y,b}\left(\delta_{\lambda_a+\lambda_b, 0}t + (1 - \delta_{\lambda_a+\lambda_b, 0})\frac{\mathrm{e}^{(\lambda_a + \lambda_b)t} -1 }{\lambda_a + \lambda_b}\right)\right]
    \end{aligned}
\end{equation}

\subsection{Scaling of Current Variance\label{sec:VarianceScaling}}

The mean current squared is
\begin{align}
    \langle \hat{J}_t\rangle^2 =& \left(\sum_{\stackrel{x\neq y}{a}}d_{xy}r_{xy}G_{x,a}\left(\delta_{a0}t + (1-\delta_{a0})\frac{\mathrm{e}^{\lambda_a t} - 1}{\lambda_a}\right)\right)^2\nonumber\\
    =&t^2\left(\sum_{x\neq y}d_{xy}r_{xy}p_x^\mathrm{s}\right)^2 - 2t\sum_{x\neq y}d_{xy}r_{xy}p_x^\mathrm{s}\sum_{i\neq y}\sum_{a>0}d_{ij}r_{ij}\frac{G_{ia}}{\lambda_a}\nonumber\\
    &+ \left(\sum_{x\neq y}\sum_{a>0}d_{xy}r_{xy}\frac{G_{x,a}}{\lambda_a}\right)^2 + \mathcal{O}(\mathrm{e}^{\lambda_1t})\label{eq:appJ2}
\end{align}
Similarly, the second moment of the current is
\begin{equation}
    \begin{aligned}[b]
        \langle \hat{J}_t^2\rangle =& t^2\left(\sum_{x\neq y}d_{xy}r_{xy}p_x^\mathrm{s}\right)^2\\
        &+ t \sum_{x\neq y} d_{xy}r_{xy}\left[d_{xy}p_x^\mathrm{s} - 2p_x^\mathrm{s}\sum_{i\neq j}\sum_{a>0}r_{ij}d_{ij}\psi^R_{i,a}\psi^L_{a,y}\frac{1}{\lambda_a} - 2\sum_{i\neq j}\sum_{b>0}d_{ij}r_{ij}p_i^\mathrm{s}\underbrace{\psi^R_{b,x}\sum_n \psi^L_{b,n}p_n(0)}_{=G_{x,b}}\frac{1}{\lambda_b}\right]\\
        &+ \left(-\sum_{x\neq y}\sum_{a>0}d_{xy}^2r_{xy}\frac{G_{x,a}}{\lambda_a}\right)+ 2\sum_{\stackrel{x\neq y}{i\neq j}}\sum_{a,b}d_{ij}d_{xy}r_{ij}r_{xy}H_{x,i,j,a,b}\left[ \delta_{ab}(1-\delta_{a0})\frac{1}{\lambda_a^2} \right.\\
        &+ \left.(1-\delta_{ab})\frac{1}{\lambda_b-\lambda_a}\left[
        -(1 - \delta_{b0})\frac{1}{\lambda_b}  + (1-\delta_{a0})\frac{1}{\lambda_a}\right]\right] + \mathcal{O}(\mathrm{e}^{\lambda_1t})
    \end{aligned}
    \label{eq:appJ3}
\end{equation}
where we use that $H_{i,x,y,0,0} = p_i^\mathrm{s}p_x^\mathrm{s}$, $H_{i,x,y,a,0} =\psi^R_{i,a}\psi^L_{a,y}p_x^\mathrm{s}$, and $H_{i,x,y,0,b} =p_i^\mathrm{s}\psi^R_{x,b}\sum_n \psi^L_{b,n}p_n(0)$. Further simplifications can be made when identifying $H_{i,x,y,a,b} =\psi^R_{i,a}\psi^L_{a,y}G_{x,b}$. Adding Eq.~\eqref{eq:appJ2} to Eq.~\eqref{eq:appJ3}, we see that the $t^2$ terms cancel. Hence, the variance is
\begin{align}
    \mathrm{var}(\hat{J})_t =& t \underbrace{\sum_{x\neq y} d_{xy}r_{xy}\left[d_{xy}p_x^\mathrm{s} - 2p_x^\mathrm{s}\sum_{i\neq j}\sum_{a>0}r_{ij}d_{ij}\psi^R_{i,a}\psi^L_{a,y}\frac{1}{\lambda_a}\right]}_{\equiv \kappa}\nonumber\\
    &+ \sum_{x\neq y}\sum_{a}d_{xy}r_{xy}\left[-(1-\delta_{a0})\frac{G_{x,a}}{\lambda_a}\left\{d_{xy} + \sum_{i\neq j}\sum_{b>0}d_{ij}r_{ij}\frac{G_{i,b}}{\lambda_b}\right\}\right.
    \nonumber\\&+2\sum_{{i\neq j}}\sum_{b}d_{ij}r_{ij}\psi^R_{i,a}\psi^L_{a,y}G_{x,b}\left\{ \delta_{ab}(1-\delta_{a0})\frac{1}{\lambda_a^2} \right.\nonumber\\
    &+ \left.\left.(1-\delta_{ab})\frac{1}{\lambda_b-\lambda_a}\left[
    -(1 - \delta_{b0})\frac{1}{\lambda_b}  + (1-\delta_{a0})\frac{1}{\lambda_a}\right]\right\}\right]\nonumber\\
    &+ \mathcal{O}(\mathrm{e}^{\lambda_1t}).\label{eq:varJSPEC}
\end{align}
In other words, the current variance scales as
\begin{align}
    \mathrm{var}(\hat{J})_t = \kappa t + \gamma + \mathcal{O}(\mathrm{e}^{\lambda_1t})\,,
\end{align}
where
\begin{align}
    \kappa \equiv& \sum_{x\neq y} d_{xy}r_{xy}\left[d_{xy}p_x^\mathrm{s} - 2p_x^\mathrm{s}\sum_{i\neq j}\sum_{a>0}r_{ij}d_{ij}\psi^R_{i,a}\psi^L_{a,y}\frac{1}{\lambda_a}\right]\,,\nonumber\\
    \gamma \equiv&\sum_{x\neq y}\sum_{a}d_{xy}r_{xy}\left[-(1-\delta_{a0})\frac{G_{x,a}}{\lambda_a}\left\{d_{xy} + \sum_{i\neq j}\sum_{b>0}d_{ij}r_{ij}\frac{G_{i,b}}{\lambda_b}\right\}\right.
    \nonumber\\&+2\sum_{{i\neq j}}\sum_{b}d_{ij}r_{ij}\psi^R_{i,a}\psi^L_{a,y}G_{x,b}\left\{ \delta_{ab}(1-\delta_{a0})\frac{1}{\lambda_a^2} \right.\nonumber\\
    &+ \left.\left.(1-\delta_{ab})\frac{1}{\lambda_b-\lambda_a}\left[
    -(1 - \delta_{b0})\frac{1}{\lambda_b}  + (1-\delta_{a0})\frac{1}{\lambda_a}\right]\right\}\right]
    \label{eq:LeadingCoefficientRelaxation}
\end{align}

\printbibliography

\end{document}